\documentclass[11pt]{article}
\usepackage{mymacros}

\title{\centering Deforming the Double-Scaled SYK \& Reaching the Stretched Horizon From Finite Cutoff Holography}
\author[a]{Sergio E. Aguilar-Gutierrez\orcidlink{0000-0003-0308-0061}}
\affiliation[a]{Qubits and Spacetime Unit, Okinawa Institute of Science and Technology Graduate University, Onna, Okinawa 904 0495, Japan}
\emailAdd{sergio.ernesto.aguilar@gmail.com}
\abstract{We study the properties of the double-scaled SYK (DSSYK) model under chord Hamiltonian deformations based on finite cutoff holography for general dilaton gravity theories with Dirichlet boundaries. The formalism immediately incorporates a lower-dimensional analog of $\TT(+\Lambda_2)$ deformations, denoted $T^2(+\Lambda_1)$, as special cases. In general, the deformation mixes the chord basis of the Hilbert space in the seed theory, which we order through the Lanczos algorithm. The resulting Krylov complexity for the Hartle-Hawking state represents a wormhole length at a finite cutoff in the bulk. We study the thermodynamic properties of the deformed theory; the growth of Krylov complexity; the evolution of $n$-point correlation functions with matter chords; and the entanglement entropy between the double-scaled algebras of the DSSYK model for a given chord state. The latter, in the triple-scaling limit, manifests as the minimal codimension-two area in the bulk following the Ryu-Takayanagi formula. By performing a sequence of $T^2$ and $T^2+\Lambda_1$ deformations in the upper tail of the energy spectrum in the deformed DSSYK, we concretely realize the cosmological stretched horizon proposal in de Sitter holography by Susskind \cite{Susskind:2021esx}. We discuss other extensions with sine dilaton gravity, end-of-the-world branes, and the Almheiri-Goel-Hu model.}

\begin{document}

\maketitle

\section{Introduction}\label{sec:intro}
\paragraph{\texorpdfstring{$\TT$}{} Deformations and Their Bulk Interpretation}
Explicit examples of holography outside the anti-de Sitter (AdS)/ conformal field theory (CFT) correspondence \cite{Maldacena:1997re} remain, arguably, not well understood and perplexing in comparison with those in conventional AdS/CFT. However, there are different approaches to bridge the gap. Some of the best examples studied have been developed through a class of tractable irrelevant deformations in the boundary theory for CFT$_2$ known as $\TT$ deformations, which were introduced in \cite{Zamolodchikov:2004ce,Smirnov:2016lqw,Cavaglia:2016oda} (reviewed in \cite{Jiang:2019epa,He:2025ppz,Guica:2025jkq}; see also App.~\ref{app:TTbar intro} for a brief introduction, and \cite{Dubovsky:2012wk,Dubovsky:2013ira, Tsolakidis:2024wut,Dubovsky:2017cnj,Dubovsky:2018bmo, Cardy:2018sdv,Aharony:2018bad,Conti:2018jho,Conti:2018tca,Tolley:2019nmm,Jafari:2019qns} for a few insightful studies in this area). Single-trace $\TT$-deformations (e.g.~\cite{Giveon:2017nie,Apolo:2019zai}) and its extensions \cite{Araujo:2018rho} modify the bulk's background metric, while double-trace $\TT$-deformations break the conformal symmetry in the dual quantum field theory (QFT). Depending on the sign of the deformation, the latter case can be interpreted in the bulk as a finite cutoff in AdS \cite{McGough:2016lol}; or gluing an auxiliary AdS path connected to the original asymptotic boundary \cite{Apolo:2023ckr,Apolo:2023vnm};
or introducing mixed boundary conditions in the asymptotic boundary \cite{Guica:2019nzm}; among others. Given that the deformation parameter introduces a length scale, it serves to regularize the deformed theory. This has motivated studying higher \cite{Hartman:2018tkw,Taylor:2018xcy,Morone:2024ffm,Cardy:2018sdv,Bonelli:2018kik,Ferko:2024yua} and lower dimensional analogs/extensions \cite{Gross:2019ach,Gross:2019uxi,Iliesiu:2020zld,Johnson:2020lns,Rosso:2020wir,Griguolo:2021wgy,Aguilar-Gutierrez:2024nst,Aguilar-Gutierrez:2024oea,Ebert:2022ehb,Ebert:2022xfh,Griguolo:2025kpi}; we collectively refer to them as $T^2$ deformations\footnote{Also referred to as 1D $\TT$ deformations in the lower-dimensional case in \cite{Gross:2019ach,Gross:2019uxi}.}. Other multitrace deformations \cite{Witten:2001ua,Parvizi:2025shq,Parvizi:2025wsg,Coleman:2020jte,Allameh:2025gsa,Sheikh-Jabbari:2025kjd,Ran:2025xas}, which modify boundary conditions in the bulk dual, are under active scrutiny. $T^2$ deformations have several exciting features, and they deserve further development in an explicitly solvable ultraviolet (UV)-finite setting, which is our aim in this work.

\paragraph{The DSSYK Model and its Holograms}Another prominent approach, lower dimensional quantum gravity offers the possibility of developing non-AdS holography from first principles in an analytically tractable manner.

In particular, the Sachdev-Ye-Kitaev (SYK) model \cite{Sachdev_1993,kitaevTalks1,kitaevTalks2,kitaevTalks3} in the double-scaling limit \cite{Cotler:2016fpe,Erdos:2014zgc}, called the DSSYK model \cite{Berkooz:2018jqr,Berkooz:2018qkz}, is a remarkable toy model to develop concepts and techniques to study lower-dimensional quantum gravity from a concrete and solvable (through chord diagram methods \cite{Berkooz:2018jqr,Berkooz:2018qkz}) boundary theory relevant for holography beyond AdS/CFT, which might carry some relevant lessons in higher dimensional holography. Key insights to understand the DSSYK model and its bulk dual include the double-scaled \cite{Lin:2022rbf,Xu:2024hoc,Cao:2025pir} and chord \cite{Lin:2023trc} von Neumann algebras; the quantum group structure \cite{Almheiri:2024ayc,Schouten:2025tvn,Belaey:2025ijg,Berkooz:2022mfk,Blommaert:2023opb,vanderHeijden:2025zkr}; Krylov complexity \cite{Rabinovici:2023yex,Ambrosini:2025hvo,Ambrosini:2024sre,Heller:2024ldz,Aguilar-Gutierrez:2025pqp,Aguilar-Gutierrez:2025mxf,Aguilar-Gutierrez:2025hty,Xu:2024hoc,Balasubramanian:2024lqk,Forste:2025gng,Aguilar-Gutierrez:2025sqh,Bhattacharjee:2022ave,Nandy:2024zcd,Anegawa:2024yia,Aguilar-Gutierrez:2024nau,Heller:2025ddj,Fu:2025kkh,Aguilar-Gutierrez:2026jjv,Xu:2024gfm} for operators \cite{Parker:2018yvk} and states \cite{Balasubramanian:2022tpr} (see recent reviews in \cite{Nandy:2024evd,Baiguera:2025dkc,Rabinovici:2025otw});\footnote{Krylov complexity in the semiclassical limit reproduces to bulk geodesic lengths \cite{Lin:2022rbf,Boruch:2023bte,Lin:2023trc,Blommaert:2024ymv,Berkooz:2022fso,Blommaert:2025avl,Griguolo:2025kpi,Verlinde:2024znh,Verlinde:2024zrh} in the bulk dual of the DSSYK model, which corresponds to an explicit realization of the volume=complexity (CV) conjecture \cite{Susskind:2014rva}.} algebraic entanglement entropy \cite{Tang:2024xgg,Aguilar-Gutierrez:2025otq}.\footnote{Holographically, the entanglement entropy between the double-scaled algebras manifests in terms of a Ryu-Takayanagi 
(RT) \cite{Ryu:2006bv,Ryu:2006ef} formula (see e.g.~\cite{Nishioka:2009un,Rangamani:2016dms,Chen:2019lcd,Harlow:2014yka} for reviews, and \cite{Hubeny:2007xt,Faulkner:2013ana,Engelhardt:2014gca} for extensions).} However, there has been active debate about the specific bulk dual to the DSSYK model. The most-discussed bulk dual proposals include sine dilaton gravity \cite{Blommaert:2024ymv,Blommaert:2023opb,Blommaert:2024whf,Blommaert:2025avl,Blommaert:2025rgw,Bossi:2024ffa,Blommaert:2025eps,Cui:2025sgy,Arundine:2025mcu}, and de Sitter (dS) space through different approaches, including Schwarzschild-dS$_3$ space \cite{Narovlansky:2023lfz,Verlinde:2024znh,Verlinde:2024zrh,Narovlansky:2025tpb,Blommaert:2025eps,Tietto:2025oxn,Aguilar-Gutierrez:2024nau}, and dS$_2$ space as a s-wave reduction of dS$_3$ space, where the DSSYK model is conjectured to live in the \textbf{stretched cosmological horizon} (illustrated in Fig.~\ref{fig:two-step_def} (d)) which is defined as a timelike codimension-one surface with Dirichlet boundary conditions somewhere in the static patch very close to the cosmological horizon, possibly a few Planck's lengths away \cite{Susskind:2021esx,Susskind:2022bia,Susskind:2023hnj,Lin:2022nss,Rahman:2022jsf,Rahman:2023pgt,Rahman:2024iiu,Rahman:2024vyg,Sekino:2025bsc,Miyashita:2025rpt}. Other interesting approaches can be found in e.g.~\cite{Milekhin:2023bjv,Milekhin:2024vbb,Narovlansky:2025tpb,Okuyama:2025hsd,Yuan:2024utc,Aguilar-Gutierrez:2025otq,Gubankova:2025gbx,Ahn:2025exp}.

At a first glance, it might seem the diversity of bulk dual proposals for the same boundary theory means that at least some of them might be incorrect. However, they are not necessarily incompatible with one another \cite{Aguilar-Gutierrez:2025hty,Blommaert:2025eps,Aguilar-Gutierrez:2024oea,Aguilar-Gutierrez:2025otq}. For instance, concrete connections between dS$_3$ space holography with sine dilaton gravity as its world-sheet theory and complex Liouville string \cite{Collier:2024kmo,Collier:2024kwt,Collier:2024lys,Collier:2024mlg,Collier:2025lux,Collier:2025pbm} and the SYK model as a collective field theory was recently investigated in \cite{Blommaert:2025eps}. To fully understand the DSSYK model, the relationship between these different proposals needs to be closely studied.

\paragraph{Addressing Susskind's Conjectures}A prominent member in this family of proposals, initiated by the work of Susskind \cite{Susskind:2021esx}, seemingly has very different features from the others \cite{Rahman:2023pgt}. Whether it fits with sine dilaton gravity and other dS$_3$ proposals, or if it is incompatible with the other proposals is not well-understood. According to \cite{Susskind:2021esx}, there are specific properties expected in dS holography that are realized by the DSSYK model at infinite (Boltzmann) temperature. Namely, \cite{Susskind:2021esx} postulates the following conjectures: 
\begin{enumerate}[label=(\roman*), nosep]
\item There is a RT formula in dS space that takes the same form as in AdS space when the asymptotic boundary is replaced by the stretched horizon. 
\item The scrambling time in the boundary theory dual to dS space, as measured by the operator size of fundamental operators in the theory \cite{Roberts:2018mnp}, which is linearly related to the out-of-time-ordered correlators (OTOCs) of the same operators \cite{Roberts:2018mnp}, is of the order of the dS length scale. 
\item There is a similar (hyper)fast growth of (some notion of) complexity associated to the exponential growth of the dS scale factor.
\item The SYK model at infinite (Boltzmann) temperature and with a scaling $p\sim N^\alpha$, where $p$ is the number of all-to-all interactions, $N$ the total number of fermions, and $0<\alpha<1$, satisfies the above properties.
\end{enumerate}
The motivation for the hyperfast properties and the infinite temperature limit comes from the fact that a surface at a finite radial distance in a gravitational spacetime experiences a Tolman temperature \cite{tolman1928extension,tolman1933thermodynamics,Tolley:2019nmm,Tolman:1930ona} with a redshift factor which diverges when the surface is close to the horizon. The temperature scale controls the rate of growth or decay of correlation functions, which would therefore lead to the hyperfast scrambling of OTOCs. This suggests that a boundary theory dual located at the cosmological stretched horizon should not be of the $k$-local type defined in \cite{Susskind:2021esx}, where $k$ is an order one parameter in the $N\gg1$ limit, and it represents the power of fundamental operators appearing in the Hamiltonian of the system. Indeed, if $k$ is fixed in the large $N$ limit, the scrambling time scales with $\log N$ \cite{Maldacena:2015waa}, and circuit complexity grows linearly in time \cite{Susskind:2014rva}, while the CV proposal in dS space \cite{Jorstad:2022mls} suggests it should grow faster than linear and it should end at a given static patch time with respect to the stretched horizon \cite{Susskind:2021esx}. In the SYK proposal of \cite{Susskind:2021esx}, we need to allow $p$ (which is the analog of $k$) to scale with $N$. In particular, when $p=N^{\alpha=1/2}$ the above conjecture suggests that the DSSYK model may be a specific example of boundary theory in stretched horizon holography \cite{Susskind:2021esx}.

The stretched horizon conjectures are seemingly \emph{in tension} with the DSSYK model being located at the asymptotic boundaries of an effective AdS$_2$ black hole in sine dilaton gravity \cite{Blommaert:2024ymv} (as well as some entropic considerations \cite{Rahman:2023pgt}). Nevertheless, there is a specific regime in sine dilaton gravity where the theory reduces Jackiw-Teitelboim (JT) \cite{JACKIW1985343,TEITELBOIM198341} gravity,\footnote{See \cite{Mertens:2022irh} for a review, and \cite{Finster:2023rkv} and references within
for connections with non-metric measures.} and to dS JT gravity \cite{Blommaert:2024whf,Heller:2025ddj}. This limit in the bulk corresponds to zooming in the UV regime in the energy spectrum of the DSSYK model. However, the boundary theory would be located at $\mathcal{I}^\pm$ when taking the appropriate limit \cite{Blommaert:2024whf,Heller:2025ddj}. In contrast, the boundary theory would be located at the stretched horizon in Susskind's proposal. The tension between them is closely related to dS$_2$/CFT$_1$ and static patch holography in dS$_2$ space.

To move the asymptotic boundary from $\mathcal{I}^\pm$ to inside the static patch, we cannot restrict ourselves to $T^2$ deformations, since they lead to a complex energy spectrum once the theory reaches a Killing horizon \cite{McGough:2016lol}. However, it was recently realized by \cite{AliAhmad:2025kki} that by applying a two-step process the boundaries can pass behind the horizon of asymptotically AdS$_3$ black holes through a combination of $\TT$ and $\TT+\Lambda_2$ deformations (which have been recently updated to $T^2+\Lambda_3$ deformations \cite{Silverstein:2024xnr}). In this process, the corresponding boundaries move from the asymptotic region to the black hole horizon through a $\TT$ deformation, and then from the horizon towards the singularity through a $\TT+\Lambda_2$ deformation. This procedure was recently applied in dS$_3$ holography by \cite{Chang:2025ays}; albeit without a concrete boundary theory in dS$_3$/CFT$_2$.

Thus, a natural way to resolve this tension is based on finite cutoff holography, by $T^2$ deforming the DSSYK model in the UV limit of its energy spectrum and implementing a lower dimensional version of $\TT+\Lambda_2$ deformations \cite{Gorbenko:2018oov,Shyam:2021ciy,Coleman:2021nor,Silverstein:2022dfj,Batra:2024kjl} to place the boundary theory near the cosmological horizon of dS$_2$ space.\footnote{Note that dS$_3$ space with Dirichlet boundaries is thermodynamically unstable, and thus also its s-wave sector; while in the dS$_{d+1\geq4}$ case, it does not lead to a well-posed initial value problem \cite{Anninos:2024wpy}.}

\paragraph{Results in This Work}
In summary, we study chord Hamiltonian deformations in the DSSYK model.\footnote{In the lower-dimensional context, this seems to be a natural choice given that the ensemble-averaged description of the SYK model is the one with a clearer bulk theory dual at the disk topology level. One could first perform the $T^2$ deformation in the physical SYK model, work out its double-scaling limit and take ensemble averaging at the end, to study the differences with respect to our approach. However, this procedure might not have a chord Hamiltonian description, or it would be non-trivially modified. This might also lead to a different holographic dual, if any. We discuss more details about this in the outlook, Sec.~\ref{ssec:Outlook}.} We focus on the boundary description with guidance from finite cutoff holography for general dilaton gravity theories; and including the infrared (IR) and UV triple-scaling limits in the DSSYK model, which are associated to JT and dS JT gravity respectively. This allows us to investigate the thermodynamic properties and dynamical observables, particularly correlation functions, in the deformed chord theory associated to a finite cutoff in the bulk dual of the DSSYK model, while allowing general counterterms in the bulk to study their role in the holographic dictionary. We deduce the Krylov basis in the deformed theory, Krylov complexity and the corresponding representation of the chord operators under a Hamiltonian deformation, as well as the entanglement entropy between the double-scaled algebras in the deformed theory. At last, these steps allow us to realize Susskind's stretched horizon proposal from the UV triple-scaling limit \cite{Aguilar-Gutierrez:2025otq}.\footnote{See \cite{Chapman:2021eyy,Jorstad:2022mls,Galante:2022nhj,Auzzi:2023qbm,Anegawa:2023wrk,Anegawa:2023dad,Baiguera:2023tpt,Aguilar-Gutierrez:2023zqm,Aguilar-Gutierrez:2023tic,Aguilar-Gutierrez:2023pnn,Aguilar-Gutierrez:2024rka,Bhattacharjee:2022ave,Rabinovici:2023yex,Aguilar-Gutierrez:2024nau,Anegawa:2024yia}, and \cite{Susskind:2021dfc,Susskind:2021esx,Shaghoulian:2021cef,Shaghoulian:2022fop,Franken:2023pni,Franken:2024ruw} for studies on quantum complexity and entanglement entropy (through bilayer vs monolayer proposals \cite{Franken:2023pni,Franken:2024ruw,Shaghoulian:2022fop}) in connection to the dS stretched horizon conjecture in \cite{Susskind:2021esx}.}

To elaborate on the summary, we first present a more complete derivation of the flow equation for the energy spectrum of the deformed boundary theory following up our companion work \cite{Aguilar-Gutierrez:2024oea} based on finite cutoff holography for dilaton gravity theories in \cite{Gross:2019ach,Gross:2019uxi,Iliesiu:2020zld}. 
For that, we consider the Hamilton-Jacobi (HJ) equation in a generic dilaton-gravity theory and we derive the flow equation encoding the energy spectrum of the dual. In contrast to the previous literature \cite{Gross:2019ach,Gross:2019uxi,Iliesiu:2020zld,Griguolo:2021wgy,Griguolo:2025kpi}, we allow the counterterm in the bulk action to be modified when the boundary radial cutoff is finite. This results in more general Hamiltonian deformations in the boundary dual, including as special cases one-dimensional $T^2$ deformations and a lower dimensional analog of $\TT+\Lambda_2$, which we refer to as $T^2+\Lambda_1$ deformations. This construction extends the holographic dictionary between the DSSYK model and a generic dilaton-gravity theory dual at finite cutoff.\footnote{The evaluations in this work are at the non-perturbative level in the deformation parameter, unless explicitly stated.} We analyze the thermodynamic properties of the deformed model from its partition function (including the density of states, microcanonical temperature, and heat capacity). A summary of the thermodynamics of the deformed model is shown in Fig.~\ref{fig:0_Thermo_summary}. 
Since the chord Hamiltonian is deformed, the chord number basis used to tridiagonalize the seed Hamiltonian does not do the same in the deformed theory. We will also derive the chord basis that is associated to the length of an Einstein-Rosen bridge in the dual AdS$_2$ black hole (which we refer to as a ``wormhole length'') at a finite cutoff, which is the Krylov basis \cite{Balasubramanian:2022tpr}, and we study the representation of operators in the chord algebra. We use these results to evaluate Krylov spread complexity \cite{Balasubramanian:2022tpr} in the HH state of the  deformed theory, which reproduces the wormhole length in the bulk at finite cutoff. We proceed evaluating $n$-point correlation functions. We find that in the semiclassical limit, they have the same functional dependence as the seed theory, but the parameters involving the ``fake'' temperature \cite{Blommaert:2024ymv} (also called ``tomperature'' \cite{Susskind:2021esx}, which is associated to the decay rate of correlation functions) contain a redshift factor, associated to the Tolman temperature at a finite distance in the bulk \cite{McGough:2016lol,Anninos:2022hqo}. Lastly, we evaluate the von Neumann entropy with respect to a density matrix associated to the chord number of the DSSYK model. The algebraic entanglement entropy is holographic, it matches the dilaton in JT gravity and dS JT gravity at a finite cutoff when we restrict to the IR and UV triple-scaling limits in the chord theory.

In particular, when applying our results in the UV limit of the DSSYK model, we realize Susskind's proposal for stretched horizon holography \cite{Susskind:2021esx}, by implementing the $T^2$ and $T^2+\Lambda_1$ deformations, and taking a limit in deformation parameter such that the boundary is placed inside the static and close to the cosmological horizon. The thermodynamic properties of the model match those in dS$_2$ JT gravity at a finite cutoff. Due to the fake temperature being infinite, the rate of growth of OTOCs and Krylov spread complexity are enhanced; the scrambling time is order one in units of the dS length scale, instead of order $\log N$ expected for a fast scrambler \cite{Susskind:2021esx} (which would diverge in the DSSYK model). The Krylov spread complexity also reproduces the minimal length between the finite Dirichlet boundaries in dS$_2$, and in the case of the $T^2+\Lambda_1$ deformed theory, it reaches a maximum value at a time of order of the dS length scale expected for hyperfast complexity growth \cite{Jorstad:2022mls}. Meanwhile, the algebraic entanglement entropy for the chord number state agrees with the RT formula for dS$_2$ JT gravity if the boundaries are in the Milne patch of dS$_2$ (Fig.~\ref{fig:two-step_def} (c)). The cosmological horizon limit reproduces the Gibbons-Hawking (GH) entropy of dS$_3$ space. However, if the boundaries are located in the static patch the RT formula in dS$_2$ does not apply, the dilaton at the minimal extremal surface becomes trivial (see Sec.~\ref{ssec:HEE dS}). However, in higher dimensions there are more possible entangling surfaces where Susskind's conjecture might apply. Our study takes the dS$_2$ geometric perspective instead of a dS$_3$ embedding, whose precise connection with the DSSYK model is actively debated \cite{Blommaert:2025eps,Rahman:2023pgt,Narovlansky:2023lfz}.

A summary comparing our results and the conjectures in \cite{Susskind:2021esx} is displayed in Tab.~\ref{tab:comparizon}.
\begin{table}[t!]
    \centering
    \begin{tabular}{|ccc|}
      \hline\textbf{Conjecture}   & \textbf{Confirmation in} & \textbf{Details} \\
      &\textbf{Deformed DSSYK}&\\\hline
        Hyperfast complexity & \bluecheck& Sec.~\ref{ssec:dS Complexity UV}\\
        growth (iii+iv)&&\\
        Hyperfast scrambling & \bluecheck & Sec.~\ref{ssec:hyperscrambling dS}\\
        (ii+iv) &&\\
        Validity of RT formula using & \redcross & Sec.~\ref{ssec:HEE dS}\\
        the stretched horizon (i+iv)&&\\\hline
    \end{tabular}
    \caption{Comparison between the (i-iv) conjectures in \cite{Susskind:2021esx}, which were explained above, and the explicit realization of the cosmological stretched horizon from the $T^2+\Lambda_1$ deformation of the UV limit of the DSSYK model, corresponding to the assumption (iv). Two out of the three remaining conjectures are satisfied in this model. Further details can be found in the respective sections.}
    \label{tab:comparizon}
\end{table}

In addition, we study other generalizations and additional technical details in the appendices. In particular, we study the consequences of interpreting the results in terms of {sine dilaton gravity} in finite cutoff holography. The energy spectrum generically becomes complex-valued since the bulk metric and dilaton at the cutoff location are complex-valued. In those cases, there is no Lorentzian evolution, rather than the deformed theory being non-unitary, so the DSSYK Hamiltonian is not isometrically dual to a canonically quantized Arnowitt–Deser–Misner \cite{Arnowitt:1959eec,Arnowitt:1960es,Arnowitt:1960zza,Arnowitt:1960zzb,Arnowitt:1960zzc,Arnowitt:1961zz,Arnowitt:1961zza} (ADM) Hamiltonian in the bulk, which is the generator of time translations. We adopt a similar prescription as in Cauchy slice holography \cite{Araujo-Regado:2022gvw,Araujo-Regado:2022jpj,Soni:2024aop,Araujo-Regado:2025elv}, where the stress tensor is allowed to be complex-valued. In contrast to previous literature, the $T^2$ deformation parameter is generically a complex number, and it can take both positive and negative values (without an auxiliary geometric extension \cite{Apolo:2023vnm,Apolo:2023ckr}). Later, we $T^2$ deform the Almheiri-Goel-Hu (AGH) model \cite{Almheiri:2024xtw}, which is a simplification of the DSSYK model in the $q\rightarrow1$ described by the Heisenberg-Weyl algebra. This integrable analog of the DSSYK model allows for analytic evaluations of correlation functions beyond the semiclassical limit.

\paragraph{Outline}
The paper is organized as follows. In Sec.~\ref{sec:flow eq}, we introduce Hamiltonian deformations in holographic quantum mechanics based on moving the HJ equation of the dual dilaton graviton theory with Dirichlet boundary conditions, and we study its thermodynamic properties. In Sec.~\ref{sec:Chord Hilbert}, we specialize the results in the DSSYK model to deduce a chord basis associated with a bulk geodesic in finite cutoff holography, and the representation of the chord operators after the deformation. We also evaluate the Krylov spread complexity of the HH state. 
In Sec.~\ref{sec:correlators} we evaluate correlation functions, including its semiclassical and its triple-scaling limits. 
In Sec.~\ref{sec:EE}, we investigate the entanglement entropy between the double-scaled algebras in the deformed DSSYK model, and their holographic interpretation. In Sec.~\ref{sec:dS holography} we apply our previous results to realize the stretched horizon proposal of Susskind. We then conclude in Sec.~\ref{sec:discussion} with a summary of our findings, and some relevant future directions. 

In addition, App.~\ref{app:Notation} contains a summary of the notation used in the manuscript. We include a brief introduction to $\TT$ deformations in App.~\ref{app:TTbar intro}, which is not required to follow the main text. In App.~\ref{app:geodesics} we provide more details about the derivation of spacelike and timelike extremal geodesic lengths connecting the asymptotic boundaries of an effective AdS$_2$ black hole. App.~\ref{app:sine dilaton gravity finite cutoff} discusses reality conditions in sine dilaton gravity at a finite cutoff. In App.~\ref{app:alternative chord number} we investigate an additional chord number basis, different from the Krylov basis, which has an interpretation in finite cutoff holography. In App.~\ref{eq:Def ETW Brane Wormhole} we explore chord Hamiltonian deformations in the presence of ETW branes and Euclidean wormholes in the bulk. In App.~\ref{app:AGH} we $T^2$-deform the AGH model to study its thermodynamics and correlation functions. At last, in App.~\ref{sec:1st law thermo} we discuss the first law of thermodynamics in the deformed DSSYK.

\section{Dilaton Gravity at Finite Cutoff and its Boundary Dual}\label{sec:flow eq}
Before specializing in the DSSYK model, we generalize previous studies of $T^2$ deformations for holographic (0+1)-dimensional models \cite{Gross:2019ach,Gross:2019uxi,Zhang:2023rub} assuming a dual dilaton gravity model with Dirichlet boundary conditions. In this section, we are particularly interested in deriving the energy flow equation in holographic quantum mechanics with a Hamiltonian deformation, and to study its semiclassical thermodynamics. In particular, one recovers a lower-dimensional analog of $\TT$ and $\TT+\Lambda_2$ deformations as special cases. 

\paragraph{Outline}In Sec.~\ref{ssec:flow eq}, we study HJ equation in dilaton gravity theories and the associated flow equation. In Sec.~\ref{ssec:TTbar+Lambda_1} we study its solutions, including one-dimensional $T^2$ deformations and we propose a ``$T^2+\Lambda_1$'' deformation in the DSSYK model, which can be interpreted as pushing the finite cutoff inside the black hole horizon/ outside the cosmological horizon. In Sec.~\ref{sec:thermo} we study the thermodynamic properties of the deformed theories, specializing to the DSSYK model for concreteness. In addition, the reader is referred to App.~\ref{app:sine dilaton gravity finite cutoff} for details on the reality conditions to implement the chord Hamiltonian deformations from sine dilaton gravity at a finite cutoff.

\subsection{Hamiltonian Deformations from Finite Cutoff Holography}\label{ssec:flow eq}
Our first goal is to derive the flow equation controlling the evolution of the energy spectrum of a (0+1)-dimensional system dual to a generic dilaton gravity theory of the form
\begin{equation}\label{eq:Dilaton-gravity theory}
\begin{aligned}
    I_{\rm E}\equiv&-\frac{\Phi_0}{16\pi G_N}\qty(\int_{\mathcal{M}}\rmd^2x\sqrt{g}\mathcal{R}+2\int_{\partial\mathcal{M}}\rmd x\sqrt{{h}}K)\\
    &-\frac{1}{16\pi G_N}\qty(\int_{\mathcal{M}}\rmd^2x\sqrt{g}\qty(\Phi\mathcal{R}+U(\Phi))+2\int_{\partial\mathcal{M}}\rmd x\sqrt{h}\qty(\Phi_{B} K-\sqrt{G(\Phi_B)}))~,
\end{aligned}
\end{equation}
where the first line is a topological term, $U(\Phi)$ and $G(\Phi_B)$ are generic $C^\infty$ functions; $G_N$ in being the two-dimensional Newton's constant; $\mathcal{M}$ is the spacetime manifold, $\partial\mathcal{M}$ is the location of the radial cutoff boundary where we impose Dirichlet boundary conditions on the dilaton and the induced metric,
\begin{equation}
    \partial\mathcal{M}:\quad\Phi=\Phi_B~,\quad h_{\mu\nu}\quad \text{fixed}~.
\end{equation}
This allows us to generalize the flow equation in \cite{Gross:2019ach} for dilaton-gravity theories without AdS boundary conditions, which is motivated by the proposed duality between sine dilaton gravity and the DSSYK model \cite{Blommaert:2023opb}. One should note that a canonical choice of the counterterm is
\begin{equation}\label{eq:Gphi special}
    G(\Phi_B)^{\rm(others)}=\int^{\Phi_B}\rmd\Phi ~U(\Phi)~,
\end{equation}
which guarantees the on-shell action is finite when $\Phi_B\gg1$; however, we will not make this restriction in deriving the flow equation below, since we investigate cases where $\Phi_B$ is a finite cutoff, instead of a radial regulator of the asymptotic boundary.\footnote{If one were to follow \eqref{eq:Gphi special}, the $\Phi_B$ variation of the boundary action leads to the on-shell mean-curvature at the radial boundary
\begin{equation}\label{eq:K symme}
    K=\frac{U(\Phi_B)}{2\sqrt{G(\Phi_B)}}~.
\end{equation}}
Also, note that in the following one may allow $g_{\mu\nu}$ and $\Phi\in\mathbb{C}$; which is the case of interest for sine dilaton gravity at finite cutoff. We will discuss this case in detail in App.~\ref{app:sine dilaton gravity finite cutoff}.

The strategy proposed in \cite{Hartman:2018tkw} is to recover the $T^2$ flow equation from the HJ equation controlling radial evolution along the flow.\footnote{A different but related approach is taken in \cite{Iliesiu:2020zld}, who instead study the radial quantization of the Wheeler-DeWitt (WDW) \cite{Wheeler:1968iap,DeWitt:1967yk}  equation (at the reduced phase space level), and the path integral dilaton gravity of the form (\ref{eq:Dilaton-gravity theory}); without a counterterm, (see e.g.~(2.20) \cite{Iliesiu:2020zld}). After properly introducing the corresponding counter term, one can then recover a relation equivalent to \cite{Gross:2019ach}, which agrees with results (see \eqref{eq:flow eq}) after rescaling.} For this reason, we will consider the dilaton-gravity theory as the input for defining the deformed DSSYK model.\footnote{An alternative way to define it would be to keep the asymptotic boundary in the same location but modify the boundary conditions instead of using Dirichlet \cite{Guica:2019nzm}. This type of procedure was considered by \cite{Gross:2019ach} in JT gravity. We reserve future directions to study if such modifications can be adapted to this context (see more comments in Sec.~\ref{ssec:Outlook}).} {Note that while we implement a similar strategy as \cite{Hartman:2018tkw,Gross:2019ach} to define the Hamiltonian deformations in the boundary theory, the crucial difference with previous approaches in the derivation of the flow equation from finite cutoff holography is that we do not require AdS asymptotics in the bulk, which is the reason for allowing the general counterterm $G(\Phi)$ in \eqref{eq:Dilaton-gravity theory}, leading to a larger class of deformations than the $T^2$ case.}

We begin considering the general variation of an action of the form (\ref{eq:Dilaton-gravity theory}), given by
\begin{equation}\label{eq:delta I_E}
\begin{aligned}
    \delta I_{\rm E}=&-\frac{1}{16\pi G_N}\int_{\mathcal{M}}\rmd^2x\sqrt{g}\qty[\mathcal{E}^{\mu\nu}\delta g_{\mu\nu}+\mathcal{E}_{\Phi}\delta \Phi]+\int_{\partial\mathcal{M}}\rmd\tau\sqrt{h}\qty[\frac{1}{2}\tilde{T}^{ab}\delta h_{ab}+\mathcal{O}_{\Phi_{B}}\delta{\Phi_{B}}]~,
\end{aligned}
\end{equation}
where
\begin{align}
    \mathcal{E}_{\mu\nu}&\equiv\nabla_\mu\nabla_\nu\Phi-g_{\mu\nu}\nabla^2\Phi+\frac{1}{2}g_{\mu\nu}U(\Phi)~,\\
    \mathcal{E}_{\Phi}&\equiv\mathcal{R}+\partial_\Phi U(\Phi)~.
\end{align}
Here we have introduced the Brown-York (BY) \cite{Brown:1992br} stress tensor and the scalar operator conjugate to $\Phi$, which we denote respectively as:
\begin{subequations}
        \begin{align}
    \tilde{T}^{ab}&\equiv\frac{2}{\sqrt{h}}\frac{\delta I_E}{\delta h_{ab}}=\frac{2}{\sqrt{h}}\qty(\pi^{ab}+\frac{\sqrt{h}}{16\pi G_N}\sqrt{G(\Phi_{B})}h^{ab})~,\label{eq:BY stress tensor}\\
        \mathcal{O}_{\Phi_{B}}&\equiv\frac{1}{\sqrt{h}}\frac{\delta I_E}{\delta {\Phi_{B}}}=\frac{1}{\sqrt{h}}\qty(\pi_{\Phi_{B}}+\frac{\sqrt{h}~G'({\Phi_{B}})}{16\pi G_N\sqrt{G({\Phi_{B}})}})~,\label{eq:source Phi}
    \end{align}
\end{subequations}
    where $G'({\Phi_{B}})\equiv\dv{G({\Phi_{B}})}{\Phi_{B}}$; and $\pi_{\Phi_{B}}$ and $\pi_{ab}$ are the canonical momenta of $h_{ab}$ and ${\Phi_{B}}$, defined as
    \begin{align}
        \pi^{ab}&\equiv-\frac{\sqrt{h}}{16\pi G_N}h^{ab}n^c\nabla_c{\Phi_{B}}~,\quad\pi_{\Phi_{B}}\equiv-\frac{\sqrt{h}}{8\pi G_N}K~.
    \end{align}
We can now study the HJ equation controlling the radial equation of the on-shell solutions. Using the general variation (\ref{eq:delta I_E}) evaluated on-shell (i.e.~$\mathcal{E}_{\mu\nu}=0$, $\mathcal{E}_\Phi=0$), {in the strict large $N\gg1$ limit in the boundary theory, $G_N\rightarrow0$ in the bulk,} we recover
    \begin{equation}\label{eq:derivative I E}
        \partial_{\Phi_{B}}I_{\rm E}^{\rm (on)}=\int_{\partial\mathcal{M}}\rmd\tau\sqrt{h}\qty[\frac{1}{2}\tilde{T}^{ab}\partial_{\Phi_{B}}h_{ab}+\mathcal{O}_{\Phi_{B}}]~,
    \end{equation}
    and we have to implement the ADM Hamiltonian constraint $\mathcal{H}$ for general dilaton gravity theories (\ref{eq:Dilaton-gravity theory}), which we evaluate at the cutoff location, and it is given by \cite{Grumiller:2007ju},\footnote{{One can recover the constraint by performing variations with respect to the lapse, represented $\tilde{N}$, in the ADM decomposition of the metric 
\begin{equation}
    \rmd s^2=\tilde{N}\rmd t^2+\rme^{2\sigma(x)}\qty(\rmd x+N_\bot\rmd t)^2~,
\end{equation}
where $N_\bot$ is the shift, $\rme^{2\sigma(x)}$ the boundary metric and $x \sim x + 1$.}}
    \begin{equation}\label{eq:HJ eq}
    \begin{aligned}
        \mathcal{H}&\equiv16\pi G_N\pi_{\Phi_{B}}\pi^{\tau\tau}-\frac{U(\Phi_{B})}{16\pi G_N}=0~.
    \end{aligned}
    \end{equation}
This constraint leads to the following relation for the dilaton source in \eqref{eq:source Phi}:
\begin{equation}\label{eq:O phi}
    \mathcal{O}_{\Phi_{B}}=\frac{U(\Phi_{B})+\qty(\frac{8\pi G_N}{G(\Phi_B)}\tilde{T}^\tau_\tau-1)G'({\Phi_{B}})}{16\pi G_N\sqrt{G({\Phi_{B}})}\qty(\frac{8\pi G_N}{\sqrt{G({\Phi_{B}})}}\tilde{T}^\tau_\tau-1)}~.
\end{equation}
Following the procedure of \cite{Gross:2019uxi}, we need to identify the scaling of the different terms in (\ref{eq:derivative I E}) with respect to $r_{B}$. Similar to \cite{Hartman:2018tkw,Gross:2019ach}, we recognize that at the finite boundary, $\Phi_{B}$, we should have the following scaling for the metric and the stress tensor given the counterterm (\ref{eq:Dilaton-gravity theory}),
\begin{equation}\label{eq:dictionary T}
    \sqrt{h}=\sqrt{G(\Phi_{B})}\sqrt{\gamma}~,\quad \tilde{T}_{\tau\tau}=\sqrt{G(\Phi_{B})}~T_{\tau\tau}~,
\end{equation}
which is a choice of normalization to match the boundary metric in the bulk to the metric of the boundary theory.\footnote{I thank Simon Lin for comments about this.} \eqref{eq:dictionary T} corresponds to a natural general generalization of the AdS$_2$ boundary conditions (with the corresponding $G(\Phi_B)=\Phi_B^2$). Our proposal for extending the holographic dictionary at finite cutoff beyond AdS \eqref{eq:dictionary T} agrees with \cite{Griguolo:2025kpi} using a different method that results in the same flow equation from general dilaton gravity theories at finite cutoff, which appeared after our results in \cite{Aguilar-Gutierrez:2024oea}.

Next, we use that in 1D we can express the following relation \cite{Gross:2019ach},
\begin{equation}\label{eq:IE as stress tensor}
    I_E=\int\rmd\tau~\sqrt{\gamma}T_\tau^\tau=\int\rmd\tau~\sqrt{\gamma}E_y~,
\end{equation}
where $\gamma_{ij}$ is the background metric of the 1D quantum mechanical theory, and $T_{ij}$ the corresponding boundary stress tensor, and in the last equality we use the fact that in (0+1)-dimensions, $T^\tau_\tau\equiv E_y$ which denotes the energy spectrum of the deformed theory, $y$ denotes a deformation parameter, which we define below. This allows us to express the HJ equation (\ref{eq:HJ eq}) as
\begin{equation}\label{eq:pre flow eq}
    \partial_{\Phi_{B}}\int_{\partial\mathcal{M}}\rmd\tau\sqrt{\gamma}~T^\tau_\tau=\int_{\partial\mathcal{M}}\rmd\tau\sqrt{\gamma}\frac{G'(\Phi_B)\qty(\qty(\frac{8\pi G_N}{G(\Phi_B)}T^\tau_\tau)^2-1)+U(\Phi_{B})}{2\qty(\frac{8\pi G_N}{G(\Phi_B)}~T^\tau_\tau-1)}~.
\end{equation}
Using the relation \eqref{eq:IE as stress tensor}, we identify that (\ref{eq:pre flow eq}) can then be expressed as a flow equation for the spectrum of the boundary theory
\begin{equation}\label{eq:flow eq}
    \boxed{\partial_{y} E_y=\frac{(E_y)^2+(\eta-1)/y^2}{2(1-y E_y)}~,}
\end{equation}
where we have defined the parameters 
\begin{equation}\label{eq:def parameter}
    y\equiv8\pi G_N/G(\Phi_{B})~,\quad  \eta\equiv U(\Phi_B)/G'(\Phi_B)~.
\end{equation}
Note that while $E_y$ depends on both $y$ and $\eta$, where the latter is not completely determined by $y$. The functional dependence on the parameter $\eta$ will be implicit in all expressions involving $y$ to lighten the notation. 

The above relation can be interpreted as a flow equation (\ref{eq:flow eq}) for the energy spectrum of the boundary dual to the dilaton gravity theory at finite cutoff. Indeed, (\ref{eq:flow eq}) takes the same form as other places in the literature (see e.g.~\cite{Gross:2019uxi,Iliesiu:2020zld}) which are restricted to $\eta=U(\Phi_B)/G'(\Phi_B)=1$. The latter choice of counterterm $G'(\Phi_B)=U(\Phi_B)$ leads to a finite on-shell action in general dilaton-gravity theories with asymptotic Dirichlet boundaries \cite{Grumiller:2007ju}. Meanwhile, the case $\eta=-1$ can be used when the cutoff surface is located at some finite $r_B$, so there is no issue with the renormalization of the on-shell action, while keeping the same holographic identification for the deformation parameter (\ref{eq:def parameter}). We remark that the deformation parameter (\ref{eq:def parameter}) and the flow equation \eqref{eq:flow eq} generalize the results in \cite{Gross:2019uxi} which are a special case of this formulation, corresponding to a class of asymptotically AdS dilaton gravity theories.

\paragraph{Deformed Boundary Hamiltonian}We promote \eqref{eq:flow eq} to an operator relation in the quantum theory as
\begin{equation}\label{eq:flow eq operator form}
    \partial_y \hH_y=\hat{\Theta}(\hH_y,y)~,\quad \hat{\Theta}=\frac{1}{2}\qty(\hH_y^2+y^{-2}(\eta-1))\qty(1-y \hH_y)^{-1}
\end{equation}
where $\hat{\Theta}(\hH_y,y)$ is the $\TT$ deformation operator \cite{Kruthoff:2020hsi}, and $\hH_y$ is the Hamiltonian deformation of the boundary theory.

\subsection{Solutions to the Flow Equation}\label{ssec:sol flow eq}
First, notice from \eqref{eq:def parameter} that $\eta=\eta(y)$ generically. As mentioned in the introductions, when we deformed the boundary theory starting at the asymptotic boundary in the bulk, we need to set $G(\Phi_B)$ as in \eqref{eq:Gphi special}. However, once we reach a finite cutoff value $\Phi_B$, we can allow $G(\Phi_B)$ to be more arbitrary, which defines a set of possible deformed theories. Below we specialize in one of them due to its applications for studying the stretched horizon conjecture of Susskind \cite{Susskind:2021esx}.

We will be interested in the case $\eta=\pm1$, as this allows us to introduce an analogous type of deformation as $\TT(+\Lambda_2)$ \cite{Silverstein:2024xnr,Batra:2024kjl,Gorbenko:2018oov,Coleman:2021nor} deformations, which we define below as
\begin{equation}
    T^2(+\Lambda_1)~{\rm deformations}:\quad U(\Phi)\quad\text{fixed}~,\quad G(\Phi_B)\rightarrow\pm G(\Phi_B)~,
\end{equation}
resulting in $\eta=\pm1$ in \eqref{eq:flow eq}. We note that the solutions to the flow equation when $\eta$ is a fixed parameter are
\begin{equation}\label{eq: E lambda}
    E_y=\frac{1}{y}\qty(1+\varepsilon\sqrt{\eta-2y f(E)})~.
\end{equation}
where $f(E)$ is a function determined by initial conditions, and $\varepsilon=\pm1$. 

\begin{figure}
    \centering
    \subfloat[]{\includegraphics[width=0.42\linewidth]{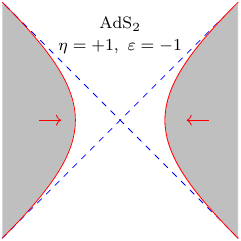}}\hspace{2cm}\subfloat[]{\includegraphics[width=0.42\linewidth]{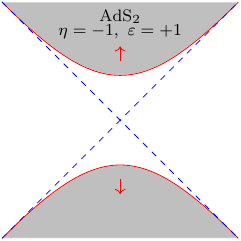}\label{sfig:eta-1_AdS}}\\
    \subfloat[]{\includegraphics[width=0.42\linewidth]{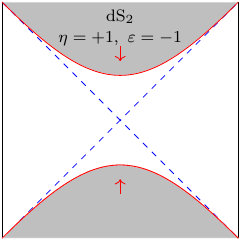}}\hspace{2cm}\subfloat[]{\includegraphics[width=0.42\linewidth]{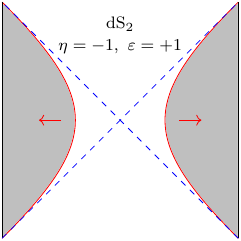}\label{sfig:eta-1_dS}}
    \caption{Bulk interpretation of the $T^2$ and $T^2+\Lambda_1$ flow in the boundary theory describing (a,b) an AdS$_2$ black hole, and (c, d) dS$_2$ space based on the IR/UV triple-scaling limits of the deformed DSSYK in Sec.~\ref{sec:Chord Hilbert}. The gray regions represent the parts of the bulk that are cut out. (a) By implementing the $T^2$ deformation, with spectrum \eqref{eq:E(theta) eta+1}, there is a finite boundary cutoff (red solid curve) in the Rindler-AdS$_2$ patch til it reaches the horizon. (b) We implement the $T^2+\Lambda_1$ deformation \eqref{eq:E(theta) eta-1} which moves the boundary inside the black hole horizon. A similar procedure applies for dS$_2$ space for (c) $T^2$ and (d) $T^2+\Lambda_1$ deformations. In the latter, the cosmological stretched horizon \cite{Susskind:2021esx} is realized when the Dirichlet boundaries are very close to the dS$_2$ horizon.}
    \label{fig:two-step_def}
\end{figure}

\paragraph{\texorpdfstring{$T^2$}{} deformation} The best studied case from the lower dimensional perspective corresponds to $\eta=+1$. The deformed energy spectrum in terms of the seed theory takes the form
\begin{equation}\label{eq:E(theta) eta+1}
    E^{\eta=+1}_y(\theta)=\frac{1}{y}\qty(1+\varepsilon\sqrt{1-2y E})~.
\end{equation}
The solution with $\varepsilon=-1$ (Fig.~\ref{fig:two-step_def} (a, c)) represents the energy spectrum at a finite cutoff in the holographic dual (see e.g.~\cite{McGough:2016lol,Hartman:2018tkw}), as seen from the on-shell action (\ref{eq:on shell action}), {which is smoothly connected to the seed theory in the limit $y\rightarrow0$.} Meanwhile, the solution with $\varepsilon=+1$ represents the complementary spacetime region with respect to the finite cutoff. 

\paragraph{\texorpdfstring{$\TT+\Lambda_1$}{} deformations}\label{ssec:TTbar+Lambda_1}
When $\eta=-1$, this corresponds to a s-wave reduction of $\TT+\Lambda_2$ deformations \cite{Gorbenko:2018oov,Coleman:2020jte,Silverstein:2022dfj,Batra:2024kjl,AliAhmad:2025kki}, which we refer to as $T^2+\Lambda_1$ deformations. In this case $\eta=-1$ means that at some point along the flow, we modify the boundary counterterm in the action \eqref{eq:Dilaton-gravity theory} with $G(\Phi_B)\rightarrow-G(\Phi_B)$,\footnote{One may recover the same sign as the $\TT+\Lambda_2$ analogue in \cite{AliAhmad:2025kki} by setting $U(\Phi)=0$, however, we do not analyze this case.} while retaining the same dilaton potential $U(\Phi)\rightarrow U(\Phi)$.\footnote{Previous literature has approached the bulk interpretation of the $\TT+\Lambda_2$ deformation in CFT$_2$ in different ways. \cite{Gorbenko:2018oov,Coleman:2020jte,Silverstein:2022dfj,Batra:2024kjl} gives evidence that this deformation corresponds to changing the bulk from an AdS black hole to dS$_3$ space. Meanwhile \cite{AliAhmad:2025kki} shows that one can probe the black hole interior by changing timelike and spacelike boundaries in the $\TT$ and $\TT+\Lambda_2$ flow respectively. The latter interpretation corresponds to the case we explore; the former interpretation of $\eta=-1$ can also be studied in our set-up by modifying the dilaton potential $U(\Phi)\rightarrow-U(\Phi)$ while keeping $G(\Phi_B)\rightarrow G(\Phi_B))$ at the transition.}

In particular, we consider a transition between the perturbative branch \eqref{eq:E(theta) eta+1} in the $\eta=+1$ case to the $\eta=-1$ case when the square-root term in \eqref{eq:E(theta) eta+1} vanishes, corresponding to
\begin{equation}\label{eq:transition TTbar+Lambda1 DSSYK}
    y=y_0\equiv1/(2E)~.
\end{equation}
By demanding continuity in \eqref{eq: E lambda}, the energy spectrum then takes the form
\begin{equation}\label{eq:E(theta) eta-1}
    E^{\eta=-1}_y=\frac{1}{y}\qty(1+\varepsilon\sqrt{2y E-1})~.
\end{equation}
As before, there are two relative signs, $\varepsilon=\pm1$; both of them are continuously connected to the $T^2$ deformation since the square root factor in either case vanishes for the transition point of the deformation parameter \eqref{eq:transition TTbar+Lambda1 DSSYK}. We will specialize in $\varepsilon=+1$ (Fig.~\ref{fig:two-step_def} (b, d)) for concreteness, and to probe the cosmological patch in the dS$_2$ space limit of the bulk. However, it is equally valid for $\varepsilon=-1$ sign;\footnote{If one requires the functional expression for the deformed energy spectrum $E_y(E(\theta))$ \eqref{eq:flow eq} for energies beyond a given seed theory spectrum $E(\theta)$, then there is a prescription in \cite{AliAhmad:2025kki} for only studying the $-$ sign solution. Since we are only concerned with the flow of the seed theory spectrum, we do not follow the prescription in \cite{AliAhmad:2025kki} .} similar to the $T^2$ case, the different signs simply correspond to different bulk regions \cite{AliAhmad:2025kki}. In particular, $\varepsilon=+1$ in the $\eta=-1$ in the \eqref{eq:E(theta) eta-1} spectrum corresponds to the bulk region connecting the asymptotic boundary to the finite cutoff inside the horizon, while the $\varepsilon=-1$ case in \eqref{eq:E(theta) eta-1} corresponds to the complement. The $\varepsilon=+1$ solution in the $\eta=-1$ case is illustrated in Figs.~\ref{sfig:eta-1_AdS} and \ref{sfig:eta-1_dS}.

Note that since there are no modifications in the dilaton potential in the $\eta=-1$ case, it describes the same bulk metric as its $\eta=+1$ analog, similar to \cite{AliAhmad:2025kki,Chang:2025ays} (in contrast to other approaches \cite{Coleman:2021nor,Batra:2024kjl,Gorbenko:2018oov,Shyam:2021ciy,Silverstein:2024xnr} which can be realized by changing the dilaton potential).\footnote{The approaches in \cite{AliAhmad:2025kki} and \cite{Coleman:2021nor} differ in terms of the sign of the cosmological constant deformation term; however, they both give rise to the same spectrum of the deformed CFT. Our approach suggests that the overall sign of the corresponding deformation is not relevant, but how this term arises in the derivation of the flow equation, which in our case corresponds to either changing the dilaton potential, $U(\Phi)\rightarrow-U(\Phi)$, in the $T^2$ to $T^2+\Lambda_1$ transition, or the sign of term $G(\Phi_B)\rightarrow-G(\Phi_B)$.} In fact, the BY stress tensor evaluated at the radial boundary $\Phi_B$ ($\tilde{T}_{\tau}^\tau=G(\Phi_B)T^\tau_\tau$ \eqref{eq:dictionary T}) changes sign during the transition between $\eta=+1$ and $\eta=-1$. This means that the normal vector to the constant $r_B$ surface changes from spacelike to timelike as explained in \cite{AliAhmad:2025kki}. 
Thus, the finite cutoff crosses the horizon once we turn on the $T^2+\Lambda_1$ deformation. We illustrate the two-step bulk interpretation of the $T^2$ and $T^2+\Lambda_1$ deformation in Fig.~\ref{fig:two-step_def}.\footnote{Note that both asymptotic boundaries move at once in the $N\gg1$ limit \cite{McGough:2016lol}.}

\subsection{Thermodynamics}\label{sec:thermo}
In this subsection, we specialize the previous results to study the semiclassical thermodynamics of the deformed DSSYK model from its partition function to evaluate its energy spectrum, thermodynamic entropy, density of states, and heat capacity as a function of its temperature. This leads to conditions when the deformed theory is thermodynamically stable. We do not assume a specific dual dilaton gravity theory (\ref{eq:Dilaton-gravity theory}) to the DSSYK model in this section (see App.~\ref{app:sine dilaton gravity finite cutoff} for considerations in sine dilaton gravity).

\paragraph{Energy spectrum, entropy and temperature}Consider the partition function of the deformed DSSYK model which we evaluate in the semiclassical limit as\footnote{There is a proposal for improvement in the partition function in finite cutoff holography in \cite{Griguolo:2025kpi} where one includes the perturbative and non-perturbative branches of the one-dimensional T$^2$ deformation with an appropriate contour of integration with a cutoff in the energy integration to regularize the integral. The contour is subtle for periodic potentials such as in sine dilaton gravity. Moreover, this prescription is not necessary for the DSSYK model whose partition function is UV finite, so it will not be discussed.}
\begin{equation}\label{eq:partition DSSYK}
	Z_y(\beta)\equiv\bra{\Omega}\rme^{-\beta \hH_y}\ket{\Omega}\eqlambda\int\rmd E(\theta)~\rme^{-\beta E_y(\theta)}\rho(\theta)~,
\end{equation}
where $E(\theta)$ is the energy spectrum of the seed theory;  $\rho(\theta)=\rme^{S(\theta)}$ is the semiclassical density of states, and $S(\theta)$ is the thermodynamic entropy, given by \cite{Goel:2023svz},\footnote{Note that the entropy is proportional to $1/\lambda=N/(2p^2)$ at all temperature scales, where $N$ is the number of degrees of freedom in the SYK and p the number of all-to-all interacting Majorana fermions. While it has been debated (see e.g.~\cite{Rahman:2024vyg,Rahman:2024iiu}) if the DSSYK model at infinite temperature should behave like any other system of $N$ qubits at infinite temperature, the entropy would instead scale as $S\propto N$. The double scaling introduces the 1/$p^2$ factor (as seen explicitly in partition function in Sec.~\ref{sec:thermo}). I thank Takanori Anegawa and Jiuci Xu for useful discussions related to this point.}
\begin{equation}\label{eq:thermodynamic entropy}
	E(\theta)\equiv-\frac{2J~\cos(\theta)}{\sqrt{\lambda(1-q)}}~,\quad S(\theta)\equiv S_0+\frac{2\theta}{\lambda}\qty(\pi-\theta)~,\quad \theta\in[0,\pi]~,
\end{equation}
where $J$, $\lambda\equiv-\log q\geq0$ are constant parameters of the theory, which will be specified in Sec.~\ref{ssec:review DSSYK}, and we have introduced $S_0$, resulting from an overall normalization of the partition function. The microcanonical inverse temperature in this theory is given by\footnote{This is a Boltzmann temperature, in contrast to the temperature scale in the correlation functions, which we discuss in Sec.~\ref{sec:correlation functions}.}
\begin{equation}\label{eq:microcanonical inverse temperature}
	\beta_y(\theta)\equiv\dv{S}{E_y}=\beta(\theta)\dv{E}{E_y}~, \quad \beta(\theta)\equiv \frac{2\pi-4\theta}{J\sin\theta}~,
\end{equation} 
where there is a redshift factor is
\begin{equation}\label{eq:redshift factor}
    \dv{E}{E_y}=\begin{cases}
        \sqrt{1-2y E(\theta)}~,&\eta=+1~,\\
        \sqrt{2y E(\theta)-1}~,&\eta=-1~.
    \end{cases}
\end{equation}
Given that $\beta_y$ is the periodicity of the thermal circle, the $T^2$ deformation effectively shrinks it. One can similarly recast (\ref{eq:partition DSSYK}) in the form:
\begin{subequations}
    \begin{align}
    Z_y(\beta)=&\int\rmd E_y(\theta)~\rme^{-\beta E_y(\theta)}\rho_y(\theta)~,\label{eq:def partition function}\\
    \rho_y(\theta)&\equiv\rho(\theta)\dv{E(\theta)}{E_y(\theta)}~,\label{eq:DOS TTbar}
\end{align}
\end{subequations}
where $\rho_y$ can be interpreted as the density of states in the deformed theory \cite{Gross:2019ach}. Note that since the redshift factor is $\mathcal{O}(1)$, it does not modify the thermodynamic entropy which is $\mathcal{O}(\lambda^{-1})$ in the limit where $\lambda\rightarrow0$.

\begin{figure}
    \centering
    \subfloat[]{\includegraphics[width=0.49\textwidth]{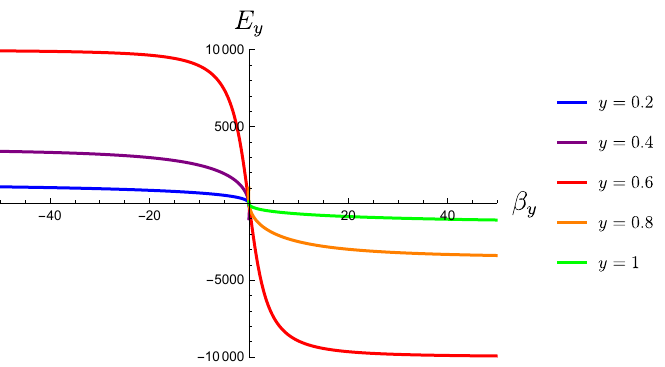}}\hfill\subfloat[]{\includegraphics[width=0.44\textwidth]{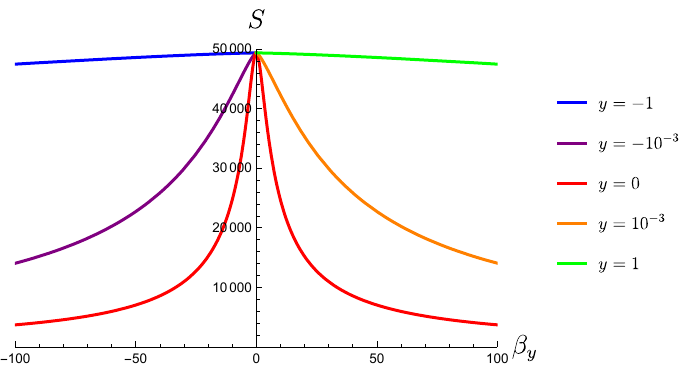}}\\
    \subfloat[]{\includegraphics[width=0.49\textwidth,valign=t]{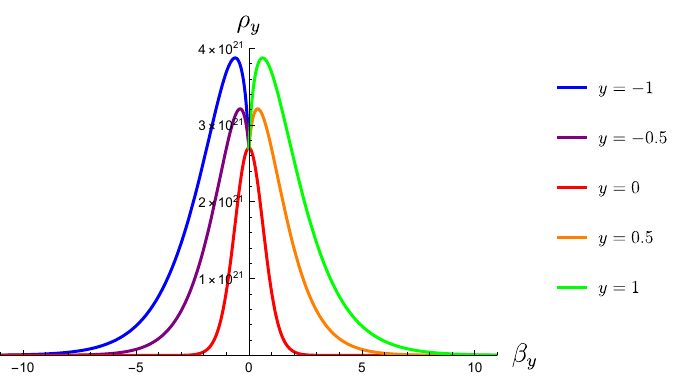}}\hfill\subfloat[]{\includegraphics[width=0.49\textwidth,valign=t]{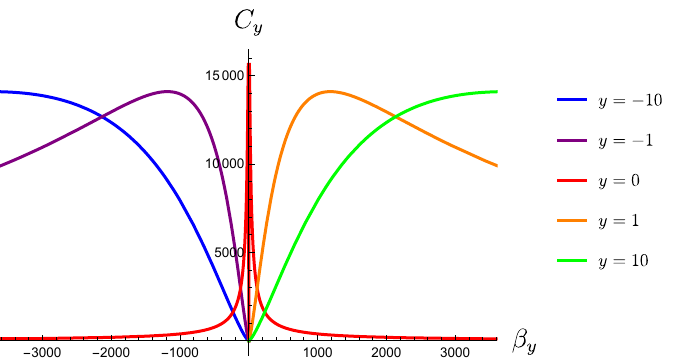}}
    \caption{(a) The energy spectrum (\ref{eq:energy spectrum}), (b) thermodynamic entropy (\ref{eq:thermodynamic entropy}), (c) density of states (\ref{eq:DOS TTbar}), and (d) heat capacity (\ref{eq:heat capacity}) for generic values of the deformation parameter $y$ (\ref{eq:def parameter}) (shown in the legends) and fixing $\eta=+1$. We use $\lambda=10^{-4}$ in all subfigures, except in (c) where $\lambda=0.1$ for visualization purposes.}
    \label{fig:0_Thermo_summary}
\end{figure}

We include Fig.~\ref{fig:0_Thermo_summary} to exemplify the resulting thermodynamic properties of the $T^2$ deformed DSSYK model for generic values of the deformation parameter $y$ appearing in the flow equation (\ref{eq:flow eq}) for $\eta=+1$.\footnote{One can repeat the same semiclassical thermodynamic analysis in sine-dilaton gravity at finite cutoff, which has the same density of states \cite{Blommaert:2023opb}; one needs to replace the deformed energy spectrum for the BY one.}

\paragraph{First law of thermodynamics} The previous results can be used to express variations of the energy spectrum with respect to the parameters of the model. We include details about the relevant calculations in App.~\ref{sec:1st law thermo}.

\paragraph{Heat capacity}Next, we use the previous results to determine the thermal stability of the system, using the definition of the heat capacity
\begin{equation}\label{eq:heat capacity}
		C_{y}(\theta)\equiv-\beta_y(\theta)^2\dv{E_y(\theta)}{\beta_y(\theta)}~.
  \end{equation}
  For concreteness, we consider the main cases of interest where $\eta=\pm1$ so that $E_y$ is given by \eqref{eq:energy spectrum} ($\eta=+1$) and \eqref{eq:E(theta) eta-1} ($\eta=-1$) respectively, so that \eqref{eq:heat capacity} becomes,\footnote{Note we are using the $\varepsilon=-1$ sign solution in the T$^2$ deformed energy spectrum \eqref{eq:E(theta) eta+1} and $\varepsilon=+1$ in the $T^2+\Lambda_1$ case \eqref{eq:E(theta) eta-1}; the signs in the heat capacity reverses.}
  \begin{equation}\label{eq:heat capacity2}
C_{y}(\theta)=\frac{\eta(J\sin\theta\beta_y(\theta))^2}{2Jy\sin\theta(\pi-2\theta)+4Jy\cos\theta(2+(\pi-2\theta)\cot\theta)+2(2+(\pi-2\theta)\cot\theta\sqrt{\lambda(1-q)})}~.
  \end{equation}
  The argument in the numerator is positive definitive when $y\in\mathbb{R}$, $\lambda>0$ and $\theta\in[0,\pi]$. Note that the case where $\eta=+1$ has a positive heat capacity while $\eta=-1$ a negative one assuming that $E_y$ can be interpreted as the conserved energy spectrum of the deformed theory.

It is important to distinguish when $C_{y}(\theta)$ physically corresponds to heat capacity when performing the $T^2$ and $T^2+\Lambda_1$ flow. For instance in dS$_2$, one performs the $T^2$ deformation, associated with $E_y^{\eta=+1}$, using the Dirichlet boundaries in the Milne patch which are spacelike \cite{Chang:2025ays}; while during the $T^2+\Lambda_1$ flow, associated to $E_y^{\eta=-1}$, the finite cutoff boundaries are located in the static patch of dS$_2$ and they are timelike, so that $E_y^{\eta=-1}$ can indeed be associated with a quasilocal BY energy \cite{Svesko:2022txo}. Then, \eqref{eq:heat capacity2} is indeed physically a heat capacity and the system can thus be thermodynamically unstable, as found in different parts of the literature, see e.g.~Sec \ref{ssec:comparison dS2}. As additional comparison, when $\TT$ deforming CFTs, one finds negative heat capacities at high energies for $T^2$-deformed CFTs and quantum mechanical systems in \cite{Barbon:2020amo}; while in this system, the energy spectrum is bounded. To make more contrast, opposite observations appear in the AdS$_2$ case, one performs the $T^2$ deformation when the boundaries are timelike, and thus the system is stable since \eqref{eq:heat capacity2} indeed corresponds to a possitive heat capacity, and $T^2+\Lambda_1$ when the boundaries are spacelike, similar to \cite{AliAhmad:2025kki}. 

\paragraph{Finite cutoff Holographic Interpretation}We now provide a bulk interpretation of \eqref{eq:microcanonical inverse temperature} from finite cutoff holography. Let us consider the dilaton gravity theory \eqref{eq:Dilaton-gravity theory}, where the metric solution to the equations of motions without matter sources is
\begin{equation}\label{eq:metric}
    \rmd s^2=F(\Phi)\rmd\tau+\frac{\rmd\Phi}{F(\Phi)}~,\quad F(x)=\int_{r_h}^x~ U(\Phi)\rmd \Phi~,
\end{equation}
where $r_h$ is the horizon radius. This can be used to evaluate the on-shell action from \eqref{eq:Dilaton-gravity theory}, resulting in
\begin{equation}\label{eq:on shell action}
    I_{\rm E}^{\rm (on)}=-\frac{1}{16\pi G_N}\qty(2\pi(\Phi_0+r_h)+\beta \qty(F(\Phi_{B})-\sqrt{F(\Phi_{B})~G(\Phi_{B})}))~,
\end{equation}
where we have used the fact that $\mathcal{M}$ has a disk topology when $\Phi_0\gg\abs{\Phi}$ everywhere. 
From \eqref{eq:on shell action} we find:
\begin{equation}\label{eq:beta_T E_rB}
    \beta_T E_{\rm BY}=\qty(\beta\sqrt{\frac{F(\Phi_{B})}{G(\Phi_B)}})\qty(\frac{{G(\Phi_B)}}{16\pi G_N}\qty(1-\sqrt{\frac{F(\Phi_{B})}{G(\Phi_{B})}}))~,
\end{equation}
where we identified the inverse the {BY quasi-local energy} and \emph{Tolman temperature} (i.e.~the respective energy and temperature that an observer in a bulk spacetime would associate with a given system at a finite distance $\Phi_B$ from it \cite{Brown:1992br}) as 
\begin{equation}\label{eq:EBY beta_Tolman}
    E_{\rm BY}\equiv\frac{1}{16\pi G_N}\qty(\sqrt{G(\Phi_{B})}-\sqrt{F(\Phi_{B})})~,\quad \beta_{\rm T}\equiv\beta\sqrt{F(\Phi_{B})}~.
\end{equation}
Comparing \eqref{eq:EBY beta_Tolman} with the microcanonical temperature of the deformed theory in (\ref{eq:microcanonical inverse temperature}) due to (\ref{eq:def parameter}), we identify the parameters:
\begin{equation}\label{eq:E_lambda beta_lambda}
    E_y=\sqrt{G(\Phi_{B})}~E_{\rm BY}~,\quad \beta_y=\frac{\beta_{\rm T}}{\sqrt{G(\Phi_{B})}}~.
\end{equation}
For these relations to hold, we have that
\begin{equation}\label{eq:relation lambda kappa}
    \lambda=8\pi G_N~,
\end{equation}
which is consistent with \cite{Blommaert:2025eps}.\footnote{\label{fnt:conventions}See e.g.~(2.32) \cite{Blommaert:2025eps} where $\pi {\rm \textbf{b}}^2=4G_N$ in our notation.}

Thus, we find a holographic dictionary at the level of finite cutoff thermodynamics. In contrast with previous literature, we include a more general counterterm $G(\Phi_{B})$ in the above expressions, which permits the analysis for non-trivial $\eta$ in the flow equations \eqref{eq:flow eq}.

Moreover, it follows from \eqref{eq:E_lambda beta_lambda} that the specific heat in the bulk at a fixed finite radial cutoff can be expressed as
\begin{equation}
    C_{r_B}\equiv-\beta_{\rm T}^2\dv{E_{\rm BY}}{\beta_{\rm T}}=-(\beta_y)^2\dv{E_y}{\beta_y}=C_y~.
\end{equation}
This means that there is a direct relation between the heat capacity in the boundary theory at a fixed deformation parameter $y$ (as well as $\eta$) \eqref{eq:heat capacity} with respect to the heat capacity in the bulk. Similar results were recovered in related settings by \cite{Aguilar-Gutierrez:2024nst}.

\paragraph{Hagedorn growth}
Before closing the section, we make contrast with other results in the literature, we analyze the semiclassical partition function with $\eta=-1$ \eqref{eq:E(theta) eta-1} 
\begin{equation}
    \int_{-\frac{2J}{\lambda}}^{\frac{2J}{\lambda}}\rmd E(\theta)~\rme^{S(\theta)-\beta E^{\eta=-1}_y(\theta)}=\int_{-\frac{2J}{\lambda}}^{\frac{2J}{\lambda}}\rmd E(\theta)~\rme^{S(\theta)-\frac{\beta}{y}\qty(1+\sqrt{2y E(\theta)-1})}~.
\end{equation}
Notably, there no divergence associated to Hagedorn growth for the $T^2+\Lambda_1$ deformed theory, in contrast to other studies \cite{AliAhmad:2025kki}.
The reason that there is no divergence even when $\beta\rightarrow0$ is that the integration is performed over a finite energy range and the factor $\rme^{S(\theta)}$ is finite, thus the model remains UV finite. It was interpreted by \cite{AliAhmad:2025kki}, that one reaches a divergence once the boundary in the dual theory reaches a curvature singularity, which is not present in JT gravity, seen as a s-wave reduction of higher dimensional charged black hole, nor in sine dilaton gravity \cite{Blommaert:2023opb}.

\section{Deforming the DSSYK Model: Chord Basis, Bulk Length \& Complexity}\label{sec:Chord Hilbert}
Following up the last subsection, we specialize in chord Hamiltonian deformations of the DSSYK model. We study the Krylov basis and Krylov spread complexity for the HH state in the deformed theory and show that in the semiclassical limit, it reproduces a wormhole length in finite cutoff holography.\footnote{There is ongoing work by a different group following up \cite{Griguolo:2025kpi}, which has some overlap with the deformed DSSYK chord Hamiltonian in this section from a bulk perspective. I thank the authors for pointing it out.}

\paragraph{Outline} In Sec.~\ref{ssec:review DSSYK} we briefly review the DSSYK model, including its Hilbert space extension with matter chords.  Sec.~\ref{ssec:spread} presents the evaluation of Krylov spread complexity for the HH state from a path integral formulation of the model. It reproduces a wormhole length in finite cutoff holography. We discuss the IR/UV triple-scaling limits of the deformed theory.

In addition, we construct a chord number basis related to finite cutoff holography in  App.~\ref{app:alternative chord number}, as an alternative to the Krylov basis in this section that leads to similar results.

\subsection{Review of the DSSYK model}\label{ssec:review DSSYK}
In this subsection, we briefly review the DSSYK model.

The SYK model \cite{Sachdev_1993,kitaevTalks1,kitaevTalks2,kitaevTalks3,Maldacena:2016hyu} is a system with $N$ Majorana fermions in $(0+1)$-dimensions, and $p$-body all-to-all interactions. The theory is given by
\begin{equation}\label{eq:SYK Hamiltonian}
    \hat{H}_{\rm SYK}=\rmi^{p/2}\sum_{I}J_{I}\hat{\psi}_{I}~,
\end{equation}
where $I=i_1,\dots,~i_p$ is a collective index, $1\leq i_1<\dots<i_p\leq N$; and similarly $\hat{\psi}_I\equiv\hat{\psi}_{i_1}\dots\hat{\psi}_{i_p}$ is a string of Majorana fermions, obeying the Clifford algebra
\begin{equation}\label{eq:Majo fer}
    \qty{\hat{\psi}_{i},~\hat{\psi}_{j}}=2\delta_{ij}\mathbbm{1}~;
\end{equation}
and the coupling constants $J_I=J_{i_1\dots i_p}$ follow a Gaussian distribution
\begin{equation}\label{eq:GEA J}
J_{I}=0~,\quad\expval{J_{I}J_{J}}=\begin{pmatrix}
        N\\
        p
    \end{pmatrix}^{-1}\frac{J^2N}{2p^2}\delta_{IJ}~,
\end{equation}
with $J\in\mathbb{R}$ being a constant, and the expectation values denote annealed averaging  over the couplings. We also introduce matter operators, which have the form:
\begin{align}\label{eq:matter ops DSSYK}
\hat{\mathcal{O}}_\Delta&\equiv\rmi^{\frac{p'}{2}}\sum_{I'}K_{I'}\hat{\psi}_{I'}~.
    \end{align}
Here $I'=i_1,\dots,i_{p'}$; and $K_{I'}$ represents Gaussian random couplings that are independent of $J_{I}$. We also define $\Delta\equiv p'/p$, so that we can express
\begin{equation}\label{eq:ensemble av operators}
        \expval{K_{I}}=0~,\quad \expval{K_{I}J_{I'}}=0~,\quad \expval{(K_{I})^2}=\frac{\mathcal{K}^2}{\Delta^2\lambda}\begin{pmatrix}
        N\\
        p'
    \end{pmatrix}^{-1}~,
    \end{equation}
where $\mathcal{K}\in\mathbb{R}$. In the following, we will set $\mathcal{K}=1$, which can be restored back with dimensional analysis. Importantly for us, once we consider the double scaling regime, where:
\begin{equation}\label{eq:double scaling}
    N,~ p \rightarrow \infty~,\quad q=\rme^{-\lambda} \equiv \rme^{-\frac{2p^2}{N}}~ \text{fixed}~.
\end{equation}
The SYK model becomes analytically solvable even away from the low energy regime; its ensemble averaged Hamiltonian moments can be described with chord diagrams techniques \cite{Berkooz:2018jqr,Berkooz:2018qkz,Cotler:2016fpe,Erdos:2014zgc},\footnote{Other models in the same universality class as the DSSYK model can be solved using the same type of chord counting techniques \cite{Parisi_1994,Berkooz:2024evs,Berkooz:2024ofm,Almheiri:2024xtw,Gao:2024lem}.} whose auxiliary Hilbert space without matter is
\begin{equation}
\mathcal{H}_0={\rm span}\qty{\ket{n}~|~~ n\in\mathbb{Z}_{\geq0}}={\rm L}^2(\theta\in[0,\pi],\rmd\mu(\theta))~,
\end{equation}
where $n$ denotes the number of open chords at a given time slice in the chord diagram (see e.g.~\cite{Berkooz:2024lgq} for a review). From now on we will denote the zero-chord number state as
\begin{equation}\label{eq:Omega state}
    \ket{n=0}\equiv\ket{\Omega}~.
\end{equation}
The auxiliary quantum mechanical theory, described by a chord Hamiltonian $\hH$, contains the thermal information about the physical system, as captured by the Hamiltonian moments
\begin{equation}\label{eq:H chord thermal}
    \expval{\tr(\hH_{\rm SYK}^k)}=\bra{\Omega}\hH^k\ket{\Omega}~,
\end{equation}
where the trace is taken over the Hilbert space states. By an appropriate orthogonal transformation, $\qty{\ket{n}}$ can be orthonormalized (i.e.~$\bra{n}\ket{m}=\delta_{nm}$), such that the Hamiltonian takes a symmetric form (see e.g.~\cite{Berkooz:2018jqr})
\begin{align}
    \hH\ket{n}&=-\frac{J}{\sqrt{\lambda}}\qty(\sqrt{[n]_q}\ket{n-1}+\sqrt{[n+1]_q}\ket{n+1})~,\label{eq:Transfer matrix}
\end{align}
where $[n]_q\equiv \frac{1-q^n}{1-q}$. We can now define the chord number operator $\hat{n}$ and its conjugate momentum, $\hat{p}$:
\begin{equation}\label{eq: Oscillators 1}
    \rme^{\pm\rmi \hat{p}}\ket{n}=\ket{n\mp1}~,\quad \hat{n}\ket{n}=n\ket{n}~,
\end{equation}
which obey the commutation relation:
\begin{equation}\label{eq:chord numb op}
\begin{aligned}
    &\qty[\hat{n},~\rme^{\rmi \hat{p}}]=\rme^{\rmi \hat{p}}~,\quad\qty[\hat{n},~\rme^{-\rmi \hat{p}}]=-\rme^{-\rmi \hat{p}}~.
    \end{aligned}
\end{equation}
The Hamiltonian can be then expressed as
\begin{equation}\label{eq:chord Hamiltonian zero particle}
\begin{aligned}
\hH=&-\frac{J}{\sqrt{\lambda}}\qty(\rme^{\rmi \hat{p}}\sqrt{[\hat{n}]_q}+\sqrt{[\hat{n}]_q}\rme^{-\rmi \hat{p}})~.
\end{aligned}
\end{equation}
One can also find the energy basis $\ket{\theta}$ of $\hH$, whose eigenvalues are given as \cite{Berkooz:2018jqr,Berkooz:2018qkz}
\begin{align}
    &\hH\ket{\theta}=E(\theta)\ket{\theta},\quad E(\theta)\equiv-\frac{2J~\cos(\theta)}{\sqrt{\lambda(1-q)}}~,\quad\theta\in[0,~\pi]~.\label{eq:energy spectrum}
\end{align} 
Solving the eigenvalue problem in \eqref{eq:chord Hamiltonian zero particle} one can then find that the inner product between the energy basis $\qty{\ket{\theta}}$ and the chord number $\qty{\ket{n}}$ is given by
\begin{equation}
    \label{eq: proj E0 theta}
    \bra{\theta}\ket{n}={\frac{H_n(\cos\theta|q)}{\sqrt{(q;q)_n}}}~,
\end{equation}
with $(a;~q)_n$ the q-Pochhammer symbol:
\begin{equation}\label{eq:Pochhammer}
    (a;~q)_n\equiv\prod_{k=0}^{n-1}(1-aq^k)~,\quad (a_0,\dots, a_N;q)_n\equiv\prod_{i=1}^N(a_i;~q)~,
\end{equation}
and $H_n(x|q)$ is the q-Hermite polynomial:
\begin{equation}\label{eq:H_n def}
    H_n(\cos\theta|q)\equiv\sum_{k=0}^n\begin{bmatrix}
        n\\
        k
    \end{bmatrix}_q\rme^{\rmi(n-2k)\theta}~,\quad
    \begin{bmatrix}
        n\\
        k
    \end{bmatrix}_{q}\equiv\frac{(q;~q)_{n}}{(q;~q)_{n-k}(q;~q)_{k}}~.
\end{equation}
The $\ket{\theta}$ basis is normalized such that
\begin{align}
\bra{\theta}\ket{\theta_0}&=\frac{1}{\mu(\theta)}\delta(\theta-\theta_0)~,\quad\mathbb{1}=\int_0^\pi\rmd\theta\mu(\theta)\ket{\theta}\bra{\theta}=\sum_{n=0}^\infty\ket{n}\bra{n}~,\label{eq:norm theta}\\
    \mu(\theta)&=\frac{(q,~\rme^{\pm 2 \rmi \theta};q)_\infty}{2\pi}\equiv \frac{1}{2\pi}(q,~q)_\infty(\rme^{2\rmi\theta},~q)_\infty(\rme^{-2\rmi\theta},~q)_\infty~.\label{eq:identity theta}
\end{align}
For later convenience, we introduce a simple generalization when considering that the DSSYK model with matter insertions \cite{Lin:2022rbf}, which we denote as
\begin{equation}\label{eq:Fock space with matter}
  \mathcal{H}_{\rm full}\equiv\bigoplus_{m=0}^\infty\mathcal{H}_m~,\quad  \mathcal{H}_{m}\equiv{\rm span}\qty{\ket{\tilde{\Delta},n_0,n_1,\cdots,n_m}}~,
\end{equation}
where $\tilde{\Delta}=\qty{\Delta_1,\dots,\Delta_m}$ is a set of conformal dimensions, and $\Delta_i$ characterizes each matter chord operator $\hat{\mathcal{O}}_{\Delta_i}$ which is the ensemble averaged analog of the SYK operator \eqref{eq:matter ops DSSYK}. 

It is also useful to define a total chord number operator
\begin{equation}\label{eq:total chord number basis}
   \hat{N}\ket{\tilde{\Delta},n_0,\cdots,n_m}=\sum_{i=0}^mn_i\ket{\tilde{\Delta},n_0,\cdots,n_m}~,
\end{equation}
which will be applied in later sections to evaluate correlation functions.

Next, we consider the two-sided chord Hamiltonian with $m$ operator insertions:
\begin{subequations}\label{eq:Hmultiple}
    \begin{align}\label{eq:two-sided Hamiltonian}
    &\hH_{L/R}=-\frac{J}{\sqrt{\lambda}}\qty(\hat{a}_{L/R}+\hat{a}^\dagger_{L/R})~,\quad \text{where}\\
&\hat{a}^\dagger_L=\hat{a}^\dagger_0~,\quad \hat{a}_L=\sum_{i=0}^m\hat{\alpha}_i{[\hat{n}_i]_q}q^{\hat{n}_i^<}~,\quad\text{with}\quad\hat{n}_i^<=\sum_{j=0}^{i-1}\qty(\hat{n}_j+\Delta_{j+1})~,\label{eq:aLdagger,aL}\\
&\hat{a}^\dagger_R=\hat{a}^\dagger_m~,\quad\hat{a}_R= \sum_{i=0}^m\hat{\alpha}_i{[\hat{n}_i]_q}q^{\hat{n}_i^>}~,\quad\text{with}\quad\hat{n}_i^>=\sum_{j={i+1}}^{m}\qty(\hat{n}_j+\Delta_{j})~,\label{eq:aRdagger,aR}
\end{align}
\end{subequations}
where the creation and annihilation operators for the chord number sectors \eqref{eq:Hmultiple} in (\ref{eq:Fock space with matter}) act as,
\begin{subequations}\label{eq:Fock Hm}
    \begin{align}\label{eq:Fock Hm 1}
    \hat{a}^\dagger_{i}\ket{\tilde{\Delta};n_0,\dots n_i,\dots, n_m}&=\ket{\tilde{\Delta};n_0,\dots, n_i+1,\dots n_m}~,\\\label{eq:Fock Hm 2}
    \hat{\alpha}_{i}\ket{\tilde{\Delta};n_0,\dots n_i,\dots, n_m}&=\ket{\tilde{\Delta};n_0,\dots, n_i-1,\dots n_m}~.
\end{align}
\end{subequations}
For instance, in terms of the above basis, the one-particle chord Hamiltonian, corresponding to $m=1$ in \eqref{eq:Hmultiple} becomes
\begin{equation}\label{eq:pair DSSYK Hamiltonians 1 particle}
\begin{aligned}
    \hH_{L/R}=\frac{J}{\sqrt{\lambda(1-q)}}\qty(\rme^{-\rmi \hat{P}_{L/R}}+\rme^{\rmi \hat{P}_{L/R}}\qty(1-\rme^{-\hat{\ell}_{L/R}})+q^{\Delta}\rme^{\rmi \hat{P}_{R/L}}\rme^{-\hat{\ell}_{L/R}}\qty(1-\rme^{-\hat{\ell}_{R/L}}))~,
    \end{aligned}
\end{equation}
where we expressed
\begin{equation}\label{eq:non-conjugate ops many particles}
\hat{a}_{i}^\dagger=\frac{\rme^{-\rmi \hat{P}_{i}}}{\sqrt{1-q}}~,\quad \hat{\alpha}_{i}=\sqrt{1-q}\rme^{\rmi \hat{P}_{i}}~,\quad q^{\hat{n}_{i}}=\rme^{-\hat{\ell}_{i}}~, \quad i=L,R~.
\end{equation}
At last, there is an energy basis conjugate to the one-particle chord number basis, diagonalizing the two-sided Hamiltonians \eqref{eq:two-sided Hamiltonian}, i.e.
\begin{equation}
    \hH_{L/R}\ket{\Delta;\theta_{L},\theta_R}=E(\theta_{L/R})\ket{\Delta;\theta_{L},\theta_R}~,
\end{equation}
where $E(\theta)$ appears in \eqref{eq:energy spectrum}. 

For later convenience, when discussing the IR and UV triple-scaling limits of the DSSYK model \cite{Lin:2022rbf,Aguilar-Gutierrez:2025otq}, we zoom in on the edges of the energy spectrum, i.e.
\begin{equation}\label{eq:seed triple scaling limits}
    {\rm IR}:~\theta\sim \mathcal{O}(\lambda)~,\quad {\rm UV}:~\pi-\theta\sim \mathcal{O}(\lambda)~,
\end{equation}
and we define the zero-point energy-subtracted and rescaled chord Hamiltonians in the seed theory,
\begin{subequations}\label{eq:H(dS)JT}
        \begin{align}
   \hH_{\rm IR/UV}&\equiv \eval{\frac{\sqrt{\lambda(1-q)}}{2J}\qty(\pm\hH+\frac{2J}{\sqrt{\lambda(1-q)}})}_{\mathcal{O}(\lambda),{\rm IR/UV}}~,\label{eq:H IR UV}\\
   \label{eq:deformed spectrum IR/UV}
        E_{\rm IR/UV}&\equiv\eval{\pm\frac{\sqrt{\lambda(1-q)}}{J}E(\theta)+1}_{\mathcal{O}(1),IR}=\frac{\theta_{\rm IR/UV}^2}{2}~,
        \end{align}
\end{subequations}
where $\theta_{IR/UV}\sim\mathcal{O}(\lambda)$ in the corresponding triple-scaling limit; the relative $\varepsilon=+1$ sign corresponds to the IR, while $\varepsilon=-1$ for the UV.

We will apply the above definitions to evaluate different properties and observables of the theory under a Hamiltonian deformation in the remainder of the manuscript.

\subsection{Krylov Spread Complexity}\label{ssec:spread}
We now explore the structure of the chord Hilbert space after a generic Hamiltonian deformation
\begin{equation}\label{eq:def Hamiltonian}
    \hH\rightarrow \hH_y\equiv f_y(\hH)\ket{\Omega}~,
\end{equation}
for some analytic function $f_y$.\footnote{For instance, in the particular case of a $T^2$ deformation \cite{Gross:2019ach}
\begin{equation}\label{eq:T^2 def f}
    f_y(x)=\frac{1}{y}\qty(1-\sqrt{1-2yx})~.
\end{equation}}
While the deformed and seed theories have the same Hilbert space, the chord basis can take different forms, and the Hamiltonian \eqref{eq:def Hamiltonian} can have different representations depending on the basis that we choose. For instance, consider the chord number basis where the seed theory Hamiltonian takes the form \eqref{eq:chord Hamiltonian zero particle}. The deformed chord Hamiltonian \eqref{eq:def Hamiltonian} for $y\neq0$ is in general not diagonal in the chord number basis \eqref{eq:chord Hamiltonian zero particle}. Since the Hamiltonian deformation scrambles the chord basis, we search for a meaningful ordering provided by the Krylov basis. This basis allows us to minimize a so-called ``cost function'' associated with a general quantum state, as we explain below based on the seminal work by \cite{Balasubramanian:2022tpr}, which defines Krylov spread complexity. Moreover, recent developments indicate that the Krylov spread complexity in the seed theory DSSYK model can be interpreted in terms of minimal geodesic lengths between asymptotic boundaries in the bulk dual \cite{Rabinovici:2023yex,Lin:2022rbf,Heller:2024ldz}. We search for the corresponding interpretation of Krylov spread complexity in the finite cutoff holographic dictionary.

\paragraph{Cost Function}In general, following \cite{Balasubramanian:2022tpr} and \cite{Erdmenger:2023wjg} we refer to the cost function of an ordered basis $\qty{\ket{B_0},\ket{B_1},\dots,\ket{B_n},\dots}$ with respect to a reference state $\ket{\psi(\tau)}$ as
\begin{equation}\label{eq:cost}
    \mathcal{C}\equiv\sum_{n=0}^\infty c_n\frac{\abs{\bra{B_n}\ket{\psi(\tau)}}^2}{\bra{\psi(\tau)}\ket{\psi(\tau)}}~,
\end{equation}
where $\qty{c_n\in\mathbb{R}}$ is a monotonically increasing sequence, and we allow that the reference state evolves in complex-valued time $\tau$ as \cite{Erdmenger:2023wjg}
\begin{equation}\label{eq:liouv}
    \ket{\psi(\tau)}\equiv\rme^{-\tau\hat{\mathcal{L}}}\ket{\psi(0)}~,
\end{equation}
where $\tau\equiv\frac{\beta}{2}+\rmi t$, with $\beta$ and $t\in {\mathbb{R}}$, leading to standard Schrödinger evolution when $\beta=0$, and $\hat{\mathcal{L}}$ is the corresponding generator of evolution in terms of the parametrization $\tau$. Spread complexity in \cite{Balasubramanian:2022tpr} is defined by minimizing the cost function \eqref{eq:cost} over all possible basis sets, which occurs when $\qty{\ket{B_n}}$ is the Krylov basis, defined below.

\paragraph{Krylov complexity}We now specialize in Krylov spread complexity \cite{Balasubramanian:2022tpr} in deformed DSSYK, so that $\hat{\mathcal{L}}=\hH_y$ in \eqref{eq:liouv}. We seek to build a Krylov basis $\qty{\ket{K_n}}$ with $\ket{K_0}=\ket{\psi(\tau=0)}$ as the initial state in the basis, which we may choose to be $\ket{\Omega}$. The other elements in the Krylov basis are obtained recursively through the Lanczos algorithm,
\begin{subequations}\label{eq:Krylov basis deformed theory}
\begin{align}
    \ket{K^{(y)}_n}&\equiv f_n\qty(\hH_y)\ket{\Omega}~,\\
    x f_n(x)&=b_{n+1}(y) f_{n+1}(x)+b_n(y) f_{n-1}(x)+a_n(y)f_{n}(x)~,\label{eq:lanczos new}
\end{align}
\end{subequations}
which is initialized by $b_0=0$, $f_{0}(x)=1$; and $b_n(y)$ and $a_n(y)$ are the Lanczos coefficients. The deformed chord Hamiltonian in this basis then takes the form
\begin{equation}\label{eq:H Krylov basis}
    \hH_y=\hat{b}_n(y)\rme^{-\rmi\hat{p}_y}+\rme^{-\rmi\hat{p}_y}\hat{b}_n(y)+\hat{a}_n(y)~,
\end{equation}
where the action of the operators in the Krylov basis is defined as
\begin{equation}\label{eq:op relations}
    \rme^{\mp\rmi\hat{p}_y}\ket{K^{(y)}_m}=\ket{K^{(y)}_{m\pm 1}}~,\quad \hat{a}_n(y)\ket{K^{(y)}_m}={a}_m(y)\ket{K^{(y)}_m}~,\quad\hat{b}_n(y)\ket{K^{(y)}_m}={b}_m(y)\ket{K^{(y)}_m}~.
\end{equation}
One can then use the above algorithm to find the explicit coefficients in terms of Hamiltonian moments (see e.g.~\cite{Caputa:2023vyr})
\begin{equation}\label{eq:Lanczos coeff}
\begin{aligned}
    a_0(y)&=m_1~,\quad b_1(y)=m_2-m_1^2~,\quad a_1(y) = \frac{m_3+m_1^3-2m_1m_2}{m_2-m_1^2}~,\\
    b_2(y)&=\frac{m_2m_4-m_2^3 - m^2_3 - m^2_1m_4 - 2m_1^2m_2m_4 }{(m_2-m^2_1)^2}
\end{aligned}
\end{equation}
where
\begin{equation}
    m_n=(-1)^n\eval{\pdv[n]{\mathcal{Z}_y(\beta)}{\beta}}_{\beta=0}~,\quad \mathcal{Z}_y(\beta)\equiv\frac{\bra{\psi(t=0)}\rme^{-\beta\hH_y}\ket{\psi(t=0)}}{\bra{\psi(t=0)}\ket{\psi(t=0)}}~.
\end{equation}
Then, based on \eqref{eq:cost} with $\qty{B_n}=\qty{\ket{K^{(y)}_n}}$ and $c_n=n$, Krylov spread complexity is simply defined as
\begin{equation}\label{eq:spread Complexity}
    \mathcal{C}_{\rm S}=\sum_{n=0}^\infty n\frac{\abs{\bra{K^{(y)}_n}\ket{\psi(t)}}^2}{\bra{\psi(t)}\ket{\psi(t)}}~,\quad \ket{\psi(t)}=\rme^{-(\frac{\beta}{2}+\rmi t)\hH_y}\ket{\Omega}~.
    \end{equation}
\paragraph{Semiclassical Approximation}
To derive the explicit form of the Krylov basis from the Lanczos coefficients \eqref{eq:Lanczos coeff}, we search for simplifications in the semiclassical limit,
\begin{equation}\label{eq:rescaled chord number krylov basis}
 \ell_y\equiv\begin{cases}
     \lambda \mathcal{C}_{\rm S}~,&  0\leq y\leq y_0=\frac{1}{2E(\theta)}~,\\
     -\rmi  \lambda \mathcal{C}_{\rm S}~,& y\geq y_0~,
 \end{cases}\quad \text{fixed as~}\lambda\rightarrow0~,
\end{equation}
which corresponds to $n\sim\mathcal{O}(1/\lambda)$, with $n$ being the order in the Lanczos basis $\ket{K_n}$, as the dominant contribution in the evaluation of Krylov complexity, which will allow us to evaluate $a_n$ and $b_n$ for $n\gg 1$.

In this limit, the Hamiltonian on the Krylov basis becomes \eqref{eq:lanczos new}
\begin{equation}\label{eq:def H Krylov basis cont limit}
    H_y={2}b(\ell_y)\cos{p_y}+a(\ell_y)~,
\end{equation}
where $p_y$ is the conjugate momentum to $\ell_y$ in the $\lambda\rightarrow0$ limit. This allows us to express the path integral of the theory as
\begin{equation}
   \int[\rmd \ell_y][\rmd p_y]\exp[\int\rmd\tau\qty( \rmi p_y\partial_\tau\ell_y-H_y)]~.
\end{equation}
This path integral prepares the HH state of the deformed theory $\rme^{-\tau\hH_y}\ket{\Omega}$, with $\tau=\frac{\beta}{2}+\rmi t$, so that the Krylov basis is built with $\ket{K_0}=\ket{\Omega}$.

We require that the measure of time scales as $t\sim\mathcal{O}(1/\lambda)$ to perform the saddle point approximation since the energy spectrum of the Hamiltonian scales with $y\propto\lambda^{-1}$. Then in the $\lambda\rightarrow0$ limit, one then finds the equations of motion
\begin{subequations}\label{eq:EOM krylov basis def}
\begin{align}
    \dv{\ell_y}{t}&=-2\sin p_y~b(\ell_y),\qquad -\dv{p_y}{t}=2\cos {p_y}\dv{b}{\ell_y}+\dv{a}{\ell_y}~,\\
   \frac{1}{2} \dv[2]{\ell_y}{t}&=\dv{\ell_y}\qty(b(\ell_y)^2)+\dv{a}{\ell_y}{b(\ell_y)\cos p_y=\dv{\ell_y}\qty(b(\ell_y)^2)-\frac{1}{2}\dv{a}{\ell_y}\qty(E_y-a(\ell_y))~,}\label{eq:2nd ODE}
\end{align}
\end{subequations}
where {we applied conservation of the deformed Hamiltonian $H_y$, with an energy spectrum
\begin{equation}\label{eq:spectrum deformed}
    E_y\equiv\begin{cases}
    \frac{1}{y}(1-\sqrt{1-2y E(\theta)})~,&0\leq y\leq y_0=\frac{1}{2E(\theta)}~,\\
    \frac{1}{y}(1+\sqrt{2y E(\theta)-1})~,&y\geq y_0~,
    \end{cases}
\end{equation}
in the last relation}. The equations of motion \eqref{eq:EOM krylov basis def} are supplemented with initial conditions
\begin{equation}\label{eq:new init cond}
    \ell_y(t=0)=\ell_*~,\quad \eval{\dv{\ell_y}{t}}_{t=0}=0~,
\end{equation}
where $\ell_*$ is a constant, and by considering that the expectation values of the chord number are taken with respect to the HH state of the deformed theory $\rme^{-\tau\hH_y}\ket{\Omega}$, the second condition follows from the h
Here, we expanded the chord number $\hat{n}_y$ in the Krylov basis and defined 
\begin{equation}
 \psi_n(\beta,y)\equiv\bra{\Omega}\rme^{-\frac{\beta}{2}\hH_y}\ket{K_n^{(y)}}~,\quad    \psi'_n(\beta,y)=\rmd \psi_n(\beta,y)/\rmd \beta~,
\end{equation}
while assuming that $\hH_y$ is Hermitian. Thus, we recover the second initial condition in \eqref{eq:new init cond}.

In particular, from the second initial condition in \eqref{eq:new init cond} we recover that $p_y(t=0)=0$, and thus, by conservation of the energy spectrum of the deformed Hamiltonian represented in \eqref{eq:def H Krylov basis cont limit}, we have
\begin{equation}
    E_y=\frac{1}{y}\qty(1+\varepsilon\sqrt{\eta(1-2y E(\theta))})=2b(\ell_*)+a(\ell_*)~.
\end{equation}
where $\eta=\pm$ and the $\varepsilon=\pm1$ in front of the square root indicate the perturbative, non-perturbative solutions in $T^2(+\Lambda_1)$ deformations. Then, to reproduce the conserved energy spectrum of the deformed theory we make an ansatz such that \eqref{eq:2nd ODE} takes the form
\begin{equation}\label{eq:2nd order Ly ODE}
    \dv[2]{\ell_y}{t_y}=\frac{1}{y^2}\tanh(\frac{\ell_y}{2})\sech^2\qty(\frac{\ell_y}{2})~,
\end{equation}
which means that in this ansatz
\begin{equation}\label{eq:a b cont}
    \dv{a}{\ell_y}=0~,\quad \dv{b(\ell_y)^2}{\ell_y}=\frac{1}{2y^2}\dv{\ell_y}\qty(\tanh^2(\ell_y/2)-1)~,
\end{equation}
where we implemented a rescaling in time parameter of \eqref{eq:2nd order Ly ODE} with $R(y)t\rightarrow 2y t$ such that the deformed Hamiltonian has the energy spectrum in \eqref{eq:spectrum deformed}.

Using the initial conditions \eqref{eq:new init cond}, one finds that the solutions of \eqref{eq:a b cont} are
\begin{equation}
a(\ell_y)=\frac{1}{y}~,\quad b(\ell_y)=\pm \frac{1}{2y}\tanh(\ell_y/2)~,
\end{equation}
where $\pm$ corresponds to the 
perturbative and non-perturbative  branches of the $T^2$ and $T^2+\Lambda_1$ deformations, which were discussed in Sec.~\ref{ssec:sol flow eq}. Thus, specializing to $T^2(+\Lambda_1)$ deformations, we find that
\begin{equation}
\begin{aligned}
    y\in[0,y_0]:~\sinh(\frac{\ell_*}{2})&=\sqrt{\frac{1}{2yE(\theta)}-1}~,\\
    y\in[y_0,\infty]:~\sin(\frac{\rmi\ell_*}{2})&=\sqrt{1-\frac{1}{2yE(\theta)}}~.
\end{aligned}
\end{equation}
The deformed chord Hamiltonian in the continuum limit \eqref{eq:def H Krylov basis cont limit} then becomes
\begin{equation}\label{eq:classical H}
    H_y=\begin{cases}
    \frac{1}{y}\qty(1-\tanh(\frac{\ell_y}{2})\cos(p_y))~,&0\leq y\leq y_0=\frac{1}{2E(\theta)}~,\\
    \frac{1}{y}\qty(1+\tan(\frac{{\ell}_y}{2})\cos({p}_y))~,&y\geq y_0~.
    \end{cases}
\end{equation}
From \eqref{eq:2nd ODE} and \eqref{eq:new init cond} we recover the Krylov spread complexity \eqref{eq:spread Complexity} of the HH state in the deformed theory as:
\begin{subequations}\label{eq:Krylov complexity deformed DSSYK}
    \begin{align}
        \mathcal{C}_{\rm S}=
        \begin{cases}
    \frac{\ell_y}{\lambda}=\frac{2}{\lambda}{\rm arcsinh}\qty(\sqrt{\frac{1}{2yE(\theta)}-1}~\cosh(\frac{\sin\theta ~\tilde{t}}{2}))~,& 0\leq y\leq y_0=\frac{1}{2E(\theta)}~,\\
    \frac{\ell_y}{\rmi\lambda}=\frac{2}{\lambda}{\rm arcsin}\qty(\sqrt{1-\frac{1}{2yE(\theta)}}~\cosh(\frac{\sin\theta ~\tilde{t}}{2}))~,& y\geq y_0~,
        \end{cases}
    \end{align}
\end{subequations}
where $\tilde{t}\equiv\csc\theta \sqrt{2E(\theta)/y}~t$. This rescaled time is convenient for comparison with the geodesic length between the asymptotic boundaries in the bulk (see App.~\ref{app:geodesics}).

\paragraph{Interpretation}
From (\ref{eq:H Krylov basis}, \ref{eq:classical H}) it follows that the chord Hamiltonian when $n\sim \mathcal{O}(1/\lambda)$ can be expressed in the Krylov basis as,
\begin{subequations}
    \begin{align}
        \hH_y&=\frac{1}{y}\qty(1\pm\frac{1}{2}\qty(\tanh\qty(\frac{\lambda \hat{n}_y}{2})\rme^{-\rmi\hat{p}_y}+\rme^{\rmi\hat{p}_y}\tanh\qty(\frac{\lambda \hat{n}_y}{2})))~,\label{eq:def H Krylov basis}\\
        \hat{n}_y&\equiv\sum_nn\ket{K^{(y)}_n}\bra{K^{(y)}_n}~,\label{eq:n_y}
    \end{align}
\end{subequations}
where the action of the above operators in the Krylov basis was specified in \eqref{eq:op relations}, and the $\pm$ signs correspond to the cases in \eqref{eq:classical H}. Crucially, while we used an ansatz to solve the differential equation for the evolution of the chord number in the semiclassical limit, the specific form of the Hamiltonian is determined from the conserved energy spectrum in the deformed theory $E_y$ such that the deformed Hamiltonian in \eqref{eq:def H Krylov basis} is tridiagonal and symmetric. Thus, the resulting Hamiltonian is expressed in the (unique) Krylov basis with the corresponding Lanczos coefficients,\footnote{Given that $a_n=$constant at leading order when $\lambda\rightarrow0$, it means that the weight $w(E_y-a_n)$ in the integration measure $\rmd\mu(E_y)\equiv\rmd E_y~w(E_y)$ obeys the property $w(E_y-a_n)=w(-(E_y-a_n))$ \cite{ismail2005classical} at leading order in a $\lambda\rightarrow0$ expansion.}
\begin{equation}
   \begin{aligned}
       \lambda\rightarrow0~,\quad n\sim\mathcal{O}(\lambda^{-1}):~&\qty{\eval{\ket{K^{(y)}_n}}\quad\bra{K^{(y)}_m}\ket{K^{(y)}_n}=\delta_{nm}~,\quad \ket{K^{(y)}_0}=\ket{\Omega}}~,\\
       &a_n(y)=y^{-1}~,\quad b_n(y)=\pm \frac{1}{2y}\tanh(\frac{\lambda n}{2})~.
   \end{aligned}
\end{equation}
Therefore, the relation between Krylov spread complexity and the wormhole length (i.e.~the extremal codimension-one volume) in the bulk \cite{Rabinovici:2023yex,Heller:2024ldz} still holds, corresponding to an explicit manifestation of the CV conjecture \cite{Susskind:2014rva,Stanford:2014jda,Susskind:2014jwa} in finite cutoff holography \cite{FarajiAstaneh:2024fpv}, as expected by \cite{Griguolo:2025kpi}.

\begin{figure}
    \centering
    \subfloat[]{\includegraphics[width=0.5\linewidth]{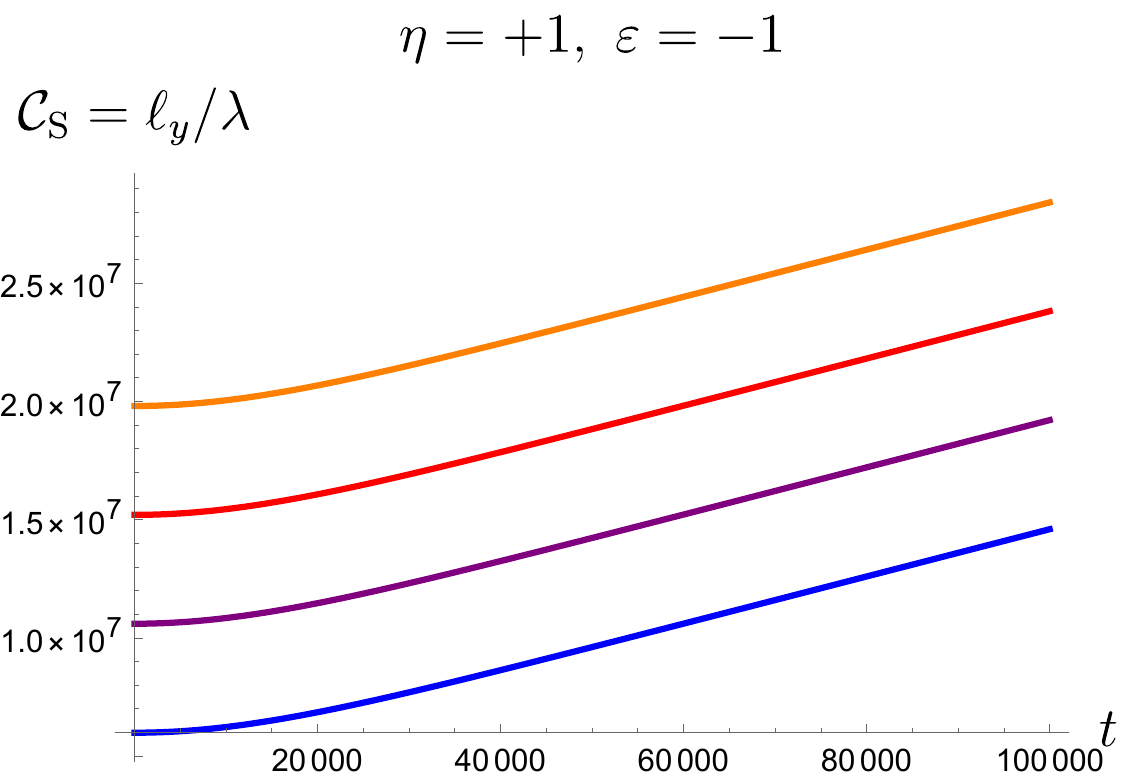}} \subfloat[]{\includegraphics[width=0.5\linewidth]{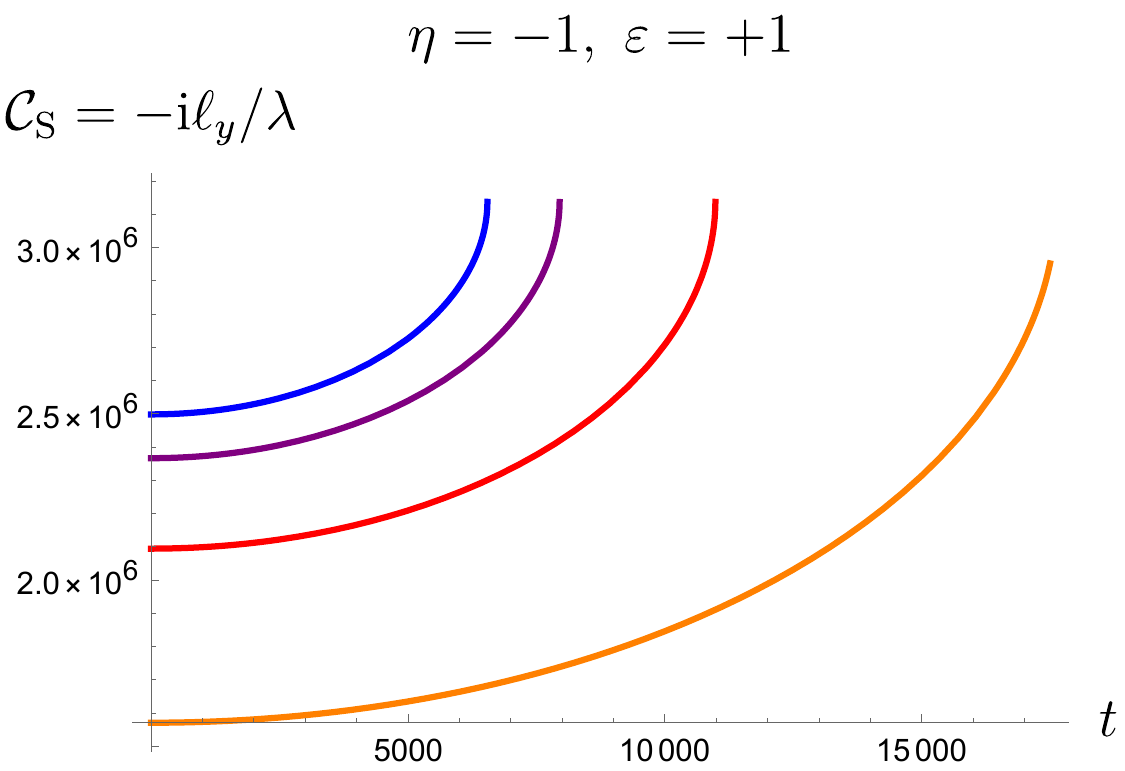}}\\
    \subfloat[]{\includegraphics[width=0.42\linewidth]{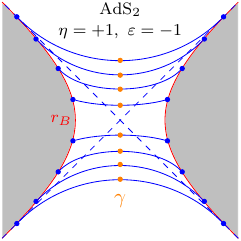}}  \hspace{1.5cm}  \subfloat[]{\includegraphics[width=0.42\linewidth]{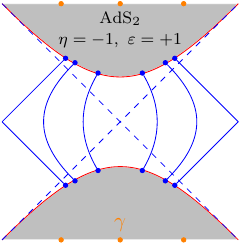}}
    \caption{\textit{Top}: Krylov spread complexity of the HH state in the IR limit \eqref{eq:Krylov complexity deformed DSSYK} after (a) a $T^2$ ($\eta=+1$) and (b) $T^2+\Lambda_1$ ($\eta=-1$) deformation, with $\theta_{\rm IR}=10^{-4}$. \textit{Bottom}: Corresponding bulk interpretation in terms of (c) spacelike and (d) timelike extremal Einstein-Rosen bridges (blue solid curves) connecting the finite Dirichlet boundaries (red solid curve), located at a constant $r=r_B$ in the Rindler-AdS$_2$ coordinates \eqref{eq:netric more gen}). We marked the minimal extremal codimension-two area (measured by the dilaton) surfaces $\gamma$ (orange dots) subject to the homology constraint to the entangling surface (blue dots at the Dirichlet boundaries). Increasing the deformation parameter $y$ \eqref{eq:def parameter} enhances the rate of growth of the Krylov spread complexity for the $\eta=+1$ case. In finite cutoff holography, this moves the location of the boundaries towards the corresponding black hole horizon, which increases the Tolman temperature, resulting in an enhancement of the growth of the length with respect to the boundary time. Meanwhile, for the $\eta=-1$ case, increasing $y$ moves the boundary cutoff away from the horizon in the bulk, leading to a decrease in the Tolman temperature. However, there is no eternal growth unlike the $\eta=+1$ case, since the timelike geodesic eventually reaches $r\rightarrow\infty$ in Rindler-AdS$_2$ space \eqref{eq:AdS blackening} as one increases the spacelike coordinate $t$.}
    \label{fig:spread_time}
\end{figure}

\paragraph{Triple-Scaling Limit}In addition, we observe from the $\varepsilon=-1$ case in \eqref{eq:classical H} that the deformed chord Hamiltonian takes the same form as the canonically quantized ADM Hamiltonian of JT gravity at a finite cutoff found by \cite{Griguolo:2021wgy} (7.6) where the parameter $r_b$ in the JT gravity case corresponds to $y$ in the boundary theory \eqref{eq:classical H}. This implies that we can define the IR/UV triple-scaling limits to recover the generators of time and spatial translations along the finite cutoff boundaries in (dS) JT gravity by simply focusing on the energy spectrum of the seed theory \eqref{eq:seed triple scaling limits}:
\begin{equation}\label{eq:Explicit Krylov basis deformed DSSYK}
    {\rm IR}:~\theta\equiv \theta_{\rm IR}\ll1~,\quad {\rm UV}:~\pi-\theta\equiv \theta_{\rm UV}\ll1~.
\end{equation}
Note that the triple-scaling limit in the deformed theory shows some sharp differences with respect to the one in the seed theory limit ($y=0$), where one instead defines a regularized length operator $\lambda\hat{n}+2\log{\lambda}$ \cite{Lin:2022rbf}. Meanwhile in \eqref{eq:classical H} the finite cutoff in the bulk dual implies that the corresponding lengths in the bulk do not need additional regularization. In contrast, \cite{Griguolo:2025kpi} mentions that the canonically quantized Hamiltonian in JT gravity \eqref{eq:classical H} may be isometrically dual to the Hamiltonian in a matrix model. Thus, our work contains a different interpretation, where (dS) JT gravity at the disk level is holographically described by the IR/UV tails of DSSYK model. In particular, while the bulk length is spacelike or time-like depending on whether the cutoff boundaries are located outside or inside the black hole or the static patch, the Krylov spread complexity \eqref{eq:Krylov complexity deformed DSSYK} with the corresponding substitution \eqref{eq:Explicit Krylov basis deformed DSSYK} for each limit remains real-valued. We display the growth of Krylov spread complexity in the IR limit and its bulk interpretation in Fig.~\ref{fig:spread_time}.

Furthermore, in this limit one can simplify the critical value of the deformation parameter $y$ \eqref{eq:def parameter} $y_0$ in \eqref{eq:transition TTbar+Lambda1 DSSYK} where the spectrum of the Hamiltonians corresponds to $\theta_{\rm IR/UV}^2/2$. That is,
\begin{equation}\label{eq:y0}
    {\rm IR}:~y_0={\theta_{\rm IR}}^{-2}~,\quad {\rm UV}:~y_0={\theta_{\rm UV}}^{-2}~.
\end{equation}
In particular, the UV case with the $T^2+\Lambda_1$ deformation describes the static patch of dS$_2$, realizing the cosmological stretched horizon proposal by Susskind \cite{Susskind:2021esx} (detailed in Sec.~\ref{sec:dS holography}).

\section{Correlation Functions}\label{sec:correlators}
In this section, we study $n$-point correlation functions of the deformed DSSYK model. They are directly defined in the chord theory with Hamiltonian deformations. We evaluate them in the semiclassical limit. Consistent with the literature, the semiclassical results are modified only by a redshift factor associated with the Tolman temperature in the bulk. 

\paragraph{Outline}In Sec.~\ref{ssec:n pnt correlation} we define $n$-point correlation functions in the chord theory after the deformation. In Sec.~\ref{sec:correlation functions} we perform the semiclassical evaluation of the correlation functions from a path integral argument, which leads to the expected redshift parameter. At last, in Sec.~\ref{eq:TSL G4} we discuss the triple-scaling limits of the expressions.

\subsection{n-Point Correlation Functions}\label{ssec:n pnt correlation}
In the following, we generalize the original flow equation \eqref{eq:flow eq} for the chord Hamiltonian with matter \eqref{eq:two-sided Hamiltonian},
\begin{equation}\label{eq:flow eq with matter}
    \partial_y\hH_{L/R,\,y}=\frac{\hH^2_{L/R,\,y}+(\eta-1)/y^2}{2(1-y \hH_{L/R,\,y})}~.
\end{equation}
The holographic dictionary entry relating the deformation parameter $y$ with a finite Dirichlet cutoff in the bulk, such as in \eqref{eq:def parameter}, may be modified since we need to account for matter in the dilaton gravity action \eqref{eq:Dilaton-gravity theory}. In this case, finite cutoff holography is less understood in general \cite{Hartman:2018tkw}; and one in principle may use a different holographic interpretation of \eqref{eq:flow eq with matter} related to mixed boundary conditions \cite{Guica:2019nzm}. Regardless of the bulk interpretation, we will study the solutions of \eqref{eq:flow eq with matter} as a deformed chord theory by itself.

While correlation functions are originally defined by taking the double-scaling limit of operators in the physical SYK model, we work directly in chord space since we are interested in evaluating correlation functions in the ensemble averaged theory. For a simple illustration, we define a (normalized) two-point function by simply deforming the chord Hamiltonian, while keeping the matter operators unchanged
\begin{equation}\label{eq:corr one particle}
    G^\Delta\equiv Z_y(\beta)^{-1}\bra{\Omega}\hmO_\Delta\rme^{-\tau^*\hH_{L,y}}\rme^{-\tau\hH_{L,y}}\hmO_\Delta\ket{\Omega}~,
\end{equation}
where $\hmO_\Delta$ represents the chord matter operator $\hmO_\Delta^{L}$ \eqref{eq:Fock space with matter}, and $\tau\equiv \frac{\beta}{2}+\rmi t$.

Note that the above definition has the same structure as the two-point function of the seed theory \cite{Berkooz:2018jqr}, where we simply replace the corresponding Hamiltonians as $\hH_{L/R}\rightarrow \hH_{L/R,y}$.

To evaluate the correlation function in the one-particle space \eqref{eq:corr one particle} we can use the resolution of the identity
\begin{equation}\label{eq:norm theta 1p}
    \mathbb{1}=\int\rmd\theta_L\rmd\theta_R~\mu(\theta_L)\mu(\theta_R)\hat{\mathcal{F}}_\Delta^\dagger\qty(\ket{\theta_L}\otimes\ket{\theta_R})\qty(\bra{\theta_L}\otimes\bra{\theta_R})\hat{\mathcal{F}}_\Delta~,
\end{equation}
where we introduce a chord intertwiner \cite{Aguilar-Gutierrez:2025mxf,vanderHeijden:2025zkr}:
\begin{equation}\label{eq:isometric factorization}
    \hat{\mathcal{F}}_\Delta\ket{\Delta;\theta_L,\theta_R}=\sqrt{\bra{\theta_L}q^{\Delta\hat{n}}\ket{\theta_R}}\ket{\theta_L}\otimes\ket{\theta_R}~.
\end{equation}
Then, \eqref{eq:corr one particle} becomes
\begin{equation}\label{eq:two-point function}
\begin{aligned}
    G^\Delta&=Z_y(\beta)^{-1}\prod_{i=1}^2\int\rmd\theta_i\mu(\theta_i)\rme^{-\tau^*E_y(\theta_1)-\tau E_y(\theta_2)}\bra{\theta_1}q^{\Delta\hat{n}}\ket{\theta_2}\\
    &=Z_y(\beta)^{-1}{\bra{\Omega}\rme^{-\tau^*\hH_y}q^{\Delta\hat{n}}\rme^{-\tau\hH_y}\ket{\Omega}}~.
\end{aligned}
\end{equation}
Therefore, since the chord number in the original chord basis without deformation $\hat{n}=\sum_{n=0}^\infty n\ket{n}\bra{n}$ appears in the above expressions, we can perform the evaluation of the correlation functions with path integral methods using the representation of the Hamiltonian  in the original basis. This implies that correlation functions in the deformed theory defined above might not capture the behavior of geodesic lengths in finite cutoff holography.  As we will see, this is consistent with the commonly employed definitions of correlation function in the literature on $T^2$ deformations and its extensions.\footnote{It might be interesting to use alternative definitions for the correlation functions, for instance, by promoting the chord number $\hat{n}$ in \eqref{eq:two-point function} to $\hat{n}_y$ \eqref{eq:n_y}.}

Next, applying \eqref{eq:two-point function}, it follows that higher-point functions can be expressed \cite{Lin:2022rbf,Lin:2023trc}
\begin{equation}\label{eq:G2m}
\begin{aligned}
    &{\bra{\Omega}\rme^{-\tau_m^*\hH_{L,y}}\cdots\hmO_\Delta\rme^{-\tau_2^*\hH_{L,y}}\hmO_\Delta\rme^{-\tau_1^*\hH_{L,y}-\tau_1\hH_{R,y}}\hmO_\Delta\rme^{-\tau_2\hH_{L,y}}\hmO_\Delta\cdots \rme^{-\tau_m\hH_{L,y}}\ket{\Omega}}\\
    &={\bra{\Omega}\rme^{-\tau_m^*\hH_{L,y}}\cdots\hmO_\Delta\rme^{-(\tau_1+\tau_2)^*\hH_{L,y}}q^{\Delta\hat{N}}\rme^{-(\tau_1+\tau_2)\hH_{L,y}}\hmO_\Delta\cdots \rme^{-\tau_m\hH_{L,y}}\ket{\Omega}}
\end{aligned}
\end{equation}
where
\begin{equation}\label{eq:taui}
    \tau_i=\frac{\beta_i}{2m}+\rmi t_i~,\quad t_i\in\mathbb{R}~,\quad 2\sum_{i=1}^m\beta_i=\beta~,
\end{equation}
with $\beta$ being the periodicity of the thermal circle, and $\hat{N}$ is the total chord number operator \eqref{eq:total chord number basis}.

\subsection{Semiclassical Evaluation}\label{sec:correlation functions}
Having found that correlation functions in the deformed theory can be treated in terms of the chord number in the Krylov basis, we will now proceed to evaluate them in the semiclassical limit \cite{Lin:2022rbf}, i.e.~$\lambda n$ fixed as $\lambda\rightarrow0$.

\begin{figure}
    \centering
    \includegraphics[width=0.43\linewidth]{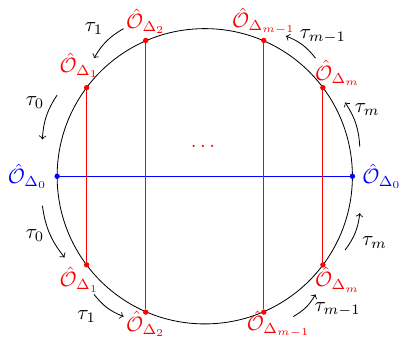}
    \caption{$(2m+2)$-correlation function in \eqref{eq:G m operators} with $\hmO_{\Delta_0}$ (blue) being light operators and $\hmO_{\Delta_{1\leq i\leq m}}$ (red) at different locations, $\tau_{1\leq i\leq m}$, within the thermal circle.}
    \label{fig:m_correlation}
\end{figure}

Based on \eqref{eq:G2m} the $(2m+2)$-correlation function of interest is,\footnote{In the bulk, this correlation function is associated with a bulk propagator being that of a probe massive scalar coupled in the effective geometry as \cite{Blommaert:2024ymv,Bossi:2024ffa},\footnote{While in this section we take a boundary approach to the deformation, more generally, there are issues with introducing matter in finite cutoff holography \cite{Gross:2019ach,Kraus:2018xrn,Hartman:2018tkw} which require adding the contribution of the matter stress tensor in the flow equations by hand.} 
\begin{equation}
    I_{\rm matter}=\int\rmd^2x\sqrt{g_{\rm eff}}\qty(g_{\rm eff}^{\mu\nu}\partial_\mu\tilde{\phi}\partial_\nu\tilde{\phi}+m^2\tilde{\phi}^2)~,
\end{equation}
$\beta_y(\theta=\pi)\rightarrow\infty~.$ It would be interesting to do an explicit matching between the finite cutoff sine dilaton gravity with the $\TT$ deformed DSSYK model; or more general multitrace deformations for the DSSYK model corresponding to other modifications in the boundary conditions in the bulk.}
\begin{subequations}
    \begin{align}\label{eq:G m operators}
        &G_{\tilde{\Delta}}^{\Delta_0}=\frac{\bra{\Psi^y_{\tilde{\Delta}}(\tau_0,\tau_1,\dots,\tau_m)}q^{\Delta_0\hat{N}}\ket{\Psi^y_{\tilde{\Delta}}(\tau_0,\tau_1,\dots,\tau_m)}}{\bra{\Psi^y_{\tilde{\Delta}}(\tau_0,\tau_1,\dots,\tau_m)}\ket{\Psi^y_{\tilde{\Delta}}(\tau_0,\tau_1,\dots,\tau_m)}}~,\\
        &\ket{\Psi_{\tilde{\Delta}}(\tau_0,\tau_1,\dots,\tau_m)}\equiv\rme^{-\tau_m\hH_{L,y}}\hmO_{\Delta_m}\cdots\rme^{-\tau_1\hH_{L,y}}\hmO_{\Delta_1}\rme^{-\tau_0\hH_{L,y}}\ket{\Omega}~,\label{eq:HH state tL tR}
    \end{align}
\end{subequations}
with $\hat{N}$ in \eqref{eq:total chord number basis}, $\tau_i\in\mathbb{C}$ and $\sum_{i=0}^m\tau_i=\beta$, and $\tilde{\Delta}=\qty{\Delta_1,\dots,\Delta_m}$. Meanwhile, $\hmO_{\Delta_1},\dots,\hmO_{\Delta_m}$ are general matter chord operators, which may be light (i.e.~$\Delta_i\mathcal{O}(1)$ as $\lambda\rightarrow0$) or heavy (i.e.~$\lambda\Delta_{1\leq i\leq m}\sim\mathcal{O}(1)$ as $\lambda\rightarrow0$). We display the representation of the correlation function \eqref{eq:G m operators} in Fig.~\ref{fig:m_correlation} (based on \cite{Aguilar-Gutierrez:2025mxf}),

To evaluate the general correlation function \eqref{eq:G m operators}, we are interested in the semiclassical regime where
\begin{equation}\label{eq:semiclassical exp}
    \expval{f\qty(\hat{N})}\eqlambda f\qty(\expval{\hat{N}})=f\qty(\frac{\ell}{\lambda})~,\quad \ell\equiv\sum_{i=0}^m\ell_i~.
\end{equation}
Here, we denote $f$ as a generic function; the expectation value is taken in a generic state (which in our case of interest corresponds to \eqref{eq:G m operators}); and at last we defined 
\begin{equation}\label{eq:exp n}
  \ell_{0\leq i\leq m}\equiv \lambda\expval{\hat{n}_{i}}  ~,
\end{equation}
as expectation values of the rescaled chord number for each chord subsector, defined in \eqref{eq:total chord number basis}; and note that the quantum fluctuations for the parameter $\hat{N}$ are suppressed in the $\lambda\rightarrow0$ limit.\footnote{This argument can also be used to evaluate a generalized notion of Krylov complexity \cite{Balasubramanian:2022tpr}
\begin{equation}\label{eq:more gen complexity}
\tilde{\mathcal{C}}\equiv\frac{\bra{\psi(t)}f(\hat{n})\ket{\psi(t)}}{\braket{\psi(t)}{\psi(t)}}=\sum_{n} f(n) \frac{\abs{\bra{n}\ket{\psi(t)}}^2}{\braket{\psi(t)}{\psi(t)}}~.
\end{equation}
where $f(n)=c_n$ in the notation of the cost function \eqref{eq:cost}; $\hat{n}$ the Krylov complexity operator, which can be extended with operator insertions \cite{Aguilar-Gutierrez:2025pqp,Aguilar-Gutierrez:2025mxf,Ambrosini:2024sre,Ambrosini:2025hvo}. In the semiclassical limit, quantum fluctuations are suppressed \eqref{eq:semiclassical exp} and \eqref{eq:more gen complexity} becomes
\begin{equation}\label{eq:more general C}
\tilde{\mathcal{C}}=f(\ell/\lambda)~.
\end{equation}
Holographically \eqref{eq:more general C} corresponds to a general function of the wormhole length.}

We now evaluate the expectation value of the length from the path integral of the DSSYK model with matter as
\begin{equation}\label{eq:Hamilton PI}
    \int\prod_{i=0}^m[\rmd \ell_i][\rmd P_i]\exp\qty[\int\rmd\tau_L\rmd\tau_R\qty(\frac{\rmi}{\lambda}\sum_{i=0}^m\qty(P_i(\partial_{\tau_L}+\partial_{\tau_R})\ell_i)+H_{L,~y}+H_{R,~y})]~,
\end{equation}
where $H_{L/R,~y}$ are the deformed chord Hamiltonians with one particle insertion \eqref{eq:pair DSSYK Hamiltonians 1 particle} $m=1$
\begin{subequations}
\begin{align}
     &\qquad\qquad H_{L/R,y}\equiv f_y(H_{L/R})~,\label{eq:f_y def}\\ H_{L/R}=\frac{J}{\sqrt{\lambda(1-q)}}&\qty(\rme^{-\rmi P_{L/R}}+\rme^{\rmi P_{L/R}}\qty(1-\rme^{-{\ell}_{L/R}})+q^{\Delta}\rme^{\rmi {P}_{R/L}}\rme^{-{\ell}_{L/R}}\qty(1-\rme^{-{\ell}_{R/L}}))~.
\end{align}
\end{subequations}
In the $\lambda\rightarrow0$ limit, we evaluate the path integral through saddle points that split into the left/right chord sectors
\begin{align}\label{eq:evol many}
\frac{1}{\lambda}\pdv{\ell}{t_{L/R}}=\sum_{i=0}^m\pdv{H_{L/R,~y}}{P_{i}}~,\qquad
\frac{1}{\lambda}\pdv{P_{i}}{t_{L/R}}=-\pdv{H_{L/R,~y}}{\ell_{i}}~,
\end{align}
given that $\hH_{L/R,~y}$ are the translation generators in $t_{L/R}$.

Using \eqref{eq:f_y def} it follows that \eqref{eq:evol many} can be expressed as
\begin{align}\label{eq:EOM new}
\frac{1}{\lambda}\pdv{\ell}{t_{L/R}}=f'_y(\hH_{L/R})\sum_{i=0}^m\pdv{H_{L/R}}{P_{i}}~,\qquad
\frac{1}{\lambda}\pdv{P_{i}}{t_{L/R}}=-f'_y(\hH_{L/R})\pdv{H_{L/R}}{\ell_{i}}~,
\end{align}
It is clear from this derivation of the saddle points of the path integral \eqref{eq:Hamilton PI} that $f'_y(\hH_{L/R})$ are just overall constant factors since the two-sided Hamiltonian $H_{L/R}$ is conserved \eqref{eq:energy spectrum}. This means that the correlation function \eqref{eq:G m operators} for the deformed theory is the same the seed correlation function after rescaling the $t_{L/R}\rightarrow f'_y(E_{L/R})t_{L/R}$. Given that the temperature is the only energy scale in the correlation functions, the effect of the deformation is equivalent to rescaling the temperature by an appropriate redshift factor. This means 
\begin{equation}\label{eq:resclaing time temp}
\begin{aligned}
\frac{t}{\beta}\rightarrow\frac{t}{\beta_y}~,\quad\text{when}\quad t\rightarrow f'_y(E)t~,
\end{aligned}
\end{equation}
where $\beta_y=f'_y(E)\beta$ \eqref{eq:microcanonical inverse temperature} as we defined in the semiclassical thermodynamic analysis in Sec.~\ref{sec:thermo}. This means that the rate of growth is modified by the Hamiltonian deformation. A similar finding was reported in T$\overline{\text{T}}$ deformations to study OTOCs in the shockwave BTZ black hole geometry \cite{Basu:2025exh,AliAhmad:2025kki}. We should note that \eqref{eq:resclaing time temp} is consistent with finite cutoff holography \cite{McGough:2016lol}, given that the factor $\beta\rightarrow\beta_y$ corresponds to a change in the Tolman temperature in the bulk, as we discussed at the end of Sec.~\ref{sec:thermo}.

Using the above results we can immediately evaluate all the correlation functions of the type \eqref{eq:Hamilton PI} in the deformed theory based on those in the seed theory. This follows from solving $\ell$ based on the specific equations of motion in \eqref{eq:EOM new} with appropriate boundary conditions and then applying \eqref{eq:semiclassical exp} to evaluate correlation functions, as we specify below.

\paragraph{Two-Point Functions}Consider \eqref{eq:G m operators} for $m=1$ with an analytic continuation in the parameter 
\begin{equation}\label{eq:analytic continuation}
    \tau=\frac{\beta(\theta)}{2}+\rmi t~,
\end{equation}
where $\beta(\theta)$ is the semiclassical temperature \eqref{eq:microcanonical inverse temperature} to compute Lorentzian correlators in the microcanonical ensemble. Based on our results from the saddle point analysis above, and using the semiclassical seed theory two-point correlation function at finite temperature in \cite{Goel:2023svz}, the corresponding correlation function in the deformed theory is given by
\begin{equation}
    \label{eq:correlator Euclidean}
\begin{aligned}
    G_2(t,\theta)&\equiv\eval{G^{\Delta_0}(\tau)}_{\tau=\frac{\beta(\theta)}{2}+\rmi t}=\eval{\frac{\bra{\Omega}\rme^{-\tau^*\hH_y}q^{\Delta_0\hat{n}}\rme^{-\tau\hH_y}\ket{\Omega}}{\bra{\Omega}\rme^{-\beta(\theta)\hH_y}\ket{\Omega}}}_{\tau=\frac{\beta(\theta)}{2}+\rmi t}\\
&\eqlambda\frac{\exp(-\Delta_0\ell)}{\bra{\Omega}\rme^{-\beta(\theta)\hH_y}\ket{\Omega}}=\qty(\sin\theta~\sech\frac{\pi~t}{\tilde{\beta}_y(\theta)})^{2\Delta_0}~,
\end{aligned}
\end{equation}
where $\tilde{\beta}_{y=0}$ above is termed as fake temperature \cite{Blommaert:2024ymv}, which in our case contains a redshift factor
\begin{equation}\label{eq:fake temperature}
    \tilde{\beta}_y(\theta)\equiv \frac{2\pi}{J\sin\theta}\dv{E}{E_y}=\begin{cases}
        \frac{2\pi}{J\sin\theta}\sqrt{1-2yE(\theta)}&\eta=+1~,\\
        \frac{2\pi}{J\sin\theta}\sqrt{2yE(\theta)-1}&\eta=+1~.
    \end{cases}~,
\end{equation}
and $\ell\equiv\lambda\bra{\Omega}\rme^{-\tau^*\hH_y}\hat{n}\rme^{-\tau\hH_y}\ket{\Omega}$ as in \eqref{eq:semiclassical exp}; while the last step comes from solving the equations of motion for $\ell$ in \eqref{eq:EOM new} (in the case $m=0$), which follows analogously to \cite{Aguilar-Gutierrez:2025pqp} with the fake temperature in \eqref{eq:fake temperature}.

\begin{figure}
    \centering
    \subfloat[]{\includegraphics[width=0.48\linewidth]{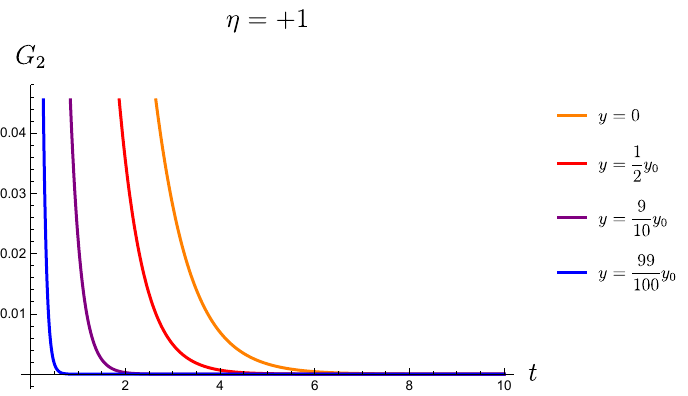}}\hspace{0.5cm}\subfloat[]{\includegraphics[width=0.48\linewidth]{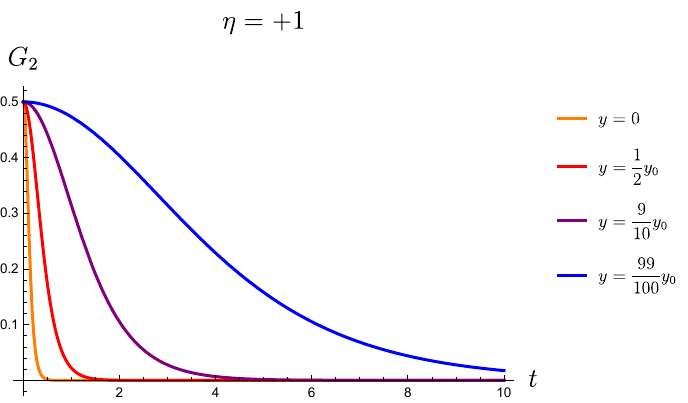}}
    \caption{Evolution of the two-point correlation function \eqref{eq:correlator Euclidean} for (a) $T^2$, and (b) $T^2+\Lambda_1$ deformations. As the deformation parameter reaches the critical value $y_0$ \eqref{eq:y0} the fake temperature \eqref{eq:fake temperature} increases resulting in faster decay of the correlation function. The parameters in this evaluation are: $\lambda=10^{-4}$, $J=1$, $\Delta=1$ and $\theta=3\pi/4$.}
    \label{fig:G2_new}
\end{figure}

We illustrate the behavior of the above expressions in Fig.~\ref{fig:G2_new}.
The fake temperature, which corresponds to the inverse rate of decay of two-point functions, is argued to represent the black hole temperature in the bulk so that the system is only maximally chaotic with respect to $\tilde{\beta}_y$ \cite{Blommaert:2024ymv}. In particular, Fig.~\ref{fig:G2_new} indicates that as the deformation parameter reaches its critical value $y_0$ \eqref{eq:y0} the boundary cutoff in the bulk approaches the location of a black hole horizon, which has infinite Tolman temperature \eqref{eq:EBY beta_Tolman} in accordance with finite cutoff holography. In contrast, the system is submaximally chaotic with respect to the physical temperature as defined from the partition function $\beta_y$ \eqref{eq:microcanonical inverse temperature}, as we find below.

\paragraph{Crossed Four-Point Functions}
Similarly, we can evaluate explicitly \eqref{eq:G m operators} for $m=2$ to recover a Lorentzian crossed four-point function using the analytic continuation \eqref{eq:analytic continuation} when $\lambda\rightarrow0$ based on the seed theory result for the crossed four-point function of matter chord operators in \cite{Aguilar-Gutierrez:2025pqp} and the time rescaling in \eqref{eq:resclaing time temp}. Including the fake temperature \eqref{eq:fake temperature}, the previous procedure gives:
\begin{align}\label{eq:generating function 1 particle}
&\begin{aligned}
    &G_{4}(t_L,t_R;\theta_L,\theta_R)\equiv\eval{G^{\Delta_0}_{{\Delta_1}}(\tau_L,\tau_R)}_{\tau_{L/R}=\frac{\beta(\theta_{L/R})}{2}+\rmi t_{L/R}}\\
&\eqlambda\qty(\frac{\rme^{-\ell_*(q^{\Delta_1},\theta_L,\theta_R)/2}}{\cosh(\frac{\pi t_L}{\tilde \beta_{L,y}})\cosh(\frac{\pi t_R}{\tilde \beta_{R,y}})+\frac{\rme^{-\ell_*(q^{\Delta_1},\theta_L,\theta_R)}q^{{\Delta_1}}}{\sin\theta_L\sin\theta_R}\sinh(\frac{\pi t_L}{\tilde \beta_{L,y}})\sinh(\frac{\pi t_R}{\tilde \beta_{R,y}})})^{2\Delta_0}~,
\end{aligned}
\end{align}
where $\tilde \beta_{L/R,y}\equiv\tilde{\beta}_y(\theta_{L/R})$ which appears in \eqref{eq:fake temperature}, while
\begin{equation}
    \begin{aligned}\label{eq:ell cond thetaLR}
    &\rme^{-\ell_*(q^{\Delta_1},\theta_L,\theta_R)}=\frac{q^{-2\Delta_{1}}}{2}\biggl(1+q^{2\Delta_{1}}-2q^{\Delta_{1}}\cos\theta_L\cos\theta_R\\
    &\quad-\sqrt{1+q^{4\Delta_{1}}+2 q^{2\Delta_{1}}(1+\cos(2\theta_L)+\cos(2\theta_R))-4q^{\Delta_{1}}(1+q^{2\Delta_{1}})\cos\theta_L\cos\theta_R}\biggr)~.
\end{aligned}
\end{equation}
In particular, when $t_L=-t_R:=t$ and $\beta_L=-\beta_R=\beta(\theta)/2$ then \eqref{eq:generating function 1 particle} takes the form
\begin{equation}
\begin{aligned}
    \label{eq:G4 crossed}
    G_{4}(-t,t;\theta,\pi-\theta)&\eqlambda\qty(\frac{\rme^{-\ell_0}}{\cosh^2\qty(\frac{\pi t}{\tilde{\beta}_y})-\frac{q^{{\Delta_1}}\rme^{-\ell_0}}{\sin^2\theta }\sinh^{2}\qty(\frac{\pi t}{\tilde{\beta}_y})})^{2\Delta_0}~,
\end{aligned}
\end{equation}
where we denote $\rme^{-\ell_0}\equiv\rme^{-\ell_*(q^{\Delta_1}, \theta,\pi-\theta)}$.\footnote{Note that \eqref{eq:G4 crossed} in the $y\rightarrow0$ limit corresponds to a crossed four-point correlation function in the physical SYK model \cite{Berkooz:2018jqr}
\begin{subequations}
\begin{align}
        G_4&=\expval{\Tr[\hmO_{\Delta_1}(t)\rme^{-\frac{\beta(\theta)}{4}}\hmO_{\Delta_0}(0)\rme^{-\frac{\beta(\theta)}{4}}\hmO_{\Delta_1}(t)\rme^{-\frac{\beta(\theta)}{4}}\hmO_{\Delta_0}(0)\rme^{-\frac{\beta(\theta)}{4}}]}~,\\
    &\qquad\qquad\hmO_{\Delta_0}(t)\equiv\rme^{\rmi t \hH_{\rm SYK}}\hmO_{\Delta}\rme^{-\rmi t \hH_{\rm SYK}}~,
\end{align}
\end{subequations}
which is evaluated in the double-scaling limit and with annealed ensemble averaging in \cite{Berkooz:2018jqr}.}

The expression \eqref{eq:G4 crossed} can then be approximated as
\begin{equation}\label{eq:G4 new}
   G_{4}(-t,t;\theta,\theta) \simeq\begin{cases}
    \rme^{-2\Delta_0\ell_0}\qty(1-2\Delta_0\qty(1-\frac{q^{\Delta_1\rme^{-\ell_0}}}{\sin^2\theta})\qty(\frac{\pi t}{\tilde{\beta}_y})^2)~,&t\ll\tfrac{\tilde{\beta}_y}{2\pi}~,\\
    \qty(\frac{2\sin^2\theta\rme^{-\ell_0}}{\sin^2\theta+q^{\Delta_1}\rme^{-\ell_0}})^{2\Delta_0}\qty(1-\Delta_0\frac{\sin^2\theta-q^{\Delta_1}\rme^{-\ell_0}}{\sin^2\theta+q^{\Delta_1}\rme^{-\ell_0}}\rme^{\frac{2\pi t}{\tilde{\beta}_y}})~,&\tfrac{\tilde{\beta}_y}{2\pi}\ll t\ll t_{\rm sc}~,\\
    \qty(\frac{\rme^{-\ell_0}}{\sin^2\theta-q^{\Delta_1}\rme^{-\ell_0}})^{2\Delta_0}\rme^{-\frac{4\pi\Delta_0t}{\tilde{\beta}_y}}~,&t\gg t_{\rm sc}~,
    \end{cases}
\end{equation}
where in the second line we approximated 
\begin{equation}
 \sinh^2\frac{\pi}{\tilde\beta_y}\simeq \frac{1}{4}\rme^{\frac{2\pi t}{\tilde\beta_y}}-\frac{1}{2}~,\quad  \rme^{\frac{2\pi t}{\tilde\beta_y}}\gg1~, 
\end{equation}
up to the scrambling time,
\begin{align}
\label{eq:scrambling time}
 t_{\rm sc}&\equiv
 \frac{\tilde{\beta}_y}{2\pi}\log\frac{2\qty({\sin^2\theta}+q^{\Delta_1}\rme^{-\ell_0})}{{\sin^2\theta}-q^{\Delta_1}\rme^{-\ell_0}}~,
\end{align}
which is the time scale of validity for the Taylor expansion used to derive the exponential growth in \eqref{eq:G4 new}.

Note that the chaos bound \cite{Maldacena:2015waa} in the deformed theory is satisfied in terms of the fake temperature $\tilde{\beta}_y$ since the OTOC decays by a rate $2\pi/\tilde{\beta}_y$ fixed by the Tolman temperature in the bulk dual theory (as we have discussed in Sec.~\ref{sec:thermo}). In contrast, if we compare the rate of growth with respect to the physical temperature of the deformed DSSYK model, given by $1/\beta_y$, then the system is submaximally chaotic given that $\tilde{\beta}_y(\theta)\geq\beta_{y}(\theta)$, as seen from (\ref{eq:fake temperature}, \eqref{eq:microcanonical inverse temperature}), when $\theta\in[0,\pi]$. This result has been observed in the seed theory \cite{Aguilar-Gutierrez:2025mxf}.

\begin{figure}
    \centering
    \subfloat[]{\includegraphics[width=0.475\linewidth]{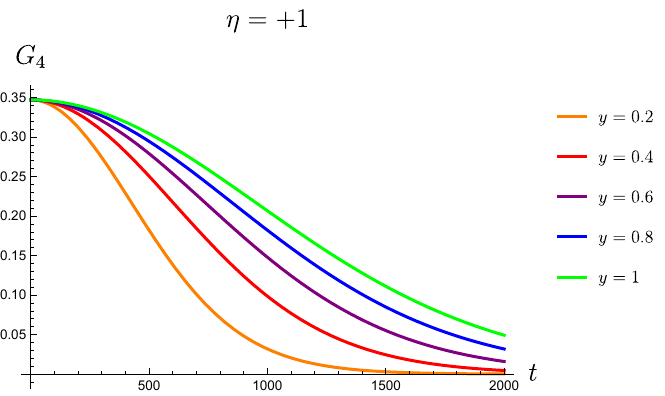}}\hspace{0.5cm}\subfloat[]{\includegraphics[width=0.475\linewidth]{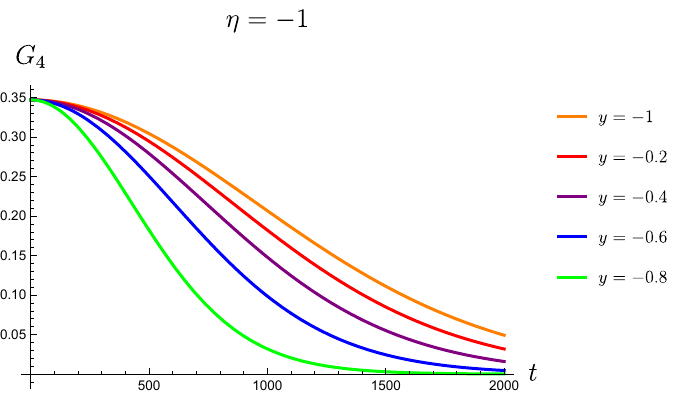}}
    \caption{Evolution of the OTOC \eqref{eq:G4 crossed} based on the energy spectrum of the deformed theory $E_y$ for (a) $\eta=+1$ \eqref{eq:E(theta) eta+1}, and (b) $\eta=-1$ \eqref{eq:E(theta) eta-1}. The parameters applied in the plot are $\theta=\pi/4$, $\lambda=10^{-6}$, $J=1$, $\Delta_0=1$, $\lambda\Delta_1=1$. We observe that the rate of decay is slower as the deformation parameter increases.}
    \label{fig:G4}
\end{figure}

 To exemplify the results, we display the growth of the OTOCs using the energy spectrum of the deformed theory $E_y$ in \eqref{eq:E(theta) eta-1} ($\eta=-1$) and \eqref{eq:E(theta) eta+1} ($\eta=+1$) in Fig.~\ref{fig:G4}. Note that we recover the expected behavior as the solution smoothly connected to the $y=0$ solution when $y\geq0$ for $\eta=+1$ and $y\leq0$ for $\eta=-1$. The reason for this is that the redshift factor $\rmd E_y/\rmd E$ (determined from \eqref{eq:E(theta) eta+1} ($\eta=+1$) and \eqref{eq:E(theta) eta-1} ($\eta=-1$)) can become imaginary when $y\leq0$ for $\eta=+1$ and $y\geq0$ for $\eta=-1$, which modifies the evolution (as exponential functions become trigonometric ones). When the redshift factor is complex-valued, the assumptions that $\hH_{L/R,~y}$ has a conserved energy spectrum are not valid, so that one has to modify the analysis.
 
As we elaborate further in App.~\ref{app:sine dilaton gravity finite cutoff}, when $y\geq0$ and $\eta=+1$ \eqref{eq:def parameter} one describes timelike boundaries in the bulk at finite cutoff, as well as for $y\geq0$ and $\eta=-1$, when $\hH_{L/R,~y}$ can be interpreted as a generator of time translations. Furthermore, we note from the results that the decay rate decreases as the magnitude of the deformation parameter $y$ increases in both cases.

Similarly, for higher point functions in \eqref{eq:G m operators} one can  perform the evaluation using the time rescaling by the redshift factor in \eqref{eq:resclaing time temp} and the semiclassical correlation functions for the seed theory in \cite{Aguilar-Gutierrez:2025mxf}. However, since this is not necessary for the subsequent analysis.

\subsection{Triple-Scaling Limit of Crossed Four-Point Functions}\label{eq:TSL G4}
For later convenience, we analyze an extension of the UV and IR triple-scaling limits \eqref{eq:Explicit Krylov basis deformed DSSYK}  in the canonical variables of the one-particle Hamiltonian \eqref{eq:pair DSSYK Hamiltonians 1 particle}
\begin{subequations}
    \begin{align}
         {\rm IV}:~\rme^{\rmi\hat{P}_{L/R}}&\rightarrow\rme^{\rmi\lambda\hat{P}_{\rm IR}^{(L/R)}}~,\quad q^{\hat{n}_{L/R}}\rightarrow\lambda\rme^{-\hat{L}_{\rm IR}^{(L/R)}}~,\\
    {\rm UV}:~\rme^{\rmi\hat{P}_{L/R}}&\rightarrow\rme^{-\hat{P}_{\rm UV}^{(L/R)}}~,\quad q^{\hat{n}_{L/R}}\rightarrow\lambda\rme^{\rmi\hat{L}_{\rm UV}^{(L/R)}}
    \end{align}
\end{subequations}
where the eigenvalues of the operators $\hat{L}_{\rm IR}^{(L/R)}$, $\hat{L}_{\rm UV}^{(L/R)}$, $\hat{P}_{\rm IR}^{(L/R)}$, $\hat{P}_{\rm UV}^{(L/R)}$, and the conformal dimension of matter chord $\Delta_1$, are $\mathcal{O}(1)$ as $\lambda\rightarrow0$. This means that the one-particle chord Hamiltonian \eqref{eq:pair DSSYK Hamiltonians 1 particle} can be expressed similarly to \eqref{eq:H(dS)JT} as
\begin{subequations}\label{eq:H LR triple scaled}
    \begin{align}
    \hH^{(L/R)}_{\rm IR}&\equiv\frac{\sqrt{\lambda(1-q)}}{2J}\qty(\hH_{L/R}+\frac{2J}{\sqrt{\lambda(1-q)}}\mathbb{1})~,\\
    \hH^{(L/R)}_{\rm UV}&\equiv\frac{\sqrt{\lambda(1-q)}}{2J}\qty(-\hH_{L/R}+\frac{2J}{\sqrt{\lambda(1-q)}}\mathbb{1})~,
    \end{align}
\end{subequations}
The corresponding correlation functions in the IR/UV triple-scaling limit can be solved based on the same procedure leading to \eqref{eq:generating function 1 particle} for the crossed four-point function, and, for instance, for the IR case one finds 
\begin{equation}\label{eq:IR OTOC}
    G_{4}(t_L,t_R;\theta_L,\theta_R)\underset{\rm IR}{\rightarrow}\qty(\frac{\rme^{-\ell_{0}/2}}{\cosh(\frac{\pi t_L}{\tilde\beta^{L}_{{\rm IR},y}})\cosh(\frac{\pi t_R}{\tilde\beta^{R}_{{\rm IR},y}})+\frac{2\rme^{-\ell_0}}{\tilde\theta_{\rm IR}^{(L)}\tilde\theta_{\rm IR}^{(R)}}\sinh(\frac{\pi t_L}{\tilde\beta^{L}_{{\rm IR},y}})\sinh(\frac{\pi t_R}{\tilde\beta^{R}_{{\rm IR},y}})})^{2\Delta_0}~,
\end{equation}
where $\tilde\theta_{\rm IR}^{(L/R)}\equiv\theta_{L/R}/\lambda$ is a fixed constant in the triple-scaling limit; while the fake temperature \eqref{eq:fake temperature} in the IR limit takes the form
\begin{equation}
    \tilde\beta^{L/R}_{{\rm IR},y}\underset{{\rm IR}}{\rightarrow}\frac{2\pi}{J\tilde\theta_{\rm IR}^{(L/R)}}\begin{cases}
        {\sqrt{1-y\qty(\tilde\theta_{\rm IR}^{(L/R)})^2}}~,&\eta=+1~,\\
        {\sqrt{y\qty(\tilde\theta_{\rm IR}^{(L/R)})^2-1}}~,&\eta=+1~.
    \end{cases}
\end{equation}
with the deformation parameter $y$ \eqref{eq:def parameter} (with $16\pi G_N=1$ from \eqref{eq:H LR triple scaled}) being related to the radial cutoff in JT gravity by $y=r_B^{-2}$. Meanwhile, the constant $\ell_0$ is determined from energy conservation in the corresponding triple-scaled Hamiltonian \eqref{eq:H LR triple scaled}. To recover real solutions in the triple-scaling limit, we implement a homogeneous energy condition after the operator insertion (such as that used in shockwave geometries \cite{Shenker:2013yza}) $\tilde\theta_{\rm IR}^L=\tilde\theta_{\rm IR}^R\equiv\tilde\theta_{\rm IR}$, which leads to
\begin{equation}\label{eq:new e ell0}
  \rme^{-\ell_0}=\frac{1}{2}\qty(\tilde\theta_{\rm IR}^2+\Delta_1^2-\Delta_1\sqrt{2\tilde\theta_{\rm IR}^2+\Delta_1^2})~.
\end{equation}
The $y\rightarrow0$ limit of the above results has appeared previously in \cite{Xu:2024gfm}. Similar outcomes are recovered in the UV limit, which we discuss in Sec.~\ref{ssec:hyperscrambling dS} in the context of the stretched horizon, i.e.~when $y\rightarrow y_0$ \eqref{eq:y0}. This leads to an infinite temperature limit (already noticed in \cite{McGough:2016lol}) enhancing the rate of growth of the OTOC \eqref{eq:IR OTOC}.

\paragraph{Summary}Our findings show that all the correlation functions at the semiclassical level experience the same evolution as the correlation functions in the seed theory; where the temperature is affected by a redshift factor, in agreement with the existing literature on finite cutoff holography. The results also highlight the analytic control on the DSSYK model. In CFT$_{d\geq 2}$ real-time correlation functions at finite temperature are known only in terms of perturbative expansions in the deformation parameter \cite{Guica:2025jkq}. In contrast, in this specific system we evaluate the correlation functions at a non-perturbative order in the deformation parameter $y$.

Additionally, we discuss how to apply the results in this section to evaluate ETW brane partition functions and correlation functions, including trumpet geometries, in App.~\ref{eq:Def ETW Brane Wormhole}.

\section{Holographic Entanglement Entropy from the Double-Scaled Algebra}\label{sec:EE}
In this section, we study the entanglement entropy between the double-scaled algebras of the DSSYK \cite{Lin:2022rbf,Xu:2024hoc}, defined below, given a chord state and we analyze its finite cutoff holographic interpretation. The analysis is based on recent developments in von Neumann algebras and generalized entropies in quantum gravity \cite{Witten:2021unn,Chandrasekaran:2022cip,Chandrasekaran:2022eqq,Jensen:2023yxy,Kudler-Flam:2023qfl}; see \cite{Witten:2018zxz,Sorce:2023fdx,Casini:2022rlv,Liu:2025krl} for modern reviews. Note that while a $\TT$ deformation generates spatial non-locality in higher-dimensional theories, we consider a theory without spatial dimensions, which remains local, at least from the boundary theory perspective. This might be a reason that there is a precise match between the boundary and bulk evaluations; even non-perturbatively in terms of the deformation parameter.

\paragraph{Outline}In Sec.~\ref{ssec:algebraic entropy} we discuss the double-scaled algebra of the (deformed) DSSYK model and define entanglement entropy between the algebras given a pure global state. We illustrate the definitions considering the HH state. In Sec.~\ref{ssec:EE} we then evaluate entanglement entropy using the chord number state in the basis that reproduces a wormhole length with finite cutoff, particularly its IR triple-scaling limit. In Sec.~\ref{ssec:HEE AdS} we show that the expressions have a bulk interpretation.

\subsection{Double-Scaled Algebras \& Entanglement Entropy}\label{ssec:algebraic entropy}
As argued in the previous sections, in general the chord operators in the deformed theory belong to the operator algebra of the seed theory since we study chord Hamiltonian deformations encoded in the flow equation \eqref{eq:flow eq operator form}. This means the double-scaled algebras \cite{Xu:2024hoc,Lin:2022rbf} can be expressed in terms of the (un)deformed Hamiltonian
\begin{equation}\label{eq:DS algebra}
    \mathcal{A}_{L/R}=\qty{\hH_{L/R,~y},~\hat{\mathcal{O}}^{(L/R)}_\Delta}~,
\end{equation}
where $\hat{\mathcal{O}}^{(L/R)}_\Delta$ are matter chord operators. Moreover, we know that $\mathcal{A}_{L/R}$ is a type II$_1$ algebra \cite{Xu:2024hoc}, where  the $\ket{\Omega}$ is the cyclic separating state in both the deformed and seed theories. This allows us to define traces
\begin{equation}\label{eq:trace def}
    \Tr(\hat{w}_{L/R})=\bra{\Omega}\hat{w}_{L/R}\ket{\Omega}~,\quad \forall\hat{w}_{L/R}\in{\mathcal{A}}_{L/R}~.
\end{equation}
We may then define the density matrix in $\hat{\rho}_{L/R}\in\mathcal{A}_{L/R}$ associated to a state $\ket{\Psi}\in\mathcal{H}_{\rm chord}$ as \cite{Witten:2021unn}
\begin{equation}\label{eq:def density}
    \Tr(\hat{\rho}_{L/R}~\hat{w}_{L/R})=\bra{\Psi}\hat{w}_{L/R}\ket{\Psi}~,\quad \forall\hat{w}_{L/R}\in\mathcal{A}_{L/R}~,
\end{equation}
Next, from the Gelfand–Naimark–Segal construction \cite{Segal1947IrreducibleRO,gelfand1943imbedding}, we can express any state in the chord Hilbert space by applying a string of the operators in the double-scaled algebras to the cyclic separating state,
\begin{equation}\label{eq:psi gen}
    \ket{\psi}={f}\qty(\hH_{L/R,y},\hat{\mathcal{O}}_{\Delta}^{(L/R)})\ket{\Omega}\in\mathcal{H}_{\rm chord}~,
\end{equation}
where $f$ is a generic function, then combining \eqref{eq:def density} and \eqref{eq:psi gen} we identify
\begin{equation}\label{eq:density matrix formal}
    \hat{\rho}_{L/R}=f\qty(\hH_{L/R,~y},\hat{\mathcal{O}}_{\Delta}^{(L/R)})f\qty (\hH_{L/R,~y},\hat{\mathcal{O}}_{\Delta}^{(L/R)})^\dagger~,
\end{equation}
where we applied cyclicity of the trace \eqref{eq:trace def}.

Given that $\ket{\Omega}$ is the tracial state in the double-scaled algebra, the von Neumann entropy of \eqref{eq:density matrix formal} can be defined as
\begin{equation}\label{eq:def entanglement entropy}
    S\equiv\log\bra{\Omega}\hrho\ket{\Omega}-\frac{\bra{\Omega}\hrho\log\hrho\ket{\Omega}}{\bra{\Omega}\hrho\ket{\Omega}}~,
\end{equation}
where we have suppressed the $L/R$ index since both density matrices lead to the same result. For this reason, we associate \eqref{eq:def entanglement entropy} as a notion of algebraic entanglement entropy \cite{Penington:2023dql}.

Note that while we can evaluate entanglement entropy between the double-scaled algebras in the seed and deformed theory in the same way, the main difference is the physical interpretation of the entropies which depend on the form that $\hrho$ takes, given that it can either involved the (un)deformed chord Hamiltonian.

\paragraph{Thermodynamic Entropy}
For instance, there is a thermodynamic entropy that can be computed from the partition function of the deformed theory
\begin{equation}\label{eq:entanglement entropy HH state}
    S=(1-\beta\partial_\beta)\log Z_y(\beta)\quad Z_y(\beta)=\bra{\Omega}\rme^{-\beta\hH_{y}}\ket{\Omega}
\end{equation}
where $\rme^{-\frac{\beta}{2}\hH_y}\ket{\Omega}$ is the HH state in the zero-particle chord space.

The above result can also be interpreted as entanglement entropy between the double-scaled algebra using the definition of the density matrix:
\begin{equation}
    Z_y(\beta)=\Tr(\hat{\rho}_{\rm HH})~,\quad \hat{\rho}_{\rm HH}=\rme^{-\beta\hH_y}~.
\end{equation}
Note that the density matrix is not normalized. The von Neumann entropy in the corresponding state \eqref{eq:def entanglement entropy} is thus the thermal entropy in the deformed theory \eqref{eq:entanglement entropy HH state}. Holographically, since the effect of a $T^2$ deformation represents a finite cutoff in the bulk \cite{McGough:2016lol}, the above expressions would have a natural geometric interpretation in terms of a finite cutoff AdS$_2$ black hole effective geometry, similar to the partition functions computed in finite cutoff holography for JT gravity \cite{Iliesiu:2020zld,Griguolo:2025kpi}.\footnote{Additionally, there are simple extensions to evaluate the algebraic entanglement entropy \eqref{eq:def entanglement entropy} using particle insertions in the HH state, which follow similarly to our discussion, and which were connected to the partially-entangled thermal state \cite{Goel:2023svz,Goel:2018ubv} in the DSSYK model in \cite{Aguilar-Gutierrez:2025otq}.}

\subsection{Boundary Perspective: Entropy From the Chord Number State}\label{ssec:EE}
In this subsection, we will investigate the holographic nature of the algebraic entropy \eqref{eq:def entanglement entropy} by considering the chord number state in the Krylov basis and its corresponding density matrix \eqref{eq:density matrix formal} in the double-scaled algebra \eqref{eq:DS algebra}.

We begin with the state
\begin{equation}
    \ket{\Psi}=\ket{K_n^{(y)}}=\tilde{f}_n(\hH_y)\ket{\Omega}~,\quad\hat{\rho}=\tilde{f}_n(\hH_y)^2~,
\end{equation}
The algebraic entanglement entropy \eqref{eq:def entanglement entropy} for the $\ket{\Psi}$ state is then
\begin{equation}\label{eq:full entropy n DSSYK}
    S\equiv-\int_{0}^{\pi}\rmd \theta~\mu(\theta){\tilde{f}_n(E_y(\theta))^2}\log{\tilde{f}_n(E_y(\theta))^2}~,
\end{equation}
with $\mu(\theta)$ in \eqref{eq:norm theta}, and $E_y(\theta)\equiv f_y(E(\theta))$ \eqref{eq:def Hamiltonian}. Next, we compute the integral \eqref{eq:full entropy n DSSYK} using the Wentzel-Kramers-Brillouin (WKB) approximation \cite{wentzel1926verallgemeinerung,kramers1926wellenmechanik,brillouin1926mecanique} approximation. We introduce a change of variables adapted for the (IR) triple-scaling limit \eqref{eq:seed triple scaling limits}, $\theta\equiv\lambda~k$, so that
\begin{equation}\label{eq:S k}
   S\underset{\rm IR}{\rightarrow} -2\lambda\int_0^\infty\rmd k~p(k,\ell_y)\log\frac{p(k,\ell_y)}{\mu(\lambda k)}~,
\end{equation}
where we defined
\begin{equation}
    p(k,\ell_y\equiv\lambda n)\propto\eval{\mu(\lambda k)\tilde{f}_n(\lambda k)^2}_{\rm IR}~,
\end{equation}
where the proportionality constant is chosen such that $p(k,\ell_y)$ represents a probability distribution in terms of the parameter $k$,
\begin{equation}\label{eq:p normalization}
\int_0^\infty\rmd k~p(k,\ell_y)=1~.
\end{equation}
\eqref{eq:S k} then transforms in
\begin{equation}\label{eq:new entropy}
    S\underset{\rm IR}{\rightarrow}2\lambda\int_0^\infty\rmd k~ p(k,\ell_y)\log\mu(\lambda k)+\mathcal{O}(\lambda)~.
\end{equation}
where we use the fact that $p(k,\ell_y)\sim\mathcal{O}(1)$ as $\lambda\rightarrow0$ while \cite{Goel:2023svz}
\begin{equation}
    \lambda\log\mu(\lambda k)=-\frac{\pi^2}{2}+\mathcal{O}(\lambda\log\lambda)~.
\end{equation}
To find the explicit functional form of $p(k,\ell_y)$, we use the WKB approximation \cite{wentzel1926verallgemeinerung,kramers1926wellenmechanik,brillouin1926mecanique}, so that the probability distribution takes the form
\begin{equation}\label{eq:prob k}
    p(k,\ell_y)\simeq\begin{cases}
        A\exp(-2\int\rmd\ell ~\tilde{p}_y)~,&k\lesssim k_{\rm crit}~,\\
        B\sin^2(\int\rmd\ell ~{p}_y)~,&k> k_{\rm crit}~
    \end{cases}
\end{equation}
where we denote the critical value of $k$ where the particle description of the system in the WKB approximation has vanishing momenta (i.e.~$p_y=0$ in \eqref{eq:classical H}) as,\footnote{We are using $k=\theta_{\rm IR}/\lambda$ and the spectrum in \eqref{eq:E(theta) eta+1}, \eqref{eq:E(theta) eta-1}:
\begin{equation}
    E_{{\rm IR},y}=\begin{cases}
        \frac{1}{y}\qty(1-\sqrt{1-y k^2})~,&\eta=+1~,\\
        \frac{1}{y}\qty(1-\sqrt{y k^2-1})~,&\eta=-1~.
    \end{cases}
\end{equation}
}
\begin{equation}
    k_{\rm crit}=\begin{cases}
        \frac{1}{\sqrt{y}}\sech\frac{\ell_y}{2}~,&\eta=+1~,\\
        \frac{1}{\sqrt{y}}\sec\frac{\rmi\ell_y}{2}~,&\eta=-1~,
    \end{cases}
\end{equation}
and $A$, $B$ are normalization constants, such that \eqref{eq:p normalization} is satisfied; i.e.
\begin{align}\label{eq:norm consts}
A=\frac{1}{{2\pi}}\begin{cases}
    \sqrt{\frac{y k^2}{ \sech^2\frac{\ell_{y}}{2}-y k^2}}~,&\eta=+1\\
    \sqrt{\frac{y k^2}{ \sec^2\frac{\rmi\ell_{y}}{2}-y k^2}}~,&\eta=-1
\end{cases}~,\quad
B={\frac{2}{\pi}}\begin{cases}
    \sqrt{\frac{y k^2}{y k^2-\sech^2\frac{\ell_{y}}{2}}}~,&\eta=+1\\
    \sqrt{\frac{y k^2}{ y k^2-\sec^2\frac{\rmi\ell_{y}}{2}}}~,&\eta=-1
\end{cases}~.
\end{align}
Meanwhile, $p_y$ and $\tilde{p}_y$ in \eqref{eq:prob k} are the momenta in the allowed and disallowed regions of the corresponding Schrödinger equations for the DSSYK chord Hamiltonian in the Krylov chord number basis \eqref{eq:classical H} in the IR limit, namely,
\begin{subequations}\label{eq:different p}
    \begin{align}
         \sin^2p_y&=-\sinh^2\tilde{p}_y=\begin{cases}
         \frac{yk^2-\sech^2(\ell_y/2)}{\tanh^2(\ell_y/2)}~,&\eta=+1~,\\
         \frac{\sec^2(\rmi\ell_y/2)-yk^2}{\tan^2(\rmi\ell_y/2)}~,&\eta=-1~,\\
         \end{cases}
    \end{align}
\end{subequations}
where we used that $\sqrt{1-y k^2}=\tanh(\ell_y/2)\cos p_y$ for $\eta=+1$.
We work in the regimes where
\begin{subequations}\label{eq:careful limit}
    \begin{align}
      \eta=+1:\quad  \sech\frac{\ell_y}{2}\gg yk~~{\rm for}~~ k\lesssim k_{\rm crit}~;\quad \sech\frac{\ell_y}{2}\ll yk~~{\rm for}~~ k\gtrsim k_{\rm crit}~,\\
        \eta=-1:\quad\sec\frac{\rmi\ell_y}{2}\gg yk~~{\rm for}~~ k\lesssim k_{\rm crit}~;\quad \sec\frac{\rmi \ell_y}{2}\ll yk~~{\rm for}~~ k\gtrsim k_{\rm crit}~.
    \end{align}
\end{subequations}
Then, we recover from \eqref{eq:different p} that
\begin{equation}
    \sin^2 p_y\simeq\begin{cases}
    yk^2~,&\eta=+1~,\\
    \csc^2\frac{\rmi \ell_y}{2}~,&\eta=-1~,
    \end{cases}
    \quad
    \sinh^2\tilde{p}_y\simeq\begin{cases}
        \csch^2\frac{\ell_y}{2}~,&\eta=+1~,\\
    yk^2~,&\eta=-1~.
    \end{cases}
\end{equation}
Thus, \eqref{eq:prob k} with normalization constants in \eqref{eq:norm consts} and the limit \eqref{eq:careful limit} becomes
\begin{equation}\label{eq:new p ell k}
    p(k,\ell_y)\simeq\begin{cases}
        0~,&k\lesssim k_{\rm crit}~,\\
        \frac{2}{\pi}\sin^2(\ell_y\arcsin(yk))\simeq\frac{1}{\pi}~,&k>  k_{\rm crit}~.
    \end{cases}
\end{equation}
The algebraic entanglement entropy \eqref{eq:new entropy} with the explicit probability distribution \eqref{eq:new p ell k} results in,\footnote{The result \eqref{eq:new S ell y} in the $y\rightarrow0$ limit reproduces the renormalized entropy in (dS) JT gravity \cite{Tang:2024xgg,Aguilar-Gutierrez:2025otq}.}
\begin{equation}\label{eq:new S ell y}
    S+S_{\rm ren}\underset{\rm IR}{\rightarrow}\pi\int_0^{k_{\rm crit}}\rmd k=\frac{1}{\pi\sqrt{y}}\begin{cases}
        \sech(\frac{\ell_y}{2})~,&\eta=+1\\
        \sec(\rmi\frac{\ell_y}{2})~,&\eta=-1~,
    \end{cases}
\end{equation}
where $S_{\rm ren}\equiv\pi\int_0^\infty\rmd k$, corresponding to a regularization of the entanglement entropy in momentum space due to the IR energy limit taken in this approximation.

Next, we analyze the bulk dual of the above result.

\subsection{Bulk Perspective: Ryu-Takayanagi Formula}\label{ssec:HEE AdS}
In the following, we search for the minimal area surface $\gamma$ in the (A)dS$_2$ space, using the finite cutoff boundary as the entangling surface to evaluate holographic entanglement entropy. The extremal surfaces in two-dimensional quantum gravity are point-like. We investigate the different configurations depending on the location of boundary cutoff surfaces in the corresponding spacetime.

\paragraph{Timelike Boundaries in an AdS Black Hole}
First, considering timelike finite cutoff boundaries, as illustrated in Fig.~\ref{fig:spread_time} (c) as the entangling surface to evaluate the RT formula in JT gravity, which corresponds to \eqref{eq:Dilaton-gravity theory} with $U(\Phi)=2\Phi$,
\begin{equation}\label{eq:RT surface dS}
    S_{\rm AdS}\equiv \frac{{\Phi_{\rm AdS}(\gamma)}}{4G_N}~,
\end{equation}
with $G_N$ in \eqref{eq:Dilaton-gravity theory}, $\Phi_{\rm AdS}$ is the dilaton in JT gravity, and $\gamma$ the extremal surface. To locate the latter, we use global and Rindler-AdS$_2$ coordinates,
\begin{equation}\label{eq:rBH}
    \rmd s^2=-(1+X^2)\rmd T^2+\frac{\rmd X^2}{1+X^2}=-(r^2-r_{\rm BH}^2)\rmd t_{L/R}^2+\frac{\rmd r^2}{r^2-r_{\rm BH}^2}~,
\end{equation}
where $t_{L/R}$ corresponds to the time asymptotic boundary time in the left/right patches of Rindler-AdS$_2$; the dilaton in JT gravity is given by
\begin{align}\label{eq:dilaton global AdS}
    \Phi_{\rm AdS}=r_{\rm BH}\sqrt{1+X^2}\cos T~.
\end{align}
Due to symmetry, $\gamma$ is located at $X=0$ in global AdS$_2$ coordinates for a given Cauchy slice, as seen by minimizing the dilaton \eqref{eq:dilaton global AdS} ($\partial_X\Phi_{\rm AdS}=0$ with $T$ fixed).

Let us then relate the  time $t$ and $T$ at the location of the entangling region, $r=r_B$, using the following coordinate transformation between global and Rindler-AdS coordinates,
\begin{equation}
    \tan T=\frac{r_{\rm BH}}{r}\sqrt{\frac{r^2}{r_{\rm BH}^2}-1}\sinh(\frac{r_{\rm BH}t}{2})~,\quad r<r_{\rm BH}~,
\end{equation}
where we are fixing the gauge $t_L=t_R\equiv t/2$ in \eqref{eq:rBH}.

Next, from the explicit form of the wormhole length in terms of Rindler-AdS$_2$ time $t$ and the metric in \eqref{eq:rBH} (see App.~\ref{app:geodesics}),
\begin{equation}\label{eq:wormhole length finite cutoff2}
    L\equiv\int\rmd\xi\sqrt{g_{\mu\nu}\dv{x^\mu}{\xi}\dv{x^\nu}{\xi}}=2~{\rm arcsinh}\qty(\sqrt{\frac{r_B}{r_{\rm BH}}-1}\cosh\frac{r_{\rm BH}t}{2})~,
\end{equation}
we have that the RT formula \eqref{eq:RT surface dS} from the dilaton at $X=0$ is
\begin{equation}\label{eq:JT gravity Sbulk}
    S_{\rm AdS}=\frac{r_B}{4G_N}\sech\frac{L}{2}~.
\end{equation}
Thus, we find exactly the same result as the boundary theory \eqref{eq:new S ell y} following the holographic dictionary between DSSYK and sine dilaton gravity \cite{Blommaert:2025eps}, including the relationship between the rescaled chord number $\ell_y$ with the geodesic length $L$ \eqref{eq:wormhole length finite cutoff2} and the deformation parameter $y$ and $r_B$ in \eqref{eq:def parameter} for the counterterm $G(\Phi_B)=\Phi_B^2$ \eqref{eq:Gphi special} in JT gravity
\begin{equation}
   \ell_y=L~,\quad \lambda=8\pi G_N~,\quad \frac{1}{\sqrt{y}}=r_B~,
\end{equation}
which is consistent with \eqref{eq:relation lambda kappa}.

In particular, the limit $r_{B}\rightarrow r_{\rm BH}$ leads to the Bekenstein-Hawking (BH) entropy of a Bañados-Teitelboim-Zanelli (BTZ) black hole,
\begin{equation}
    S_{\rm BH}=\frac{2\pi r_{\rm BH}}{4G_3}~,
\end{equation}
where the two and three-dimension Newton's constants are related by $G_3=4\pi G_N$ \cite{Svesko:2022txo}, and we can recognize the factor $2\pi r_{\rm BH}$ as the codimension-two area of the BTZ black hole, corresponding to the perimeter of a circle with radius $r_{\rm BH}$. This can be interpreted from the boundary theory answer \eqref{eq:JT gravity Sbulk} when $y\rightarrow y_0$.

\paragraph{Spacelike Boundaries in an AdS Black Hole}Next, we return to the JT gravity case with spacelike cutoff boundaries inside the black hole interior, illustrated in Fig.~\ref{fig:spread_time} (d), interpreted as the bulk dual to a $T^2+\Lambda_1$ deformed DSSYK model in the IR limit. 

An attempt might be made to evaluate the holographic entanglement entropy of \eqref{eq:RT surface dS} in a similar way to the previous case. However, if we extremize the dilaton \eqref{eq:dilaton global AdS} with $T$ fixed and varying $X$ to obey the homology constraint ($\partial_T\Phi=0$) one would find $T=0$ as the location for the RT surface. This corresponds to a maximum dilaton value instead of a minimum one. In fact, the minimum value is reached when $T=\pm \pi/2$, where $\Phi=0$ for any $X$. However, the dilaton in the minimal area surface cannot describe holographic entanglement entropy, it would not even be smoothly connected to the extremal area when the timelike boundaries approach the black hole horizon \eqref{eq:JT gravity Sbulk}. This means that the RT formula does not apply when the boundaries are located in the black hole interior. Thus, while we expect that the algebraic entanglement entropy for a chord number in the $T^2+\Lambda_1$ deformed DSSYK \eqref{eq:new S ell y} has a bulk interpretation, it does not correspond to the minimal area in the RT formula measured by the dilaton at the corresponding extremal surface.

\section{Stretched Horizon Holography From the Deformed DSSYK Model}\label{sec:dS holography}
We collect the lessons from the previous sections to explicitly realize the stretched horizon conjecture by Susskind \cite{Susskind:2021esx}. We perform the $T^2$ and $T^2+\Lambda_1$ deformation procedure in the DSSYK chord Hamiltonian in the UV triple-scaling limit \eqref{eq:H(dS)JT}, which is isomorphic to the dS JT gravity generator of spatial translations along $\mathcal{I}^\pm$. After the $T^2+\Lambda_1$ deformation, the finite cutoff boundary moves inside the static patch of dS$_2$ space, which takes the role of the cosmological stretched horizon \cite{Susskind:2021esx}. We evaluate the rate of growth of the Krylov spread complexity for the HH state defined in Sec.~\ref{ssec:spread}; OTOCs (based on Sec.~\ref{sec:correlation functions}), as well as the algebraic entanglement entropy for the chord number state in Sec.~\ref{ssec:EE}. We show that the Krylov complexity and OTOCs display hyperfast growth and hyperfast scrambling respectively, as the boundary approaches the cosmological horizon. Meanwhile, the entanglement entropy matches the dilaton in dS JT gravity at finite cutoff if the boundaries are spacelike, and it reproduces the expected formula in the Gibbon-Hawking entropy in dS$_3$ space in the stretched horizon limit.

\paragraph{Outline}In Sec.~\ref{ssec:comparison dS2} we compare general aspects and the thermodynamic properties between the UV limit of the deformed DSSYK model after $T^2$ and $T^2+\Lambda_1$ deformations with dS$_2$ JT gravity with a stretched horizon in the static patch. In Sec.~\ref{ssec:dS Complexity UV} we discuss Krylov complexity of the deformed HH state and its bulk interpretation as a geodesic length; particularly its hyperfast growth. In Sec.~\ref{ssec:hyperscrambling dS} we evaluate the OTOCs in the UV triple-scaling limit and its Lyapunov exponent. At last, in Sec.~\ref{ssec:HEE dS} we evaluate algebraic entanglement entropy for a chord number state in the Krylov basis; we match it to the dilaton in dS$_2$ JT gravity, confirming the RT formula.

A summary of results in this section in connection with the stretched horizon holographic proposal \cite{Susskind:2021esx} is shown in Tab.~\ref{tab:comparizon}.

\subsection{UV Limit of the DSSYK Model vs dS Space at Finite Cutoff}\label{ssec:comparison dS2}
To realize the stretched horizon proposal in the UV limit \eqref{eq:H(dS)JT}, as explained in Sec.~\ref{ssec:spread}, we apply the $T^2$ deformation into the UV chord Hamiltonian \eqref{eq:H(dS)JT}. Once the deformation parameter reaches the critical value $y=y_0$ \eqref{eq:y0}, we apply the $T^2+\Lambda_1$ deformation. We thus specialize in \eqref{eq:def H Krylov basis} in the UV triple scaling limit \eqref{eq:H IR UV},
\begin{equation}\label{eq:H UV deformed}
    \hH_{{\rm UV},y}\equiv\begin{cases}
   \frac{1}{y}\qty(1-\frac{1}{2}\qty(\tanh\qty(\frac{\lambda \hat{n}_y}{2})\rme^{-\rmi\hat{p}_y}+\rme^{\rmi\hat{p}_y}\tanh\qty(\frac{\lambda \hat{n}_y}{2})))&0\leq y\leq y_0~,\\
    \frac{1}{y}\qty(1+\frac{1}{2}\qty(\tanh\qty(\frac{\lambda \hat{n}_y}{2})\rme^{-\rmi\hat{p}_y}+\rme^{\rmi\hat{p}_y}\tanh\qty(\frac{\lambda \hat{n}_y}{2})))&y\geq y_0~,
    \end{cases}
\end{equation}
where $y_0\equiv\theta_{\rm UV}^{-2}$ in the regime of interest. We may now interpolate between dS$_2$/CFT$_1$ holography at finite cutoff and static patch holography. In particular, when $y= y_0+\delta$ for $0<\delta\ll 1$ we recover the stretched horizon conjecture.\footnote{This resonates with similar ideas presented in \cite{Chang:2025ays}, which views $\TT+\Lambda_2$ deformations in dS$_3$/CFT$_2$ as a way to unify it with static patch holography.}

As emphasized in \cite{Aguilar-Gutierrez:2025otq}, the UV limit of the DSSYK model is an explicit example of a unitary theory which contrasts the dS/CFT correspondence \cite{Strominger:2001pn,Witten:2001kn,Spradlin:2001pw,Bousso:2001mw,Maldacena:2002vr} in higher dimensions, where a non-unitary CFT dual is expected to live at future/past infinity $\mathcal{I}^\pm$.\footnote{There have been explicit realizations of the dS/CFT correspondence in different approaches, including higher spin gravity \cite{Vasiliev:1999ba,Vasiliev:1990en} where dS$_4$ space \cite{Anninos:2011ui,Ng:2012xp,Anninos:2012ft} is dual to Sp($N$) CFT$_3$s; as well as a  dS$_3$ space higher spin proposal being dual to a SU($2$) Wess-Zumino-Witten model \cite{Hikida:2021ese,Hikida:2022ltr,Chen:2022ozy,Chen:2022xse} from an analytic continuation of Euclidean AdS$_3$ space; and from transitions between Euclidean AdS$_4$ space to dS$_4$ space \cite{Hartle:2012tv,Hertog:2011ky,Bobev:2022lcc}. A different approach involves geometries interpolating from asymptotically AdS$_2$ and dS$_2$ spacetimes, often referred to as centaur geometries \cite{Anninos:2017hhn,Anninos:2018svg,Anninos:2020cwo,Chapman:2021eyy,Anninos:2022ujl,Anninos:2022hqo,Aguilar-Gutierrez:2023odp}, whose holographic dual has been recently investigated in \cite{Anninos:2022qgy,Chapman:2024pdw}. See \cite{Galante:2023uyf} for a review and references therein.} Work in this area indicates that the renormalization group (RG) flow in the dual CFT at $\mathcal{I}^\pm$ allows for an emergent notion of time evolution from the bulk \cite{Strominger:2001gp} (recently debated by \cite{Araujo-Regado:2025elv}). Similarly, the $T^2$ flow in \eqref{eq:H UV deformed} indicates one can associate an emergent global time generated by the $T^2$ deformation.

Next, we apply our results in Sec.~\ref{sec:flow eq} to compare the thermodynamics in the deformed boundary theory \eqref{eq:H UV deformed} to those in the bulk at a finite cutoff, discussed below.

First, consider the geometry of the dS$_2$ static patch (resulting from the s-wave reduction of dS$_3$ space), which is given by:
\begin{equation}\label{eq:dS2 metric}
	\rmd s^2=-f(r)\rmd t_{L/R}^2+\frac{\rmd r^2}{f(r)}~,\quad f(r)={r_{\rm CH}^2-r^2}~,\quad \Phi=r\in[0,~r_{\rm CH}]~,
\end{equation}
where $r_{\rm CH}$ is a parametrization of the cosmological horizon radius.\footnote{This parameter is associated with the mass of SdS$_3$ space \cite{Balasubramanian:2001nb}, where a value $r_{\rm CH}\neq1$ implies the existence of a conical deficit.} We illustrate the geometry of dS$_2$ and the cosmological stretched horizon in Fig.~\ref{fig:two-step_def} (d). Note that in writing \eqref{eq:dS2 metric} we may assume a gauge fixing where $t_L=t_R\equiv t/2$. Adopting a gauge-fixing choice for the coordinate system is not necessary for the evaluation of diffeomorphism invariant observables, all other gauge choices are equivalent.

Let us now evaluate the Euclidean on-shell action of the half-reduction dS$_2$ JT gravity model with a finite time-like Dirichlet surface, $r=r_{B}$, using (\ref{eq:dS2 metric}, \ref{eq:on shell action}) with $U(\Phi)=-2\Phi$, and a counterterm $G(\Phi_B)=\Phi_B^2$ for the dual $T^2$ deformation, or $G(\Phi_B)=-\Phi_B^2$ for $T^2+\Lambda_1$,\footnote{Note that this is the same boundary term choice as \cite{Aguilar-Gutierrez:2024nst}, which can be motivated for defining a notion of ADM mass for SdS$_3$ space with respect to $\mathcal{I}^+$, as elaborated in \cite{Balasubramanian:2001nb}, and it has also been applied in $\TT+\Lambda_2$ deformations in \cite{Coleman:2021nor,Batra:2024kjl}.} which, like \eqref{eq:on shell action}, results in
\begin{equation}
    I_{\rm E}=S_{\rm dS}-\beta_T E_{\rm BY}(r_B)~,
\end{equation}
where $S_{\rm dS}=r_{\rm CH}/(4G_N)$ is the Gibbons-Hawking (GH) entropy \cite{Gibbons:1976ue,Gibbons:1977mu} (which does not include a topological contribution, unlike the term $\Phi_0$ in (\ref{eq:thermodynamic entropy})). Meanwhile the BY \cite{Brown:1992br} quasi-local energy with respect to the Dirichlet wall is given by
\begin{equation}\label{eq:E BY dS}
E_{\rm BY}=\begin{cases}
\frac{1}{8\pi G_N}\qty(r_{B}+\sqrt{r_{\rm CH}^2-r_{B}^2})~,&{\rm when}~r_{\rm CH}>r_{B}~,\\
\frac{1}{8\pi G_N}\qty(r_{B}+\sqrt{r_{B}^2-r_{\rm CH}^2})~,&{\rm when}~r_{\rm CH}<r_{B}~,
\end{cases}
\end{equation}
which is directly related to the energy spectrum of the triple-scaled SYK \eqref{eq:deformed spectrum IR/UV} as expected from the holographic dictionary at a finite cutoff \eqref{eq:EBY beta_Tolman}.
 
The corresponding inverse Tolman temperature is then given by
\begin{equation}\label{eq:tomperature new}
\beta_T=\beta\begin{cases}
    \sqrt{{r_{B}^2}-{r_{\rm CH}^2}}~,&{\rm when}~r_{\rm CH}>r_{B}~,\\
    \sqrt{{r_{\rm CH}^2}-{r_{B}^2}}~,&{\rm when}~r_{\rm CH}<r_{B}~,
\end{cases}\quad\beta=\frac{2\pi}{r_{\rm CH}}~.
\end{equation}
Performing a similar analysis of the heat capacity with a Dirichlet wall with a fixed radial location $r=r_{B}$, as in (\ref{eq:heat capacity}), results in
\begin{equation}\label{eq:heat capacity dS}
    C_{r_B}=-\frac{r_B}{8\pi G_N}\frac{4\pi^2\beta_T^2}{(4\pi^2-\beta_T^2)^{3/2}}~.
\end{equation}
Notice that the heat capacity for fixed location $r_{B}$ is always negative, which corresponds to the $\eta=-1$ case in (\ref{eq:heat capacity}).

Thermodynamic properties of dS$_2$ space are indeed in agreement with the analysis of the UV behavior of the deformed boundary theory in Sec.~\ref{ssec:spread}. Particularly, given $G(\Phi_B)=\pm\Phi_B^2$ then the dictionary entry for the deformation parameter (\ref{eq:def parameter}) 
\begin{equation}\label{eq:def par dS2}
y=\pm {1}/{r_B^2}~,    
\end{equation}
allow us to explicitly check of the dictionary \eqref{eq:EBY beta_Tolman} relating BY quasi-local energy \eqref{eq:E BY dS} and Tolman temperature \eqref{eq:tomperature new} to the energy spectrum of the $T^2$ deformed \eqref{eq:E(theta) eta+1} and $T^2+\Lambda_1$ deformed \eqref{eq:E(theta) eta-1} theories and microcanonical temperature \eqref{eq:microcanonical inverse temperature} respectively, as well as the corresponding heat capacity at fixed $y$ \eqref{eq:heat capacity2} with that at fixed $r_B$ \eqref{eq:heat capacity dS}. Thus, there is an exact match at the level of thermodynamic properties. We will extend this to dynamical observables and holographic entanglement entropy in the rest of the section.

\subsection{Hyperfast Krylov Spread Complexity Growth}\label{ssec:dS Complexity UV}
\begin{figure}
    \centering
    \subfloat[]{\includegraphics[height=0.27\linewidth]{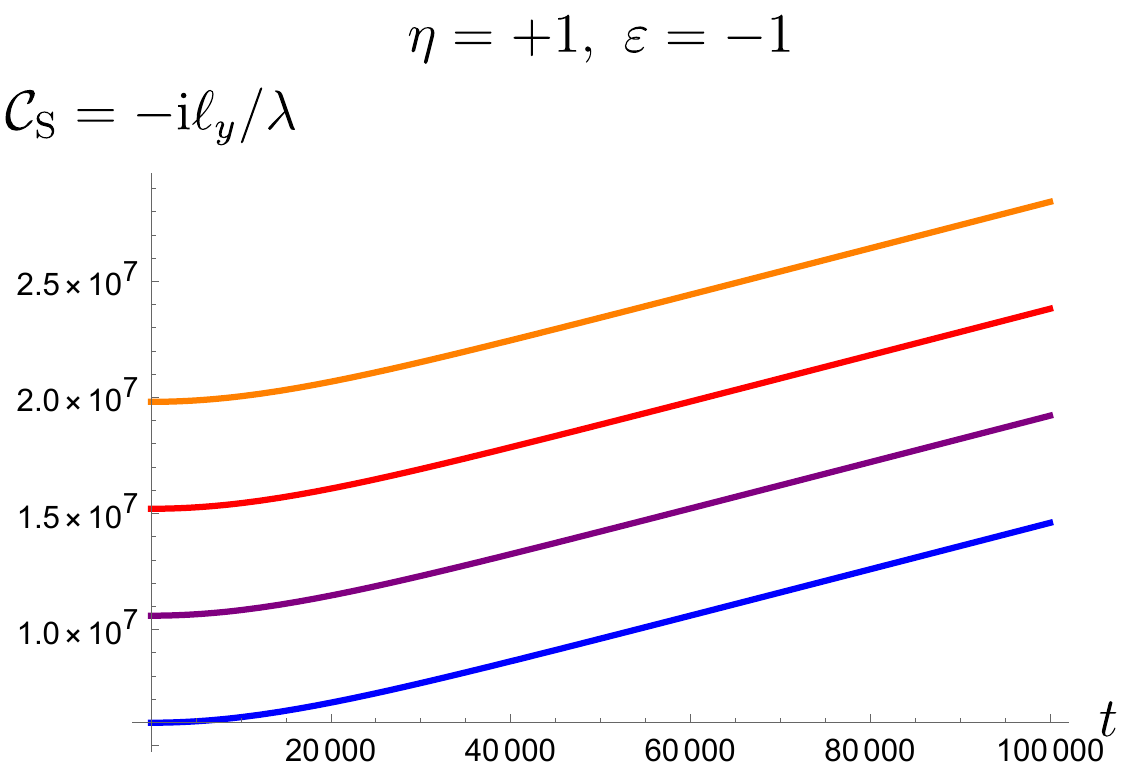}}  \hfill \subfloat[]{\includegraphics[height=0.27\linewidth]{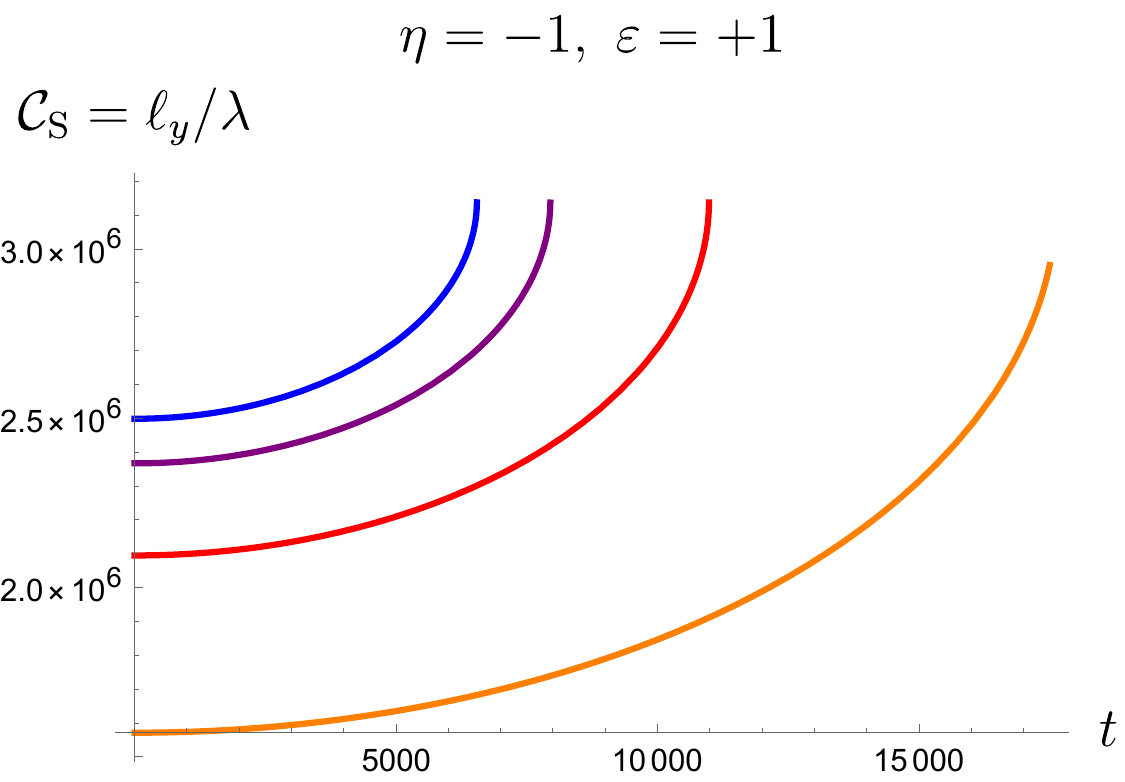}}\\
    \subfloat[]{\includegraphics[width=0.42\linewidth]{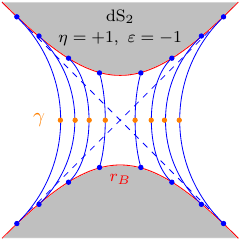}}  \hspace{1.5cm}  \subfloat[]{\includegraphics[width=0.425\linewidth]{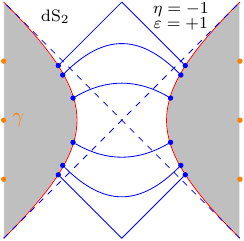}}
    \caption{\textit{Top}: UV triple-scaling limit of Krylov spread complexity \eqref{eq:Krylov complexity deformed DSSYK} in the HH state from (a) $T^2$, and (b) $T^2+\Lambda_1$ deformations of the DSSYK model using $\theta_{\rm UV}=10^{-4}$. \textit{Bottom}: Corresponding bulk interpretation of the $T^2$ and $T^2+\Lambda_1$ deformations in terms of (c) timelike and (d) spacelike extremal length geodesics at a finite cutoff in dS$_2$ space. We marked minimal codimension-two area (measured by the dilaton) points $\gamma$ homologous to the finite boundaries as orange dots, and the entangling surfaces as blue dots in the corresponding boundaries.}
\label{fig:Length_dS}
\end{figure}

Let us now discuss the expectation value of the Krylov spread complexity \eqref{eq:Krylov complexity deformed DSSYK} in the HH state in the UV limit \eqref{eq:Explicit Krylov basis deformed DSSYK} in the microcanonical ensemble , which we repeat for the convenience of the reader,
\begin{equation}\label{eq:LUV hyperfast}
\begin{aligned}
\mathcal{C}_{\rm S}(t)&\equiv\eval{\sum_{n=0}^\infty n\frac{\abs{\bra{\Omega}\rme^{-\tau^*\hH_y}\ket{K_n^{(y)}}}^2}{Z_y(\beta)}}_{\tau=\frac{\beta(\theta)}{2}+\rmi t}\\
&\underset{\rm UV}{\rightarrow}\begin{cases}
    \frac{2}{\lambda}{\rm arcsinh}\qty(\sqrt{\frac{1}{y{\theta_{\rm UV}}^2}-1}~\cosh(\frac{\theta_{\rm UV}t}{2}))~,&0\leq y\leq y_0~,\\
    \frac{2}{\lambda}\arcsin\qty(\sqrt{1-\frac{1}{y{\theta_{\rm UV}}^2}}~\cosh(\frac{\theta_{\rm UV}t}{2}))~,&y\geq y_0~.
    \end{cases}
    \end{aligned}
\end{equation}
We display the growth of this observable in Fig.~\ref{fig:Length_dS}.
Similar to the IR case there is eternal growth in the $\eta=+1$ case, and a limit in the evolution for the $\eta=-1$. The first observation is in agreement with \cite{Heller:2025ddj} (for the seed theory case), who finds that the expectation value of the chord number in the regime we have considered as the Krylov spread complexity of the corresponding state. The authors interpret this as a dual observable to a geodesic length in dS$_2$ space, which is thus a concrete manifestation of the CV conjecture \cite{Susskind:2014rva}. It eternally evolves in terms of the spacelike coordinate $t$ in \eqref{eq:dS2 metric}. Our results can be seen as a finite cutoff realization. Meanwhile, by performing the $T^2+\Lambda_1$ deformation, the rescaled chord number has a finite range of time evolution. In the bulk interpretation, this means that the corresponding geodesic curves become spacelike instead of timelike. This indicates that the cutoff surface has crossed the cosmological horizon in the bulk dual, as depicted in Fig.~\ref{fig:Length_dS}, which is consistent with the $\TT+\Lambda_2$ interpretation \cite{AliAhmad:2025kki,Chang:2025ays} in one lower spacelike dimension, as well as our findings in the AdS$_2$ case (Fig.~\ref{fig:spread_time}). However, note that there is no enhanced growth as the cutoff surface approaches the cosmological horizon since we work in the canonical ensemble which does not include the corresponding redshift factor. Additionally, one finds from \eqref{eq:LUV hyperfast} that the maximal range of evolution is:
\begin{equation}\label{eq:tcrit}
t\in[-t_{\rm crit},~t_{\rm crit}]~,\quad t_{\rm crit}\equiv\frac{1}{\theta_{\rm UV}}{\rm arccosh}\qty(\sqrt{1-\frac{1}{y\theta_{\rm UV}^2}})~.
\end{equation}
The result agrees with the hyperfast complexity growth conjecture (iii) and (iv) in Sec.~\ref{sec:intro}; the Krylov spread complexity reaches its maximum at a critical time $t_{\rm crit}$ which is order one in units of the dS length scale.

In the bulk interpretation, \eqref{eq:tcrit} is consistent with previous results in the CV conjecture in dS$_{d+1\geq 3}$ with a cosmological stretched horizon \cite{Jorstad:2022mls}. While the previous proposal does not discuss an explicit boundary dual, it was noticed that the extremal volume surfaces do not evolve eternally since there is a maximum value for the timelike coordinate, similar to $t=t_{\rm crit}$ above. At this point, the extremal surfaces reach $\mathcal{I}^\pm$, as seen in Fig.~\ref{fig:two-step_def} which diverges in higher dimensions. Thus, our results provide an explicit confirmation of the CV proposal in dS$_2$ space, and it displays the hyperfast growth expected by \cite{Susskind:2021esx}.

\subsection{Hyperfast Scrambling From OTOCs}\label{ssec:hyperscrambling dS}
The crossed four-point function of the deformed DSSYK model in the semiclassical limit was calculated in \eqref{eq:G4 crossed} and its scrambling time in \eqref{eq:scrambling time}. Now, we are interested in its UV triple-scaling limit \eqref{eq:H(dS)JT}, where $\tilde\theta_{\rm UV}\equiv(\pi-\theta)/\lambda$ is fixed as $\lambda\rightarrow0$. Since the analysis follows similarly to the IR limit of the crossed four-point function \eqref{eq:generating function 1 particle} with two operators of conformal dimension $\Delta_0$ and two with dimension $\Delta_1$, we specialize in the OTOC case, where $t_L=-t_R\equiv t$, as explained in \eqref{eq:G4 crossed}. Using the corresponding canonical variables, the UV triple-scaling limit of \eqref{eq:generating function 1 particle} becomes
\begin{align}\label{eq:OTOC UV}
&\begin{aligned}
    &G_{4}(t,-t;\theta,\pi-\theta)\underset{\rm UV}{\rightarrow}\qty(\frac{\rme^{-\ell_0/2}}{1+\qty(1-\frac{2\rme^{-\ell_0}}{{\tilde\theta_{\rm UV}}^2})\sinh^2(\frac{\lambda_L}{2} t)})^{2\Delta_0}~,
\end{aligned}
\end{align}
where $\lambda_L\equiv2\pi/\tilde{\beta}_y$ is the corresponding Lyapunov exponent in the scrambling phase of the OTOC \eqref{eq:G4 crossed}, and similar to \eqref{eq:new e ell0}, the constant $\ell_0$ above is
\begin{equation}
    \begin{aligned}
    \rme^{-\ell_0}\equiv\rme^{-\ell_*(q^{\Delta_1}, \theta,\pi-\theta)}\underset{\rm UV}{\rightarrow}&~\frac{1}{2}\qty(\tilde\theta_{\rm UV}^2+\Delta_1^2-\Delta_1\sqrt{2\tilde\theta_{\rm UV}^2+\Delta_1^2})~,
\end{aligned}
\end{equation}
with $\ell_*$ defined in
\eqref{eq:ell cond thetaLR}. The scrambling time is obtained by Taylor expanding \eqref{eq:OTOC UV} in powers of $\rme^{\lambda_L t}$ by
\begin{equation}
    G_{4}\simeq\qty(\frac{2\tilde\theta_{\rm UV}^2\rme^{-\ell_0}}{\tilde\theta_{\rm UV}^2+2\rme^{-\ell_0}})^{2\Delta_0}\qty(1-\Delta_0\frac{\tilde\theta_{\rm UV}^2-2\rme^{-\ell_0}}{\tilde\theta_{\rm UV}^2+2\rme^{-\ell_0}}\rme^{\lambda_Lt})~,\quad t\ll t_{\rm sc}~,
\end{equation}
where
\begin{subequations}
\begin{align}
    \lambda_L&\equiv\frac{2\pi}{\tilde\beta_y}\underset{\rm UV}{\rightarrow}\begin{cases}
        \frac{\tilde\theta_{\rm UV}}{\sqrt{1- y{\tilde\theta_{\rm UV}}^2}}~,&0\leq y_0\leq y_0~,\label{eq:Lyapunov exp}\\
        \frac{\tilde\theta_{\rm UV}}{\sqrt{y{\tilde\theta_{\rm UV}}^2-1}}~,&y\geq y_0~,
    \end{cases}\\
    t_{\rm sc}&\underset{\rm UV}{\rightarrow}
 \frac{1}{\lambda_L}\log\frac{2\qty({\tilde\theta_{\rm UV}}^2+2\rme^{-\ell_0})}{{\tilde\theta_{\rm UV}}^2-2\rme^{-\ell_0}}~.\label{eq:t scr dS}
\end{align}
\end{subequations}
Note that the latter expression is consistent with (3.83) in \cite{Ambrosini:2024sre}. 
Manifestly, the result \eqref{eq:t scr dS} agrees with the hyperfast scrambling conjecture (ii) and (iv) in Sec.~\ref{sec:intro}. The scrambling time of the OTOC is order one, corresponding to the dS scale, instead of order $t_{\rm sc}\sim\log S\propto \log N$ \cite{Goel:2023svz,Aguilar-Gutierrez:2025otq} the expected scrambling time for fast scrambler, which diverges in the double-scaling limit.
Next, we see that the presence of the redshift factor in $\lambda_L$, the limit $y\rightarrow y_0$ reproduces the infinite temperature limit, confirming the hyperfast conjecture in \cite{Susskind:2021esx}.

\subsection{Holographic Entanglement Entropy and dS Space}\label{ssec:HEE dS}
In this section, we discuss the algebraic entanglement entropy \eqref{eq:def entanglement entropy} of the DSSYK model in the UV limit and its holographic interpretation. Similar to the previous discussion, we focus on the chord number state $\ket{K_n^{(y)}}$. 

\paragraph{Chord Number State}
From the boundary side, the algebraic entanglement entropy given the state $\ket{K_n^{(y)}}$ in the UV triple-scaling limit \eqref{eq:H(dS)JT} can be evaluated using very analogous steps as in Sec.~\ref{ssec:EE}  with the corresponding limit of the Hamiltonian \eqref{eq:H IR UV} using the symmetry in the integration measure $\mu(\theta)$ \eqref{eq:norm theta} around $\theta=\pi/2$. Specifically, \eqref{eq:new S ell y} extends to the UV case as
\begin{equation}\label{eq:dS WKB Delta S}
       S\underset{\rm UV}{\rightarrow}\frac{2\pi}{\lambda\sqrt{y}} \sec\frac{\ell_{y}}{2}~.
\end{equation}
Now, we discuss its bulk interpretation, including its limit as the GH entropy in the stretched horizon limit.

\paragraph{RT Surfaces From Spacelike Boundaries}
We are interested in evaluating the RT formula in dS JT gravity, i.e.~\eqref{eq:Dilaton-gravity theory} with $U(\Phi)=-2\Phi$ as
\begin{equation}\label{eq:dS RT formula}
    S_{\rm dS}\equiv \frac{{\Phi_{\rm dS}(\gamma)}}{4G_N}~,
\end{equation}
where the finite cutoff boundaries are spacelike, as displayed in Fig.~\ref{fig:Length_dS} (c).

First, consider global coordinates, where the metric and the dilaton take the form
\begin{equation}\label{eq:global coord}
    \rmd s^2=-\rmd T^2+\cosh^2 T~\rmd \varphi^2~,\quad \Phi_{\rm dS}=r_{\rm CH}\sin\varphi\cosh T~.
\end{equation}
To map to static patch coordinates in \eqref{eq:dS2 metric} (with gauge fixing $t_L=t_R\equiv t/2$) describing the Milne patch (i.e.~outside the cosmological horizon in Fig.~\ref{fig:two-step_def} (d)) we need to use $r_{\rm CH}~t\rightarrow r_{\rm CH}~t+\rmi \frac{\pi}{2}$ \cite{Faruk:2023uzs}, so that
\begin{subequations}\label{eq:coordinate change}
    \begin{align}
    \sqrt{\frac{r^2}{r_{\rm CH}^2}-1}\cosh\frac{r_{\rm CH} t}{2}&=\sinh T~,\quad {\rm when}~r>r_{\rm CH}~,\\
    \sqrt{\frac{r^2}{r_{\rm CH}^2}-1}\sinh\frac{r_{\rm CH} t}{2}&=\cos\varphi~\cosh T~.
\end{align}
\end{subequations}
From symmetry of dS$_2$ space the finite cutoff boundaries are located at $T=0$ in global coordinates, as displayed in Fig.~\ref{fig:Length_dS} (c). Following the same derivation as in App.~\ref{app:geodesics} for the dS$_2$ space  with $r=r_B>r_{\rm CH}$ for geodesics anchored at spacelike boundaries in the Milne patch (Fig.~\ref{fig:two-step_def}), we find
\begin{equation}\label{eq:LdS def}
    L_{\rm dS}\equiv\int\rmd\xi~\sqrt{g_{\mu\nu}\dv{x^\mu}{\xi}\dv{x^\nu}{\xi}}=-2\rmi~{\rm arcsinh}\sqrt{\qty(\frac{r_B}{r_{\rm CH}}-1)\cosh\frac{r_{\rm CH}t}{2}}~.
\end{equation}
Then, using the relation between the coordinates in \eqref{eq:coordinate change} with $r=r_B>r_{\rm CH}$ for the entangling surface we get
\begin{equation}\label{eq: eval dS dilaton}
    \eval{\Phi_{\rm dS}}_{T=0}=r_B\sec\frac{L_{\rm dS}}{2}~.
\end{equation}
Then, the evaluation of the RT formula in dS$_2$ space \eqref{eq:dS RT formula} gives
\begin{equation}\label{eq:final dilaton }
    S_{\rm dS}=\frac{r_B}{4G_N}\sec\frac{L_{\rm dS}}{2}~.
\end{equation}
Note that the only difference with respect to the JT gravity setting \eqref{eq:JT gravity Sbulk} is in replacing $L\rightarrow-\rmi L_{\rm dS}$, which is consistent with the boundary evaluation. In particular, note that in the stretched horizon limit, i.e.~$r_B\rightarrow r_{\rm CH}$, we recover the GH entropy of dS$_3$ space, since $L_{\rm dS}\rightarrow0$ \eqref{eq: eval dS dilaton}
\begin{equation}
    S_{\rm dS}=\frac{2\pi r_{\rm CH}}{4 G_3}~,
\end{equation}
where we again used that $2\pi G_N=G_3$ \cite{Svesko:2022txo}, and $2\pi r_{\rm CH}$ is the corresponding codimension-two area in dS$_3$ space. Due to the general correspondence with the entanglement entropy in the deformed theory \eqref{eq:dS WKB Delta S} the GH entropy can indeed be interpreted as entanglement entropy from the perspective of the microscopic theory. The BH entropy in the $y\rightarrow 0$ limit can also be recovered as explained in \cite{Aguilar-Gutierrez:2025otq}.

\paragraph{RT Surfaces From Timelike Boundaries}
At last, we search to evaluate the RT formula for dS JT gravity  \eqref{eq:Dilaton-gravity theory} (with $U(\Phi)=-2\Phi$) in \eqref{eq:dS RT formula} using global coordinates \eqref{eq:global coord} when the entangling surface are points along the finite cutoff boundaries that reside in the static patch (e.g.~the cosmological stretched horizon case) in Fig.~\ref{fig:Length_dS} (d).\footnote{Note that while we use entangling surface points in the two-dimensional context, in higher dimensions one can adopt spatial subregions, which may lead to a larger variety of the minimal codimension-two area with boundaries in the static patch \cite{Doi:2022iyj,Chang:2025ays}.} The relationship between static patch coordinates and global dS$_2$ coordinates in the case of interest is similar to \eqref{eq:coordinate change} without the analytic continuation between Milne and static patch, namely
\begin{subequations}
    \begin{align}
    \sqrt{1-\frac{r^2}{r_{\rm CH}^2}}\sinh\frac{r_{\rm CH} t}{2}&=\sinh T~,\quad {\rm when}~r<r_{\rm CH}\\
    \sqrt{1-\frac{r^2}{r_{\rm CH}^2}}\cosh\frac{r_{\rm CH} t}{2}&=\cos\varphi~\cosh T~.
\end{align}
\end{subequations}
Analogous to the AdS case in Sec.~\ref{ssec:HEE AdS}, the homology constraint with $r$ timelike boundaries implies that the dilaton in dS JT gravity \eqref{eq:global coord} is extremized in the $\varphi$ direction ($\partial_\varphi \Phi_{\rm dS}=0$) and for a fixed $T$ coordinate. This leads to $\varphi=\pi/2$, which is a maximum value for the dilaton \eqref{eq:global coord}, while the minimum value is reached when $\varphi=0$ or $\pi$, corresponding to the poles of the static patch. However, since the dilaton is trivial for this minimum, this would not even be connected to the previous result the extremal value of the dilaton with spacelike boundaries \eqref{eq:final dilaton }, and neither to the boundary result for the algebraic entanglement entropy. This indicates that the RT formula does not apply after the $T^2+\Lambda_1$ deformation, and one might need to search for a different geometric quantity dual to the algebraic entanglement entropy found in \eqref{eq:dS WKB Delta S}.

Thus, the result differs from Susskind’s conjectures (i) and (iv) in Sec.~\ref{sec:intro}, which expected the RT formula to hold when the corresponding boundary is inside the static patch instead of the Milne patch. However, in the strict limit where the boundary is at the cosmological horizon one recovers the expected GH entropy. Meanwhile, in higher dimensions, one can have subregions in the stretched horizon and develop a different form of the RT formula in \cite{Susskind:2021esx}, which is not accessible from the two-dimensional analysis.

\section{Discussion}\label{sec:discussion}
In summary, we developed chord Hamiltonian deformations based on general dilaton gravity theories with Dirichlet boundaries and different boundary counterterms. In particular, this recovers $T^2$ deformations in (0+1)-dimension, as well as a lower-dimensional analog of $\TT+\Lambda_2$ deformations (``$T^2+\Lambda_1$'') as special cases. The latter deformation might clarify the relationship and differences in the bulk interpretation of $\TT+\Lambda_2$ by \cite{Coleman:2021nor} and \cite{AliAhmad:2025kki}. Namely, one recovers the same deformation by either changing the dilaton potential, or the counterterm in the dilaton gravity action \eqref{eq:Dilaton-gravity theory}, which modifies the location of the finite cutoff boundaries, crossing the location of the corresponding horizon \cite{AliAhmad:2025kki}. While the procedure can be applied in general (0+1)-dimensional quantum mechanical systems with a bulk dual, we specialized most of the study in the DSSYK model for concreteness. 

We evaluated the semiclassical thermodynamic properties of the deformed model from its partition function. We investigated the Krylov spread complexity of the HH state in the deformed theory and matched it to a wormhole length in finite cutoff holography. We also deduced expressions for general and semiclassical correlation functions in the deformed theory. Remarkably, the algebraic entanglement entropy between the double-scaled algebras of the DSSYK given the new chord number basis precisely matches the dilaton at the extremal surface in the JT and dS JT gravity once we restrict to the IR and UV triple-scaling limits respectively. By combining the previous results in the UV limit of the DSSYK chord spectrum and applying a sequence of $T^2$ and $T^2+\Lambda_1$ deformations, we explicitly realize the stretched horizon proposal by Susskind \cite{Susskind:2021esx}, including its characteristic hyperfast growth for dynamical observables. This provides a direct link between Susskind's conjectures \cite{Susskind:2021esx} with sine dilaton gravity, which recovers dS$_2$ space in a specific limit \cite{Blommaert:2024whf} associated with our UV triple-scaling limit, and therefore also with Schwarzschild-dS$_3$ holography \cite{Blommaert:2025eps}. This leads to a more precise connection between different proposals for the bulk dual of the DSSYK model.

Now, we discuss directions for future investigation.

\subsection{Outlook}\label{ssec:Outlook}
\paragraph{Krylov operator complexity}\label{ssec:Kcomplexity}
In the main text, we discussed Krylov spread complexity, which has a bulk interpretation in terms of bulk geodesics at finite cutoff in (A)dS$_2$ space. This connects with literature on holographic complexity with a finite cutoff in the bulk, e.g.~\cite{Jorstad:2022mls,Baiguera:2023tpt,Anegawa:2023wrk,Anegawa:2023dad,Auzzi:2023qbm,Baiguera:2024xju,Aguilar-Gutierrez:2023zqm,Aguilar-Gutierrez:2023pnn,Aguilar-Gutierrez:2024rka,FarajiAstaneh:2024fpv}. In the main text, we also explored the effect of the deformations in the semiclassical correlation function, which are only affected by a modification in the inverse temperature to be the Tolman temperature. It is then natural to investigate what would be the effect of the Hamiltonian deformation in Krylov operator complexity.

For concreteness, consider the one-particle HH state in \eqref{eq:HH state tL tR} for $m=1$ which can be expressed as
\begin{equation}\label{eq:Lanczos eta}
    \ket{\Psi_{\Delta_1}(\tau_L=\kappa\tau,\tau_R=\tau)}=\rme^{-\hat{\mathcal{L}}_\kappa\tau}\ket{\Delta_1;0,0}=\sum_n\Psi_n^{(\kappa)}(\tau)\ket{K^{(\kappa)}_n}~,
\end{equation}
where 
\begin{equation}\label{eq:Liou}
\hat{\mathcal{L}}_\kappa\equiv\hH_{R,y}+\kappa\hH_{L,y}~,\quad \kappa\equiv\tau_L/\tau_R~.
\end{equation}
is a Liouvillian operator. Note that 
\begin{subequations}
\begin{align}
\kappa=1:\quad &\ket{\Psi_{\Delta_1}(\tau_L=\tau,\tau_R=\tau)}=\rme^{-\tau(\hH_L+\hH_R)}\ket{\Delta_1;0,0}~,\\
\kappa=-1:\quad &\ket{\Psi_{\Delta_1}(\tau_L=-\tau,\tau_R=\tau)}=\rme^{-\frac{\beta}{2}\hH_{L,y}}\hmO_{\Delta_1}(t)\rme^{-\frac{\beta}{2}\hH_{L,y}}\ket{\Omega}~.
\end{align}
\end{subequations}
where 
\begin{equation}
\hmO_{\Delta_1}(t)\equiv \rme^{\rmi t\hH_L}\hmO_{\Delta_1}\rme^{-\rmi t\hH_L}~,\quad \tau\equiv\frac{\beta}{2}+\rmi t~.
\end{equation}
The $\kappa=+1$ case corresponds to Schrödinger evolution with respect to a total Hamiltonian $\hH_L+\hH_R$ in complex-valued time $\tau$; while $\kappa=-1$ corresponds to Heisenberg evolution with respect to the single-sided Hamiltonian $\hH_L$. For this reason, based on the Liouvillian evolution operator $\hat{\mathcal{L}}_\kappa$, we can evaluate Krylov complexity for states and operators depending on the value of $\kappa$. Other choices of $\kappa$ are associated with generalized notions of Krylov complexity \cite{Aguilar-Gutierrez:2025pqp}. We will focus on $\kappa=\pm 1$.

Meanwhile, we refer to $\qty{\ket{K^{(\kappa)}_n}}$ as the Krylov basis for the reference state $\ket{K_0^{(\kappa)}}\equiv\ket{\Delta_1;0,0}$, which are defined to satisfy a Lanczos algorithm
\begin{equation}\label{eq:Liouville Lanczos}
    \hat{\mathcal{L}}_\eta\ket{K_n^{(\kappa)}}=a_n^{(\kappa)}\ket{K^{(\kappa)}_n}+b^{(\kappa)}_{n+1}\ket{K_{n+1}^{(\kappa)}}+b^{(\kappa)}_{n}\ket{K_{n-1}^{(\kappa)}}~,
\end{equation}
with $b_0^{(\kappa)}=0$, $a_0^{(\kappa)}=\bra{K_0^{(\kappa)}}\ket{K_0^{(\kappa)}}$, while $a_n^{(\kappa)}$ and $b_n^{(\kappa)}$ are called Lanczos coefficients. We define Krylov operator complexity as
\begin{equation}\label{eq:complexity eta}
    \mathcal{C}(t)\equiv \eval{\frac{\bra{\Psi_{\Delta}(-\tau,\tau)}\hat{\mathcal{C}}^{(\kappa)}\ket{\Psi_{\Delta}(-\tau,\tau)}}{\bra{\Psi_{\Delta}(-\tau,\tau)}\ket{\Psi_{\Delta}(-\tau,\tau)}}}_{\tau=\frac{\beta}{2}+\rmi t}~,\quad \hat{\mathcal{C}}^{(\kappa)}\equiv \sum_nn\ket{K_n^{(\kappa)}}\bra{K_n^{(\kappa)}}~.
\end{equation}
Similar to the spread complexity case, the effect of the deformation in the Liouvillian operator \eqref{eq:Liou} is to scramble the Krylov basis $\ket{K_n^{(\kappa)}}$. It would be interesting to deduce the semiclassical Krylov operator and spread complexity with a one-particle insertion. In particular, this could be useful to learn about these deformations in higher dimensional examples. For instance, it was reported in \cite{Chattopadhyay:2024pdj} that the exponential growth of Krylov operator complexity for CFTs might not agree with the OTOC chaos bound \cite{Maldacena:2015waa} when the deformation parameter has the sign associated to superluminal motion \cite{Cooper:2013ffa}, unlike in finite cutoff holography.

It would also be interesting to compare the standard definition of Krylov operator complexity above with the work by \cite{Bhattacharyya:2025gvd}, where the authors use a notion of Krylov operator complexity in a microcanonical window in JT gravity. They find that at leading order in the approximation, the Lanczos coefficients display linear growth, and Krylov complexity grows exponentially with a simple modification in the inverse temperature, which includes the corresponding redshift factor, similar to our findings for the semiclassical correlation functions. This suggests that one might find a redefinition of Krylov operator complexity \eqref{eq:complexity eta} in a microcanonical window which can be expressed as the first derivative of the OTOC at finite temperature in \eqref{eq:G4 crossed} based on the seed theory result \cite{Aguilar-Gutierrez:2025pqp} where one would incorporate the fake temperature with the corresponding redshift factor in \eqref{eq:fake temperature}. This redefinition of Krylov operator complexity would be associated to the wormhole length connecting the asymptotic boundaries of an AdS$_2$ black hole instead of the one with finite Dirichlet boundary \eqref{eq:wormhole length finite cutoff2} even when $\Delta\rightarrow0$. For this reason, it might also be fruitful to develop in detail the definition of Krylov operator complexity in \eqref{eq:complexity eta} without fixing an energy window, which might be more closely related to finite cutoff holography.

\paragraph{Generalized Boundary Conditions in Finite Cutoff Holography}
Our work has focused on the finite cutoff interpretation of Hamiltonian deformations, when the bulk dual has Dirichlet boundary conditions. There is literature regarding how to extend holography between Dirichlet through multitrace deformations, which is expected to modify the boundary conditions in the bulk \cite{Parvizi:2025shq,Parvizi:2025wsg,Sheikh-Jabbari:2025kjd,Witten:2001ua,Adami:2025pqr,Li:2025lpa}. In particular, there has been significant interest in developing conformal boundary conditions and its extensions \cite{Anderson_2008,Witten:2018lgb,An:2021fcq,Coleman:2020jte,Anninos:2023epi,Anninos:2024wpy,Anninos:2024wpy,Batra:2024qju,Banihashemi:2024yye,Galante:2025tnt,Banihashemi:2025qqi,Galante:2025emz,Anninos:2025zgr,Allameh:2025gsa,Anninos:2025zgr,Liu:2024ymn} to obtain well-posed solutions in general relativity when Dirichlet boundary conditions are ill-posed. Finite cutoff in AdS space with conformal boundary conditions was investigated in \cite{Coleman:2020jte,Allameh:2025gsa,Parvizi:2025shq,Parvizi:2025wsg}. Given that the DSSYK model is a useful analytically solvable model at different energy scales, it might be convenient to generalize our study, incorporating cutoff holography with conformal boundary conditions instead of Dirichlet in the dilaton-gravity theories, possibly along the lines of \cite{Galante:2025tnt}, and analyze how the thermodynamics properties of the model and its correlation functions are modified. Given the connection between the deformations in the DSSYK model and stretched horizon holography, this study could help us gain general lessons for static patch holography, since, as we saw, the theory is otherwise thermodynamically unstable. It is also relevant to study the perturbative stability of the solutions under background fluctuations \cite{Anninos:2024wpy}, which could be incorporated following previous studies in the dilaton-gravity context by \cite{Ishii:2019uwk,Okumura:2020dzb}.

Additionally, it should be possible to adapt our study in the mixed boundary condition interpretation where the DSSYK would stay at the original asymptotic boundary location (i.e.~$r_{B}$). The development of this interpretation in the Schwarzian of JT gravity has been previously performed in \cite{Gross:2019uxi}. We hope to come back to this point in future work to implement this interpretation in sine dilaton gravity. 

\paragraph{Entanglement entropy in \texorpdfstring{$T^2+\Lambda_1$}{} Deformation}
As emphasized in the main text, the algebraic entanglement entropy in the double-scaled algebras associated to a chord number state (which reproduces bulk geodesics at a finite cutoff) leads to a precise matching between the bulk entanglement entropy in the $T^2$ deformed theory and the dilaton in JT and dS JT gravity at the extremal surface predicted by the RT formula. However, once we move to $T^2+\Lambda_1$ deformations, the RT formula does not longer apply as the minimal value of the dilaton becomes trivial, as explained in Secs.~\ref{ssec:HEE AdS} and \ref{ssec:HEE dS}. It would be interesting to find a geometric quantity that one can associate with the corresponding algebraic entanglement entropy \eqref{eq:def entanglement entropy} given the chord number state in the Krylov basis $\ket{K_n^{(y)}}$ in the $T^2+\Lambda_1$ case, although the system is not even thermodynamically stable. For instance, a central part in Susskind's conjectures  are rules to implement RT formula calculations in stretched horizon holography in the bilayer vs monolayer proposals \cite{Shaghoulian:2022fop,Susskind:2021esx,Franken:2023pni,Franken:2024ruw}. These rules rely on angular dependence in dS$_3$ space, and they are thus inaccessible from the dS$_2$ perspective which is dual to the UV limit of the DSSYK model. In principle one may connect these additional RT proposals with the dS$_3$ embedding of sine dilaton gravity \cite{Blommaert:2025eps}.

Additionally, we suspect that one does not need to limit the evaluations to a triple-scaling limit to be able to interpret the algebraic entanglement entropy in the chord number state holographically. We would like to extend this result for the semiclassical regime in the DSSYK model \cite{Lin:2022rbf}, which is expected to have a bulk interpretation in sine dilaton gravity. One might derive further results in this direction by developing a Lewkowycz-Maldacena path integral derivation of the corresponding RT formula in the bulk \cite{Lewkowycz:2013nqa}, or possibly through a factorization map \cite{Lin:2018xkj,Lin:2017uzr,Blommaert:2018iqz,Lin:2021tlr,Jafferis:2019wkd,Mertens:2022ujr}.

\paragraph{Deforming Before Averaging}
Throughout our analysis, we deformed the DSSYK model after ensemble averaging since the Hamiltonian deformation can be interpreted as a radial bulk flow in the sine dilaton gravity theory. However, one might investigate the differences in the theory by applying first the deformation and then the averaging. This would mean a transformation in the SYK Hamiltonian \eqref{eq:SYK Hamiltonian}
\begin{equation}
    \hH_{\rm SYK}\rightarrow \hH_{{\rm SYK},y}=\frac{1}{y}\qty(1-\sqrt{1-2y \hH_{\rm SYK}})~.
\end{equation}
The ensemble average evaluation is more complicated due the square root factor, although one can make much progress by working in the $G\Sigma$-formalism, as in \cite{Gross:2019uxi}. It would be interesting to consider in this alternative case how the different thermodynamic properties and dynamical observables in our study are modified, such as Krylov spread complexity and correlation functions, and to understand the corresponding bulk geometry.

\paragraph{Verifying Sine Dilaton at Finite Cutoff}
Our work has mostly taken a boundary perspective without requiring a specific bulk dual to the DSSYK model beyond its IR and UV triple-scaling limits, as long as it corresponds to a dilaton gravity theory. Nevertheless, we found explicit results that suggest a natural bulk interpretation in terms of sine dilaton gravity, where Krylov spread complexity for the HH state in the deformed DSSYK in the semiclassical limit \eqref{eq:Krylov complexity deformed DSSYK} describes a wormhole geodesic at finite cutoff \eqref{eq:wormhole length finite cutoff2}, as well as the RT formula verification for the UV and IR triple-scaling limits in Sec.~\ref{ssec:spread}; see App.~\ref{app:sine dilaton gravity finite cutoff} for further exploration. As a next step, one could perform a bulk analysis of correlations functions to provide a concrete bulk dual interpretation of some of our other results. A simple start would be studying the functional dependence of correlation functions in the microcanonical ensemble and an overall redshift factor associated to the Tolman temperature in the semiclassical limit, as in the existing literature in JT gravity and higher dimensional cutoff holography \cite{AliAhmad:2025kki,Basu:2025exh,Aguilar-Gutierrez:2024nst,Iliesiu:2020zld,Gross:2019ach,Gross:2019uxi}. However, it is important to make further checks outside the AdS$_2$ regime of the bulk dual theory. For instance, in sine dilaton gravity, one could carry out a gravitational path integral calculation in the bulk, or study solutions to the WDW equation in the bulk at finite cutoff (similar to \cite{Iliesiu:2020zld}) to verify our boundary theory results for the partition functions and correlation functions.

Additionally, it was previously pointed out by \cite{Blommaert:2024ymv} that there is a defect interpretation of the thermodynamics in sine dilaton gravity related to an infinite degeneracy in the black hole and cosmological horizons due to the periodicity in the dilaton potential. Meanwhile, the finite cutoff interpretation of the Hamiltonian deformations in this work might allow to remove the degeneracy by placing the boundaries in appropriate locations in the complex-valued geometry. It would be interesting to investigate how the properties of the conical defect associated with the removed horizons can be matched with the exact deformed partition function in (\ref{eq:partition DSSYK}).

Moreover, while the practical computations are semiclassical, one could repeat our analysis of correlation functions and partition functions beyond the on-shell level by including one-loop corrections as in \cite{Goel:2023svz} for the $T^2$ deformed DSSYK, and comparing them with recent developments \cite{Bossi:2024ffa,Griguolo:2025kpi}. It is also important to extend our investigation with non-perturbative series expansions of our results in terms of Borel resummation and instanton processes following analogous studies in JT gravity \cite{Griguolo:2021wgy} to better understand the spectrum of the deformed theory and the putative bulk dual at a finite cutoff.

At last, a related interesting extension would be to explore finite cutoff holography using the chord Hamiltonian \eqref{eq:chord Hamiltonian zero particle}  when $q=\rme^{\rmi b}$ (where $b\in\mathbb{R}$) (instead of $q\in[0,1)$ in the main text) which is argued to be dual to a $\sinh\Phi$ dilaton gravity theory \cite{Belaey:2025kiu,Blommaert:2023wad}, and it is closely connected with Liouville gravity \cite{Mertens:2020hbs,Fan:2021bwt,Kyono:2017pxs,Suzuki:2021zbe,Goel:2020yxl,Collier:2023cyw}; see \cite{Griguolo:2025kpi} for recent progress.

\paragraph{Narovlansky-Verlinde Model}
Another interesting proposal for the bulk dual of the DSSYK model was made by V.~Narovlansky and H.~Verlinde in \cite{Narovlansky:2023lfz} (see also \cite{Verlinde:2024znh,Verlinde:2024zrh,Tietto:2025oxn,Blommaert:2025eps}),  who developed an approach towards dS$_3$ holography by considering a two-copy SYK model with an equal energy constraint. In the double-scaling limit and after ensemble-averaging, the system can be described by the zero-particle chord Hamiltonian \eqref{eq:chord Hamiltonian zero particle}. 

In the bulk proposal of \cite{Narovlansky:2023lfz}, the dual dilaton gravity theory has a potential (\ref{eq:potential}) of the form:
\begin{equation}\label{eq:NV potential}
    U(\Phi)=-2\cosh(\Phi)~.
\end{equation}
The $T^2$ deformation parameter is given by (\ref{eq:def parameter}) for the potential with the counterterm (\ref{eq:Gphi special}) is
\begin{equation}\label{eq:lambda NV}
    y=-\frac{\lambda}{\sinh \Phi_B}~.
\end{equation}
It would be interesting to test our boundary theory findings with finite cutoff holography in the NV model to check if it leads to manifestly different thermodynamics or correlation functions with respect to those in sine dilaton gravity at finite cutoff and compare both with our boundary theory results. In particular, our results about entanglement entropy at a finite cutoff have a clear geometric interpretation for (A)dS space when we zoom in the UV and IR triple-scaling limits. These limits are naturally compatible with sine dilaton gravity; it would be interesting to find if there are similar limit that can be associated to the dilaton gravity theory described by \eqref{eq:NV potential}, or whether one can show there are none unless $\Phi\rightarrow\rmi(\Phi)$ which would recover an equivalent theory to sine dilaton gravity. It would also be interesting to relate the NV model at finite cutoff to a dimensionally reduced version of the $\TT+\Lambda_2$ deformations \cite{Coleman:2021nor,Batra:2024kjl} similar to our proposal for $T^2+\Lambda_1$ deformations of the DSSYK model in the triple-scaling limit \eqref{eq:H IR UV}.

\paragraph{Relational Holographic Flow}In the future, we would like to study the relational interpretation in the $T^2$ flow, focusing on the dressed observables and the finite cutoff holographic dictionary. It has been noticed that quantum reference frame (QRF) transformation will significantly affect the thermodynamics, and algebraic operator structure of general systems \cite{DeVuyst:2024pop,Hoehn:2023ehz} (see also \cite{Susskind:2023rxm}). Given the explicit knowledge of the solution for the DSSYK model, it would be intriguing to study QRF transformations in this system explicitly and how the semiclassical thermodynamics studied here will be modified. More generally, it would be interesting to study the algebra of observables for the $T^2$-deformed DSSYK model with matter, as a microscopic realization of \cite{Silverstein:2022dfj}.

If one interprets the boundary in terms of an observer to define expectation values of gravitationally dressed operators in the bulk algebra, we might view that as we change the location of the cutoff boundary, the observer also changes due to the Hamiltonian deformation of the boundary theory. For instance, based on recent work by \cite{Blommaert:2025eps}, we could re-express sine dilaton gravity theory by a pair of Liouville fields; and we study how the flow in the Schrödinger evolution is captured by relational observables. The effect of the $\TT$ deformation is to change the Hamiltonian but not the proper time measured by a boundary observer. From the relational interpretation of the Liouville-CFT formulation of sine dilaton gravity this means that the collective field description of the DSSYK is modified, but not the time parameter itself corresponding to the observer's clock in the Lioville theory. It would thus be interesting to explore how the $\TT$ deformation changes the Page-Wootters mechanism \cite{Page:1983uc,Wootters:1984wfv} in the bulk theory. We expect there is a relational description in the bulk in terms proper time in sine dilaton gravity. In contrast, the boundary theory just experiences Schrödinger evolution since it only has access to gauge-invariant data of the bulk.

\paragraph{Deforming Matrix Models}$T^2$ deformations in an ensemble of matrix models have been previously considered in the JT gravity context by \cite{Rosso:2020wir}; which allowed them to study the energy spectrum of the deformed model for all real values of the deformation parameter without recurring to a mixed boundary condition interpretation in the bulk. It would be interesting to generalize the previous work of \cite{Rosso:2020wir} to explore how this type of interpretation works in a matrix model dual to the sine dilaton gravity model with a higher genus expansion, which is not well developed. It might also be interesting to perform the $T^2$ deformation of the two-matrix model of \cite{Jafferis:2022uhu,Jafferis:2022wez} that reproduces certain DSSYK amplitudes (and which has been further developed in \cite{Okuyama:2023kdo,Okuyama:2023yat,Okuyama:2024eyf}), to compare with our findings on semiclassical thermodynamics in Sec.~\ref{sec:thermo}, and possibly to investigate recent connections between JT gravity at finite cutoff with matrix models \cite{Griguolo:2025kpi}.

\paragraph{Charged DSSYK model and J$\overline{\text{T}}$ deformations}
In line with our analysis of $T^2$ deformations in the DSSYK model, there also exist J$\overline{\text{T}}$ deformations \cite{Guica:2017lia} in the literature \cite{Chakraborty:2020xwo}, where J is a chiral U(1) current. It would be interesting to investigate the nature of the bulk dual to the DSSYK model with a U(1) charge \cite{Berkooz:2020uly} (see also \cite{Davison:2016ngz,Sachdev:2015efa,Gu:2019jub}), which has been recently discussed in \cite{Arundine:2025mcu,Gubankova:2025gbx}. One could then introduce J$\overline{\text{T}}$ deformations, or alternatively a combination of $T^2$, J$\overline{\text{T}}$ deformations in the DSSYK model and study the thermodynamic properties and correlation functions we have encountered in this work. In particular, it would be interesting to evaluate the corresponding Krylov spread complexity (away from the vanishing deformation parameter $y\rightarrow0$ limit, discussed by \cite{Forste:2025gng}). A related possibility in this direction is using the $\mathcal{N}=2$ extension of the DSSYK model \cite{Berkooz:2020xne} which counts with an R-charge. It would be very interesting to extend the results deforming the theory and finding the corresponding bulk length at finite cutoff, particularly for the BPS sector \cite{Lin:2022zxd}, its interpretation as chord number \cite{Boruch:2023bte} and in terms of Krylov spread complexity \cite{Aguilar-Gutierrez:2025sqh}.

\paragraph{Stretched Horizon in Higher Dimensions}
Recently, \cite{Silverstein:2024xnr} found an extension of $\TT+\Lambda_2$ deformations in higher dimensions, which we denote $T^2+\Lambda_d$ deformations. Explicitly, consider deformation the action of a CFT$_d$ as
\begin{equation}\label{eq:action deformed theory}
\dv{I_{\rm bdry}}{y}=\int d^{d-1}x\,\sqrt{-h} \left[ \frac{\pi y}{d} :\left(\tilde T_{\mu\nu} \tilde T^{\mu\nu}- \frac{1}{d-1} (\tilde T^\mu_\mu)^2 \right):-\frac{d^2(d-1)}{4\pi} \frac{\eta-1}{y}+ \frac{d}{4\pi} \left(\frac{C_d^2}{y^{d-2}} \right)^{1/d} R^{(d)}\right]\,.    
\end{equation}
with $h_{\mu\nu}$ being the metric of the boundary theory in $d$ space-time dimensions, with an associated Ricci scalar $R^{(d)}$; $y$ is the deformation parameter and the stress tensor is $\tilde T_{\mu\nu}=T_{\mu\nu}+\tilde T^{\rm CT}_{\mu\nu}$ where $:\;\; :$ represents large-$N$ normal ordering as described in \cite{Hartman:2018tkw}, where CT denotes the counterterm contribution, and $\eta$ is a  parameter corresponding to $T^2$ ($\eta=+1$) and $T^2+\Lambda_d$ ($\eta=-1$) deformations.

The authors associate the deformed theory \eqref{eq:action deformed theory} to Einstein gravity in (A)dS$_{d+1}$ space at finite cutoff:
\begin{equation}\label{eq-action-Lorentzian}
\begin{aligned}
    I_{\rm bulk} =& \frac{1}{16\pi G_N} \int_{\mathcal{M}} d^{d+1}x\,\sqrt{-g} \left(R^{(d+1)} + \frac{d(d-1)}{\ell^2}\eta\right)\\
    &+\frac{1}{8\pi G_N} \int_{\partial \mathcal{M}} d^dx\,\sqrt{-h} \left(K +{\rm counterterm} \right)  ~,
\end{aligned}
\end{equation}
where $\mathcal{M}$ represents the spacetime manifold, $G_N$ Newton's constant, $g_{\mu\nu}$ the bulk metric, $R^{(d+1)}$ the Ricci scalar, and $K$ the mean curvature.

The authors find an associated (dimensionless) energy spectrum for the deformed theory, which takes the form
\be\label{eq:dressed-gen}
\mathcal{E}_y=\frac{d(d-1)}{2\pi y} \left[\sqrt{1+ \Omega^{2/(d-1)} (C_d y)^{2/d}}\Big|_{y^{ (\lceil d-2 \rceil)/d }} \mp \sqrt{\eta+\Omega^{2/(d-1)} (C_d y)^{2/d}-\frac{4\pi}{d(d-1)}y \mathcal{E}_0}\; \right]~,
\ee
where $C_d$ represents constants determined by the counterterms and $\mathcal{E}_0$ is the (dimensionless) energy spectrum of the seed theory. Using the available tools to repeat the analysis in this work in higher dimensional settings, it would be interesting to verify our findings on the stretched horizon, such as the hyperfast scrambling in the OTOCs and holographic entanglement entropy from the algebraic perspective in the higher dimensional settings. One difficulty is to identify a specific boundary theory in the dS$_{d+1}$ case.

\section*{Conﬂict of Interest}None of the authors have a conﬂict of interest to disclose.

\section*{Data Availability Statement}The data that support the findings of this study are available from the corresponding author upon reasonable request.

\section*{Keywords}Hamiltonian deformation, gauge/gravity correspondence, double-scaled SYK model, anti-de Sitter black holes, de Sitter space, lower-dimensional quantum gravity.

\section*{Acknowledgments}
I thank Andrew Svesko, and Manus Visser for previous collaboration on \cite{Aguilar-Gutierrez:2024nst}, which partially motivated this work; to Gonçalo Araújo-Regado, Philipp A. Höhn, Francesco Sartini, Jiayue Yang and Ming Zhang for collaboration in other works with related themes; to Takanori Anegawa, Andreas Blommaert, Klaas Parmentier, and Jiuci Xu for useful comments on an earlier version of the draft; and to Dio Anninos, Luis Apolo, Micha Berkooz, Arghya Chattopadhyay, Chuanxin Cui, Victor Franken, Damián A. Galante, Eleanor Harris, Taishi Kawamoto, Simon Lin, Masamichi Miyaji, Sangjin Sin, Yuan Sun, Andrew Rolph, and Gopal Yadav for insightful discussions/comments. I also thank the organizers of the ``V Siembra-HoLAGrav Young Frontiers Meeting'', the ``Quantum Extreme Universe: Matter, Information and Gravity 2024'' workshop in Okinawa Institute of Science and Technology Graduate University (OIST), ``New Avenues in Quantum Many-body Physics and Holography'' in Asia-Pacific Centre (APCTP) for Theoretical Physics, the ``Young Researchers of Quantum Gravity'' online seminars, ``Strings 2025'' in New York University Abu Dhabi; the ``QISS 2025 Conference'' in Vienna University, and ``2025 East Asia Joint Workshop on Fields and Strings'' in OIST, and the joint Israeli seminars at Neve Shalom where part of the work was presented in different formats; to the ``XX Modave summer school'' for allowing relevant discussions with other participants; to the high energy physics groups at Rikkyo University  and the Weizmann Institute of Science for hospitality while this work was in finishing stages. SEAG acknowledges the APCTP, the Office of Scholarships, and the QISS consortium for travel support.

\section*{Statement of Funding}
SEAG was partially supported by the FWO Research Project G0H9318N and the inter-university project iBOF/21/084 when this project started. SEAG is supported by the Okinawa Institute of Science and Technology Graduate University. This work was made possible through the support of the WOST, WithOut SpaceTime project (\hyperlink{https://withoutspacetime.org}{https://withoutspacetime.org}), supported by Grant ID\# 63683 from the John Templeton Foundation (JTF), and ID\#62312 grant from the JTF, as part of the ‘The Quantum Information Structure of Spacetime’ Project (QISS), as well as Grant ID\# 62423 from the JTF. The opinions expressed in this work are those of the author(s) and do not necessarily reflect the views of the John Templeton Foundation. 

\appendix
\section{Notation}\label{app:Notation}
\paragraph{Acronyms}
\begin{itemize}[noitemsep]
\item ADM: Arnowitt–Deser–Misner.
    \item (A)dS: (Anti-)de Sitter.
\item AGH: Almheiri-Goel-Hu
\item ASC: Al-Salam Chihara.
\item BTZ: Bañados-Teitelboim-Zanelli.
\item BH: Bekenstein-Hawking.
\item BY: Brown-York.
    \item CFT: Conformal field theory.
     \item CV: Complexity=volume.
    \item (DS)SYK: (Double-scaled) Sachhev-Ye-Kitaev
    \item ETW: End-Of-The-World
    \item GH: Gibbons-Hawking.
    \item HH: Hartle-Hawking
    \item IR: Infrared
    \item JT: Jackiw-Teitelboim
    \item OTOC: Out-of-time-order correlations.
    \item QFT: Quantum field theory
    \item SC: Square correlators.
    \item RT: Ryu-Takayanagi.
\item WKB: Wentzel-Kramers-Brillouin.
    \item UV: Ultraviolet
    \item WDW: Wheeler–DeWitt
\end{itemize}
\paragraph{Notation}
\begin{itemize}[noitemsep]
\item $F(r)$ \eqref{eq:metric}: Blackening factor.
    \item $G(\Phi_B)$ \eqref{eq:Dilaton-gravity theory}: Boundary counterterm, where $\Phi_B(\equiv r_B)$ denotes the dilaton at the finite or asymptotic boundary.
     \item $h_{\mu\nu}$ \eqref{eq:Dilaton-gravity theory}: Induced metric.
    \item $\gamma_{ab}$ \eqref{eq:IE as stress tensor}: Boundary metric.
    \item $U(\Phi)$ \eqref{eq:Dilaton-gravity theory}: Dilaton potential.
    \item $\mathcal{H}$ \eqref{eq:HJ eq}: Hamiltonian constraint.
    \item $\tilde{T}_{ab}$ \eqref{eq:BY stress tensor}: Bulk matter stress tensor.
    \item $T_{ab}$ \eqref{eq:dictionary T}: Boundary matter stress tensor.
    \item $\mathcal{O}_{\Phi_B}$ \eqref{eq:source Phi}: Dilaton source.
    \item $K$, $\mathcal{R}$ \eqref{eq:Dilaton-gravity theory}: Mean curvature, and Ricci scalar.
    \item $Z_y(\beta)$ \eqref{eq:def partition function}: Deformed partition function without matter.
    \item $y$ and $\eta$ \eqref{eq:def parameter}: Deformation parameters.
    \item $E(\theta)=\frac{-2J~\cos(\theta)}{\sqrt{\lambda(1-q)}}$ (\ref{eq:thermodynamic entropy}): DSSYK energy spectrum, where $\theta$ is a parametrization.
    \item $E_y$ \eqref{eq:flow eq}: Deformed energy spectrum.
    \item $\hH_y=f_y(\hH)$ \eqref{eq:flow eq operator form}: Deformed Hamiltonian, with $f_y$ a generic analytic function.
\item $S(\theta)$ (\ref{eq:thermodynamic entropy}): Thermodynamic entropy.
\item $\rho_y$ \eqref{eq:DOS TTbar}: Density of states of the deformed theory.
\item $\beta_y$ \eqref{eq:microcanonical inverse temperature}: Microcanonical inverse temperature.
\item $C_y$ \eqref{eq:heat capacity}: Heat capacity for fixed deformation parameter.
\item $N$, and $p$ \eqref{eq:SYK Hamiltonian}: Number of Majorana fermions and all-to-all interactions respectively.
    \item $q\in[0,1)$ \eqref{eq:double scaling}: q-deformation parameter, the penalty factor associated to DSSYK Hamiltonian interceptions in chord diagrams. 
\item $\lambda=2N/p^2\equiv-\log q$: Parameter in the double-scaling limit (\ref{eq:double scaling}).
\item $\mathcal{H}_m$ and $\mathcal{H}_{\rm chord}$ \eqref{eq:identity theta}: Chord space with $m$ operators, and an all-operator extension.
    
    \item $\ket{K_n^{(y)}}$ \eqref{eq:Explicit Krylov basis deformed DSSYK}: Krylov basis in the deformed DSSYK model.
    \item $\hat{n}_y$ \eqref{eq:n_y}: Chord number operator in the deformed DSSYK model.
    \item $\ket{K^{(\kappa)}_n}$, $a_n^{(\kappa)}$, $b_n^{(\kappa)}$ \eqref{eq:Liouville Lanczos}: Krylov basis and Lanczos coefficients, where $\kappa$ is a parametrization for one-particle states.
\item $\mathcal{C}$ \eqref{eq:cost}: Cost function.
\item $\mathcal{C}_{\rm S}$ \eqref{eq:spread Complexity}, $a_n$, $b_n$ \eqref {eq:lanczos new}: Krylov spread complexity in the zero-particle chord space.
\item $\ell_y$ \eqref{eq:rescaled chord number krylov basis}: Expectation value of the rescaled chord number in the HH state.
\item $L$ \eqref{eq:length functional}, $L_{\rm dS}$ \eqref{eq:LdS def}: Minimal geodesic lengths between the asymptotic boundaries in the bulk.
\item $\ket{n}$, $\ket{\theta}$ \eqref{eq: proj E0 theta}: Chord number and energy basis for zero-particle states respectively.
    \item $\ket{\Omega}$ \eqref{eq:Omega state}: Tracial (zero chord number) state.
\item $[n]_q=\frac{1-q^n}{1-q}$ \eqref{eq:Transfer matrix}: q-deformed integer.

\item $(a;q)_n=\prod_{k=0}^{n-1}(1-aq^k)$ \eqref{eq:Pochhammer}: q-Pochhammer symbol.

\item $(a_1,a_2,\dots a_m;q)_n=\prod_{i=1}^N(a_i;~q)_n$ \eqref{eq:Pochhammer}.

\item $\tilde{\beta}_y=\frac{\pi}{J\sin\theta}$ (\ref{eq:fake temperature}): Fake temperature of the deformed theory.

\item $H_n(x|q)$ (\ref{eq:H_n def}): q-Hermite polynomials.

\item $\mu(\theta)$ (\ref{eq:norm theta}): Integration measure in the energy parametrization basis $\theta$.

\item $\hat{\mathcal{O}}_\Delta$ (\ref{eq:matter ops DSSYK}): Matter chord operator with conformal dimension $\Delta$.

\item $\hat{n}$ (\ref{eq: Oscillators 1}): Chord number operator in the zero-particle sector $\mathcal{H}_0$.

\item $\rmd E/\rmd E_y$ \eqref{eq:redshift factor}: Redshift factor.

\item $\hH_{\rm IR/UV}$ \eqref{eq:H(dS)JT}: Chord Hamiltonian in the IR and UV triple-scaling limit respectively.
\item $\theta_{\rm IR}\equiv\theta\ll1$ and $\theta_{\rm UV}\equiv\pi-\theta\ll1$ \eqref{eq:deformed spectrum IR/UV}: IR and UV triple-scaled energy parametrizations respectively in the seed theory.
    \item $\hat{a}^\dagger_{i}$ and $\hat{\alpha}_{i}$ (\ref{eq:Fock Hm 1}): Non-Hermitian conjugate creation and annihilation operators in $\mathcal{H}_{m}$.
\item $\tau_{L/R}=\frac{\beta_{L/R}}{2}+\rmi t_{L/R}$ \eqref{eq:taui}: Complex-valued time, with $\beta_{L/R}$ the inverse temperature and $t_{L/R}$ as real time.

\item $\ket{\Psi_{\Delta}(\tau_L,\tau_R)}$ (\ref{eq:HH state tL tR}): Generalization of the HH state with matter chord insertions.

\item $\tilde{\Delta}=\qty{\Delta_1,\dots,\Delta_m}$ \eqref{eq:Fock space with matter}: Collective index of conformal dimensions.

\item $\hat{N}$ (\ref{eq:total chord number basis}): Total chord number with one or more particle chords.

\item $\hat{n}_i^>\equiv\sum_{j={i+1}}^{m}\qty(\hat{n}_j+\Delta_{j})$ and $\hat{n}_i^<\equiv\sum_{j=0}^{i-1}\qty(\hat{n}_j+\Delta_{j+1})$ (\ref{eq:Hmultiple}).

\item $\hH_{L/R,y}$ \eqref{eq:flow eq with matter}: Deformed two-sided chord Hamiltonian 

\item $G_{2m}$ \eqref{eq:G2m}: ($2m$)-point correlation function.
\item $\ell_*$ \eqref {eq:ell cond thetaLR}, $\ell_0$ \eqref {eq:new e ell0}: Constants in the crossed-correlation function in the double-scaling and triple-scaling limits respectively.
\item $\lambda_L$ \eqref{eq:Lyapunov exp}: Lyapunov exponent.
\item $t_{\rm sc}$ \eqref{eq:scrambling time}: Scrambling time.
\item $\mathcal{A}_{L/R}\equiv\qty{\hH_{L/R},\,\hmO_\Delta^{L/R}}$ 
\eqref{eq:DS algebra}: Double-scaled algebras.
    \item $\Tr$ \eqref{eq:trace def}: Type II algebraic trace.
    \item $S$ \eqref{eq:def entanglement entropy}: Entanglement entropy between double-scaled algebras given a chord state.
\item $\hat{\rho}$ \eqref{eq:density matrix formal}: Reduced density matrix.
\item $r_{\rm BH}$ \eqref{eq:rBH}, $r_{\rm CH}$ \eqref{eq:dS2 metric}): AdS black hole and cosmological horizon radial locations.
  \item  $\Phi_{\rm (A)dS}$ \eqref{eq:RT surface dS}: Dilaton
$\Phi$ \eqref{eq:Dilaton-gravity theory} in (dS) JT gravity.
\item $\gamma$ \eqref{eq:RT surface dS}: RT surface.
\item $\hat{\varphi}_i$ (\ref{eq:symmetries general2}): Constraints selecting symmetry sectors in the chord Hilbert space.
\item $\hH_{\rm ASC}$ (\ref{eq:ASC Hamiltonian}): ASC Hamiltonian. $X$, $Y$: Parameters in the ASC Hamiltonian.
\item $\tilde{\mu}(\theta)$ \eqref{eq:completeness relation}: Integration measure factor in energy basis in systems with ETW branes.
\item $Q_l(\cos\theta|X,Y;q)$ (\ref{eq:ASC pol}): ASC polynomials.
\item $Z^{(y)}_{\rm ETW}(\beta|X,Y)$ (\ref{eq:disk part ETW}): Disk partition function with ETW brane.
    \item $\chi$ \eqref{eq:trajectory real line}: Stretched horizon parameter.
\end{itemize}

\section{Brief Review of \texorpdfstring{$\TT(+\Lambda_2)$}{} Deformations}\label{app:TTbar intro}
This appendix is a short review of $\TT$ ($+\Lambda_2$) deformations, mostly based on \cite{Aguilar-Gutierrez:2024nst}.

\subsection{\texorpdfstring{$\text{T}\overline{\text{T}}$}{} deformations}
$\text{T}\overline{\text{T}}$ deformations are a special class of irrelevant deformations of CFT$_2$s \cite{Smirnov:2016lqw,Cavaglia:2016oda}. Here, one introduces a term in the Euclidean action of the would-be CFT resulting in family of local quantum field theories of the form
\beq I_{\text{QFT}}=I_{\text{CFT}}+\mu\int \rmd^{2}x\sqrt{\gamma}\,\text{T}\overline{\text{T}}(x)\;,\label{eq:defQFTaction}\eeq
where $\mu$ is a coupling term parametrizing the deformation, $I_{\text{CFT}}$ is the `seed' CFT, and $\text{T}\overline{\text{T}}(x)$ an irrelevant, local, composite product of the stress components  $T_{ij}$,
\beq \text{T}\overline{\text{T}}(x)\equiv\frac{1}{8}\left[T^{ij}(x)T_{ij}(x)-(T^{i}_{i}(x))^{2}\right]\;.\label{eq:TTbardef}\eeq
Also, denoting $z\equiv x+i\tau$ in complex Cartesian coordinates (where the CFT metric is $\rmd s^2=\rmd \tau^2+\rmd x^2$), one has that $8\text{T}\overline{\text{T}}=T_{zz}T_{\bar{z}\bar{z}}-(T_{z\bar{z}})^{2}=\text{det}(T_{ij})$, with trace $T^{i}_{i}=4T_{z\bar{z}}$. It is conventional to define 
\beq
T\equiv T_{zz}~,\quad \overline{T}\equiv T_{z\bar{z}}\equiv\Theta~,
\eeq
such that $8\text{T}\overline{\text{T}}=T\bar{T}-\Theta^{2}$.

Crucially, the expectation value of the $\TT$ operator in any translation invariant satisfies \cite{Zamolodchikov:2004ce}
\beq \langle \text{T}\overline{\text{T}}\rangle= \langle T\rangle\langle \overline{T}\rangle-\langle \Theta\rangle^{2}\;,\label{eq:factorizationTT}\eeq
for any finite $\mu$.

Consider that the QFT lives
on a cylinder with circumference $L$, where we denote
\begin{subequations}
    \begin{align}
\text{Quantized~energy}~E_{n}&\equiv L\langle n|T_{\tau\tau}|n\rangle~,\\
\text{Pressure}~\partial_{L}E_{n}&\equiv\langle n|T_{xx}|n\rangle~,\\
\text{Momentum}~iP_{n}&\equiv L\langle n|T_{\tau x}|n\rangle~.
\end{align}
\end{subequations}
Noticing that the QFT generating function $Z_{\text{QFT}}$ obeys 
\beq\label{eq:log Z} \partial_{\mu}\log Z_{\text{QFT}}(\mu)=-\int \rmd^{2} x\sqrt{\gamma}\langle \text{T}\overline{\text{T}}\rangle\;,\eeq
with $Z_{\text{QFT}}(0)=Z_{\text{CFT}}$, then it follows that $\partial_{\mu}\langle H\rangle=L\langle \text{T}\overline{\text{T}}\rangle$ (while $\partial_{\mu}P_{n}=0$), which means that \eqref{eq:log Z} with (\ref{eq:factorizationTT}) is equivalent to the forced inviscid Burgers equation (which has applications in turbulent fluid dynamics),
\beq\label{eq:Burges} 4\partial_{\mu}E_{n}+E_{n}\partial_{L}E_{n}+\frac{P_{n}^{2}}{L}=0\;.\eeq
Next, we can introduce 
\begin{equation}\label{eq:def lambda}
y\equiv\pi\mu/L^{2}~,    
\end{equation}
so that \eqref{eq:Burges} becomes
\beq \label{eq:curly En}\frac{4\pi}{L}\partial_{y}E_{n}-2yE_{n}\partial_{y}E_{n}-E_{n}^{2}+P_{n}^{2}=0\;.\eeq
The solution of the differential equation \eqref{eq:curly En} subject to initial conditions dictated by the seed CFT (i.e.~at $y=0$), which we denote
\beq
E_{n}(y=0,L)=2\pi M_{n}/L~,\quad P_{n}(y=0)\equiv2\pi J_{n}/L
\eeq
becomes \cite{Smirnov:2016lqw,Cavaglia:2016oda,Zamolodchikov:2004ce}
\beq\label{eq:En sols} E_{n}= \frac{2\pi}{y L}\left(1-\sqrt{1-2y M_{n}+y^{2}J_{n}^{2}}\right)\;.\eeq

\paragraph{Dimensional reduction to one-dimensions} We perform a dimensional reduction of $\text{T}\overline{\text{T}}$ deformations. Similar to \cite{Gross:2019ach,Gross:2019uxi}, we assume the QFT metric is $\rmd s^2=\rmd\theta^2+\rmd\tau^2$ where $\theta+2\pi\sim\theta$. Solving for $T^{\theta}_{\;\theta}$, the QFT action (\ref{eq:defQFTaction}) can be recast as
\beq \partial_{\mu}I_{\text{QFT}}=\int \rmd^{2}x\sqrt{\gamma}\left(\frac{(T^{\tau}_{\;\tau})^{2}+T_{\tau\theta}T^{\tau\theta}}{4-2\mu T^{\tau}_{\;\tau}}\right)\;,\label{eq:flowv2}\eeq
which has a pole at $T^{\tau}_{\;\tau}=2/\mu$; i.e.~$E=\int d\theta T^{\tau}_{\;\tau}=LT^{\tau}_{\;\tau}$, which is reflected in (\ref{eq:En sols}). To perform the dimensional reduction of (\ref{eq:flowv2})~\cite{Gross:2019ach}, we can use the angular invariance to set $T_{\tau\theta}=0$, i.e.~$P_{n}=J_{n}=0$ in (\ref{eq:En sols}), i.e.
\beq\partial_{\mu}(E/L)=\frac{(E/L)^{2}}{4-2\mu E/L}\;\;\Longrightarrow \;\; E(y)=\frac{2\pi}{y L}\qty(1-\sqrt{1-2yM_n})\;,\label{eq:defen1d}\eeq
where the spectrum becomes complex if $2y M_n>1$. The energy spectrum of the deformed theory (\ref{eq:defen1d}) agrees with the quasi-local energy of its dual $\text{AdS}_{2}$ black hole \cite{Gross:2019ach}, as expected in finite cutoff holography \cite{McGough:2016lol}. For instance, the flow equation of the deformed theory can be derived from the Hamilton-Jacobi equation (see Sec.~\ref{ssec:flow eq}) which depends on the Wheeler-DeWitt constraint in the wavefunction of three-dimensional gravity \cite{McGough:2016lol}. In more general dimensions, we denote $T^2$ deformations the analog to $\TT$ deformations as explained in Sec.~\ref{sec:intro}.

\subsection{\texorpdfstring{$\text{T}\overline{\text{T}}+\Lambda_2$}{} deformations}

Now, we move on to $\text{T}\overline{\text{T}}+\Lambda_2$ deformations. It has been emphasized that this type of solvable irrelevant deformation holographically corresponds to connecting an  $\text{AdS}_{3}$ patch to a $\text{dS}_{3}$ patch \cite{Gorbenko:2018oov,Lewkowycz:2019xse,Silverstein:2024xnr,Batra:2024kjl,Shyam:2021ciy,Coleman:2021nor}; while it can also be interpreted as probing the black hole interior when modifying the sign of the deformation. We also present its two-dimensional analog via dimensional reduction similar to the last subsection, based on \cite{Aguilar-Gutierrez:2024nst}. 

First, consider adding a two-dimensional `cosmological constant' $\Lambda_{2}$ \cite{Gorbenko:2018oov,Lewkowycz:2019xse} in (\ref{eq:defQFTaction}),
\beq I_{\text{QFT}}=I_{\text{CFT}}+\mu\int \rmd^{2}x\sqrt{\gamma}\text{T}\overline{\text{T}}+\frac{(1-\eta)}{\mu}\int \rmd^{2}x\sqrt{\gamma}\;,\label{eq:TTbarL2act}\eeq
where $\eta=\pm1$, so that $\eta=+1$ corresponds to the $\text{T}\overline{\text{T}}$ deformed theory. The corresponding modification of \eqref{eq:log Z} is
\beq \partial_{\mu}\log Z_{\text{QFT}}(\mu,\eta)=-\int\rmd^{2}x\sqrt{\gamma}\langle\text{T}\overline{\text{T}}\rangle+\frac{(1-\eta)}{\mu^{2}}\int\rmd^{2}x\sqrt{\gamma}\;,\label{eq:flowZL2}\eeq
where $Z_{\text{QFT}}(\mu=0,\eta=1)=Z_{\text{CFT}}$. Similarly, the Burgers equation (\ref{eq:curly En}) becomes \cite{Gorbenko:2018oov}
\beq \frac{4\pi}{L}\partial_{y}E_{n}-2y E_{n}\partial_{y} E_{n}-E^{2}_{n}+P_{n}^{2}+\left(\frac{2\pi}{yL}\right)^{\hspace{-1mm}2}(1-\eta)=0\;,\label{eq:BurgeqnEL2}\eeq
with $y$ in (\ref{eq:def lambda}). The solution is
\beq E_{n}(y,\eta)=\frac{2\pi}{y L}\left(1\pm\sqrt{\eta-\frac{4C_{1}y}{2\pi^{2}}+y^{2}J_{n}^{2}}\right)\;,\label{eq:gensolL2}\eeq
with $C_{1}$ an integration constant which depends on the $\pm$ solution branch.

As explained in \cite{Gorbenko:2018oov}, and explained at length in Sec.~\ref{ssec:sol flow eq} in the lower-dimensional analog case, (\ref{eq:flowZL2}) can be used piecewise; (a) starting with a $\text{T}\overline{\text{T}}$ deformation (i.e.~at $\eta=1$), up to a critical value in the deformation parameter $y=y_{0}$ \eqref{eq:y0}, and (b) for $y>y_{0}$ one can switch to $\eta=-1$ while imposing continuity in the energy spectrum respect to the $\text{T}\overline{\text{T}}$ deformation at $y=y_0$. The motivation for considering this deformation comes from the three-dimensional Einstein gravity with cosmological constant $2\Lambda_{3}\equiv-{2\eta}/{\ell^{2}_{3}}$, 
\beq
    I_{\text{EH}}=\frac{1}{16\pi G_{3}}\int_{\mathcal{M}}\rmd^3 x\sqrt{-g}\qty(R+\frac{2\eta}{\ell^{2}_{3}})+\frac{1}{8\pi G_{3}}\int_{\partial\mathcal{M}}\rmd^2 x\sqrt{h}\left(K-\frac{1}{\ell_{3}}\right)\;.
\label{eq:3D action}\eeq
One might then expect that the flow in the dual theory between $\eta=+1$ to $\eta=-1$ corresponds to a transition from bulk spacetimes with $\Lambda_{3}<0$ to those with $\Lambda_{3}>0$, so that the dual theory of dS space might be connected to that in AdS thought an appropriate deformation with a matching condition at $y=y_0$. This limit corresponds to connecting the near-horizon region of the BTZ black hole at its Hawking-Page phase transition point (where the black hole horizon radius matches the scale of AdS space, i.e.~$r_{h}=\ell_{3}$) and the near-horizon of the $\text{dS}_{3}$ static patch \cite{Coleman:2021nor}. The matching at $y=y_0$ implies that the deformed energy spectrum \eqref{eq:gensolL2} for $J_n =0$ is
\beq E_{n}(\lambda,\eta)=\frac{2\pi}{\lambda L}\left(1-\sqrt{\eta(1-2M_{n}\lambda)}\right)\;.\label{eq:micenergyHP}\eeq
This indeed matches the quasi-local Brown-York energy in the cosmological patch of $\text{dS}_{3}$ \cite{Banihashemi:2022htw} with Dirichlet boundary conditions at the radial cutoff $r_B=\sqrt{4G_3\ell_3/y}$. Its role in the microscopic interpretation of the Gibbons-Hawking entropy has been discussed in \cite{Coleman:2021nor}. In \cite{Batra:2024kjl} the $\text{T}\overline{\text{T}}+\Lambda_{2}$ prescription was generalized to further account for bulk matter fields, going beyond the model-independent pure (semi-classical) gravity sector. In \cite{Silverstein:2024xnr}, the flow was extended to higher dimensions without matter.

\paragraph{Dimensional to one-dimensions}
Similarly to the previous subsection, applying spherical dimensional reduction of the 2D $\text{T}\overline{\text{T}}+\Lambda_2$ deformation so that it corresponds to a flow for a one-dimensional quantum theory. The dimensional reduction of the 3D bulk action (\ref{eq:3D action}) leads to a 2D bulk action of the form (\ref{eq:Dilaton-gravity theory}) without a topological term \cite{Svesko:2022txo},
\begin{equation}\label{eq:dim reduction to Dilaton gravity}
    I_E=-\frac{1}{16\pi G_N}\int_{\mathcal{M}} \rmd^2x\sqrt{g}(R\Phi+V(\Phi))-\frac{1}{8\pi G_N}\int_{\partial\mathcal{M}}\rmd \tau\sqrt{h}\,\Phi \qty(K-\tfrac{1}{\ell})~,
\end{equation}
with 
\begin{equation}\label{eq:eta potential 2d}
        V(\Phi)=\frac{2\eta}{\ell^2}{\Phi}~,
\end{equation} 
and where $\ell_3 = \ell.$ Then the background solution of the theory becomes
\begin{align}\label{eq:blackening}
    \rmd s^2=-N(r)\rmd t^2+\frac{\rmd r^2}{N(r)}~,\quad N(r)=\eta\frac{r^2-r_H^2}{\ell^2}\,,
\end{align}
where the blackening factor corresponds to ($\eta=+1$) an AdS$_{2}$ black hole, or ($\eta=-1$) the reduction of dS$_3$ space with a conical deficit when $r_H\neq\ell$, which has been discussed in \cite{Aguilar-Gutierrez:2024nst} in connection with interpolating models \cite{Anninos:2017hhn,Anninos:2018svg}. The deformed energy spectrum with the dilaton potential (\ref{eq:eta potential 2d}) in the flow equation (\ref{eq:flow eq}) is then
\begin{equation}\label{eq:E spectrum TTbar+Lambda2}
    E(y)=\tfrac{1}{y}\left(1\pm\sqrt{\eta(1-{2y}E_0)} \right)~,
\end{equation}
with $\mathcal{E}_0$ corresponding to $\lambda=0$ and $\eta=+1$ in the higher-dimensional $\TT+\Lambda_2$ flow, where the `$-$' sign corresponds to the smoothly connected solution to the seed theory, and `$+$' for describing the cosmic patch of the dimensionally reduced dS$_2$ space (see Fig.~\ref{fig:two-step_def}).

\section{Matching Time/Space-like Geodesics with \texorpdfstring{\eqref{eq:Krylov complexity deformed DSSYK}}{}}\label{app:geodesics}
In this appendix, we show that the Krylov spread complexity of the HH state in the deformed DSSYK model in \eqref{eq:Krylov complexity deformed DSSYK} describes both spacelike and timelike geodesic lengths connecting the finite cutoff boundaries of AdS$_2$ black hole, which are illustrated in Fig.~\ref{fig:spread_time}.

We define geodesic lengths as
\begin{equation}\label{eq:length functional}
    L=\int\rmd\xi~\mathcal{L}(\xi)~,\quad\mathcal{L}\equiv\sqrt{g_{\mu\nu}\dot{x}^\mu\dot{x}^\nu}=\sqrt{-f(r)\dot{v}_{L/R}^2+2\dot{v}_{L/R}\dot{r}}~,
\end{equation}
where $\xi$ is a general parameterization, $\dot{x}\equiv\dv{x}{\xi}$, the index $L/R$ indicates left/right patches, while
\begin{equation}\label{eq:netric more gen}
    \rmd s^2\equiv g_{\mu\nu}\rmd{x}^\mu\rmd{x}^\nu=-f(r)\rmd{v}_{L/R}^2+2\rmd{v}_{L/R}\rmd{r}=-{f(r)}{\rmd t_{L/R}^2}+\frac{\rmd r^2}{f(r)}~.
\end{equation}
Here we will be interested in the blackening factor
\begin{equation}\label{eq:AdS blackening}
    f(r)=r^2-r_{\rm BH}^2~,
\end{equation}
although, we keep some expressions with $f(r)$ general unless otherwise specified.

There is a conserved charge
\begin{equation}\label{eq:cond2}
    P=\frac{\delta \mathcal{L}}{\delta \dot{v}_{L/R}}=\frac{-f(r)\dot{v}_{L/R}+\dot{r}}{\mathcal{L}}~.
\end{equation}
Note that $\mathcal{L}$ is a diffeomorphism invariant function, and it can be thus gauged fixed to any value. In the literature, one commonly chooses $\mathcal{L}=\pm 1$ where $+1$ corresponds to the spacelike case, and $-1$ to timelike \cite{Chapman:2022mqd}. However, given that $\mathcal{L}$ is diffeomorphism invariant, it is not necessary to make any particular gauge-choice, all will give the same answer, and thus, regardless of the geodesic being spacelike or timelike, they will have the same functional expression, as we show below by combining $\mathcal{L}$ in \eqref{eq:netric more gen} with \eqref{eq:cond2}. This gives
\begin{equation}\label{eq:dot r}
    \dot{r}^2=\qty(P^2+f(r))\mathcal{L}^2~.
\end{equation}
Then, we can evaluate the geodesic length functional as
\begin{equation}\label{eq:length new}
    L=\int_{\xi_*}^{\xi_B}\mathcal{L}~\rmd\xi=\int^{r_B}_{r_*}\frac{\rmd r}{\sqrt{P^2+f(r)}}~,
\end{equation}
where $\xi_B$ is the parametrization of the boundary cutoff value where $r(\xi_B)=r_B$, and $r(\xi_*)=r_*$ represents the turning-point i.e.~when $\dot{r}=0$ in \eqref{eq:dot r}
\begin{equation}
    r_*=f^{-1}\qty(-P^2)~,
\end{equation}
where we are using that $P$ is a conserved charge. Meanwhile, using the relation between $t$ and $v$ in \eqref{eq:netric more gen}, we have that $f(r)\dot{t}_{L/R}=-\mathcal{L}P$, i.e.
\begin{equation}\label{eq:time new}
    t=-P\int_{\xi_*}^{\xi_B}\frac{\rmd r}{f(r)\sqrt{P^2+f(r)}}~,
\end{equation}
where we are gauge-fixing $t_{L}=t_R\equiv t$.

We can now specialize in the cases of interest in the AdS$_2$ black hole \eqref{eq:AdS blackening}, where \eqref{eq:length new} and \eqref{eq:time new} gives
\begin{subequations}\label{eq:parts L t}
\begin{align}
     L&=2{\rm arccosh}\frac{r_B}{r_*}~,\label{eq:part1}\\
     \frac{r_{\rm BH}t}{2}&=\arctan\frac{r_B\sqrt{r_*^2-r_{\rm BH}^2}}{r_{\rm BH}\sqrt{r_B^2-r_{*}^2}}~.\label{eq:part2}
\end{align}
\end{subequations}
Note that in the case of timelike geodesics ($r_*< r_B$) $L$ \eqref{eq:part1} becomes complex-valued, while $t$ remains real-valued; while both $L$ and $t$ remain real-valued for spacelike geodesics, where $r_*> r_B$. One can simplify the above expressions by inverting $r_*(t)$ from \eqref{eq:part1}, plugging it back in \eqref{eq:part2} and using hyperbolic trigonometric identities, which gives
\begin{equation}
    L(t)=2~{\rm arcsinh}\qty(\sqrt{\frac{r_B}{r_{\rm BH}}-1}~\cosh\frac{r_{\rm BH}t}{2})~.
\end{equation}
This indeed agrees with Krylov spread complexity in \eqref{eq:Krylov complexity deformed DSSYK}, where time includes the corresponding redshift factor. The geodesic length at finite cutoff has been previously noticed for the spacelike case in e.g.~\cite{Chapman:2021eyy,Griguolo:2025kpi}.

\section{Sine Dilaton Gravity at Finite Cutoff}
\label{app:sine dilaton gravity finite cutoff}
In this appendix, we explore the putative duality between the deformation of the DSSYK model and sine dilaton gravity at the disk topology level from finite cutoff holography. We are interested in the reality properties of the theory. Recent progress relating sine dilaton gravity and the DSSYK chord model can be found in \cite{Blommaert:2024ymv,Blommaert:2024whf,Blommaert:2023opb,Blommaert:2023wad,Cui:2025sgy,Heller:2024ldz,Heller:2025ddj} among others. 

Sine dilaton gravity (in Euclidean-like signature) is described by \eqref{eq:Dilaton-gravity theory} where the dilaton potential takes the form
\begin{equation}
    \label{eq:potential}
    U(\Phi)=2\sin\Phi~.
\end{equation}
Since the metric and the dilaton in this model are complex-valued, the holographic dictionary implies that the deformation parameter in the deformed DSSYK model can be complex. Thus, we are particularly interested in the conditions for the energy spectrum of the deformed theory to be real, and its corresponding interpretation in the Lorentzian signature bulk spacetime. {We also search for locations in the complex plane where the model has positive and near constant curvature, corresponding to first implementing the deformation and then taking the UV triple-scaling limit of the DSSYK model, which we applied in previous sections.}

\paragraph{Reality Conditions}One might be concerned that the Euclidean path integral of the theory might not converge when the energy spectrum is generically complex-valued. For instance, consider the partition
\begin{equation}\label{eq:partition E}
\int\rmd E ~\rme^{S(E)-\beta E_y}~,
\end{equation}
where $S(\theta)=S_0-\frac{(\pi-2\theta)^2}{\lambda}$ \cite{Goel:2023svz}, $S_0$ is a constant; and $E(\theta)$ appears in \eqref{eq:energy spectrum}. In principle, due to the complex spectrum \eqref{eq:partition E} could diverge along the flow. After all, this what happens for instance in JT gravity (and other set-ups) when the finite cutoff reaches the black hole horizon in the bulk, where instanton effects need to be taken into account \cite{Griguolo:2021wgy}. However, the model that we study is UV finite, and \eqref{eq:partition E} is well-defined even when the exponent of the integral is complex.\footnote{If the Hamiltonian has a complex-valued spectrum, one needs to include its left and right-eigenstates, in contrast to the past sections where we considered the eigenbasis for a Hermitian Hamiltonian.}

\begin{figure}
    \centering
   \subfloat[]{\includegraphics[width=0.52\textwidth]{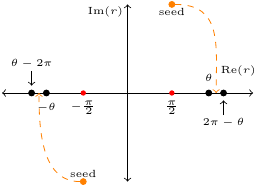}}\hfill\subfloat[]{\includegraphics[width=0.46\textwidth]{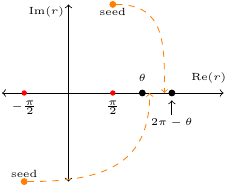}}
    \caption{Moving the boundaries in the complex $r$-plane by $T^2$ deforming the DSSYK model according to the dictionary entry (\ref{eq:def parameter}). The seed DSSYK model is initially located at $r\rightarrow\pm(\frac{\pi}{2}+\rmi\infty)$, and it can be moved in the indicated contours (dashed orange lines) close to the horizons at (a) $r=\pm\theta$, and (b) $r=\theta,~2\pi-\theta$. Both ways can be used to probe the positive curvature regions between the black hole and cosmological horizons when $\theta\simeq\pi$ (see the respective Penrose diagrams in Fig.~\ref{fig:Euclidean_Lorentzian_interpretation}).}
    \label{fig:Contour_TTbar}
\end{figure}

For the above reasons, we now analyze the constraints to recover a unitary energy spectrum (\ref{eq: E lambda}) by considering that the radial bulk flow is described by the HJ equation (\ref{eq:HJ eq}). A diagram of the paths generating these flows is shown in Fig.~\ref{fig:Contour_TTbar},\footnote{\label{fnt:extended analysis}As mentioned in Sec.~\ref{sec:intro}, this is not the only possible way to place the seed theories; for instance, one could consider placing the boundaries in between the black hole region; or between several black hole and cosmological patches (similar to \cite{Aguilar-Gutierrez:2024rka} in the extended SdS spacetime). Given that our focus is to probe the positive curvature region between the black hole and cosmological horizons, we leave these other possibilities for future directions (Sec.~\ref{ssec:Outlook}).} where we choose as the final radial location for the boundary theories to be the black hole and cosmological stretched horizon in (\ref{eq:def stretched horizon}), and (\ref{eq:def stretched horizon_2}). To be more explicit about the trajectory, one can, for instance, use
\begin{equation}\label{eq:change variables}
    \Phi_B=\pm\qty(\frac{\pi}{2}+\rmi\log(\upsilon+\rmi\cos r_B))~,
\end{equation}
 to connect $r=\frac{\pi}{2}+\rmi\infty$ (the asymptotic boundary) with $r=r_B$. 
Meanwhile, the trajectory from the black hole horizon at $r=\theta$ and the boundary location $r=r_B$ can be expressed by
\begin{equation}\label{eq:trajectory real line}
r=\theta\chi+(1-\chi)r_B~,\quad \chi\in[0,~1]~.
\end{equation}
Here $\chi$ is a parametrization that allows us to interpolate between the location of the cosmological horizon (i.e.~$\chi=1$ in $r_{B}(\theta)$) and black hole horizon (i.e.~$\chi=0$ in $r_{B}(\theta)$)\footnote{$\chi=0$ implies $r^{(1)}_{B_1}=-\theta$, which is a $2\pi$ displacement (also referred to as a duplicate in \cite{Blommaert:2024ymv}) of the cosmological horizon at $r=2\pi-\theta$, so we will attribute it as a cosmological horizon as well.} in $r\in[-2\pi,~2\pi]$.

\begin{figure}
    \centering
    \subfloat[]{\includegraphics[width=0.5\textwidth]{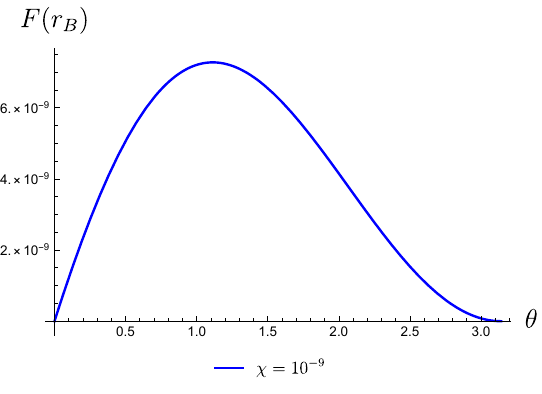}}\subfloat[]{\includegraphics[width=0.5\textwidth]{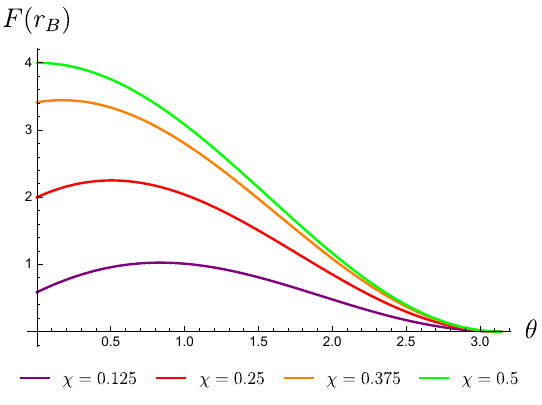}}
    \caption{Evaluation of the blackening factor (\ref{eq:blackening standard}) at the stretched horizon defined in (\ref{eq:def stretched horizon}) in terms of the parameter $\chi$ (for the different values shown in the legend). It remains positive $\forall\,\theta\in[0,~\pi]$.}
    \label{fig:blackening_positive}
\end{figure}

As we flow the boundary theory from $r_B\rightarrow\pm\qty(\frac{\pi}{2}+\rmi\infty)$ to a generic location $r_B\in\mathbb{C}$ in Fig.~\ref{fig:Contour_TTbar}, the energy spectrum (\ref{eq:E(theta) eta+1}) becomes complex; however, one should notice that the metric (\ref{eq:metric}) is also complex along this flow, and there is no Lorentzian interpretation where we could associate the complex energy spectrum, e.g.~a dissipation process.\footnote{In higher dimensions, a non-unitary energy spectrum with $T^2$ deformations has been studied in \emph{Cauchy slice holography} \cite{Caputa:2020fbc,Araujo-Regado:2022gvw,Araujo-Regado:2022jpj}. The interpretation is that if one deforms a CFT living on the boundary of a Euclidean Cauchy slice, its energy spectrum is inherently complex. Once the deformation is large enough, the extrinsic curvature $K_{\alpha\beta}$ becomes timelike, resulting in a real energy spectrum, corresponding to a Lorentzian signature in the bulk.} Once we reach the real line, we must study the locations where we can have a thermodynamic description of the system. First, notice that if we use the parametrization of the stretched horizon in (\ref{eq:def stretched horizon}) as the location of the boundary, we need to account for the functional dependence of $r_{B}(\theta)$. We can see that the blackening factor
\begin{equation}\label{eq:blackening standard}
    F(r_{B})=2(\cos\theta-\cos(\theta+2\chi(\pi-\theta)))~,
\end{equation}
remains positive for $\chi\in[0,~1]$ and $\theta\in[0,~\pi]$, as shown in Fig.~\ref{fig:blackening_positive}. 

\begin{figure}
    \centering
    \subfloat[]{\includegraphics[width=0.485\textwidth]{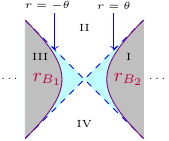}}\hfill\subfloat[]{\includegraphics[width=0.485\textwidth]{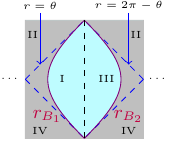}}
    \caption{Penrose diagram in the $\tilde{U}$, $\tilde{V}$ coordinates below (\ref{eq:Kruskal_coord}) for all the quadrants (Roman numerals). The energy spectrum of the dual $T^2$ deformed dual DSSYK models (\ref{eq:E(theta) eta+1}) has a relative (a) $\varepsilon=-1$, and (b) $\varepsilon=+1$ sign. We have placed the Dirichlet boundaries (purple) in (a) (\ref{eq:def stretched horizon}), and (b) (\ref{eq:def stretched horizon_2}), corresponding to the respective bulk radial flow paths. The gray region represents the part of the bulk that has been cut out in the bulk interpretation of the $T^2$ deformation, and the light blue area between the stretched horizon and the black hole and cosmological horizons (dashed blue lines) represents the bulk positive curvature region (for $\theta\simeq\pi$). The dashed black line in (b) only serves to separate the respective quadrants, and the dots indicate the $2n\pi$ periodicity of the spacetime.}\label{fig:Euclidean_Lorentzian_interpretation}
\end{figure}

In what follows, we search the reality constraints for unitary evolution in (\ref{eq: E lambda}), meaning that
\begin{equation}\label{eq:identify lambda}
    1\geq2y E(\theta)~.
\end{equation}
For concreteness, consider $\pi/2\leq\theta\leq\pi$ so $E(\theta)\geq0$. We identify different possibilities for the location $r_B$ given the condition (\ref{eq:identify lambda}), as we list below.
\begin{itemize}
    \item When $y>0$,
\begin{equation}\label{eq: cond TTbar}
    \theta<r_B~.
\end{equation}
    However, in this case, the blackening factor in the metric (\ref{eq:blackening standard}) is $F(r_{B})<0$, so that $r$ is a timelike coordinate and $\tau$ spacelike. Thus, the quantity (\ref{eq: E lambda}) would be related to a conserved momentum instead of the energy spectrum of the system. One might attempt a double Wick rotation in $\tau\rightarrow\rmi t$ and $r\rightarrow\rmi \tilde{r}$, but in that case, $F(r_{B})\in\mathbb{C}$. For this reason, we will not consider this case. Notice that if we take $r_{B}=\pi+\rmi r_0$ with $r_0\in\mathbb{R}$, we will face a similar problem.
    \item When $y\leq0$, there are no restrictions, and $F(r_{B})>0$. We can then perform the Wick rotation $\tau\rightarrow\rmi t$ in the metric (\ref{eq:metric}) so that $E_y(\theta)$ in (\ref{eq: E lambda}) along the real line indeed can be interpreted as conserved energy.
\end{itemize}
To draw the Penrose diagram, we consider the timelike Eddington–Finkelstein coordinates:
\begin{equation}
\begin{aligned}
    v&=t+r^*(r)~,\quad u=t-r^*(r)~,\\
    r^*(r)&=\int_{r_0}^{r}\frac{\rmd r'}{F(r')}=\frac{1}{2\sin\theta}\log\abs{\frac{\sin\frac{\theta-r}{2}}{\sin\frac{\theta+r}{2}}}~,
\end{aligned}
\end{equation}
where $r^*(r)$ is the tortoise coordinate, and we have chosen $r_0$ such that $r^*(r=0)=0$. We then define Kruskal coordinates
\begin{equation}\label{eq:Kruskal_coord}
U=\rme^u~,\quad V=-\rme^{-v}~,
\end{equation}
for the first quadrant of the Penrose diagram, Fig.~\ref{fig:Euclidean_Lorentzian_interpretation}, while  $U=\rme^u$, $V=\rme^{-v}$ in the second; $U=-\rme^u$, $V=\rme^{-v}$ third; and $U=-\rme^u$, $V=-\rme^{-v}$ fourth. At last, we define compact coordinates $U=\tan\tilde{U}$ and $V=\tan\tilde{V}$ to generate the Penrose diagram. Notice that the location of the horizons is $UV=0$, and the periodic nature of the diagram shares similarities with the (extended) full-reduction dS$_2$ spacetime (see e.g.~\cite{Balasubramanian:2020xqf,Levine:2022wos,Aguilar-Gutierrez:2021bns}).

\paragraph{Probing the Bulk Interior}
We now probe the positive curvature region in Fig.~\ref{fig:Euclidean_Lorentzian_interpretation} (and (\ref{fig:Contour_TTbar})) using two configurations for the \emph{cosmological} and \emph{black hole} stretched horizons where we move the boundaries into the locations:\footnote{A similar definition as (\ref{eq:def stretched horizon}) was first proposed in \cite{Aguilar-Gutierrez:2024rka} in the context of the extended Schwarzschild-de Sitter (SdS) black hole geometries. It allows us to generalize the notion of the stretched horizon in \cite{Susskind:2021esx} for pure dS space to spacetimes with multiple horizons.}
\begin{align}\label{eq:def stretched horizon}
    r_{B_1}^{(1)}&=r_{B}(-\theta)~,\quad r_{B_2}^{(1)}=r_{B}(\theta)~,\quad\text{where}~~ r_{B}(\theta)=(1-\chi)\theta+\chi(\theta-2\pi)~,\\
    r_{B_1}^{(2)}&=r_{B}(\theta)~,\quad r^{(2)}_{B_2}=r_{B}(2\pi-\theta)~.\label{eq:def stretched horizon_2}
\end{align}
As seen from Fig.~\ref{fig:Euclidean_Lorentzian_interpretation}, by setting the boundary locations to be $r_{B}=r^{(1)}_{B_1}$, $r_{B}=r^{(1)}_{B_2}$ in (\ref{eq:def stretched horizon}), the deformed DSSYK model can be located at a very close distance with respect to the cosmological ($\chi\ll1$) or black hole ($\chi\simeq1$), where the bulk thermodynamics with respect to the boundary locations can be associated with the respective black hole or cosmological horizon.

By taking $\theta\simeq \pi$, the region $r\in[-2\pi+\theta,-\theta]\cup[\theta,~2\pi-\theta]$ has a positive and near constant scalar curvature $\mathcal{R}\simeq 2$ \cite{Blommaert:2023opb}. This is equivalent to implementing the UV triple-scaling limit after performing the T$^2(+\Lambda_1)$ deformation in the DSSYK model, which we applied in previous sections, albeit without assuming the correspondence between sine dilaton gravity and the DSSYK model. The procedure in this section allows us to be more explicitly explore the structure of the finite cutoff holography with a concrete bulk dual for the range of its energy spectrum of the DSSYK model.

\section{An Alternative Chord Number Basis}\label{app:alternative chord number}
In this appendix, we define a new chord number basis, an alternative to the Krylov basis \eqref{eq:Krylov basis deformed theory} in the deformed DSSYK model, which is motivated by finite cutoff holography. Concretely, consider 
\begin{equation}\label{eq: new chord basis T^2}
    \ket{n}_y\equiv g_{n}\qty(\frac{\sqrt{\lambda(1-q)}}{-J}\hH)\ket{\Omega}~,
\end{equation}
where $\hH$ is the zero-particle chord Hamiltonian of the seed theory \eqref{eq:H chord thermal} and $g_n(x)$ is a polynomial function defined recursively
\begin{equation}\label{eq:new recursion rel T^2}
x~ g_n(x)=g_{n+1}(x)+\qty(1-R(y)^2\sech^2\frac{\lambda n}{2})g_{n-1}(x)~,
\end{equation}
with $R(y)$ a constant dependent on the deformation parameter $y$, and we impose as initial conditions,\footnote{For instance:
\begin{equation}
    g_1(x)=R(y)^2\sech^2\qty(\frac{\lambda}{2})~x~,
\end{equation}
and the higher-order polynomials, $g_{n\geq 2}$, follow iteratively.}
\begin{equation}
    g_{0}(x)=1~,\quad g_{-1}(x)=0~.
\end{equation}
There is a map between the different chord basis, given by
\begin{equation}\label{eq:n lambda}
    \ket{n}_y=g_n(\hh)\ket{\Omega}=\sum_{m=0}^\infty a_{nm}\ket{m}~,
\end{equation}
where 
\begin{equation}
a_{nm}\equiv\bra{m}\ket{n}_y=\int_0^\pi\rmd\theta~\mu(\theta)g_n(2\cos\theta)\frac{H_n(\cos\theta|q)}{\sqrt{(q;q)_n}}~.
\end{equation}
The explicit holographic dictionary between the arbitrary parameter $R(y)$ and the bulk radial cutoff will be determined below. Using the basis \eqref{eq: new chord basis T^2}, the seed theory Hamiltonian is then expressed as,\footnote{\label{fnt:imposing}Note that by imposing
\begin{equation}
    \lim_{y\rightarrow0}\frac{\cosh(L_y/2)}{R(y)}=\rme^{L/2}/2~,
\end{equation}
one recovers the corresponding chord Hamiltonian in the appropriate limit after a canonical transformation \cite{Aguilar-Gutierrez:2025pqp}. Importantly, \eqref{eq:Hamiltian cutoff canonical variables} remains as a Hermitian Hamiltonian with respect to the chord inner product \cite{Aguilar-Gutierrez:2025hty,Aguilar-Gutierrez:2025mxf}.}
\begin{equation}\label{eq:Hamiltian cutoff canonical variables}
\begin{aligned}
    \hH&=\frac{-J}{\sqrt{\lambda(1-q)}}\qty(2\cos\hat{P}_y-R(y)^2\rme^{\rmi \hat{P}_y}\sech^2\frac{\lambda \hat{n}_y}{2})~,
\end{aligned}
\end{equation}
where
\begin{equation}\label{eq:Py ny}
\rme^{\mp\rmi \hat{P}_y}\ket{n}_y\equiv\ket{n\pm 1}_y~,\quad \hat{\text{n}}_y\ket{n}_y\equiv n\ket{n}_y~.
\end{equation}
The resulting representation for the chord Hamiltonian \eqref{eq:Hamiltian cutoff canonical variables} relates to recent findings by \cite{Griguolo:2025kpi}, since, as seen below, the triple-scaling limit of the chord Hamiltonian recovers the canonically quantized JT gravity ADM Hamiltonian \cite{Griguolo:2025kpi}. However, we emphasize that the chord Hamiltonian representation \eqref{eq:Hamiltian cutoff canonical variables} only connects with finite cutoff holography once we study the deformed Hamiltonian $\hH_y$ \eqref{eq:def Hamiltonian}, either with the chord number basis $\ket{n}_y$ or a different one.\footnote{We expect that the choice of chord number basis $\ket{n}_y$ corresponds to the foliation in terms of the auxiliary parameter $y$ in the bulk dual, such that the regularized geodesic length between the asymptotic boundaries for chosen foliation reproduces the minimal length geodesic length connecting finite boundaries. We will explore this interpretation in upcoming work.} In contrast, the construction of the Krylov basis for the deformed DSSYK in Sec.~\ref{ssec:spread} does not incorporate auxiliary parameters associated to a spacetime foliation in the bulk, and indeed the Krylov spread complexity of the HH state reproduces the minimal geodesic length at finite cutoff \eqref{eq:Krylov complexity deformed DSSYK}. As we will see in the semiclassical analysis of the next subsection, the chord number in the $\ket{n}_y$ basis is related to an Einstein-Rosen bridge connecting the boundaries of an AdS$_2$ black hole with a finite radial cutoff with Dirichlet boundary conditions \cite{Griguolo:2025kpi} (7.5) (see also App.~\ref{app:geodesics}).

\paragraph{Outline}We investigate the path integral of the theory in App.~\ref{ssec:path integral} and we evaluate the expectation value of the chord number, which reproduces the wormhole length in the bulk. In App.~\ref{ssec:IR UV} we study the triple scaling limits of the chord Hamiltonian, and we show it reproduces the canonically quantized ADM Hamiltonian of JT gravity at a finite cutoff in a different ensemble with respect to the one found in the main text. At last, App.~\ref{sapp:AEE extra} contains an evaluation of algebraic entanglement entropy \eqref{eq:def entanglement entropy} using the chord number basis $\ket{n}_y$ which, in the IR/UV triple-scaling reproduces the dilaton in JT and dS JT gravity in the bulk dual, respectively.

\subsection{Path Integral Formulation}\label{ssec:path integral}
We study the path integral of the deformed DSSYK chord theory using the representation \eqref{eq:Hamiltian cutoff canonical variables}, which is
\begin{subequations}
\begin{align}
    Z&=\int[\rmd L_y] [\rmd P_y]\exp\qty[\int\rmd\tau\qty[\frac{\rmi}{\lambda} P_y\partial_\tau L_y-H_y]]~,\label{eq:parition new}\\
H_y&=f_y\qty(\frac{J}{\lambda}\qty(2\cos P_y-\frac{R(y)^2\rme^{\rmi P_y}}{\cosh^2\frac{L_y}{2}}))~.\label{eq:energy y}
\end{align}
\end{subequations}
where $\tau\equiv \frac{\beta}{2}+\rmi t$. Here, we represent $L_y$ as the semiclassical expectation value of the chord number in the deformed HH state $\rme^{-\tau\hH_y}\ket{\Omega}$, namely
\begin{equation}\label{eq:def Ly}
L_y\equiv\lambda\frac{\langle\Omega|{\rm e}^{-\tau^*\hat{H}_y}\hat{\text{n}}_y{\rm e}^{-\tau\hat{H}_y}|\Omega\rangle}{\langle\Omega|{\rm e}^{-\beta\hat{H}_{{y}}}|\Omega\rangle}~,
\end{equation}
and $P_y$ is the corresponding conjugate momenta. The saddle point solutions of \eqref{eq:parition new} are then
\begin{subequations}
\begin{align}
     \frac{1}{\lambda}\dv{L_y}{t}&=\pdv{H_y}{P_y}=f_y'(H)\pdv{H}{P_y}~,\\
 \frac{1}{\lambda}\dv{P_y}{t}&=-\pdv{H_y}{L_y}=-f_y'(H)\pdv{H}{L_y}~.
\end{align}
\end{subequations}
Note that we have used the chain rule. Using the explicit form in \eqref{eq:Hamiltian cutoff canonical variables} we recover
\begin{subequations}\label{eq:ODEs important}
\begin{align}
    \dv{L_y}{t_y}&=2\rmi\qty(\cos\theta-\rme^{\rmi P_y})~,\label{eq:dL dt}\\
     -\rmi\dv{\rme^{-\rmi P_y}}{t_y}&=R(y)^2\sech^2\qty(\frac{L_y}{2})\tanh\qty(\frac{L_y}{2})~,
\end{align}
\end{subequations}
where
\begin{equation}\label{eq:ty}
    t_y\equiv J f_y'(H)~t~.
\end{equation}
In particular for $T^2$ deformations \eqref{eq:T^2 def f} we have that the redshift factor is $f_y'(H)={1}/{\sqrt{1-2y H}}$, and for $T^2+\Lambda_1$ $f_y'(H)={1}/{\sqrt{2y H-1}}$.

The differential equations in \eqref{eq:ODEs important} are supplemented by boundary conditions
\begin{equation}\label{eq:BDRY cond cutoff}
    L_y(0)\equiv L_0~,\quad L_y'(0)=0~,
\end{equation}
Meanwhile, $L_0$ in \eqref{eq:BDRY cond cutoff} is a numerical constant determined from initial conditions and conservation of energy. Namely, consider \eqref{eq:dL dt} and the second initial condition in \eqref{eq:BDRY cond cutoff} for $t=0$ implies
\begin{equation}\label{eq:P(0)}
    \eval{\dv{L_y}{t_y}}_{t_y=0}=0~,\quad \rme^{\rmi P_y(0)}=\cos\theta~.
\end{equation}
Then, by introducing \eqref{eq:P(0)} and the first initial condition in \eqref{eq:BDRY cond cutoff} to \eqref{eq:energy y}, it follows that
\begin{equation}\label{eq:L0}
    R(y)^2\sech^2\frac{L_0}{2}=\sin^2\theta~.
\end{equation}
Combining them, we recover
\begin{equation}\label{eq:2nd order Ly ODE2}
    \dv[2]{L_y}{t_y}=2R(y)^2\tanh(\frac{L_y}{2})\sech^2\qty(\frac{L_y}{2})~.
\end{equation}
The solution to the above ordinary differential equation subject to the boundary conditions \eqref{eq:BDRY cond cutoff} is
\begin{equation}\label{eq:wormhole length finite cutoff}
    L_y=2~{\rm arcsinh}\qty(\sqrt{\frac{R(y)^2}{\sin^2\theta}-1}~\cosh(t_y\sin\theta ))~,
\end{equation}
which is precisely the spacelike geodesic length between the finite cutoff boundaries of AdS$_2$ black holes \cite{Chapman:2021eyy,Griguolo:2025kpi} with a black hole horizon at $\rho=\sin\theta$ in the coordinates
\begin{equation}\label{eq:fake AdS2}
    \rmd s^2_{\rm eff}=\qty(\rho^2-\sin^2\theta)\rmd\tau^2+\frac{\rmd \rho^2}{\rho^2-\sin^2\theta}~.
\end{equation}
In fact, we show that \eqref{eq:wormhole length finite cutoff} is also valid to describe timelike geodesics connecting finite cutoff boundaries inside the AdS$_2$ black hole \eqref{eq:fake AdS2}, see App.~\ref{app:geodesics}.

In the AdS$_2$ interpretation, $\rho=R(y)$ \eqref{eq:wormhole length finite cutoff} corresponds to the radial cutoff parameter in the effective AdS$_2$ metric. Thus, while the original chord Hamiltonian in canonical variables measure lengths/momenta with respect to the boundaries \cite{Lin:2022rbf}, we can implement the canonical transformation in the Hamiltonian \eqref{eq:Hamiltian cutoff canonical variables} to the new chord basis \eqref{eq: new chord basis T^2} to describe the deformed theory.\footnote{For example, in the specific case of the sine dilaton gravity proposal \cite{Blommaert:2024ymv}, the map between the AdS$_2$ black hole metric \eqref{eq:fake AdS2} to the geometry where we identified $y=2\lambda/G(r_B)$ based on \eqref{eq:def parameter} and \eqref{eq:relation lambda kappa} is given by
\begin{equation}
    r_B=\frac{\pi}{2}+\rmi \log (R(y)+\rmi\cos\theta)~,\quad y=\frac{-\lambda}{\cos(r_B)}~,
\end{equation}
where we are using $G(x)=-2\cos(x)$ \eqref{eq:Gphi special} for sine dilaton gravity. The above relation means that
\begin{equation}\label{eq:complicated R(y) in sine dilaton}
    \frac{\lambda}{y}=\frac{\rmi}{2}\qty(R(y)+\rmi\cos\theta-\frac{1}{R(y)+\rmi\cos\theta})~,
\end{equation}
if the DSSYK model were dual to sine dilaton gravity at the disk level.}

\paragraph{\texorpdfstring{$T^2$}{} vs \texorpdfstring{$T^2+\Lambda_1$}{} Deformations}Note that in the derivation of the semiclassical chord number \eqref{eq:wormhole length finite cutoff} we have made no assumptions about the explicit form of $f_y$; the expression is valid for both general deformations in finite cutoff holography. In particular, the difference between $T^2$ and $T^2+\Lambda_1$ deformation is contained in the factor $t_y$ \eqref{eq:ty} which depends on the explicit form of the deformation. Also, from the bulk perspective, we will be interested in the cases where $R(y)\geq\sin\theta$ for the $T^2$ case, and $R(y)\leq\sin\theta$ for $T^2+\Lambda_1$, resulting real-valued and pure imaginary lengths.

\subsection{IR/UV Triple-Scaling Limits and \texorpdfstring{$T^2(+\Lambda_1)$}{} Deformations}\label{ssec:IR UV}
To examine more closely the chord number in the $\ket{n}_y$ basis, we now define the IR/UV triple-scaling limit to further examine its properties:
\begin{subequations}\label{eq:IR UV triple scaling}
    \begin{align}
{\rm IR}:\quad{\sech^2\frac{\lambda\hat{\text{n}}_y}{2}}&\rightarrow\lambda^2 \sech^2\frac{\hat{L}_{{\rm IR}}}{2}~,\quad \hat{P}_y\rightarrow\lambda \hat{P}_{{\rm IR}}~,\\        
{\rm UV}:\quad{\sech^2\frac{\lambda\hat{\text{n}}_y}{2}}&\rightarrow\lambda^2 \sech^2\frac{\rmi\hat{L}_{{\rm UV}}}{2}~,\quad \hat{P}_y\rightarrow\pi-\rmi\lambda\hat{P}_{{\rm UV}}~,
    \end{align}
\end{subequations}
where the eigenvalues of the operators $\hat{L}_{{\rm IR}}$, $\hat{P}_{{\rm IR}}$ and $\hat{L}_{{\rm UV}}$, $\hat{P}_{{\rm UV}}$ are fixed as $\lambda\rightarrow0$.

We note that by subtracting the zero-energy contribution in the IR and UV limits of the chord Hamiltonian in the $\ket{n}_y$ basis \eqref{eq:Hamiltian cutoff canonical variables} and considering leading order terms, we recover
\begin{subequations}\label{eq:H(dS)JT2}
        \begin{align}
   \hH_{\rm IR}&\equiv \eval{\frac{1}{2J\lambda}\qty(\hH+\frac{2J}{\sqrt{\lambda(1-q)}})}_{\mathcal{O}(1),{\rm IR}}=\frac{\hP^2_{{\rm IR}}}{2}+\frac{R(y)^2}{2}\sech^2\frac{\hL_{{\rm IR}}}{2}~,\\
   \hH_{\rm UV}&\equiv \eval{\frac{1}{2J\lambda}\qty(\hH-\frac{2J}{\sqrt{\lambda(1-q)}})}_{\mathcal{O}(1),{\rm UV}}=-\frac{\hP^2_{{\rm UV}}}{2}+\frac{R(y)^2}{2}\sec^2\frac{\hL_{{\rm UV}}}{2}~.
        \end{align}
\end{subequations}
$\hH_{\rm IR}$ is isomorphic to the canonically quantized JT gravity Hamiltonian at finite cutoff in \cite{Griguolo:2025kpi} ($\hH_{\rm IR}$)  in terms of the finite cutoff length and its canonical conjugate, and $\hH_{\rm UV}$ is its dS$_2$ counterpart, which is the boundary operator dual to the generator of spatial translations along $\mathcal{I}^\pm$ in dS JT gravity \cite{Heller:2025ddj,Aguilar-Gutierrez:2025otq}, as a s-wave dimensional reduction of dS$_3$ space \cite{Heller:2025ddj}. Note that both are manifestly Hermitian, since the spectrum of $\hat{L}_{\rm UV}$ is pure imaginary.

Importantly, the need of the $\lambda^2$ factor in \eqref{eq:IR UV triple scaling} signals that in the triple-scaling limit one is defining a two-sided geodesic length between the asymptotic boundaries. In contrast, as we found in Sec.~\ref{ssec:spread}, one does not need to implement a regularization in the canonical variables describing a finite cutoff geodesic length. In a similar way, the triple-scaling limit describing geodesic lengths connecting one asymptotic boundary to a matter particle or specifically an ETW brane, one requires introducing a regularized geodesic length with a $\lambda^1$ factor \cite{Xu:2024hoc,Okuyama:2023byh,Aguilar-Gutierrez:2025hty,Cao:2025pir} in contrast to the $\lambda^2$ factor in the two-sided case. 

In either of the IR/UV limits, we can consider the sequence of a $T^2$ deformation and $T^2+\Lambda_1$ deformation once the cutoff boundary reaches the corresponding AdS$_2$ black hole and dS$_2$ cosmological horizons \eqref{eq:H IR UV}, namely
\begin{equation}\label{eq:deformed DSSYK cases}
    \hH_{{\rm IR/UV},y}=\begin{cases}
        \frac{1}{y}\qty(1-\sqrt{1-2y\hH_{\rm IR/UV}})&0\leq y\leq y_0~,\\
        \frac{1}{y}\qty(1+\sqrt{2y\hH_{\rm IR/UV}-1})&y\geq y_0~,
    \end{cases}
\end{equation}
We represent the energy spectrum introducing the parameters $\theta_{\rm IR}\equiv \theta\ll1$ and $\theta_{\rm UV}\equiv \pi-\theta\ll1$ in the corresponding triple-scaling limits, so that the deformed energy spectrum becomes
\begin{equation}\label{eq:deformed spectrum}
     E_{{\rm IR/UV},y}=\begin{cases}
        \frac{1}{y}\qty(1-\sqrt{1-{y}{~\theta^{2}_{\rm IR/UV}}})~,&0\leq y\leq y_0~,\\
        \frac{1}{y}\qty(1+\sqrt{{y}{~\theta^2_{\rm IR/UV}}-1})~,&y\geq y_0~,
    \end{cases}
\end{equation}
where $y_0$ appears in \eqref{eq:y0}.

\paragraph{Chord Number}
We also define the expectation value of the IR and UV length operators in \eqref{eq:IR UV triple scaling} in the deformed HH state $\rme^{-\tau\hH_y}\ket{\Omega}$ ($\tau=\beta/2+\rmi t$) as
\begin{equation}\label{eq:length dS exp }
\begin{aligned}
L_{\rm IR}(t)\equiv&\frac{\langle\Omega|{\rm e}^{-\tau^*\hat{H}_y}\hat{L}_{\rm IR}{\rm e}^{-\tau\hat{H}_y}|\Omega\rangle}{\langle\Omega|{\rm e}^{-\beta\hat{H}_y}|\Omega\rangle}~,\quad L_{\rm UV}(t)\equiv\frac{\langle\Omega|{\rm e}^{-\tau^*\hat{H}_y}\hat{L}_{\rm UV}{\rm e}^{-\tau\hat{H}_y}|\Omega\rangle}{\langle\Omega|{\rm e}^{-\beta\hat{H}_y}|\Omega\rangle}~,
    \end{aligned}
\end{equation}
which can be seen as the UV and IR limits of expectation value of the chord number in the $\ket{n}_y$ basis in \eqref{eq:wormhole length finite cutoff}. To carry out their evaluation, we use the explicit solution for the wormhole length $L_y$ in \eqref{eq:wormhole length finite cutoff} with $R(y)$ identified as the boundary cutoff an AdS$_2$ black hole metric \eqref{eq:fake AdS2} in the IR and UV triple-scaling limits \eqref{eq:IR UV triple scaling}, which describe JT ($U(\Phi)=2\Phi$) and dS JT gravity ($U(\Phi)=-2\Phi$) respectively. Using the counterterm $G(\Phi_B)$ \eqref{eq:Gphi special} (corresponding to $\eta=+1$) and the deformation parameter $y$ through \eqref{eq:def parameter} we have that
\begin{equation}\label{eq:Ry explicit IR}
   \eta=+1:\quad {\rm IR}: y=R(y)^{-2}~,\quad  {\rm UV}: y=-R(y)^{-2}~.
\end{equation}
Meanwhile, for $T^2+\Lambda_1$ deformations, we invert the sign of $G(\Phi_B)$ in both the IR and UV cases \eqref{eq:E(theta) eta-1}, resulting in
\begin{equation}\label{eq:Ry explicit UV}
   \eta=-1:\quad {\rm IR}: y=-R(y)^{-2}~,\quad  {\rm UV}: y=R(y)^{-2}~.
\end{equation}
It follows that the wormhole length \eqref{eq:wormhole length finite cutoff} in each of the corresponding triple-scaling limits \eqref{eq:length dS exp } becomes\footnote{Note that in \eqref{eq:LUV} we are using the fact that the conformal transformation between Rindler-AdS$_2$ \eqref{eq:fake AdS2} and static patch coordinates is given by a $-$ sign, instead applying the map in sine dilaton gravity of \eqref{eq:complicated R(y) in sine dilaton}.}
\begin{subequations}\label{eq:both lengths}
\begin{align}\label{eq:LIR}
    L_{\rm IR}(t)&=\begin{cases}
        2{\rm arcsinh}\qty(\sqrt{\frac{1}{y{\theta_{\rm IR}}^2}-1}~\cosh(\frac{J\theta_{\rm IR}t}{\sqrt{1-y{\theta_{\rm IR}}^2}}))~,&0\leq y\leq y_0~,\\
2\rmi~{\rm arcsin}\qty(\sqrt{1-\frac{1}{y{\theta_{\rm IR}}^2}}~\cosh(\frac{J\theta_{\rm IR}t}{\sqrt{y{\theta_{\rm IR}}^2-1}}))~,&y\geq y_0~,
    \end{cases}\\
     L_{\rm UV}(t)&=\begin{cases}
         2\rmi~{\rm arcsinh}\qty(\sqrt{\frac{1}{y{\theta_{\rm UV}}^2}-1}~\cosh(\frac{J\theta_{\rm UV}t}{\sqrt{1-y{\theta_{\rm UV}}^2}}))~,&0\leq y\leq y_0~,\\
         2{\rm arcsin}\qty(\sqrt{1-\frac{1}{y{\theta_{\rm UV}}^2}}~\cosh(\frac{J\theta_{\rm UV}t}{\sqrt{y{\theta_{\rm UV}}^2-1}}))~,&y\geq y_0~.\end{cases}\label{eq:LUV}
\end{align}
\end{subequations}
Note that $L_{\rm IR}$ and $L_{\rm UV}$ are real and pure imaginary respectively for the $T^2$ deformation, and they exchange to pure imaginary and real for the $T^2+\Lambda_1$ flow. 
The results show that the rescaled chord number in the IR triple-scaling limit $L_{\rm IR}$ grows eternally in the $\eta=+1$ case, while in the $\eta=-1$, there is a finite growth in terms of the $t$ parametrization. In both cases, the deformation parameter $y$ modifies the rate of growth of $L_{\rm IR}$. This reflects that the Tolman temperature in the bulk is the thermal scale accompanying $t$. In the $\eta=+1$ case as we approach $y\rightarrow y_0$ the rate of growth increases since the cutoff surface approaches the black hole horizon, increasing the Tolman temperature. Meanwhile, in the $\eta=-1$ case, corresponding to the interior of the black hole in the AdS$_2$ description, the imaginary value of $L_{\rm IR}$ can be interpreted in terms of the growth of the timelike geodesics. Note the growth of $L_{\rm IR}$ stops when the argument in arcsin of \eqref{eq:LIR} is greater than $1$ (which we discuss further in Sec.~\ref{ssec:dS Complexity UV}). This signals that the geodesics in the bulk dual reach $r\rightarrow\infty$ in Rindler-AdS$_2$ space, as this corresponds to the maximum value of the renormalized length in the AdS$_2$ description in Fig.~\eqref{fig:spread_time}. Similar findings occur in dS$_2$ space, as discussed in Sec.~\ref{ssec:hyperscrambling dS}.

Next, one sees that $L_{\rm IR/UV}$ (\ref{eq:both lengths}) takes the same form as $\ell_y$ \eqref{eq:Krylov complexity deformed DSSYK}, with a rescaling in the time parameter
\begin{equation}\label{eq:subst time}
 \frac{J\theta_{\rm IR/UV} t}{\sqrt{1-y \theta_{\rm IR/UV}^2}}\rightarrow\frac{\theta_{\rm IR/UV}~ \tilde{t}}{2}~,
\end{equation}
where the different time parametrizations have been interpreted in the JT gravity case in terms of different ensembles in \cite{Griguolo:2025kpi}. Similar to our analysis, the left-hand side in \eqref{eq:subst time} corresponds to a microcanonical ensemble incorporating the redshift factor due to the fixed energy \eqref{eq:fake temperature}, while the right-hand side to a canonical ensemble, where the redshift factor is not present in the Krylov complexity growth. Thus, if one instead evaluates Krylov spread complexity in the microcanonical ensemble, then one needs to incorporate the redshift factor in the rescaled time.

\subsection{Algebraic Entanglement Entropy}\label{sapp:AEE extra}
In this subappendix, we show that the entanglement entropy between the double-scaled algebras \eqref{eq:def entanglement entropy} for the chord number state $\ket{n}_y$ leads to the same result as using the Krylov basis in the deformed DSSYK model in the triple-scaling limit. Thus, our starting point is
\begin{equation}\label{eq:rho n new}
    \ket{\Psi}=\ket{n}_y=g_n(\hh)\ket{\Omega}~,\quad\hat{\rho}=g_n(\hh)^2~,
\end{equation}
where $g_n$ appears in \eqref{eq:new recursion rel T^2}.

From the general expression for algebraic entanglement entropy \eqref{eq:def entanglement entropy}, we get
\begin{equation}\label{eq:full entropy n DSSYK2}
    S(L)\equiv-\int_{0}^{\pi}\rmd \theta~\mu(\theta)\frac{g_n(2\cos\theta)^2}{\bra{\Psi}\ket{\Psi}}\log\frac{g_n(2\cos\theta)^2}{\bra{\Psi}\ket{\Psi}}~,
\end{equation}
where the energy measure appears in \eqref{eq:norm theta}. 

For analytic progress, we study the triple-scaling limit of $S(L)$; see \cite{Tang:2024xgg} for similar considerations, and \cite{Aguilar-Gutierrez:2025otq}.\footnote{We expect that the evaluation in the $\lambda\rightarrow0$ limit leads to a RT formula in sine dilaton gravity at finite cutoff, which has not been addressed even in the seed theory case.} For concreteness, we focus on the IR triple-scaling limit \eqref{eq:IR UV triple scaling}; very similar arguments can be implemented for the UV limit, which we discuss in Sec.~\ref{ssec:HEE dS}.

To study the IR triple-scaling limit of the different terms in \eqref{eq:full entropy n DSSYK2}, let us first find solutions to the inner product $\bra{k}\ket{L}$, where $\ket{L}$ represents the eigenstates of $\hat{L}_{\rm IR}$, and $\ket{k}\equiv\ket{\theta=\lambda k}$ the eigenstates of $\hH_{\rm IR}$, i.e.
\begin{equation}
\hat{L}_{\rm IR}\ket{L}=L_{\rm IR}\ket{L}~,\quad \hH_{\rm IR}\ket{k}=\frac{k^2}{2}\ket{k}~,
\end{equation}
Plugging these states in the corresponding Hamiltonian in \eqref{eq:H(dS)JT2},
\begin{equation}\label{eq:Schrödinger triple scaled York}
\begin{aligned}
    \bra{k}\hH_{{\rm IR}}\ket{L}&=\qty(\frac{-1}{2}\partial_{L_{\rm IR}}^2+\frac{R(y)^2}{2}\sech^2\frac{L_{\rm IR}}{2})\bra{k}\ket{L}=\frac{k^2}{2}\bra{k}\ket{L}~,
\end{aligned}
\end{equation}
where we used $\hat{P}_{\rm IR}=-\rmi\partial_{L_{\rm IR}}$, and $\theta\equiv k\lambda$ in the IR triple-scaling limit \eqref{eq:IR UV triple scaling}, so that \eqref{eq:Schrödinger triple scaled York} can be seen as a Schrödinger equation. The solutions are associated Legendre functions of first ($P_\nu^\mu(x)$) and second ($Q_\nu^\mu(x)$) kind,
\begin{equation}
    \bra{k}\ket{L}=c_1~P_{\frac{1}{2} \left(\sqrt{1-16 R(y)^2}-1\right)}^{2 \rmi k}\left(\tanh \left(\frac{L_{\rm IR}}{2}\right)\right)+c_2~Q_{\frac{1}{2} \left(\sqrt{1-16 R(y)^2}-1\right)}^{2 \rmi k}\left(\tanh \left(\frac{L_{\rm IR}}{2}\right)\right)~,
\end{equation}
where $c_1$ and $c_2$ are constants. We are interested in solutions that are square integrable, so we keep $c_2=0$ and take $c_1$ such that
\begin{equation}
    c_1=\qty(\int_0^\infty\rmd k\rho(k)\abs{P_{\frac{1}{2} \left(\sqrt{1-16 R(y)^2}-1\right)}^{2 \rmi k}\left(\tanh \left(\frac{L_{\rm IR}}{2}\right)\right)}^2)^{-1/2}~,
\end{equation}
where $\rho(k)$ is the density of states of JT gravity
\begin{equation}
    \rho(k)\equiv 4k\sin(2\pi k)~.
    \end{equation}
Then, \eqref{eq:full entropy n DSSYK2} becomes
\begin{equation}\label{eq:triple scaled entropy}
S\underset{\rm IR}{\rightarrow}-\lambda\int_0^\infty \rmd k~p(L_{\rm IR},k)\log \frac{p(L_{\rm IR},k)}{\mu(\theta\equiv\lambda k)}~.
\end{equation}
where we use that
\begin{subequations}
    \begin{align}
        \mu(\theta\equiv \lambda k)&\underset{\rm IR}{\rightarrow}\frac{\lambda^{1/2}}{\sqrt{2\pi}}\rme^{-\frac{\pi^2}{2\lambda}}(\rho(k)+\mathcal{O}(\lambda))~,\label{eq:case1}\\
        \mu(\theta)\frac{g_n(2\cos\theta)^2}{\bra{\Psi}\ket{\Psi}}&\underset{\rm IR}{\rightarrow}p(L_{\rm IR},k)~.\label{eq:case2}
    \end{align}
\end{subequations}
Here, \eqref{eq:case1} is known in \cite{Lin:2022rbf}, and \eqref{eq:case2} follows the fact that $\bra{\theta}\ket{n}_y$ satisfies the recurrence relation \eqref{eq:new recursion rel T^2} where $x=-\lambda E(\theta)/(2J)$, while $\bra{k}\ket{L}$ satisfies the IR triple-scaled version of the same equation, namely \eqref{eq:Schrödinger triple scaled York}.

To make more analytic progress, we apply the WKB approximation, where we look for semiclassical solutions to the Schrödinger equation \eqref{eq:Schrödinger triple scaled York}. We consider energy scales away from the one where the kinetic energy vanishes, meaning $\hat{P}_{\rm IR}^2=0$ in \eqref{eq:H(dS)JT2}. This can be expressed in terms of a cutoff scale
\begin{equation}\label{eq:critical E}
    k_{\rm crit}\equiv R(y)\sech\frac{L_{\rm IR}}{2}~,
\end{equation}
so that the wavefunction in \eqref{eq:Schrödinger triple scaled York} can be approximately described by a particle trapped in a potential given by the $R(y)^2\sech^2\frac{L_{\rm IR}}{2}$ term in the Schrödinger equation \eqref{eq:Schrödinger triple scaled York},
resulting in
\begin{align}\label{eq:psi E L WKB}
        &p(L_{\rm IR},k)\simeq\\
    &\begin{cases}
        \frac{1}{{2\pi}}\sqrt{\frac{k^2}{R(y)^2\sech^2\frac{L_{\rm IR}}{2}-k^2}}\exp(-2\int_{L_{\rm IR}}^{L_0}\rmd L\sqrt{R(y)^2\sech^2\frac{L}{2}-k^2})&k\lesssim k_{\rm crit}~,\\
        {\frac{2}{\pi}}\sqrt{\frac{k^2}{k^2-R(y)^2\sech^2\frac{L_{\rm IR}}{2}}}\sin^2\qty(\int^{L_{\rm IR}}\rmd L\sqrt{k^2-R(y)^2\sech^2\frac{L}{2}})&k>k_{\rm crit}~,
            \end{cases}\nonumber
\end{align}
where
\begin{equation}
    L_0\equiv2~{\rm arccosh}\qty(\frac{R(y)}{k})
\end{equation}
results from inverting \eqref{eq:critical E}, and the proportionality constant in \eqref{eq:psi E L WKB} is chosen for normalization. By evaluating the integrals in the approximation where
\begin{equation}
    \begin{aligned}
        {R(y)\sech\frac{L_{\rm IR}}{2}}&\gg k~,\quad k\lesssim k_{\rm crit}~,\\
        {R(y)\sech\frac{L_{\rm IR}}{2}}&\ll k~,\quad k>k_{\rm crit}~,
    \end{aligned}
\end{equation}
then the corresponding probability distribution in the WKB approximation \eqref{eq:psi E L WKB} becomes
\begin{equation}\label{eq:p L k dS}
p(L_{\rm IR},k)\simeq\begin{cases}
0&k\lesssim k_{\rm crit}~,\\
\frac{2}{\pi k}\sin^2(L_{\rm IR}k)&k>k_{\rm crit}~.
\end{cases}
\end{equation}
We can now evaluate the following entropy difference based on \eqref{eq:triple scaled entropy}
\begin{equation}\label{eq:entropy difference}
    \Delta S\equiv S(L)-S(L\rightarrow\infty)~,
\end{equation}
using the triple-scaled probability distribution in the following terms inside \eqref{eq:triple scaled entropy},
\begin{equation}\label{eq:new eq york}
\begin{aligned}
\lambda\int_0^\infty\rmd k \qty(p(L_{\rm IR},k)\log p(L_{\rm IR},k)-p_\infty(k)\log p_\infty(k))=\mathcal{O}(\lambda)~,
\end{aligned}
\end{equation}
where 
\begin{equation}
    p_\infty(k)\equiv p(L_{\rm IR}\rightarrow\infty,k)=\frac{1}{\pi}~,
\end{equation}
as seen from \eqref{eq:p L k dS}; and the last equality in \eqref{eq:new eq york} follows from $p(L_{\rm IR},E)\sim \mathcal{O}(1)$.

Thus, the leading order terms contributing to the entropy difference \eqref{eq:entropy difference} are
\begin{equation}\label{eq:intermediate Delta S}
\begin{aligned}
\Delta S\simeq 2\int_0^\infty\rmd k\qty(p(L_{\rm IR},k)-p_\infty(k))\lambda\log(\mu(\theta=\lambda k))~.
\end{aligned}
\end{equation}
We can implement the following relation from \eqref{eq:case1}
\begin{equation}\label{eq:lambda mu}
\begin{aligned}
\lambda\log \mu(\theta=\lambda k)\underset{\rm UV}{\rightarrow}-\frac{1}{2}\pi^2+\mathcal{O}(\lambda\log\lambda)~.
\end{aligned}
\end{equation}
Therefore, to leading order in the semiclassical expansion, the entropy difference \eqref{eq:intermediate Delta S} with (\ref{eq:p L k dS}, \ref{eq:new eq york}, \ref{eq:lambda mu}) becomes
\begin{equation}
\begin{aligned}
\Delta S&\simeq-\pi^2\int_0^\infty\rmd k \qty(p(L_{\rm IR},k)-p_\infty(k)){\simeq}\pi\int_0^{k_{\rm crit}}{\rmd k}~.
\end{aligned}
\end{equation}
Then, the WKB approximation gives
\begin{equation}\label{eq:Delta S final}
   \Delta S\simeq \pi R(y)\sech\frac{L_{\rm IR}}{2}~.
\end{equation}
Moreover, since, in general, traces are defined up to a proportionality constant (equivalently with respect to the choice of normalization of $\ket{\Omega}$), we can perform the following rescaling
\begin{equation}
    \Tr\rightarrow\frac{2}{\lambda}\Tr~,
\end{equation}
so that \eqref{eq:Delta S final} after the rescaling becomes
\begin{equation}\label{eq:WKB entropy}
    \Delta S\underset{\rm IR}{\rightarrow}\frac{2\pi}{\lambda} R(y)\sech\frac{L_{\rm IR}}{2}~.
\end{equation}
The specific choice of overall constant is motivated by the RT formula \eqref{eq:JT gravity Sbulk}.

\section{Deformations with End-Of-The-World Branes and Wormholes}\label{eq:Def ETW Brane Wormhole}
As emphasized in the main text, Hamiltonian deformations modify the energy and microcanonical temperature of the chord auxiliary theory while keeping the thermodynamic entropy the same. There is a complementary situation where the thermodynamic entropy with respect to the seed theory is modified while keeping the same energy spectrum. This occurs when applying constraints in the chord Hilbert space of the DSSYK model \cite{Aguilar-Gutierrez:2025hty}, which lead to ETW brane Hamiltonians in sine dilaton gravity \cite{Blommaert:2025avl}. In this section, we implement Hamiltonian deformations in the DSSYK model with matter  in constrained chord states, resulting in the ETW branes in the finite cutoff bulk theory,\footnote{There are higher dimensional analogs, such as in double holography \cite{Deng:2023pjs} where both effects result in intricate dynamical observables.} and we develop its connection with the double trumpet geometry in the DSSYK model and sine dilaton gravity \cite{Jafferis:2022wez,
Okuyama:2023yat,Okuyama:2023aup,Okuyama:2024eyf,Blommaert:2025avl,Cui:2025sgy,Cao:2025pir,Watanabe:2025rwp}.

\paragraph{Outline}
We begin App.~\ref{sapp:ETW} with a brief review of the ETW brane Hamiltonian from implementing Hilbert space constraints used in Dirac quantization (see \cite{dirac2013lectures,Henneaux:1994lbw} for reviews). In App.~\ref{sapp:ETW deformation} we investigate the semiclassical partition function based on the Al-Salam Chihara (ASC) polynomials and two-point correlation functions with the ETW branes in the DSSYK model. In App.~\ref{sapp:wormhole} we extend the results for trumpet geometries resulting from a Hilbert space trace.

\subsection{Matter Chords as ETW Branes from Deformed DSSYK Model}\label{sapp:ETW}
We investigate Hamiltonian deformations in the ETW brane Hamiltonian for the DSSYK model, first studied by \cite{Okuyama:2023byh}, and followed up by \cite{Blommaert:2025avl,Aguilar-Gutierrez:2025hty,Cao:2025pir,Watanabe:2025rwp}, among others. The seed theory of interest is described by the ASC Hamiltonian \cite{Blommaert:2025avl,Cui:2025sgy,Aguilar-Gutierrez:2025hty},
\begin{equation}\label{eq:ASC Hamiltonian}
    \hH_{\rm ASC}=\frac{J}{\sqrt{\lambda(1-q)}}\qty(\rme^{-\rmi \hat{P}}+(X+Y)\rme^{-\hat{\ell}}+\qty(1-XY\rme^{-\hat{\ell}})\rme^{\rmi \hat{P}}\qty(1-\rme^{-\hat{\ell}}))~,
\end{equation}
which can be derived by imposing constraints in the chord Hilbert space \eqref{eq:Fock space with matter} with one particle insertion \cite{Aguilar-Gutierrez:2025hty}, and $X$ and $Y$ are constant parameters which depend on the specific symmetry sector under consideration. Details about the gauging procedure which fixes $X$ and $Y$ in \eqref{eq:ASC Hamiltonian} can be found in \cite{Aguilar-Gutierrez:2025hty}, which we briefly summarize below for completeness. We begin considering the one particle chord state
\begin{equation}
    \mathcal{H}_1={\rm span}\qty{\ket{\Delta;n_L,n_R}}~,
\end{equation}
and the corresponding one-particle chord Hamiltonian \eqref{eq:pair DSSYK Hamiltonians 1 particle}. We may study symmetry sectors in the one-particle chord space to recover the basis $\ket{H_n}$ \eqref{eq:number op ASC} and the ASC Hamiltonian \eqref{eq:ASC Hamiltonian}. By a symmetry we refer to a transformation, $\hat{U}_i$, that leaves a set of states $\qty{\ket{\psi}}$ in the system invariant i.e.
\begin{equation}\label{eq:transformation}
    \hat{U}_i:~\ket{\psi}\rightarrow\ket{\psi}~.
\end{equation}
One generates the physical Hilbert space ${\mathcal{H}}_{\rm phys}$ from the set of states $\qty{\ket{\psi}}$ that are annihilated by the generator of the corresponding transformation \eqref{eq:transformation}, which we refer to as constraints $\hat{\varphi}_i$:
\begin{equation}\label{eq:symmetries general2}
    \hat{\varphi}_i\ket{\psi}=0~,\quad \ket{\psi}\in{\mathcal{H}}_{\rm phys}~.
\end{equation}
There are different constraints leading to a basis where the one-particle Hamiltonian adopts the ASC form \eqref{eq:ASC Hamiltonian}, which we summarize below (based on \cite{Aguilar-Gutierrez:2025hty}):
\paragraph{One-sided branes} Let $\hat{\varphi}_A:=\hat{P}_L-\delta~\hat{\tilde{P}}_L$, $\hat{\varphi}_B:=\hat{\ell}_L-\hat{\tilde{\ell}}_L/\delta$, where $\delta$ is a constant so that $[\hat{\ell}_L,\hat{P}_L]=[\hat{\tilde{\ell}}_L,\hat{\tilde{P}}_L]$ is invariant under the constraint, even under the limit $\delta\rightarrow0$; and let $\hat{P}_R=\hat{P}$, $\hat{\ell}_R=\hat{\ell}$. By applying these constraints to build the physical Hilbert space, one of the chord Hamiltonians \eqref{eq:pair DSSYK Hamiltonians 1 particle} takes the representation
    \begin{equation}\label{eq:true Okuyama}
    \hH_R\rightarrow\hH^{(\rm 1s)}_{\rm ASC}=\frac{J}{\sqrt{\lambda(1-q)}}\qty(\rme^{-\rmi\hat{P}}+\rme^{\rmi\hat{P}}\qty(1-\rme^{-\hat{\ell}})+q^\Delta\rme^{-\hat{\ell}})~,
\end{equation}
while $\hH_L$ becomes constant.
\paragraph{Two-sided branes}We can take $\hat{\varphi}_A^{(L/R)}:=\hat{P}_{L/R}-\hat{P}$, $\hat{\varphi}_B^{(L/R)}:=\hat{\ell}_{L/R}-\hat{\ell}$, so that the total Hamiltonian can be represented as
\begin{equation}\label{eq:H + ASC}
\hH_L+\hH_R\rightarrow\hH_{\rm ASC}^{(\rm 2s)}=\frac{2J}{\sqrt{\lambda(1-q)}}\qty(\rme^{-\rmi \hat{P}}+\qty(1+q^{\Delta+1}\rme^{-\hat{\ell}})\rme^{\rmi \hat{P}}\qty(1-\rme^{-\hat{\ell}}))~.
\end{equation}
\paragraph{States in the Constrained Chord space}
Next, we summarize the general structure of the states $\qty{\ket{H_n}}$ in the Hilbert space where the ASC Hamiltonian \eqref{eq:ASC Hamiltonian} acts. The operators in the Hamiltonian above can be represented by
\begin{equation}\label{eq:number op ASC}
\begin{aligned}
&\rme^{-\rmi\hat{P}}\ket{H_n}=\ket{H_{n+1}}~,\quad &&\rme^{\rmi\hat{P}}\ket{H_n}=\ket{H_{n-1}}~,\quad &&\hat{n}\ket{H_n}=n\ket{H_n}~.
\end{aligned}
\end{equation}
We can also look for a basis that diagonalizes the Hamiltonian
\begin{equation}
    \hH_{\rm ASC}\ket{\theta}=E(\theta)\ket{\theta}~,
\end{equation}
where $\theta\in[0,\pi]$ is a parametrization of the energy. Next, by acting with the basis $\qty{\ket{H_n}}$ and $\ket{\theta}$ in \eqref{eq:ASC Hamiltonian}, we recover
\begin{equation}
\begin{aligned}
    2\cos\theta ~Q_n=Q_{n+1}+(X+Y)q^n Q_n+(1-XY q^{n-1})(1-q^n)Q_{n-1}~,
\end{aligned}
\end{equation}
where 
\begin{equation}
    \bra{\theta}\ket{H_n}:=Q_n(\cos\theta|X,Y;q)~,\quad E(\theta)=\frac{2J}{\sqrt{\lambda(1-q)}}\cos\theta~,\quad \theta\in[0,\pi]~.
\end{equation}
By requiring $Q_0=1$ and $Q_{-1}=0$, the recurrence relation is solved by the ASC polynomials \cite{al1976convolutions}, which are given by
\begin{equation}\label{eq:ASC pol}
    Q_l(\cos\theta|X,Y;q)=\frac{(XY;q)_l}{X^l}\sum_{n=0}^\infty\frac{(q^{-l},X \rme^{\pm\rmi\theta};q)_n}{(XY,q;q)_n}q^n~.
\end{equation}
The ASC polynomials satisfy the relation
\begin{equation}\label{eq:ortho relations ASC}
\int_0^\pi\rmd\theta ~\tilde{\mu}(\theta)\frac{Q_{l_1}(\cos\theta,X,Y|q)}{\sqrt{(q,XY;q)_{l_1}}}\frac{Q_{l_2}(\cos\theta,X,Y|q)}{\sqrt{(q,XY;q)_{l_2}}}=\delta_{l_1,l_2}~,
\end{equation}
where
\begin{equation}\label{eq:completeness relation}
\begin{aligned}
    \tilde{\mu}(\theta)&=\mu(\theta)\frac{\qty(XY;q)_\infty}{(X\rme^{\pm\rmi\theta}{;}Y\rme^{\pm\rmi\theta};q)_\infty}~,\quad\int_0^\pi\rmd\theta~\tilde{\mu}(\theta)\ket{\theta}\bra{\theta}=\mathbb{1}~.
\end{aligned}
\end{equation}
The integration measure can be expressed as
\begin{equation}\label{eq:deff mu tilde}
   {\bra{\Omega}\ket{B_{X,Y}}=\int_0^\pi\rmd\theta~\tilde{\mu}(\theta)~,}
\end{equation}
where $\ket{B_{X,Y}}$ refers to the ETW brane state,  a generalized q-coherent state, which is defined to satisfy the relation:\footnote{See e.g.~\cite{Watanabe:2025rwp}; the dictionary relating our notation with theirs is $\beta_{\rm There}=(1-q)/((1+X)(1+Y))$ and $\delta_{\rm There}=-(1-q)XY/((1+X)(1+Y))$. The relation between $X$, $Y$ to ETW brane parameters in sine dilaton gravity has been explored in \cite{Aguilar-Gutierrez:2025hty}.}
\begin{equation}
     (\hat{a}+XY\hat{a}^\dagger)\ket{B_{X,Y}}\equiv\frac{X+Y}{\sqrt{1-q}}\ket{B_{X,Y}}~,
 \end{equation}
with $\ha$, $\ha^\dagger$ being annihilation and creation operators in the zero-particle chord space in \eqref{eq:chord numb op}
\begin{equation}
    \hat{a}\equiv\rme^{\rmi\hat{p}}\sqrt{[\hat{n}]}_q~,\quad \hat{a}^\dagger\equiv\sqrt{[\hat{n}]}_q\rme^{-\rmi\hat{p}}~.
\end{equation}

\subsection{Deformed ETW Brane theory}\label{sapp:ETW deformation}
We consider deforming the Hamiltonian after gauging the corresponding symmetry sector in the chord Hilbert space\footnote{Alternatively, one may deform the chord Hamiltonian with matter (\ref{eq:flow eq with matter}) 
\begin{equation}
     \hH_{L/R}\rightarrow \hH_{L/R,\,y}=f(\hH_{L/R})~,
\end{equation}
and then gauge the corresponding symmetry $\hH_{L/R,\,y}\rightarrow\hH_{\rm ASC}$.}
\begin{equation}\label{eq:def ASC}
\begin{aligned}
    {\rm Gauging:}\quad &\hH_{L/R}\rightarrow \hH_{\rm ASC}~,\\
     {\rm Deforming:}\quad  &\hH_{\rm ASC}\rightarrow \hH_{\rm ASC}^y=f_y(\hH_{\rm ASC})~,
\end{aligned}
\end{equation}
with $f_y$ being the Hamiltonian deformation function \eqref{eq:f_y def}.

Now, we evaluate the disk partition function with the ETW brane based on \eqref{eq:def ASC}, which is
\begin{equation}\label{eq:disk part ETW}
     \begin{aligned}
        Z^{(y)}_{\rm ETW}(\beta&|X,Y)\equiv\bra{H_0}\rme^{-\beta\hH_{\rm ASC}^y}\ket{H_0}=\int_0^\pi\rmd\theta\frac{(XY,\rme^{\pm2\rmi\theta};q)_\infty~\rme^{-\beta E_y(\theta)}}{2\pi(X\rme^{\pm\rmi\theta},Y\rme^{\pm\rmi\theta};q)_\infty}
        \\
        &\eqlambda\int_0^\pi\rmd\theta\sqrt{\frac{8\pi(1-XY)\sin^2\theta}{\lambda(1+X^2-2X\cos\theta)(1+Y^2-2Y\cos\theta)}}\rme^{S_\Delta(\theta)-\beta E_y(\theta)}~,
    \end{aligned}
\end{equation}
where the thermodynamic entropy, and microcanonical temperature  are respectively
\begin{equation}\label{eq:temperature entropy ETW one sided}
    \begin{aligned}
        S_{\rm ETW,y}(\theta)&=S(\theta)+\frac{1}{\lambda}\qty({\rm Li}_2(XY)+\frac{\pi^2}{6}+\sum_{\varepsilon=\pm}\qty({\rm Li}_2(X\rme^{\rmi\varepsilon\theta})+{\rm Li}_2(Y\rme^{\rmi\varepsilon\theta})))~,\\
    \dv{S_{\rm ETW,y}}{E_y}&=\dv{E}{E_y}\qty(\beta(\theta)+\frac{\rmi}{2J\sin\theta}\log\frac{\qty(1-X\rme^{i\theta})\qty(1-Y\rme^{i\theta})}{\qty(1-X\rme^{-\rmi\theta})\qty(1-Y\rme^{-\rmi\theta})})~{,}
    \end{aligned}
\end{equation}
and $S(\theta)$, $\beta(\theta)$ appear in \eqref{eq:thermodynamic entropy}, \eqref{eq:microcanonical inverse temperature}.

\begin{figure}
    \centering
    \subfloat[]{\includegraphics[width=0.5\linewidth]{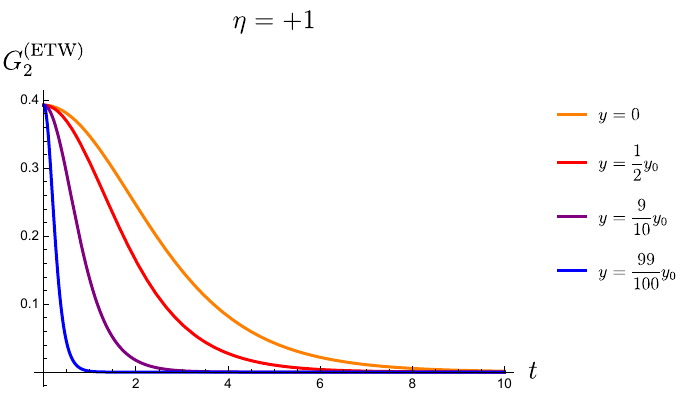}}\subfloat[]{\includegraphics[width=0.5\linewidth]{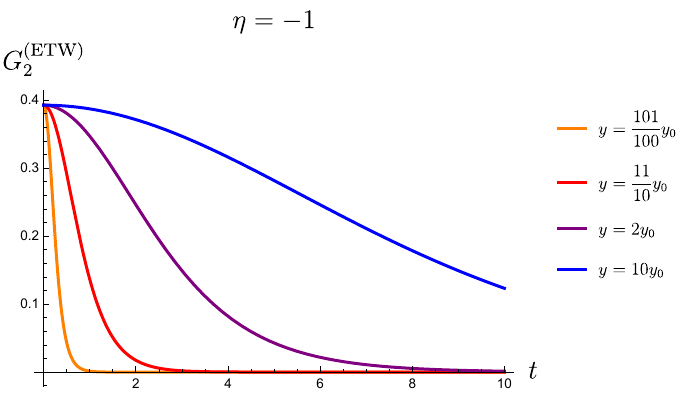}}
    \caption{Two-point correlation function \eqref{eq:correlator XY} resulting from (a) $T^2$ deformation ($\eta=+1$) and (b) $T^2+\Lambda_1$ deformation ($\eta=-1$). We display the enhanced rate of decay of the two-point function as the deformation parameter to approaches the critical value $y_0$ \eqref{eq:y0} due to the redshift factor in the fake temperature \eqref{eq:fake temperature}. The parameters in the plot are $\lambda=10^{-4}$, $J=1$, $\lambda\Delta_1=1$ $\Delta_2=1$ and $\theta=3\pi/4$ for one-sided ETW brane ($X+Y=q^{\Delta}$ and $XY=0$ in\eqref{eq:correlator XY}). Similar results are recovered for two-sided ETW branes.}
    \label{fig:G2ETW}
\end{figure}

Applying the constraints in the crossed four-point function \eqref{eq:fake temperature}, the corresponding two-point function in presence of an ETW brane contains redshift factor in \eqref{eq:resclaing time temp} gives
\begin{align}\label{eq:correlator XY}
    &G_2^{(\rm ETW)}(t,\theta)\equiv\frac{\bra{H_0}\rme^{-\qty(\frac{\beta(\theta)}{2}-\rmi t)\hH_{\rm ASC}^y}q^{\Delta\hat{n}}\rme^{-\qty(\frac{\beta(\theta)}{2}+\rmi t)\hH_{\rm ASC}^y}\ket{H_0}}{\bra{H_0}\rme^{-\beta(\theta)\hH^y_{\rm ASC}}\ket{H_0}}\\
&\eqlambda\qty(\frac{2\sin^2\theta}{1+XY-(X+Y)\cos\theta+\cosh(\frac{2\pi t}{\tilde{\beta}_y(\theta)})\prod_{X_i=X,Y}\sqrt{X_i^2-2X_i\cos\theta+1}})^{\Delta}~,\nonumber
\end{align}
with $\tilde{\beta}_y$ in \eqref{eq:fake temperature}. We display the behavior of the two-point function in Fig.~\ref{fig:G2ETW}.

From the figure, it can be seen that, as expected in finite cutoff holography \cite{McGough:2016lol}, once the deformation parameter approaches its critical value $y_0$ \eqref{eq:y0}, it will experience an enhanced decay rate, while it leads to an enhanced rate of growth for crossed-four point functions \eqref{eq:G4 crossed}. This is also expected since the growth of OTOCs is bounded by that of the two-point correlation functions \cite{Milekhin:2024vbb}. We discussed its implications in the IR and UV regimes in Sec.~\ref{eq:TSL G4} and \ref{ssec:hyperscrambling dS} respectively.

\subsection{Euclidean Wormholes from the DSSYK Model}\label{sapp:wormhole}
Now, we proceed by studying the trumpet geometries from the deformed theory perspective.

The ASC Hamiltonian \eqref{eq:ASC Hamiltonian} can be used to evaluate partition functions and correlation functions for Euclidean wormhole geometries \cite{Aguilar-Gutierrez:2025hty,Blommaert:2025avl}. In this subsection, we will study how the thermodynamics and correlation functions in the ETW brane and Euclidean wormhole theories are modified after implementing the deformations in the corresponding Hamiltonians.

\begin{figure}
    \centering
    \subfloat[]{\includegraphics[width=0.164\linewidth]{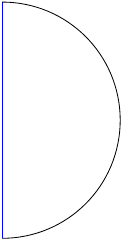}}\hspace{1cm}\subfloat[]{\includegraphics[width=0.22\linewidth]{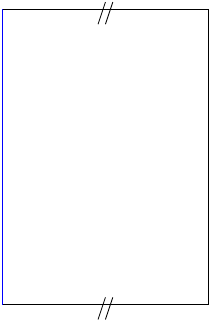}}\hspace{1cm}\subfloat[]{\includegraphics[width=0.35\linewidth]{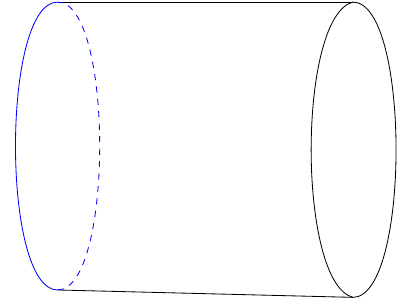}}
    \caption{Construction of the cylinder partition function \eqref{eq:half wormhole ETW}. (a) The disk topology with an ETW brane (blue solid line) is mapped to (b) a rectangle. The identification between two of its edges represents the Hilbert space trace $\sum\bra{H_n}\cdots\ket{H_n}$, which is glued together to form (c) the cylinder, which leads to the divergence in \eqref{eq:half wormhole ETW}.}
    \label{fig:trumpets}
\end{figure}

\begin{figure}
    \centering
    \subfloat[]{\includegraphics[width=0.22\linewidth]{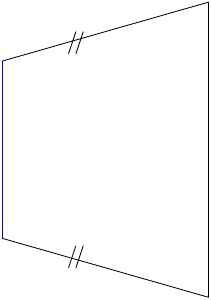}}\hspace{2cm}\subfloat[]{\includegraphics[width=0.33\linewidth]{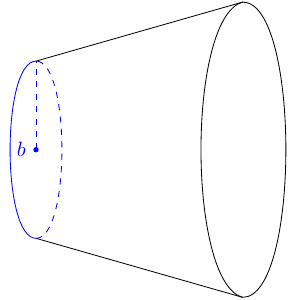}}
    \caption{The regularization implemented in the trumpet partition function \eqref{eq:trumpet amplitude2} can be seen as reducing the length of the ETW brane (solid blue line) in Fig.~\ref{fig:trumpets} to a finite size and gluing appropriately the edges of the plane.}
    \label{fig:trumpet}
\end{figure}

$T^2$ deformations in JT gravity with non-trivial topology have been recently considered in \cite{Bhattacharyya:2025gvd,Griguolo:2021wgy}. This motivates us to look at the corresponding wormhole partition functions
\begin{equation}\label{eq:half wormhole ETW}
\begin{aligned}
Z^{(y)}_{X,Y}(\beta)&\equiv\Tr[\rme^{-\beta \hH^y_{\rm ASC}}]\\
&=\sum_{n=0}^\infty\int_{0}^{\pi}\rmd\theta~\tilde{\mu}(\theta)\rme^{-\beta E_y(\theta)}\bra{H_n}\ket{\theta}\bra{\theta}\ket{H_n}=\int\rmd\theta~\delta(\theta-\theta)\rme^{-\beta E_y(\theta)}~,
\end{aligned}
\end{equation}
where we implemented $\mo=\sum_{n=0}^\infty\ket{H_n}\bra{H_n}$. This partition function can be visualized as cutting in half a partition function and gluing its ends, resulting in a trumpet geometry, as we illustrate in Fig.~\ref{fig:trumpets}.
While the above correlation function is singular, one can define a regularized version, which is the trumpet partition function 
\begin{equation}\label{eq:trumpet amplitude2}
\begin{aligned}
   Z^{(y)}_{b}(\beta)&\equiv\frac{1}{2\pi\rmi}\oint\frac{\rmd Y}{Y^{(b+1)}}Z^{(y)}_{X,Y}(\beta)\\
    &=\frac{2}{(q;q)_\infty}\sum_{n=0}^\infty\int_0^\pi\rmd\theta~2\cos (b\theta)~\rme^{-\beta E_y(\theta)}q^{nb}~,
\end{aligned}
\end{equation}
which we display in Fig.~\ref{fig:trumpet}.

The above expression is not simple to evaluate; it would be interesting to use the techniques of resurgence, similar to the JT gravity analog \cite{Griguolo:2021wgy}, to complete the evaluation. One would be able to use the result to determine the double trumpet amplitude in the bulk
\begin{equation}
    \label{eq:trumpet amplitude}
    Z^{(y)}_b(\beta):=\frac{(q;q)_\infty(1-q^b)}{2}\tilde{Z}^{(y)}_b(\beta)~,
\end{equation}
and construct a partition function for multiple-boundaries Euclidean wormhole \cite{Aguilar-Gutierrez:2025hty}
\begin{equation}
    Z_y(\beta_1,\beta_2,\dots,\beta_m):=\sum_{b=1}^\infty b~Z^{(y)}_{b}(\beta_1)Z^{(y)}_{b}(\beta_2)\cdots Z^{(y)}_{b}(\beta_m)~,
\end{equation}
which reduces to a double-trumpet partition function in the $m=2$ case. Relevant evaluations for this quantity in finite cutoff holography have recently appeared in \cite{Griguolo:2025kpi}.

We move on to a unnormalized thermal two-point correlation function in the trumpet configuration, given by gauging symmetries and implement the Hilbert space trace as 
\begin{equation}\label{eq:G cylinder}
    \begin{aligned}
   &G^{(\Delta_w)}_{\rm cylinder}=\Tr({{\hat{\mathcal{O}}_{\Delta_w}}\rme^{-\beta_1\hH^y_{\rm ASC}}{\hat{\mathcal{O}}_{\Delta_w}}\rme^{-\beta_2\hH^y_{\rm ASC}}})\\
    &={\sum_{n=0}^\infty\bra{H_n}\rme^{-\beta_2\hH^y_{\rm ASC}}q^{\Delta_w\hat{n}}\rme^{-\beta_1\hH^y_{\rm ASC}}\ket{H_n}}\\
    &=\sum_{n=0}^\infty\qty(\int_{0}^{\pi}\prod_{i=1}^2\rmd\theta_i\frac{(XY,\rme^{\pm2\rmi\theta_i}; q)_{\infty}}{(X\rme^{\pm \rmi\theta_i},Y\rme^{\pm \rmi\theta_i};q)_{\infty}}\rme^{-\beta_iE_y(\theta_i)})\bra{H_n}\ket{\theta_1}\bra{\theta_1}q^{\Delta_w\hat{n}}\ket{\theta_2}\bra{\theta_2}\ket{H_n}~.
    \end{aligned}
\end{equation}
Using \eqref{eq:ortho relations ASC} and defining $\beta=\beta_1+\beta_2$ one recovers \begin{equation}\label{eq:cylinder amplitude}
    G^{(\Delta_w)}_{\rm cylinder}(\beta)={\int_0^\pi \rmd\theta~\tilde{\mu}(\theta)\bra{\theta}q^{\Delta_w\hat{n}}\ket{\theta}\rme^{-\beta E(\theta)}}~,
\end{equation}
where $\tilde{\mu}(\theta)=\frac{(XY,\rme^{\pm2\rmi\theta}; q)_{\infty}}{(X\rme^{\pm \rmi\theta},Y\rme^{\pm \rmi\theta};q)_{\infty}}$. This can be further simplified by inserting the identity $\sum_n\ket{K_n^y}\bra{K_n^y}=\mathbb{1}$ and $\hat{n}_y\ket{K_n^y}=n_y\ket{K_n^y}$, resulting into
\begin{equation}\label{eq:correlator cylinder 2}
    G^{(\Delta_w)}_{\rm cylinder}(\beta)=\sum_{n=0}^\infty q^{\Delta_wn}{\int_0^\pi\rmd\theta~\tilde{\mu}(\theta)\abs{\bra{\theta}\ket{K_n}}^2\rme^{-\beta E_y(\theta)}}~.
\end{equation}
It might be possible to simplify the result using saddle point method where the semiclassical solution takes the form
\begin{equation}\label{eq:G cylinder saddle}
    G^{(\Delta_w)}_{\rm cylinder}(\beta)\eqlambda\sum_{n=0}^\infty q^{\Delta_wn}
    \rme^{-\beta E_y(\theta_{\rm s.p.})}\int_0^\pi\rmd\theta~\tilde{\mu}(\theta)\abs{\bra{\theta}\ket{K_n}}^2~,
\end{equation}
where $\theta_{\rm s.p.}$ denotes the corresponding saddle-point solution. This would result from a one-loop quantum corrected two-point correlation function similar to \cite{Aguilar-Gutierrez:2025hty}. However, the details of the evaluation are outside the scope of the current investigation, and we leave them for future directions.

\section{Almheiri-Goel-Hu Model}\label{app:AGH}
In this appendix, we explore the previous results on $T^2$ deformation for the model developed in \cite{Almheiri:2024xtw}, which is significantly simpler than the one in the main text. We are interested in its thermodynamic properties, correlation functions, and the evolution of the chord number.

Motivated by the high-temperature limit of the DSSYK model, \cite{Almheiri:2024xtw} studied the limit where its ensemble averaged theory can be approximately described by
\begin{equation}\label{eq:Hamiltonian AGH}
    \hH=\frac{1}{\sqrt{2}}(\hat{a}+\hat{a}^\dagger)~,
\end{equation}
with $\hat{a}^\dagger$, $\hat{a}$ being creation and annihilation operators of the harmonic oscillator.

By matching correlation functions with bulk geodesics, a (non-local like) dilaton gravity theory dual to the model was proposed by \cite{Almheiri:2024xtw}, which is described by (\ref{eq:Dilaton-gravity theory}) with a potential
\begin{equation}\label{eq:AGH potential}
    U(\Phi)=\frac{96\pi^2}{\beta^4}(1-\Phi)^{-1/3}~,
\end{equation}
where we have performed an overall rescaling in the dilaton field with respect to \cite{Almheiri:2024xtw} to make it dimensionless as in the ansatz (\ref{eq:metric}), where we now have
\begin{equation}\label{eq:AGH background}
  \rmd s^2=F(r)\rmd\tau^2+\frac{\rmd r^2}{F(r)}~,\quad  F(r)=\frac{\beta^4}{4}(1-(1-r)^{2/3})~.
\end{equation}
We assume a counter term of the type (\ref{eq:Gphi special}) to keep the on-shell action finite at the asymptotic boundary, located at $\Phi_B=1$.

\paragraph{Thermodynamics} In the rest of the section, we investigate $T^2$ deformations of the AGH model \eqref{eq:Hamiltonian AGH}. We can directly use the thermodynamic results in \cite{Almheiri:2024xtw}, where
\begin{equation}\label{eq:negative S}
    \beta=-2E~,\quad S=-E^2~,
\end{equation}
to derive the boundary inverse temperature of the $T^2$ deformed model in terms of energy spectrum (\ref{eq:energy spectrum}) and thermodynamic entropy (\ref{eq:thermodynamic entropy})
\begin{align}
    \beta_y=&E_y(-2+y E_y)({1-2y E_y})=-2\sqrt{\abs{S}}\sqrt{1-2y \sqrt{\abs{S}}}~.\label{eq:betalambda negative S}
\end{align}
where the interpretation of the negative thermodynamic entropy is similar to \cite{Almheiri:2024xtw}, namely a relative entropy with respect to the maximal mixed state of the model. Similarly, the density of states (\ref{eq:DOS TTbar}) and the heat capacity (\ref{eq:heat capacity}) can be calculated as:
\begin{equation}
    \rho_y=\rme^{S}\sqrt{1-2y\sqrt{-S}}~,
\end{equation}
\begin{equation}
    C_y=\frac{(\beta_y)^2}{2-3y E_y(2-y E_y)}~.
\end{equation}
Moreover, using the dictionary entry (\ref{eq:def parameter}), we can modify the boundary location from $r=1$ to the stretched horizon (i.e. nearby $r=0$). Note that since the potential (\ref{eq:AGH potential}) is obtained by matching geodesic lengths with two-point correlations, its boundary temperature will be modified $\beta\rightarrow\beta_y$, and the corresponding deformation parameter (\ref{eq:def parameter}) is expressed as
\begin{equation}\label{eq:deformation AGH}
    y=\frac{16\pi G_N (\beta_y)^4}{144\pi^2(1-r_B)^{2/3}}~.
\end{equation}
Note that the boundary temperature dependence is consistent with the non-local nature of the theory proposed by \cite{Almheiri:2024xtw}, and the factor $G_N$ in (\ref{eq:Dilaton-gravity theory}) allows to match ADM energy and entropy in (\ref{eq:on shell action}) with those of the DSSYK model (\ref{eq:energy spectrum}, \ref{eq:thermodynamic entropy}) when we set $16\pi G_N=\sqrt{2}$.

\begin{figure}
    \centering
    \subfloat[]{\includegraphics[width=0.49\textwidth]{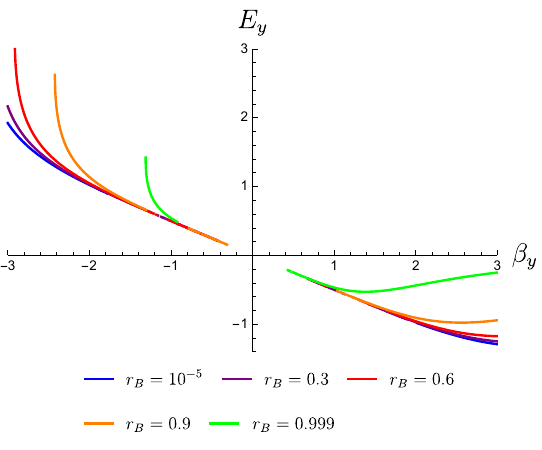}}\hfill\subfloat[]{\includegraphics[width=0.49\textwidth]{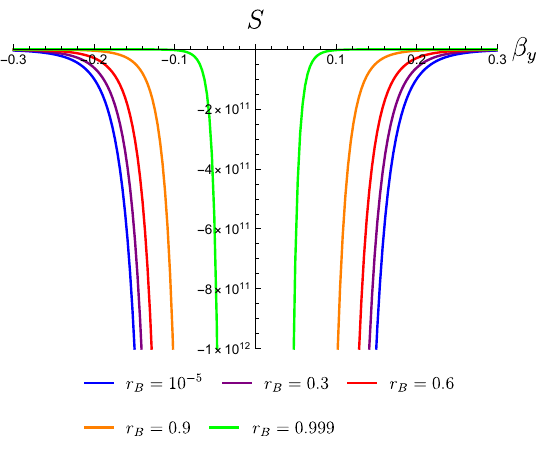}}\\
    \subfloat[]{\includegraphics[width=0.49\textwidth]{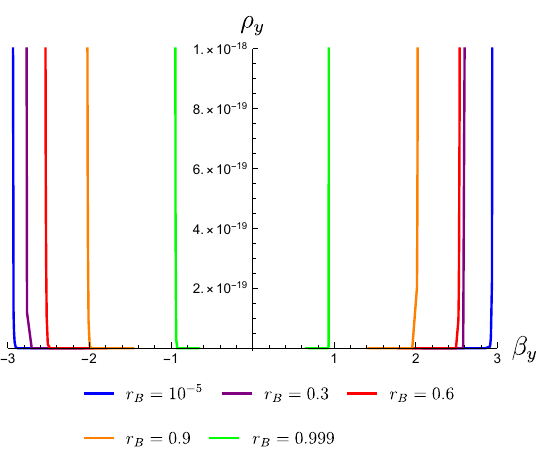}}\hfill\subfloat[]{\includegraphics[width=0.49\textwidth]{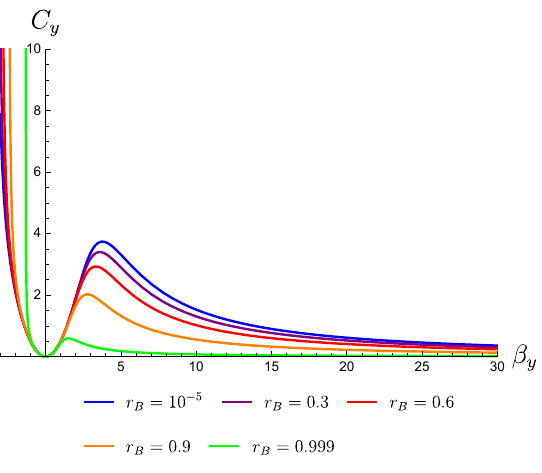}}
    \caption{(a) $T^2$ deformed ($\eta=+1$ \eqref{eq:def parameter}) energy spectrum, (b) thermodynamic entropy, (c) density of states, and (d) heat capacity of the AGH model \cite{Almheiri:2024xtw}, using the map (\ref{eq:deformation AGH}) for different values of the boundary location $r_B$.}
    \label{fig:thermo_AGH}
\end{figure}

The evaluation of (\ref{eq:matter ops DSSYK}) for different boundary locations $r_{B}\in(0,1)$ (i.e.~between the event horizon and the asymptotic boundary) used in (\ref{eq:deformation AGH}) are shown in Fig.~\ref{fig:thermo_AGH}. We {emphasize} that, for the purposes of the analysis, we are allowing the boundary inverse temperature to have a finite range of values, even though the connection with the DSSYK model occurs at high temperatures. As seen, for instance, in Fig.~\ref{fig:thermo_AGH} the energy spectrum is real only under a finite range of boundary temperatures, which do not include strictly infinite boundary temperatures. Note also, there is a non-zero inverse boundary temperature where $S$ and $C_y$ will diverge, independent of the stretched horizon limit. Meanwhile, the fact that the model has a negative thermodynamic entropy (\ref{eq:negative S}, \ref{eq:betalambda negative S}) implies that the growth of the density of states is suppressed in the regime where $S$ diverges. At last, it is seen in Fig.~\ref{fig:thermo_AGH} (d) that the system is manifestly thermodynamically stable for generic values of $r_B$. Further discussion of the results is provided at the end of this subsection.

\begin{figure}
    \centering
    \subfloat[]{\includegraphics[width=0.3\linewidth,valign=t]{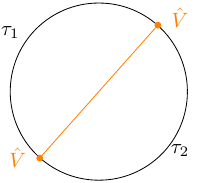}}\hspace{2cm}\subfloat[]{\includegraphics[width=0.33\linewidth,valign=t]{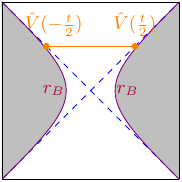}}\\
    \subfloat[]{\includegraphics[width=0.33\linewidth,valign=t]{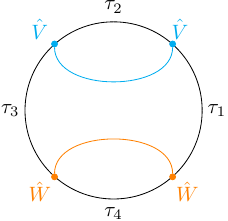}}\hspace{2cm}\subfloat[]{\includegraphics[width=0.32\linewidth,valign=t]{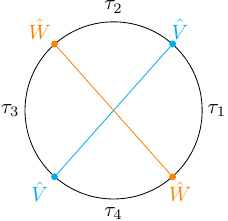}}
    \caption{Correlation functions of matter field operators $\hat{V}$ (cyan) and $\hat{W}$ (orange) (with conformal dimension $\Delta_V$ and $\Delta_W$ respectively) in the AGH model: two-point correlation functions from the (a) boundary and (b) the Lorentzian-signature bulk perspectives (where $\hat{V}(t)=\rme^{\rmi\hH t}\hat{V}(0)\rme^{-\rmi\hH t}$); as well as the (c) uncrossed, and (d) crossed four-point correlation functions.}
    \label{fig:AGH_correlators}
\end{figure}

\paragraph{Correlation functions}
We now proceed to evaluate the $T^2$ deformed thermal two-point correlation function of matter operators $\hat{V}$ with conformal dimension $\Delta_V$ and the corresponding OTOCs (see Fig.~\ref{fig:AGH_correlators}), to compare with our previous findings in Sec.~\ref{sec:correlation functions} for the sine dilaton gravity model. Note that while in Sec.~\ref{sec:correlation functions} we evaluate the correlation functions through saddle point methods in the $\lambda\rightarrow0$ regime, since we are viewing the (\ref{eq:Hamiltonian AGH}) as a model on its own; which is not necessarily related to the DSSYK model in the $\lambda\rightarrow0$ regime, we will perform the evaluation without taking any limits.
In general, these evaluations become intricate due to the $E_y(E)$ factor (except in the $y=0$ case). Fortunately for us, we notice that we can approach a stretched horizon regime for $y\ll1$ in (\ref{eq:deformation AGH}) when $\beta_y\rightarrow0$, which is the high temperature regime where the AGH model is expected to describe the DSSYK model. For this reason, in this subsection, we will be mostly interested in a perturbative $y$ expansion, in contrast to most of the work, which is at non-perturbative order in $y$.

Similar to the identity (\ref{eq:identity theta}), we can express the (Euclidean-signature) evolution of the oscillator number basis $\qty{\ket{n}}$ (i.e. where $\hat{a}^\dagger \hat{a}\ket{n}=n\ket{n}$) of (\ref{eq:Hamiltonian AGH}) in terms of a complete basis of energy eigenstates $\ket{E}$:
\begin{equation}\label{eq:evolved chord}
    \ket{\phi^{(y)}_n(\tau)}\equiv\rme^{-\hH_y\tau}\ket{n}=\int_{-\infty}^{\infty}{\rmd E}\frac{\rme^{-E^2-E_y(E)\tau}H_n(E)}{\sqrt{n!2^n\pi}}\ket{E}~,
\end{equation}
where $H_n(E)=\bra{E}\ket{n}$ is the $n$-th order Hermite polynomial. Note that
\begin{equation}\label{eq:braket AGH y=0}
    \bra{\Omega}\ket{\phi^{(y=0)}_n(\tau)}=\frac{(-\tau)^n}{\sqrt{n!2^n}}\rme^{\tau^2/4}~.
\end{equation}

\paragraph{Two-point correlation functions}
We can then evaluate the correlation function in Fig.~\ref{fig:AGH_correlators} (a) as:
\begin{equation}\label{eq:correlation function AGH}
\begin{aligned}
    \frac{1}{Z_y(\beta)}\bra{\Omega}\rme^{-\tau_2\hH_y}\hat{V}\rme^{-\tau_1\hH_y}\hat{V}\ket{\Omega}&=\bra{\Omega}\rme^{-\tau_2\hH_y}\rme^{-\Delta_V\hat{n}}\rme^{-\tau_1\hH_y}\ket{\Omega}\\
    &=\sum_n{\bra{\phi^{(y)}_n(\tau_2)}\ket{\Omega}}\bra{\Omega}\ket{\phi_n^{(y)}(\tau_1)}\rme^{-\Delta_Vn}~.
\end{aligned}
\end{equation}
To perform the evaluation above, we will first consider a perturbative $y$ expansion, where $E_y(E)=E+\frac{y}{2}E^2+\mathcal{O}(y^2)$. In this case (\ref{eq:correlation function AGH}) with (\ref{eq:braket AGH y=0}) can be expressed as
\begin{equation}
\begin{aligned}
    &\bra{\Omega}\rme^{-\tau_2\hH_y}\hat{V}\rme^{-\tau_1\hH_y}\hat{V}\ket{\Omega}\\
    &=\exp\qty[\frac{1}{4}\sum_{i=1}^2\tau_i^2\qty(1-\frac{y}{2}\tau_i)+\frac{\rme^{-\Delta_V}}{2}\tau_1\tau_2\qty(1-\frac{y}{4}(\tau_1+\tau_2))+\mathcal{O}(y^2)]~.
\end{aligned}
\end{equation}
We can then see the expansion $E_y(E)=E+\frac{y}{2}E^2+\mathcal{O}(y^2)$ as the first order correction in $y$ in the exponent of the correlation function. 

To make the analytic continuation to Lorentzian signature, we take 
\begin{equation}\label{eq:analytic cont}
\tau_1=\rmi t+\frac{\beta}{2}~,\quad\tau_2=-\rmi t+\frac{\beta}{2}~,    
\end{equation}
so that the normalized two-point correlation function becomes
\begin{equation}\label{eq:correlator AGH}
\begin{aligned}
    \mathcal{G}(t)&\equiv\frac{\bra{\Omega}\rme^{-\qty({\beta}/{2}-\rmi t)\hH_y}\hat{V}\rme^{-\qty({\beta}/{2}+\rmi t)\hH_y}\hat{V}\ket{\Omega}}{\bra{\Omega}\rme^{-\frac{\beta}{2}\hH_y}\hat{V}\rme^{-\frac{\beta}{2}\hH_y}\hat{V}\ket{\Omega}}\\
    &=\exp\qty(-\frac{t^2}{2}\qty((1-\rme^{-\Delta_V})+\frac{y\beta}{8}(1-\rme^{-2\Delta_V})+\mathcal{O}(y^2)))~.
\end{aligned}
\end{equation}
We can then see the time dependence of the function in (\ref{eq:correlator AGH}) remains unaffected, at least at the first order perturbation in $y$. Moreover, it follows from (\ref{eq:correlator AGH}) that the tomperature, defined as the rate of decay of the thermal correlation functions, is a fixed value
\begin{equation}\label{eq:def alpha AGH}
    \tilde{\beta}_y=2\pi\alpha~,\quad\text{where}\quad\alpha=\qty(1-\rme^{-\Delta_V}+\frac{y\beta}{8}(1-\rme^{-2\Delta_V}))+\mathcal{O}(y^2)~.
\end{equation}
where we notice that even without the perturbation, the system's temperature and decay of correlation functions are different in general, as in the DSSYK case \cite{Blommaert:2024ymv}.

Performing the calculation (\ref{eq:correlation function AGH}) at the non-perturbative level in the deformation parameter $y$ is in general complicated; however, to get the answer at any order in perturbation theory, we can carry out the expansion
\begin{equation}
    \rme^{-E_y(E)\tau}=\rme^{-E\tau}\sum_{m=0}^\infty \tilde{c}_m(\tau,~y)E^m~,
\end{equation}
where $\tilde{c}_m(\tau,~y)$ is a polynomial function in $\tau$ and $y$, which can be explicitly evaluated for the 1D $T^2$ deformation (\ref{eq:energy spectrum}). One can then use (\ref{eq:braket AGH y=0}) to evaluate:
\begin{equation}
    \int_{-\infty}^{\infty}\rmd E~E^{m}\rme^{-\tau E-E^2}H_n(E)=\qty(-\dv{\tau})^m\qty[\frac{(-\tau)^n}{\sqrt{\pi}}\rme^{\tau^2/4}]~,
\end{equation}
and obtain (\ref{eq:correlation function AGH}) at any order in perturbation theory in $y$. However, for the purpose of later computations, we find it more convenient to evaluate the $y$-corrections in the exponent (\ref{eq:correlator AGH}), instead of the correlation function itself. Lastly, to find the full, non-perturbative answer for (\ref{eq:correlation function AGH}), one might also attempt a numerical evaluation of the integral corresponding to $\bra{\Omega}\ket{\phi_n(\tau)}$ from (\ref{eq:evolved chord}) and performing the sum in $n$. However, we do not pursue this direction, since perturbation theory is a good approximation to study the stretched horizon regime analytically in the $\beta\rightarrow0$ regime.

\begin{figure}
    \centering
    \includegraphics[width=0.6\textwidth]{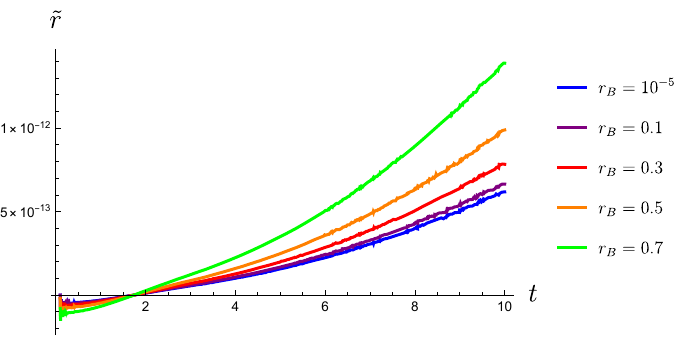}
    \caption{Real part of the relative difference between the out of time order correlator with the perturbative $y$ correction (\ref{eq:OTOC AGH}) and the one in the seed theory (\ref{eq:AGH original OTOC}), in terms of the ratio $\tilde{r}$ \eqref{eq:new ratio}, and the radial cutoff $r_B$ related to the deformation parameter through (\ref{eq:deformation AGH}) as it approaches the horizon located at $r=0$. We have taken $\Delta_V=1$, $\Delta_W=2$, $\beta=0.01$, and different values of $r_B$. The numerical noise is produced by the relative phases in (\ref{eq:OTOC AGH})}
    \label{fig:OTOC_time_AGH}
\end{figure}

\paragraph{Squared correlators}We would like to evaluate the Lorentzian-signature squared correlator defined by
\begin{equation}\label{eq:Lorentzian OTOC}
   {\rm SC}_y(t)\equiv\frac{1}{Z_y(\beta)}\bra{\Omega}\rme^{-\beta\hH_y}\qty[\hat{W}(0),~\rme^{\rmi \hH_y t}\hat{V}(0)\rme^{-\rmi \hH_y t}]^2\ket{\Omega}~,
\end{equation}
for the operators $\hat{V}$ and $\hat{W}$ with conformal dimensions $\Delta_{V}$ and $\Delta_W$ respectively. The diagrammatic techniques for the calculation have been worked out in App.~B of \cite{Almheiri:2024xtw}, so we state the results after incorporating the $T^2$ deformation at order $\mathcal{O}(y)$, which essentially amounts to the modification $\tau_i\rightarrow\tau_i(1-\frac{y}{4}\tau_i)$ inside the overall exponent. This allows us to express the uncrossed four-point correlation function (Fig.~\ref{fig:AGH_correlators} (c)):
\begin{equation}\label{eq:uncrossed_4pnt}
    \begin{aligned}
        &\log(\bra{\Omega}\rme^{-\hH_y\tau_4}V\rme^{-\hH_y\tau_3}W\rme^{-\hH_y\tau_2}W\rme^{-\hH_y\tau_1}V\ket{\Omega})\\
        &=\sum_{i=1}^4\tau_{i}^2+\left(\tau _2+\tau _4\right)\qty(\frac{\rme^{-\Delta_V}}{2}\tau _3+\frac{\rme^{-\Delta_W}}{2}\tau _1)+\frac{\rme^{-\Delta_V-\Delta_W}}{2}\tau _2 \tau _4+\frac{\tau _1 \tau _3}{2} \\
      &\quad-\frac{y}{8} \left(\rme^{-\Delta_V}\tau _1\left(\tau _2 \left(\tau _2+\tau _3\right)+\tau _4 \left(\tau _1+\tau_4\right)\right)+\left(\tau _1^3+\tau _3^3+\left(\tau _2+\tau _4\right)\left(\tau _2^2+\tau _4^2\right)\right)\right)\\
   &\quad-\frac{y}{8} \rme^{-\Delta_V-\Delta _W}\tau _3 \left(\tau _1 \left(\tau _1+\tau_3\right)+\left(\tau _2 \left(\tau _1+\tau _2\right)+\tau _4 \left(\tau_3+\tau _4\right)\right) \rme^{\Delta_V}\right)+\mathcal{O}\left(y^2\right)~,
        \end{aligned}
\end{equation}
as well as the crossed four-point correlation function (Fig.~\ref{fig:AGH_correlators} (d)):
\begin{equation}\label{eq:crossed_4pnt}
    \begin{aligned}
        &\bra{\Omega}\rme^{-\hH_y\tau_4}V\rme^{-\hH_y\tau_3}W\rme^{-\hH_y\tau_2}V\rme^{-\hH_y\tau_1}W\ket{\Omega}\\
        &=\sum_{i=1}^4\tau_{i}^2+\frac{\rme^{-\Delta_V}}{2}\left(\tau _1 \tau _2+\tau _3 \tau _4\right)+\frac{\rme^{-\Delta_W}}{2}\left(\tau _2 \tau _3+\tau _1 \tau _4\right)+\frac{\rme^{-\Delta_V-\Delta_W}}{2}\left(\tau _1 \tau _3+\tau _2 \tau _4\right) -\Delta_V\Delta_W\\
        &\quad-\frac{y}{8}\qty(\left(\tau _2 \tau _3 \left(\tau_2+\tau _3\right)+\tau _1 \tau _4 \left(\tau _1+\tau_4\right)\right) \rme^{-\Delta _W}+\left(\tau_1 \tau _2 \left(\tau _1+\tau _2\right)+\tau _3 \tau _4\left(\tau _3+\tau _4\right)\right) \rme^{-\Delta_V})\\
        &\quad+\frac{y}{8}\qty(\sum_{i=1}^4\tau_i^3-\left(\tau _1 \tau _3\left(\tau _1+\tau _3\right)+\tau _2 \tau _4 \left(\tau_2+\tau _4\right)\right) \rme^{-\Delta _V-\Delta _W})+\mathcal{O}(y^2)~.
        \end{aligned}
\end{equation}
Combining (\ref{eq:uncrossed_4pnt}) and (\ref{eq:crossed_4pnt}), and choosing the corresponding value for $\tau_i$, we can evaluate (\ref{eq:Lorentzian OTOC}) at order $\mathcal{O}(y)$ as
\begin{dmath}\label{eq:OTOC AGH}
{{\rm SC}_y(t)}\equiv\eval{\frac{1}{Z_y(\beta)}\bra{\Omega}\rme^{-\hH_y\beta}\qty[W(0),~V(t)]^2\ket{\Omega}}_{\mathcal{O}(y)}=\rme^{\frac{1}{16} \left(-8 t^2-\beta  t \rme^{-\Delta _V-\Delta _W}
   \left(\rme^{\Delta _V} (y (t (\rmi t y+\beta  y-4)-2 \rmi \beta )+8 \rmi)+2 i (\beta 
   y-4)\right)-\beta  t \rme^{-\Delta _V} (y (t (\rmi t y+\beta  y-4)-2 \rmi \beta )+8
   i)-2 \rmi \beta  t (3 \beta  y-4)-2 \beta ^3 y\right)}\\ \left(\rme^{\frac{1}{32} t^2 \rme^{-\Delta _V-\Delta _W} \left(\rme^{\Delta _V}
   \left(y \left(2 t^2 y+\beta  (\beta  y-12)\right)+32\right)+\rme^{\Delta _W}
   \left(y \left(2 t^2 y+\beta  (\beta  y-12)\right)+32\right)+y \left(2 t^2
   y+\beta  (12-\beta  y)\right)+4 (3 \beta  y-4) \rme^{\Delta _V+\Delta
   _W}-32\right)}\\ \left(1+\rme^{\frac{1}{8} \rmi \beta  t \rme^{-\Delta
   _V-\Delta _W} \left(2 \rme^{\Delta _W} \left(\rme^{\Delta _V} (3 \beta 
   y-4)+\beta  (-y)+4\right)-2 \left(\rme^{\Delta _V}-1\right) (\beta 
   y-4)\right)}\right)-\left(\rme^{\frac{1}{16} \beta  t^2 y \rme^{-\Delta
   _V} (\beta  y-4)}+\rme^{\frac{1}{16} \beta  t^2 y \rme^{-\Delta _W} (\beta 
   y-4)}\right) \rme^{\frac{1}{32} t \rme^{-\Delta _V} \left(\rme^{\Delta _V}
   (16 t+4 \rmi \beta  (3 \beta  y-4))-4 \rmi \beta  \left(e^{\Delta _V}-1\right)
   e^{-\Delta _W} (\beta  y-4)-4 \rmi \beta  (\beta  y-4)\right)}\right)~.
\end{dmath}
Note that the $y\rightarrow0$ limit reproduces one of the results in \cite{Almheiri:2024xtw}
\begin{equation}\label{eq:AGH original OTOC}
\begin{aligned}
    &{\rm SC}_{y\rightarrow0}(t)\equiv\frac{1}{Z(\beta)}\bra{\Omega}\rme^{-\beta\hH}\qty[\hat{W}(0),~\hat{V}(t)]^2\ket{\Omega}\\
    &=-2+2\cos\qty(\qty(1-\rme^{-\Delta_V})(1-\rme^{-\Delta_W}){\frac{\beta t}{2}})\rme^{-\qty(1-\rme^{-\Delta_V})(1-\rme^{-\Delta_W})(1+t^2)}~.
\end{aligned}
\end{equation}
A comparison between the evolution of (\ref{eq:OTOC AGH}) and (\ref{eq:AGH original OTOC}) is shown in Fig.~\ref{fig:OTOC_time_AGH} in terms of a ratio
\begin{equation}\label{eq:new ratio}
    \tilde{r}\equiv\Re\qty(({\rm SC}(t)-{\rm SC}_{\rm AGH}(t))/{\rm SC}_{\rm AGH}(t))
\end{equation}
We notice that the increased rate of growth of the OTOC (\ref{eq:OTOC AGH}) is not significantly impacted in the stretched horizon limit $y\rightarrow\frac{2\sqrt{2}\beta^4}{144\pi^2}$.
Moreover, one can evaluate the scrambling time in this model, defined as the time scale associated to the exponential decay in (\ref{eq:OTOC AGH}); one finds that it is not significantly different from the one corresponding to (\ref{eq:AGH original OTOC}) $t_{\rm sc}=\qty((1-\rme^{-\Delta_V})(1-\rme^{-\Delta_W}))^{-1/2}$, given that it is a $\mathcal{O}(1)$ value that remains below the time scale where ${\rm SC}$ and ${\rm SC}_{\rm AGH}$ start differing substantially (see Fig.~\ref{fig:OTOC_time_AGH}).

\paragraph{Krylov Spread Complexity}
Meanwhile, the expectation value of the chord number operator (\ref{eq:spread Complexity}) in the state $\rme^{-\hH_y \tau}\ket{\Omega}$ can be evaluated directly from (\ref{eq:correlation function AGH}) by taking a derivative in $\Delta_V$ and taking the limit $\Delta_V\rightarrow0$ (see e.g.~\cite{Heller:2024ldz}). This results in:
\begin{equation}
    \begin{aligned}
        \frac{1}{Z}\bra{\Omega}\rme^{-\hH_y\tau_2}\hat{n}\rme^{-\hH_y\tau_1}\ket{\Omega}&=\sum_{n=0}^\infty n{\bra{\phi^{(y)}_n(\tau_2)}\ket{\Omega}}\bra{\Omega}\ket{\phi_n^{(y)}(\tau_1)}\\
        &=\frac{\tau_1\tau_2}{2}\qty(1+2y(\tau_1+\tau_2))~.
    \end{aligned}
\end{equation}
Its analytic continuation to Lorentzian signature (\ref{eq:analytic cont}) allow us to evaluate the spread complexity as
\begin{equation}
    \expval{n}_y(t)=\frac{\bra{\Omega}\rme^{-\hH_y\tau_2}\hat{n}\rme^{-\hH_y\tau_1}\ket{\Omega}}{\bra{\Omega}\rme^{-\hH_y{\beta_y}}\ket{\Omega}}=\frac{1}{2}t^2(1+2y\beta)+\mathcal{O}(y^2)~.
\end{equation}
Note that the rate of growth of $\expval{n}_y(t)$ is faster when $y\rightarrow\frac{\sqrt{2}\beta^4}{144\pi^2}$ (i.e.~the stretched horizon limit), similar to our results in the main text. In contrast, the thermodynamic properties in Fig.~\ref{fig:thermo_AGH} do not display an enhancement in the same limit. Perhaps this is expected since \cite{Susskind:2021esx} associates the enhanced growth of observables in time with a hyperfast scrambling of information, which would not be exhibited in generic integrable models, such as this one. It had been noticed in \cite{Blommaert:2024ymv} that the model proposed in \cite{Almheiri:2024xtw} displays quite different observables with respect to the $\lambda\rightarrow0$ limit in the sine dilaton gravity model, which should still describe the DSSYK model at high temperatures, without the Heisenberg algebra limit (\ref{eq:Hamiltonian AGH}). Therefore, our findings in this section compared to Secs. \ref{sec:thermo} add further contrast between these two approaches, as different observables display rather different behaviors in the stretched horizon limit.

\section{First Law of Thermodynamics at Finite Cutoff}\label{sec:1st law thermo}
Motivated by recent developments on black hole thermodynamics in finite cutoff holography \cite{Zhang:2025dgm}, we study the first law of thermodynamics in the deformed DSSYK model based on our results in Sec.~\ref{sec:flow eq}.

We specialize in $T^2(+\Lambda_1)$ deformations, where the relevant deformation parameter $\eta=\pm1$ \eqref{eq:def parameter}. Let us use the semiclassical entropy in \eqref{eq:thermodynamic entropy} to express the energy spectrum in \eqref{eq:E(theta) eta+1} ($\eta=+1$) for the $T^2$ deformation, and \eqref{eq:E(theta) eta-1} ($\eta=-1$) for the $T^2+\Lambda_1$ deformation. This results in
\begin{equation}\label{eq:E new eta}
    \begin{aligned}
        E^{\eta}_y=\frac{1}{y}\qty(1-\eta\sqrt{\eta\qty(1-\frac{2y}{\pi V\sqrt{\lambda(1-q)}}\sin\sqrt{\frac{\lambda(S_0-S)}{2}})})~,
    \end{aligned}
\end{equation}
where we have defined $V\equiv\frac{2\pi}{J}$. By taking variations with respect to each of the parameters in \eqref{eq:E new eta} we can write the extended first law for the deformed DSSYK model as
\begin{align}
    \rmd E_y=&P_y\rmd V+(\beta_y)^{-1}\rmd S+\tilde{\mu}_y\rmd \lambda+\nu_y\rmd y~,\\
    P_y\equiv&-\frac{\eta ^2 \sin
   \left(\frac{\sqrt{\lambda 
   \left(S_0-S\right)}}{\sqrt{2}}\right)}{\pi  V^2 \sqrt{\lambda 
   (1-q)} \sqrt{\eta -\frac{2 \eta 
   y \sin \left(\frac{\sqrt{\lambda 
   \left(S_0-S\right)}}{\sqrt{2}}\right)}{\pi  V \sqrt{\lambda 
   (1-q)}}}}~,\\
    \tilde{\mu}_y\equiv&-\frac{\eta ^2 q \left(\frac{2
   (\lambda  q-q+1) \sqrt{\lambda 
   \left(S_0-S\right)} \sin
   \left(\frac{\sqrt{\lambda 
   \left(S_0-S\right)}}{\sqrt{2}}\right)}{q}+\frac{\sqrt{2} \lambda 
   (1-q) \left(S-S_0\right) \cos
   \left(\frac{\sqrt{\lambda 
   \left(S_0-S\right)}}{\sqrt{2}}\right)}{q}\right)}{4 \sqrt{\pi }
   \lambda  (1-q) V \sqrt{\lambda 
   (1-q)} \sqrt{\lambda 
   \left(S_0-S\right)} \sqrt{\pi 
   \eta -\frac{2 \eta  y \sin
   \left(\frac{\sqrt{\lambda 
   \left(S_0-S\right)}}{\sqrt{2}}\right)}{V \sqrt{\lambda  (1-q)}}}}~,\\
    \nu_y\equiv&\frac{\eta  \sqrt{\eta -\frac{2 \eta
    y \sin \left(\frac{\sqrt{\lambda\left(S_0-S\right)}}{\sqrt{2}}\right)}{\pi  V \sqrt{\lambda 
   (1-q)}}}-1}{y^2}+\frac{\eta ^2
   \sin \left(\frac{\sqrt{\lambda 
   \left(S_0-S\right)}}{\sqrt{2}}\right)}{\pi  V y \sqrt{\lambda 
   (1-q)} \sqrt{\eta -\frac{2 \eta 
   y \sin \left(\frac{\sqrt{\lambda 
   \left(S_0-S\right)}}{\sqrt{2}}\right)}{\pi  V \sqrt{\lambda 
   (1-q)}}}}~,
\end{align}
while $\beta_y$ appears in \eqref{eq:microcanonical inverse temperature}. One may interpret $\tilde{\mu}_y$ as a chemical potential associated with variations in the parameter $\lambda$ \eqref{eq:double scaling} for a fixed deformation parameter $y$; and similarly for $\nu_y$, where $y$ is allowed to vary instead. Note that the notation suggests that $V$ and $P_y$ are analogs to thermodynamic volumes and pressures in the bulk \cite{Zhang:2025dgm}; however, they are viewed only as definitions of different combinations of parameters from our analysis.

\bibliographystyle{JHEP}
\bibliography{references.bib}

@article{Gross:2019uxi,
    author = "Gross, David J. and Kruthoff, Jorrit and Rolph, Andrew and Shaghoulian, Edgar",
    title = "{Hamiltonian deformations in quantum mechanics, $T\bar T$, and the SYK model}",
    eprint = "1912.06132",
    archivePrefix = "arXiv",
    primaryClass = "hep-th",
    doi = "10.1103/PhysRevD.102.046019",
    journal = "Phys. Rev. D",
    volume = "102",
    number = "4",
    pages = "046019",
    year = "2020"
}

@article{Nandy:2024zcd,
    author = "Nandy, Pratik",
    title = "{Tridiagonal Hamiltonians modeling the density of states of the Double-Scaled SYK model}",
    eprint = "2410.07847",
    archivePrefix = "arXiv",
    primaryClass = "hep-th",
    month = "10",
    year = "2024"
}

@article{Strominger:2001pn,
    author = "Strominger, Andrew",
    title = "{The dS / CFT correspondence}",
    eprint = "hep-th/0106113",
    archivePrefix = "arXiv",
    doi = "10.1088/1126-6708/2001/10/034",
    journal = "JHEP",
    volume = "10",
    pages = "034",
    year = "2001"
}

@inproceedings{Witten:2001kn,
    author = "Witten, Edward",
    title = "{Quantum gravity in de Sitter space}",
    booktitle = "{Strings 2001: International Conference}",
    eprint = "hep-th/0106109",
    archivePrefix = "arXiv",
    month = "6",
    year = "2001"
}

@article{Maldacena:2002vr,
    author = "Maldacena, Juan Martin",
    title = "{Non-Gaussian features of primordial fluctuations in single field inflationary models}",
    eprint = "astro-ph/0210603",
    archivePrefix = "arXiv",
    doi = "10.1088/1126-6708/2003/05/013",
    journal = "JHEP",
    volume = "05",
    pages = "013",
    year = "2003"
}

@inproceedings{Spradlin:2001pw,
    author = "Spradlin, Marcus and Strominger, Andrew and Volovich, Anastasia",
    title = "{Les Houches lectures on de Sitter space}",
    booktitle = "{Les Houches Summer School: Session 76: Euro Summer School on Unity of Fundamental Physics: Gravity, Gauge Theory and Strings}",
    eprint = "hep-th/0110007",
    archivePrefix = "arXiv",
    pages = "423--453",
    month = "10",
    year = "2001"
}

@article{Vasiliev:1999ba,
    author = "Vasiliev, Mikhail A.",
    editor = "Shifman, Mikhail A.",
    title = "{Higher spin gauge theories: Star product and AdS space}",
    eprint = "hep-th/9910096",
    archivePrefix = "arXiv",
    reportNumber = "FIAN-TD-24-99",
    doi = "10.1142/9789812793850_0030",
    pages = "533--610",
    month = "10",
    year = "1999"
}

@article{Ng:2012xp,
    author = "Ng, Gim Seng and Strominger, Andrew",
    title = "{State/Operator Correspondence in Higher-Spin dS/CFT}",
    eprint = "1204.1057",
    archivePrefix = "arXiv",
    primaryClass = "hep-th",
    doi = "10.1088/0264-9381/30/10/104002",
    journal = "Class. Quant. Grav.",
    volume = "30",
    pages = "104002",
    year = "2013"
}

@article{Anninos:2011ui,
    author = "Anninos, Dionysios and Hartman, Thomas and Strominger, Andrew",
    title = "{Higher Spin Realization of the dS/CFT Correspondence}",
    eprint = "1108.5735",
    archivePrefix = "arXiv",
    primaryClass = "hep-th",
    doi = "10.1088/1361-6382/34/1/015009",
    journal = "Class. Quant. Grav.",
    volume = "34",
    number = "1",
    pages = "015009",
    year = "2017"
}

@article{Hartle:2012tv,
    author = "Hartle, James B. and Hawking, S. W. and Hertog, Thomas",
    title = "{Quantum Probabilities for Inflation from Holography}",
    eprint = "1207.6653",
    archivePrefix = "arXiv",
    primaryClass = "hep-th",
    doi = "10.1088/1475-7516/2014/01/015",
    journal = "JCAP",
    volume = "01",
    pages = "015",
    year = "2014"
}

@article{Hertog:2011ky,
    author = "Hertog, Thomas and Hartle, James",
    title = "{Holographic No-Boundary Measure}",
    eprint = "1111.6090",
    archivePrefix = "arXiv",
    primaryClass = "hep-th",
    doi = "10.1007/JHEP05(2012)095",
    journal = "JHEP",
    volume = "05",
    pages = "095",
    year = "2012"
}

@article{Bobev:2022lcc,
    author = "Bobev, Nikolay and Hertog, Thomas and Hong, Junho and Karlsson, Joel and Reys, Valentin",
    title = "{Microscopics of de Sitter Entropy from Precision Holography}",
    eprint = "2211.05907",
    archivePrefix = "arXiv",
    primaryClass = "hep-th",
    doi = "10.1103/PhysRevX.13.041056",
    journal = "Phys. Rev. X",
    volume = "13",
    number = "4",
    pages = "041056",
    year = "2023"
}

@article{Bousso:2001mw,
    author = "Bousso, Raphael and Maloney, Alexander and Strominger, Andrew",
    title = "{Conformal vacua and entropy in de Sitter space}",
    eprint = "hep-th/0112218",
    archivePrefix = "arXiv",
    doi = "10.1103/PhysRevD.65.104039",
    journal = "Phys. Rev. D",
    volume = "65",
    pages = "104039",
    year = "2002"
}

@article{Strominger:2001gp,
    author = "Strominger, Andrew",
    title = "{Inflation and the dS / CFT correspondence}",
    eprint = "hep-th/0110087",
    archivePrefix = "arXiv",
    doi = "10.1088/1126-6708/2001/11/049",
    journal = "JHEP",
    volume = "11",
    pages = "049",
    year = "2001"
}

@article{Vasiliev:1990en,
    author = "Vasiliev, Mikhail A.",
    title = "{Consistent equation for interacting gauge fields of all spins in (3+1)-dimensions}",
    reportNumber = "LEBEDEV-90-29",
    doi = "10.1016/0370-2693(90)91400-6",
    journal = "Phys. Lett. B",
    volume = "243",
    pages = "378--382",
    year = "1990"
}

@article{Anninos:2012ft,
    author = "Anninos, Dionysios and Denef, Frederik and Harlow, Daniel",
    title = "{Wave function of Vasiliev\textquoteright{}s universe: A few slices thereof}",
    eprint = "1207.5517",
    archivePrefix = "arXiv",
    primaryClass = "hep-th",
    reportNumber = "SU-ITP-12-19",
    doi = "10.1103/PhysRevD.88.084049",
    journal = "Phys. Rev. D",
    volume = "88",
    number = "8",
    pages = "084049",
    year = "2013"
}

@article{Tang:2024xgg,
    author = "Tang, Haifeng",
    title = "{Entanglement entropy in type II$_1$ von Neumann algebra: examples in Double-Scaled SYK}",
    eprint = "2404.02449",
    archivePrefix = "arXiv",
    primaryClass = "hep-th",
    month = "4",
    year = "2024"
}

@article{Goel:2023svz,
    author = "Goel, Akash and Narovlansky, Vladimir and Verlinde, Herman",
    title = "{Semiclassical geometry in double-scaled SYK}",
    eprint = "2301.05732",
    archivePrefix = "arXiv",
    primaryClass = "hep-th",
    doi = "10.1007/JHEP11(2023)093",
    journal = "JHEP",
    volume = "11",
    pages = "093",
    year = "2023"
}

@article{Hoehn:2023ehz,
    author = "Hoehn, Philipp A. and Kotecha, Isha and Mele, Fabio M.",
    title = "{Quantum Frame Relativity of Subsystems, Correlations and Thermodynamics}",
    eprint = "2308.09131",
    archivePrefix = "arXiv",
    primaryClass = "quant-ph",
    month = "8",
    year = "2023"
}

@article{Dubovsky:2013ira,
    author = "Dubovsky, Sergei and Gorbenko, Victor and Mirbabayi, Mehrdad",
    title = "{Natural Tuning: Towards A Proof of Concept}",
    eprint = "1305.6939",
    archivePrefix = "arXiv",
    primaryClass = "hep-th",
    doi = "10.1007/JHEP09(2013)045",
    journal = "JHEP",
    volume = "09",
    pages = "045",
    year = "2013"
}

@article{Barbon:2020amo,
    author = "Barbon, Jos\'e L. F. and Rabinovici, E.",
    title = "{Remarks on the thermodynamic stability of $T \bar T$ deformations}",
    eprint = "2004.10138",
    archivePrefix = "arXiv",
    primaryClass = "hep-th",
    doi = "10.1088/1751-8121/ab99ee",
    journal = "J. Phys. A",
    volume = "53",
    number = "42",
    pages = "424001",
    year = "2020"
}

@article{Galante:2022nhj,
    author = "Galante, Damian",
    title = "{Geodesics, complexity and holography in (A)dS$_2$}",
    doi = "10.22323/1.406.0359",
    journal = "PoS",
    volume = "CORFU2021",
    pages = "359",
    year = "2022"
}

@article{Ishii:2019uwk,
    author = "Ishii, Takaaki and Okumura, Suguru and Sakamoto, Jun-Ichi and Yoshida, Kentaroh",
    title = "{Gravitational perturbations as $T\bar{T}$-deformations in 2D dilaton gravity systems}",
    eprint = "1906.03865",
    archivePrefix = "arXiv",
    primaryClass = "hep-th",
    reportNumber = "KUNS-2765, YITP-19-48",
    doi = "10.1016/j.nuclphysb.2019.114901",
    journal = "Nucl. Phys. B",
    volume = "951",
    pages = "114901",
    year = "2020"
}

@article{Okumura:2020dzb,
    author = "Okumura, Suguru and Yoshida, Kentaroh",
    title = "{$T\bar{T}$-deformation and Liouville gravity}",
    eprint = "2003.14148",
    archivePrefix = "arXiv",
    primaryClass = "hep-th",
    reportNumber = "KUNS-2810",
    doi = "10.1016/j.nuclphysb.2020.115083",
    journal = "Nucl. Phys. B",
    volume = "957",
    pages = "115083",
    year = "2020"
}

@article{Anderson_2008,
   title={On boundary value problems for Einstein metrics},
   volume={12},
   ISSN={1465-3060},
   url={http://dx.doi.org/10.2140/gt.2008.12.2009},
   DOI={10.2140/gt.2008.12.2009},
   number={4},
   journal={Geometry \&amp; Topology},
   publisher={Mathematical Sciences Publishers},
   author={Anderson, Michael T},
   year={2008},
   month=jul, pages={2009–2045} }

@article{Witten:2018lgb,
    author = "Witten, Edward",
    title = "{A note on boundary conditions in Euclidean gravity}",
    eprint = "1805.11559",
    archivePrefix = "arXiv",
    primaryClass = "hep-th",
    doi = "10.1142/S0129055X21400043",
    journal = "Rev. Math. Phys.",
    volume = "33",
    number = "10",
    pages = "2140004",
    year = "2021"
}

@article{Tietto:2025oxn,
    author = "Tietto, Damiano and Verlinde, Herman",
    title = "{A microscopic model of de Sitter spacetime with an observer}",
    eprint = "2502.03869",
    archivePrefix = "arXiv",
    primaryClass = "hep-th",
    month = "2",
    year = "2025"
}

@article{Narovlansky:2025tpb,
    author = "Narovlansky, Vladimir",
    title = "{Towards a microscopic description of de Sitter dynamics}",
    eprint = "2506.02109",
    archivePrefix = "arXiv",
    primaryClass = "hep-th",
    month = "6",
    year = "2025"
}

@article{Silverstein:2024xnr,
    author = "Silverstein, Eva and Torroba, Gonzalo",
    title = "{Timelike-bounded $dS_4$ holography from a solvable sector of the $T^2$ deformation}",
    eprint = "2409.08709",
    archivePrefix = "arXiv",
    primaryClass = "hep-th",
    month = "9",
    year = "2024"
}

@article{Apolo:2019zai,
    author = "Apolo, Luis and Detournay, Stephane and Song, Wei",
    title = "{TsT, $T\bar{T}$ and black strings}",
    eprint = "1911.12359",
    archivePrefix = "arXiv",
    primaryClass = "hep-th",
    doi = "10.1007/JHEP06(2020)109",
    journal = "JHEP",
    volume = "06",
    pages = "109",
    year = "2020"
}

@article{Okuyama:2023byh,
    author = "Okuyama, Kazumi",
    title = "{End of the world brane in double scaled SYK}",
    eprint = "2305.12674",
    archivePrefix = "arXiv",
    primaryClass = "hep-th",
    doi = "10.1007/JHEP08(2023)053",
    journal = "JHEP",
    volume = "08",
    pages = "053",
    year = "2023"
}

@article{Giveon:2017nie,
    author = "Giveon, Amit and Itzhaki, Nissan and Kutasov, David",
    title = "{$ \mathrm{T}\overline{\mathrm{T}} $ and LST}",
    eprint = "1701.05576",
    archivePrefix = "arXiv",
    primaryClass = "hep-th",
    doi = "10.1007/JHEP07(2017)122",
    journal = "JHEP",
    volume = "07",
    pages = "122",
    year = "2017"
}

@article{Anninos:2023epi,
    author = "Anninos, Dionysios and Galante, Dami\'an A. and Maneerat, Chawakorn",
    title = "{Gravitational observatories}",
    eprint = "2310.08648",
    archivePrefix = "arXiv",
    primaryClass = "hep-th",
    doi = "10.1007/JHEP12(2023)024",
    journal = "JHEP",
    volume = "12",
    pages = "024",
    year = "2023"
}

@article{Chang:2025ays,
    author = "Chang, Jing-Cheng and He, Yang and Liu, Yu-Xiao and Sun, Yuan",
    title = "{Toward a Unified de Sitter Holography: A Composite $T\bar{T}$ and $T\bar{T}+\Lambda_2$ Flow}",
    eprint = "2511.16098",
    archivePrefix = "arXiv",
    primaryClass = "hep-th",
    month = "11",
    year = "2025"
}

@article{Chakraborty:2020xwo,
    author = "Chakraborty, Soumangsu and Mishra, Amiya",
    title = "{$ T\overline{T} $ and $ J\overline{T} $ deformations in quantum mechanics}",
    eprint = "2008.01333",
    archivePrefix = "arXiv",
    primaryClass = "hep-th",
    reportNumber = "TIFR/TH/20-28",
    doi = "10.1007/JHEP11(2020)099",
    journal = "JHEP",
    volume = "11",
    pages = "099",
    year = "2020"
}

@article{Grumiller:2007ju,
    author = "Grumiller, Daniel and McNees, Robert",
    title = "{Thermodynamics of black holes in two (and higher) dimensions}",
    eprint = "hep-th/0703230",
    archivePrefix = "arXiv",
    reportNumber = "BROWN-HET-1478, MIT-CTP-3825",
    doi = "10.1088/1126-6708/2007/04/074",
    journal = "JHEP",
    volume = "04",
    pages = "074",
    year = "2007"
}

@article{Franken:2023pni,
    author = "Franken, Victor and Partouche, Herv\'e and Rondeau, Fran\c{c}ois and Toumbas, Nicolaos",
    title = "{Bridging the static patches: de Sitter holography and entanglement}",
    eprint = "2305.12861",
    archivePrefix = "arXiv",
    primaryClass = "hep-th",
    reportNumber = "CPHT-RR018.042023",
    doi = "10.1007/JHEP08(2023)074",
    journal = "JHEP",
    volume = "08",
    pages = "074",
    year = "2023"
}

@inproceedings{Franken:2024ruw,
    author = "Franken, Victor",
    title = "{de Sitter Connectivity from Holographic Entanglement}",
    booktitle = "{23rd Hellenic School and Workshops on Elementary Particle Physics and Gravity}",
    eprint = "2403.14889",
    archivePrefix = "arXiv",
    primaryClass = "hep-th",
    reportNumber = "CPHT-PC016.032024",
    month = "3",
    year = "2024"
}

@article{He:2025ppz,
    author = "He, Song and Li, Yi and Ouyang, Hao and Sun, Yuan",
    title = "{$T\overline{T}$ Deformation: Introduction and Some Recent Advances}",
    eprint = "2503.09997",
    archivePrefix = "arXiv",
    primaryClass = "hep-th",
    month = "3",
    year = "2025"
}

@article{Svesko:2022txo,
    author = "Svesko, Andrew and Verheijden, Evita and Verlinde, Erik P. and Visser, Manus R.",
    title = "{Quasi-local energy and microcanonical entropy in two-dimensional nearly de Sitter gravity}",
    eprint = "2203.00700",
    archivePrefix = "arXiv",
    primaryClass = "hep-th",
    doi = "10.1007/JHEP08(2022)075",
    journal = "JHEP",
    volume = "08",
    pages = "075",
    year = "2022"
}

@article{Banihashemi:2022htw,
    author = "Banihashemi, Batoul and Jacobson, Ted and Svesko, Andrew and Visser, Manus",
    title = "{The minus sign in the first law of de Sitter horizons}",
    eprint = "2208.11706",
    archivePrefix = "arXiv",
    primaryClass = "hep-th",
    doi = "10.1007/JHEP01(2023)054",
    journal = "JHEP",
    volume = "01",
    pages = "054",
    year = "2023"
}

@article{Susskind:2014jwa,
    author = "Susskind, Leonard and Zhao, Ying",
    title = "{Switchbacks and the Bridge to Nowhere}",
    eprint = "1408.2823",
    archivePrefix = "arXiv",
    primaryClass = "hep-th",
    month = "8",
    year = "2014"
}

@article{Jorstad:2022mls,
    author = "J\o{}rstad, Eivind and Myers, Robert C. and Ruan, Shan-Ming",
    title = "{Holographic complexity in dS$_{d+1}$}",
    eprint = "2202.10684",
    archivePrefix = "arXiv",
    primaryClass = "hep-th",
    reportNumber = "YITP-22-15",
    doi = "10.1007/JHEP05(2022)119",
    journal = "JHEP",
    volume = "05",
    pages = "119",
    year = "2022"
}

@article{Anegawa:2023wrk,
    author = "Anegawa, Takanori and Iizuka, Norihiro and Sake, Sunil Kumar and Zenoni, Nicol\`o",
    title = "{Is action complexity better for de Sitter space in Jackiw-Teitelboim gravity?}",
    eprint = "2303.05025",
    archivePrefix = "arXiv",
    primaryClass = "hep-th",
    reportNumber = "OU-HET-1170",
    doi = "10.1007/JHEP06(2023)213",
    journal = "JHEP",
    volume = "06",
    pages = "213",
    year = "2023"
}

@article{Ryu:2006bv,
    author = "Ryu, Shinsei and Takayanagi, Tadashi",
    title = "{Holographic derivation of entanglement entropy from AdS/CFT}",
    eprint = "hep-th/0603001",
    archivePrefix = "arXiv",
    reportNumber = "NSF-KITP-06-11",
    doi = "10.1103/PhysRevLett.96.181602",
    journal = "Phys. Rev. Lett.",
    volume = "96",
    pages = "181602",
    year = "2006"
}

@article{Ryu:2006ef,
    author = "Ryu, Shinsei and Takayanagi, Tadashi",
    title = "{Aspects of Holographic Entanglement Entropy}",
    eprint = "hep-th/0605073",
    archivePrefix = "arXiv",
    reportNumber = "NSF-KITP-06-31, KUNS-2021",
    doi = "10.1088/1126-6708/2006/08/045",
    journal = "JHEP",
    volume = "08",
    pages = "045",
    year = "2006"
}

@article{Hubeny:2007xt,
    author = "Hubeny, Veronika E. and Rangamani, Mukund and Takayanagi, Tadashi",
    title = "{A Covariant holographic entanglement entropy proposal}",
    eprint = "0705.0016",
    archivePrefix = "arXiv",
    primaryClass = "hep-th",
    reportNumber = "DCPT-07-13, KUNS-2069",
    doi = "10.1088/1126-6708/2007/07/062",
    journal = "JHEP",
    volume = "07",
    pages = "062",
    year = "2007"
}

@article{Faulkner:2013ana,
    author = "Faulkner, Thomas and Lewkowycz, Aitor and Maldacena, Juan",
    title = "{Quantum corrections to holographic entanglement entropy}",
    eprint = "1307.2892",
    archivePrefix = "arXiv",
    primaryClass = "hep-th",
    doi = "10.1007/JHEP11(2013)074",
    journal = "JHEP",
    volume = "11",
    pages = "074",
    year = "2013"
}

@article{Nishioka:2009un,
    author = "Nishioka, Tatsuma and Ryu, Shinsei and Takayanagi, Tadashi",
    title = "{Holographic Entanglement Entropy: An Overview}",
    eprint = "0905.0932",
    archivePrefix = "arXiv",
    primaryClass = "hep-th",
    reportNumber = "KUNS-2207, IPMU09-0056",
    doi = "10.1088/1751-8113/42/50/504008",
    journal = "J. Phys. A",
    volume = "42",
    pages = "504008",
    year = "2009"
}

@article{Anegawa:2023dad,
    author = "Anegawa, Takanori and Iizuka, Norihiro",
    title = "{Shock waves and delay of hyperfast growth in de Sitter complexity}",
    eprint = "2304.14620",
    archivePrefix = "arXiv",
    primaryClass = "hep-th",
    reportNumber = "OU-TH-1183",
    doi = "10.1007/JHEP08(2023)115",
    journal = "JHEP",
    volume = "08",
    pages = "115",
    year = "2023"
}

@article{Baiguera:2023tpt,
    author = "Baiguera, Stefano and Berman, Rotem and Chapman, Shira and Myers, Robert C.",
    title = "{The cosmological switchback effect}",
    eprint = "2304.15008",
    archivePrefix = "arXiv",
    primaryClass = "hep-th",
    doi = "10.1007/JHEP07(2023)162",
    journal = "JHEP",
    volume = "07",
    pages = "162",
    year = "2023"
}

@article{Aguilar-Gutierrez:2023zqm,
    author = "Aguilar-Gutierrez, Sergio E. and Heller, Michal P. and Van der Schueren, Silke",
    title = "{Complexity equals anything can grow forever in de Sitter space}",
    eprint = "2305.11280",
    archivePrefix = "arXiv",
    primaryClass = "hep-th",
    doi = "10.1103/PhysRevD.110.066009",
    journal = "Phys. Rev. D",
    volume = "110",
    number = "6",
    pages = "066009",
    year = "2024"
}

@article{DeVuyst:2024pop,
    author = "De Vuyst, Julian and Eccles, Stefan and Hoehn, Philipp A. and Kirklin, Josh",
    title = "{Gravitational entropy is observer-dependent}",
    eprint = "2405.00114",
    archivePrefix = "arXiv",
    primaryClass = "hep-th",
    month = "4",
    year = "2024"
}

@article{Belaey:2025ijg,
    author = "Belaey, Andreas and Mertens, Thomas G. and Tappeiner, Thomas",
    title = "{Quantum group origins of edge states in double-scaled SYK}",
    eprint = "2503.20691",
    archivePrefix = "arXiv",
    primaryClass = "hep-th",
    month = "3",
    year = "2025"
}

@article{Aguilar-Gutierrez:2024oea,
    author = "Aguilar-Gutierrez, Sergio E.",
    title = "{$T^2$ deformations in the double-scaled SYK model: Stretched horizon thermodynamics}",
    eprint = "2410.18303",
    archivePrefix = "arXiv",
    primaryClass = "hep-th",
    month = "10",
    year = "2024"
}

@article{Xu:2024gfm,
    author = "Xu, Jiuci",
    title = "{On Chord Dynamics and Complexity Growth in Double-Scaled SYK}",
    eprint = "2411.04251",
    archivePrefix = "arXiv",
    primaryClass = "hep-th",
    month = "11",
    year = "2024"
}

@article{Aguilar-Gutierrez:2023tic,
    author = "Aguilar-Gutierrez, Sergio E. and Patra, Ayan K. and Pedraza, Juan F.",
    title = "{Entangled universes in dS wedge holography}",
    eprint = "2308.05666",
    archivePrefix = "arXiv",
    primaryClass = "hep-th",
    reportNumber = "IFT-UAM/CSIC-23-95",
    doi = "10.1007/JHEP10(2023)156",
    journal = "JHEP",
    volume = "10",
    pages = "156",
    year = "2023"
}

@article{Aguilar-Gutierrez:2023pnn,
    author = "Aguilar-Gutierrez, Sergio E.",
    title = "{C=Anything and the switchback effect in Schwarzschild-de Sitter space}",
    eprint = "2309.05848",
    archivePrefix = "arXiv",
    primaryClass = "hep-th",
    doi = "10.1007/JHEP03(2024)062",
    journal = "JHEP",
    volume = "03",
    pages = "062",
    year = "2024"
}

@article{Finster:2023rkv,
    author = "Finster, Felix and Guendelman, Eduardo and Paganini, Claudio F.",
    title = "{Modified measures as an effective theory for causal fermion systems}",
    eprint = "2303.16566",
    archivePrefix = "arXiv",
    primaryClass = "gr-qc",
    doi = "10.1088/1361-6382/ad1711",
    journal = "Class. Quant. Grav.",
    volume = "41",
    number = "3",
    pages = "035007",
    year = "2024"
}

@article{Berkooz:2018jqr,
    author = "Berkooz, Micha and Isachenkov, Mikhail and Narovlansky, Vladimir and Torrents, Genis",
    title = "{Towards a full solution of the large N double-scaled SYK model}",
    eprint = "1811.02584",
    archivePrefix = "arXiv",
    primaryClass = "hep-th",
    doi = "10.1007/JHEP03(2019)079",
    journal = "JHEP",
    volume = "03",
    pages = "079",
    year = "2019"
}

@article{Milekhin:2024vbb,
    author = "Milekhin, Alexey and Xu, Jiuci",
    title = "{On scrambling, tomperature and superdiffusion in de Sitter space}",
    eprint = "2403.13915",
    archivePrefix = "arXiv",
    primaryClass = "hep-th",
    doi = "10.1007/JHEP07(2025)272",
    journal = "JHEP",
    volume = "07",
    pages = "272",
    year = "2025"
}

@article{Adami:2025pqr,
    author = "Adami, H. and Sheikh-Jabbari, M. M. and Taghiloo, V.",
    title = "{Gravity Is Induced By Renormalization Group Flow}",
    eprint = "2508.09633",
    archivePrefix = "arXiv",
    primaryClass = "hep-th",
    month = "8",
    year = "2025"
}

@article{Li:2025lpa,
    author = "Li, Yun-Ze and Xie, Yunfei and He, Song",
    title = "{Emergent classical gravity as stress-tensor deformed field theories}",
    eprint = "2508.15461",
    archivePrefix = "arXiv",
    primaryClass = "hep-th",
    month = "8",
    year = "2025"
}

@article{Ran:2025xas,
    author = "Ran, Xi-Yang and Hao, Feng and Ouyang, Hao",
    title = "{Holography for stress-energy tensor flows}",
    eprint = "2508.12275",
    archivePrefix = "arXiv",
    primaryClass = "hep-th",
    month = "8",
    year = "2025"
}

@article{Boruch:2023bte,
    author = "Boruch, Jan and Lin, Henry W. and Yan, Cynthia",
    title = "{Exploring supersymmetric wormholes in $ \mathcal{N} $ = 2 SYK with chords}",
    eprint = "2308.16283",
    archivePrefix = "arXiv",
    primaryClass = "hep-th",
    doi = "10.1007/JHEP12(2023)151",
    journal = "JHEP",
    volume = "12",
    pages = "151",
    year = "2023"
}

@article{Berkooz:2022fso,
    author = "Berkooz, Micha and Brukner, Nadav and Ross, Simon F. and Watanabe, Masataka",
    title = "{Going beyond ER=EPR in the SYK model}",
    eprint = "2202.11381",
    archivePrefix = "arXiv",
    primaryClass = "hep-th",
    doi = "10.1007/JHEP08(2022)051",
    journal = "JHEP",
    volume = "08",
    pages = "051",
    year = "2022"
}

@article{Okuyama:2025hsd,
    author = "Okuyama, Kazumi",
    title = "{de Sitter JT gravity from double-scaled SYK}",
    eprint = "2505.08116",
    archivePrefix = "arXiv",
    primaryClass = "hep-th",
    doi = "10.1007/JHEP08(2025)181",
    journal = "JHEP",
    volume = "08",
    pages = "181",
    year = "2025"
}

@article{Anninos:2017hhn,
    author = "Anninos, Dionysios and Hofman, Diego M.",
    title = "{Infrared Realization of dS$_2$ in AdS$_2$}",
    eprint = "1703.04622",
    archivePrefix = "arXiv",
    primaryClass = "hep-th",
    doi = "10.1088/1361-6382/aab143",
    journal = "Class. Quant. Grav.",
    volume = "35",
    number = "8",
    pages = "085003",
    year = "2018"
}

@article{Anninos:2018svg,
    author = "Anninos, Dionysios and Galante, Dami\'an A. and Hofman, Diego M.",
    title = "{De Sitter horizons \& holographic liquids}",
    eprint = "1811.08153",
    archivePrefix = "arXiv",
    primaryClass = "hep-th",
    doi = "10.1007/JHEP07(2019)038",
    journal = "JHEP",
    volume = "07",
    pages = "038",
    year = "2019"
}

@article{Anninos:2022hqo,
    author = "Anninos, Dionysios and Harris, Eleanor",
    title = "{Interpolating geometries and the stretched dS$_{2}$ horizon}",
    eprint = "2209.06144",
    archivePrefix = "arXiv",
    primaryClass = "hep-th",
    doi = "10.1007/JHEP11(2022)166",
    journal = "JHEP",
    volume = "11",
    pages = "166",
    year = "2022"
}

@article{Galante:2023uyf,
    author = "Galante, Damian A.",
    title = "{Modave lectures on de Sitter space \& holography}",
    eprint = "2306.10141",
    archivePrefix = "arXiv",
    primaryClass = "hep-th",
    doi = "10.22323/1.435.0003",
    journal = "PoS",
    volume = "Modave2022",
    pages = "003",
    year = "2023"
}

@article{Chandrasekaran:2022cip,
    author = "Chandrasekaran, Venkatesa and Longo, Roberto and Penington, Geoff and Witten, Edward",
    title = "{An algebra of observables for de Sitter space}",
    eprint = "2206.10780",
    archivePrefix = "arXiv",
    primaryClass = "hep-th",
    doi = "10.1007/JHEP02(2023)082",
    journal = "JHEP",
    volume = "02",
    pages = "082",
    year = "2023"
}

@article{Jensen:2023yxy,
    author = "Jensen, Kristan and Sorce, Jonathan and Speranza, Antony J.",
    title = "{Generalized entropy for general subregions in quantum gravity}",
    eprint = "2306.01837",
    archivePrefix = "arXiv",
    primaryClass = "hep-th",
    doi = "10.1007/JHEP12(2023)020",
    journal = "JHEP",
    volume = "12",
    pages = "020",
    year = "2023"
}

@article{Taylor:2018xcy,
    author = "Taylor, Marika",
    title = "{TT deformations in general dimensions}",
    eprint = "1805.10287",
    archivePrefix = "arXiv",
    primaryClass = "hep-th",
    month = "5",
    year = "2018"
}

@article{Morone:2024ffm,
    author = "Morone, Tommaso and Negro, Stefano and Tateo, Roberto",
    title = "{Gravity and T$\overline{\rm T}$ flows in higher dimensions}",
    eprint = "2401.16400",
    archivePrefix = "arXiv",
    primaryClass = "hep-th",
    doi = "10.1016/j.nuclphysb.2024.116605",
    journal = "Nucl. Phys. B",
    volume = "1005",
    pages = "116605",
    year = "2024"
}

@article{Ferko:2024yua,
    author = "Ferko, Christian and Hou, Jue and Morone, Tommaso and Tartaglino-Mazzucchelli, Gabriele and Tateo, Roberto",
    title = "{TT{\textasciimacron}-like Flows of Yang-Mills Theories}",
    eprint = "2409.18740",
    archivePrefix = "arXiv",
    primaryClass = "hep-th",
    doi = "10.1103/PhysRevLett.134.101603",
    journal = "Phys. Rev. Lett.",
    volume = "134",
    number = "10",
    pages = "101603",
    year = "2025"
}

@article{Ebert:2022ehb,
    author = "Ebert, Stephen and Ferko, Christian and Sun, Hao-Yu and Sun, Zhengdi",
    title = "{$T\bar{T}$ in JT Gravity and BF Gauge Theory}",
    eprint = "2205.07817",
    archivePrefix = "arXiv",
    primaryClass = "hep-th",
    doi = "10.21468/SciPostPhys.13.4.096",
    journal = "SciPost Phys.",
    volume = "13",
    number = "4",
    pages = "096",
    year = "2022"
}

@article{Ebert:2022xfh,
    author = "Ebert, Stephen and Ferko, Christian and Sun, Hao-Yu and Sun, Zhengdi",
    title = "{$ T\overline{T} $ deformations of supersymmetric quantum mechanics}",
    eprint = "2204.05897",
    archivePrefix = "arXiv",
    primaryClass = "hep-th",
    doi = "10.1007/JHEP08(2022)121",
    journal = "JHEP",
    volume = "08",
    pages = "121",
    year = "2022"
}

@article{Bonelli:2018kik,
    author = "Bonelli, Giulio and Doroud, Nima and Zhu, Mengqi",
    title = "{$T \bar{T}$-deformations in closed form}",
    eprint = "1804.10967",
    archivePrefix = "arXiv",
    primaryClass = "hep-th",
    doi = "10.1007/JHEP06(2018)149",
    journal = "JHEP",
    volume = "06",
    pages = "149",
    year = "2018"
}

@article{Allameh:2025gsa,
    author = "Allameh, Kuroush and Shaghoulian, Edgar",
    title = "{Timelike Liouville theory and AdS$_3$ gravity at finite cutoff}",
    eprint = "2508.03236",
    archivePrefix = "arXiv",
    primaryClass = "hep-th",
    month = "8",
    year = "2025"
}

@article{Milekhin:2023bjv,
    author = "Milekhin, Alexey and Xu, Jiuci",
    title = "{Revisiting Brownian SYK and its possible relations to de Sitter}",
    eprint = "2312.03623",
    archivePrefix = "arXiv",
    primaryClass = "hep-th",
    doi = "10.1007/JHEP10(2024)151",
    journal = "JHEP",
    volume = "10",
    pages = "151",
    year = "2024"
}

@article{Cui:2025sgy,
    author = "Cui, Chuanxin and Rozali, Moshe",
    title = "{Splitting and gluing in sine-dilaton gravity: matter correlators and the wormhole Hilbert space}",
    eprint = "2509.01680",
    archivePrefix = "arXiv",
    primaryClass = "hep-th",
    month = "9",
    year = "2025"
}

@article{Blommaert:2025eps,
    author = "Blommaert, Andreas and Tietto, Damiano and Verlinde, Herman",
    title = "{SYK collective field theory as complex Liouville gravity}",
    eprint = "2509.18462",
    archivePrefix = "arXiv",
    primaryClass = "hep-th",
    month = "9",
    year = "2025"
}

@article{Lewkowycz:2019xse,
    author = "Lewkowycz, Aitor and Liu, Junyu and Silverstein, Eva and Torroba, Gonzalo",
    title = "{$ T\overline{T} $ and EE, with implications for (A)dS subregion encodings}",
    eprint = "1909.13808",
    archivePrefix = "arXiv",
    primaryClass = "hep-th",
    reportNumber = "CALT-TH-2019--031",
    doi = "10.1007/JHEP04(2020)152",
    journal = "JHEP",
    volume = "04",
    pages = "152",
    year = "2020"
}

@article{Gao:2024lem,
    author = "Gao, Ping and Lin, Han and Peng, Cheng",
    title = "{D-commuting SYK model: building quantum chaos from integrable blocks}",
    eprint = "2411.12806",
    archivePrefix = "arXiv",
    primaryClass = "hep-th",
    month = "11",
    year = "2024"
}

@article{Balasubramanian:2024lqk,
    author = "Balasubramanian, Vijay and Magan, Javier M. and Nandi, Poulami and Wu, Qingyue",
    title = "{Spread complexity and the saturation of wormhole size}",
    eprint = "2412.02038",
    archivePrefix = "arXiv",
    primaryClass = "hep-th",
    month = "12",
    year = "2024"
}

@article{Griguolo:2021wgy,
    author = "Griguolo, Luca and Panerai, Rodolfo and Papalini, Jacopo and Seminara, Domenico",
    title = "{Nonperturbative effects and resurgence in Jackiw-Teitelboim gravity at finite cutoff}",
    eprint = "2106.01375",
    archivePrefix = "arXiv",
    primaryClass = "hep-th",
    reportNumber = "UUITP-25/21",
    doi = "10.1103/PhysRevD.105.046015",
    journal = "Phys. Rev. D",
    volume = "105",
    number = "4",
    pages = "046015",
    year = "2022"
}

@article{Kruthoff:2020hsi,
    author = "Kruthoff, Jorrit and Parrikar, Onkar",
    title = "{On the flow of states under $T\overline{T}$}",
    eprint = "2006.03054",
    archivePrefix = "arXiv",
    primaryClass = "hep-th",
    month = "6",
    year = "2020"
}

@article{Bossi:2024ffa,
    author = "Bossi, Leonardo and Griguolo, Luca and Papalini, Jacopo and Russo, Lorenzo and Seminara, Domenico",
    title = "{Sine-dilaton gravity vs double-scaled SYK: exploring one-loop quantum corrections}",
    eprint = "2411.15957",
    archivePrefix = "arXiv",
    primaryClass = "hep-th",
    month = "11",
    year = "2024"
}

@article{Blommaert:2024whf,
    author = "Blommaert, Andreas and Levine, Adam and Mertens, Thomas G. and Papalini, Jacopo and Parmentier, Klaas",
    title = "{An entropic puzzle in periodic dilaton gravity and DSSYK}",
    eprint = "2411.16922",
    archivePrefix = "arXiv",
    primaryClass = "hep-th",
    doi = "10.1007/JHEP07(2025)093",
    journal = "JHEP",
    volume = "07",
    pages = "093",
    year = "2025"
}

@article{Wheeler:1968iap,
    author = "Wheeler, J. A.",
    editor = "Fang, Li-Zhi and Ruffini, R.",
    title = "{SUPERSPACE AND THE NATURE OF QUANTUM GEOMETRODYNAMICS}",
    journal = "Adv. Ser. Astrophys. Cosmol.",
    volume = "3",
    pages = "27--92",
    year = "1987"
}

@article{DeWitt:1967yk,
    author = "DeWitt, Bryce S.",
    editor = "Fang, Li-Zhi and Ruffini, R.",
    title = "{Quantum Theory of Gravity. 1. The Canonical Theory}",
    doi = "10.1103/PhysRev.160.1113",
    journal = "Phys. Rev.",
    volume = "160",
    pages = "1113--1148",
    year = "1967"
}

@article{Ambrosini:2024sre,
    author = "Ambrosini, Marco and Rabinovici, Eliezer and S\'anchez-Garrido, Adri\'an and Shir, Ruth and Sonner, Julian",
    title = "{Operator K-complexity in DSSYK: Krylov complexity equals bulk length}",
    eprint = "2412.15318",
    archivePrefix = "arXiv",
    primaryClass = "hep-th",
    month = "12",
    year = "2024"
}

@article{Heller:2024ldz,
    author = "Heller, Michal P. and Papalini, Jacopo and Schuhmann, Tim",
    title = "{Krylov spread complexity as holographic complexity beyond JT gravity}",
    eprint = "2412.17785",
    archivePrefix = "arXiv",
    primaryClass = "hep-th",
    month = "12",
    year = "2024"
}

@article{Mertens:2022irh,
    author = "Mertens, Thomas G. and Turiaci, Gustavo J.",
    title = "{Solvable models of quantum black holes: a review on Jackiw\textendash{}Teitelboim gravity}",
    eprint = "2210.10846",
    archivePrefix = "arXiv",
    primaryClass = "hep-th",
    doi = "10.1007/s41114-023-00046-1",
    journal = "Living Rev. Rel.",
    volume = "26",
    number = "1",
    pages = "4",
    year = "2023"
}

@article{Blommaert:2025avl,
    author = "Blommaert, Andreas and Levine, Adam and Mertens, Thomas G. and Papalini, Jacopo and Parmentier, Klaas",
    title = "{Wormholes, branes and finite matrices in sine dilaton gravity}",
    eprint = "2501.17091",
    archivePrefix = "arXiv",
    primaryClass = "hep-th",
    doi = "10.1007/JHEP09(2025)123",
    journal = "JHEP",
    volume = "09",
    pages = "123",
    year = "2025"
}

@article{Tolman:1930ona,
    author = "Tolman, Richard and Ehrenfest, Paul",
    title = "{Temperature Equilibrium in a Static Gravitational Field}",
    doi = "10.1103/PhysRev.36.1791",
    journal = "Phys. Rev.",
    volume = "36",
    number = "12",
    pages = "1791--1798",
    year = "1930"
}

@article{tolman1933thermodynamics,
  title={Thermodynamics and relativity},
  author={Tolman, Richard C},
  journal={Science},
  volume={77},
  number={1995},
  pages={291--298},
  year={1933},
  publisher={American Association for the Advancement of Science}
}

@article{Belaey:2025kiu,
    author = "Belaey, Andreas and Mertens, Thomas G. and Papalini, Jacopo",
    title = "{Probing the singularity at the holographic screen via $q$-holography}",
    eprint = "2507.13873",
    archivePrefix = "arXiv",
    primaryClass = "hep-th",
    month = "7",
    year = "2025"
}

@article{Lin:2022zxd,
    author = "Lin, Henry W. and Maldacena, Juan and Rozenberg, Liza and Shan, Jieru",
    title = "{Looking at supersymmetric black holes for a very long time}",
    eprint = "2207.00408",
    archivePrefix = "arXiv",
    primaryClass = "hep-th",
    doi = "10.21468/SciPostPhys.14.5.128",
    journal = "SciPost Phys.",
    volume = "14",
    number = "5",
    pages = "128",
    year = "2023"
}

@article{Berkooz:2020xne,
    author = "Berkooz, Micha and Brukner, Nadav and Narovlansky, Vladimir and Raz, Amir",
    title = "{The double scaled limit of Super--Symmetric SYK models}",
    eprint = "2003.04405",
    archivePrefix = "arXiv",
    primaryClass = "hep-th",
    doi = "10.1007/JHEP12(2020)110",
    journal = "JHEP",
    volume = "12",
    pages = "110",
    year = "2020"
}

@article{Araujo-Regado:2025elv,
    author = "Araujo-Regado, Goncalo and Thavanesan, Ayngaran and Wall, Aron C.",
    title = "{Holographic Cosmology at Finite Time}",
    eprint = "2511.04511",
    archivePrefix = "arXiv",
    primaryClass = "hep-th",
    month = "11",
    year = "2025"
}

@book{ismail2005classical,
  title={Classical and quantum orthogonal polynomials in one variable},
  author={Ismail, Mourad},
  volume={13},
  year={2005},
  publisher={Cambridge university press}
}

@article{al1976convolutions,
  title={Convolutions of orthonormal polynomials},
  author={Al-Salam, WA and Chihara, TS},
  journal={SIAM Journal on Mathematical Analysis},
  volume={7},
  number={1},
  pages={16--28},
  year={1976},
  publisher={SIAM}
}

@article{Watanabe:2025rwp,
    author = "Watanabe, Masataka",
    title = "{A JT/KPZ correspondence}",
    eprint = "2511.02529",
    archivePrefix = "arXiv",
    primaryClass = "hep-th",
    month = "11",
    year = "2025"
}

@article{tolman1928extension,
  title={On the extension of thermodynamics to general relativity},
  author={Tolman, Richard C},
  journal={Proceedings of the National Academy of Sciences},
  volume={14},
  number={3},
  pages={268--272},
  year={1928}
}

@article{Chapman:2024pdw,
    author = "Chapman, Shira and Demulder, Saskia and Galante, Dami\'an A. and Sheorey, Sameer U. and Shoval, Osher",
    title = "{Krylov complexity and chaos in deformed SYK models}",
    eprint = "2407.09604",
    archivePrefix = "arXiv",
    primaryClass = "hep-th",
    month = "7",
    year = "2024"
}

@article{Levine:2022wos,
    author = "Levine, Adam and Shaghoulian, Edgar",
    title = "{Encoding beyond cosmological horizons in de Sitter JT gravity}",
    eprint = "2204.08503",
    archivePrefix = "arXiv",
    primaryClass = "hep-th",
    doi = "10.1007/JHEP02(2023)179",
    journal = "JHEP",
    volume = "02",
    pages = "179",
    year = "2023"
}

@article{Gibbons:1976ue,
    author = "Gibbons, G. W. and Hawking, S. W.",
    title = "{Action Integrals and Partition Functions in Quantum Gravity}",
    reportNumber = "PRINT-76-0995 (CAMBRIDGE)",
    doi = "10.1103/PhysRevD.15.2752",
    journal = "Phys. Rev. D",
    volume = "15",
    pages = "2752--2756",
    year = "1977"
}

@article{Gibbons:1977mu,
    author = "Gibbons, G. W. and Hawking, S. W.",
    title = "{Cosmological Event Horizons, Thermodynamics, and Particle Creation}",
    doi = "10.1103/PhysRevD.15.2738",
    journal = "Phys. Rev. D",
    volume = "15",
    pages = "2738--2751",
    year = "1977"
}

@article{Aguilar-Gutierrez:2021bns,
    author = "Aguilar-Gutierrez, Sergio E. and Chatwin-Davies, Aidan and Hertog, Thomas and Pinzani-Fokeeva, Natalia and Robinson, Brandon",
    title = "{Islands in Multiverse Models}",
    eprint = "2108.01278",
    archivePrefix = "arXiv",
    primaryClass = "hep-th",
    doi = "10.1007/JHEP05(2022)137",
    journal = "JHEP",
    volume = "11",
    pages = "212",
    year = "2021",
    note = "[Addendum: JHEP 05, 137 (2022), Erratum: JHEP 05, 082 (2022)]"
}

@article{Balasubramanian:2020xqf,
    author = "Balasubramanian, Vijay and Kar, Arjun and Ugajin, Tomonori",
    title = "{Islands in de Sitter space}",
    eprint = "2008.05275",
    archivePrefix = "arXiv",
    primaryClass = "hep-th",
    doi = "10.1007/JHEP02(2021)072",
    journal = "JHEP",
    volume = "02",
    pages = "072",
    year = "2021"
}

@article{Dubovsky:2012wk,
    author = "Dubovsky, Sergei and Flauger, Raphael and Gorbenko, Victor",
    title = "{Solving the Simplest Theory of Quantum Gravity}",
    eprint = "1205.6805",
    archivePrefix = "arXiv",
    primaryClass = "hep-th",
    doi = "10.1007/JHEP09(2012)133",
    journal = "JHEP",
    volume = "09",
    pages = "133",
    year = "2012"
}

@article{Cardy:2018sdv,
    author = "Cardy, John",
    title = "{The $ T\overline{T} $ deformation of quantum field theory as random geometry}",
    eprint = "1801.06895",
    archivePrefix = "arXiv",
    primaryClass = "hep-th",
    doi = "10.1007/JHEP10(2018)186",
    journal = "JHEP",
    volume = "10",
    pages = "186",
    year = "2018"
}

@article{Aharony:2018bad,
    author = "Aharony, Ofer and Datta, Shouvik and Giveon, Amit and Jiang, Yunfeng and Kutasov, David",
    title = "{Modular invariance and uniqueness of $T\bar{T}$ deformed CFT}",
    eprint = "1808.02492",
    archivePrefix = "arXiv",
    primaryClass = "hep-th",
    doi = "10.1007/JHEP01(2019)086",
    journal = "JHEP",
    volume = "01",
    pages = "086",
    year = "2019"
}

@article{Tolley:2019nmm,
    author = "Tolley, Andrew J.",
    title = "{$ T\overline{T} $ deformations, massive gravity and non-critical strings}",
    eprint = "1911.06142",
    archivePrefix = "arXiv",
    primaryClass = "hep-th",
    reportNumber = "Imperial/TP/2019/AT/01",
    doi = "10.1007/JHEP06(2020)050",
    journal = "JHEP",
    volume = "06",
    pages = "050",
    year = "2020"
}

@article{Dubovsky:2017cnj,
    author = "Dubovsky, Sergei and Gorbenko, Victor and Mirbabayi, Mehrdad",
    title = "{Asymptotic fragility, near AdS$_{2}$ holography and $ T\overline{T} $}",
    eprint = "1706.06604",
    archivePrefix = "arXiv",
    primaryClass = "hep-th",
    doi = "10.1007/JHEP09(2017)136",
    journal = "JHEP",
    volume = "09",
    pages = "136",
    year = "2017"
}

@article{Dubovsky:2018bmo,
    author = "Dubovsky, Sergei and Gorbenko, Victor and Hern\'andez-Chifflet, Guzm\'an",
    title = "{$ T\overline{T} $ partition function from topological gravity}",
    eprint = "1805.07386",
    archivePrefix = "arXiv",
    primaryClass = "hep-th",
    doi = "10.1007/JHEP09(2018)158",
    journal = "JHEP",
    volume = "09",
    pages = "158",
    year = "2018"
}

@article{Jiang:2019epa,
    author = "Jiang, Yunfeng",
    title = "{A pedagogical review on solvable irrelevant deformations of 2D quantum field theory}",
    eprint = "1904.13376",
    archivePrefix = "arXiv",
    primaryClass = "hep-th",
    reportNumber = "CERN-TH-2019-058",
    doi = "10.1088/1572-9494/abe4c9",
    journal = "Commun. Theor. Phys.",
    volume = "73",
    number = "5",
    pages = "057201",
    year = "2021"
}

@article{Hartman:2018tkw,
    author = "Hartman, Thomas and Kruthoff, Jorrit and Shaghoulian, Edgar and Tajdini, Amirhossein",
    title = "{Holography at finite cutoff with a $T^2$ deformation}",
    eprint = "1807.11401",
    archivePrefix = "arXiv",
    primaryClass = "hep-th",
    doi = "10.1007/JHEP03(2019)004",
    journal = "JHEP",
    volume = "03",
    pages = "004",
    year = "2019"
}

@article{Parker:2018yvk,
    author = "Parker, Daniel E. and Cao, Xiangyu and Avdoshkin, Alexander and Scaffidi, Thomas and Altman, Ehud",
    title = "{A Universal Operator Growth Hypothesis}",
    eprint = "1812.08657",
    archivePrefix = "arXiv",
    primaryClass = "cond-mat.stat-mech",
    doi = "10.1103/PhysRevX.9.041017",
    journal = "Phys. Rev. X",
    volume = "9",
    number = "4",
    pages = "041017",
    year = "2019"
}

@article{Baiguera:2024xju,
    author = "Baiguera, Stefano and Berman, Rotem",
    title = "{The Cosmological Switchback Effect II}",
    eprint = "2406.04397",
    archivePrefix = "arXiv",
    primaryClass = "hep-th",
    doi = "10.1007/JHEP08(2024)086",
    journal = "JHEP",
    volume = "08",
    pages = "086",
    year = "2024"
}

@article{Auzzi:2023qbm,
    author = "Auzzi, Roberto and Nardelli, Giuseppe and Ungureanu, Gabriel Pedde and Zenoni, Nicolo",
    title = "{Volume complexity of dS bubbles}",
    eprint = "2302.03584",
    archivePrefix = "arXiv",
    primaryClass = "hep-th",
    reportNumber = "OU-HET-1171",
    doi = "10.1103/PhysRevD.108.026006",
    journal = "Phys. Rev. D",
    volume = "108",
    number = "2",
    pages = "026006",
    year = "2023"
}

@article{Penington:2023dql,
    author = "Penington, Geoff and Witten, Edward",
    title = "{Algebras and States in JT Gravity}",
    eprint = "2301.07257",
    archivePrefix = "arXiv",
    primaryClass = "hep-th",
    month = "1",
    year = "2023"
}

@article{Balasubramanian:2022tpr,
    author = "Balasubramanian, Vijay and Caputa, Pawel and Magan, Javier M. and Wu, Qingyue",
    title = "{Quantum chaos and the complexity of spread of states}",
    eprint = "2202.06957",
    archivePrefix = "arXiv",
    primaryClass = "hep-th",
    doi = "10.1103/PhysRevD.106.046007",
    journal = "Phys. Rev. D",
    volume = "106",
    number = "4",
    pages = "046007",
    year = "2022"
}

@article{Collier:2024kmo,
    author = {Collier, Scott and Eberhardt, Lorenz and M{\"u}hlmann, Beatrix and Rodriguez, Victor A.},
    title = "{Complex Liouville String}",
    eprint = "2409.17246",
    archivePrefix = "arXiv",
    primaryClass = "hep-th",
    doi = "10.1103/k74n-s63l",
    journal = "Phys. Rev. Lett.",
    volume = "134",
    number = "25",
    pages = "251602",
    year = "2025"
}

@article{Collier:2024kwt,
    author = {Collier, Scott and Eberhardt, Lorenz and M{\"u}hlmann, Beatrix and Rodriguez, Victor A.},
    title = "{The complex Liouville string: The worldsheet}",
    eprint = "2409.18759",
    archivePrefix = "arXiv",
    primaryClass = "hep-th",
    doi = "10.21468/SciPostPhys.19.2.033",
    journal = "SciPost Phys.",
    volume = "19",
    number = "2",
    pages = "033",
    year = "2025"
}

@article{Collier:2024lys,
    author = {Collier, Scott and Eberhardt, Lorenz and M{\"u}hlmann, Beatrix and Rodriguez, Victor A.},
    title = "{The complex Liouville string: The matrix integral}",
    eprint = "2410.07345",
    archivePrefix = "arXiv",
    primaryClass = "hep-th",
    doi = "10.21468/SciPostPhys.18.5.154",
    journal = "SciPost Phys.",
    volume = "18",
    number = "5",
    pages = "154",
    year = "2025"
}

@article{Collier:2024mlg,
    author = {Collier, Scott and Eberhardt, Lorenz and M{\"u}hlmann, Beatrix and Rodriguez, Victor A.},
    title = "{The complex Liouville string: Worldsheet boundaries and non-perturbative effects}",
    eprint = "2410.09179",
    archivePrefix = "arXiv",
    primaryClass = "hep-th",
    doi = "10.21468/SciPostPhys.19.2.034",
    journal = "SciPost Phys.",
    volume = "19",
    number = "2",
    pages = "034",
    year = "2025"
}

@article{Collier:2025lux,
    author = {Collier, Scott and Eberhardt, Lorenz and M{\"u}hlmann, Beatrix},
    title = "{A microscopic realization of dS$_3$}",
    eprint = "2501.01486",
    archivePrefix = "arXiv",
    primaryClass = "hep-th",
    doi = "10.21468/SciPostPhys.18.4.131",
    journal = "SciPost Phys.",
    volume = "18",
    number = "4",
    pages = "131",
    year = "2025"
}

@article{Collier:2025pbm,
    author = {Collier, Scott and Eberhardt, Lorenz and M{\"u}hlmann, Beatrix},
    title = "{The complex Liouville string: the gravitational path integral}",
    eprint = "2501.10265",
    archivePrefix = "arXiv",
    primaryClass = "hep-th",
    month = "1",
    year = "2025"
}

@article{Apolo:2023ckr,
    author = "Apolo, Luis and Hao, Peng-Xiang and Lai, Wen-Xin and Song, Wei",
    title = "{Extremal surfaces in glue-on AdS/$ T\overline{T} $ holography}",
    eprint = "2311.04883",
    archivePrefix = "arXiv",
    primaryClass = "hep-th",
    doi = "10.1007/JHEP01(2024)054",
    journal = "JHEP",
    volume = "01",
    pages = "054",
    year = "2024"
}

@article{Jafferis:2022wez,
    author = "Jafferis, Daniel Louis and Kolchmeyer, David K. and Mukhametzhanov, Baur and Sonner, Julian",
    title = "{Jackiw-Teitelboim gravity with matter, generalized eigenstate thermalization hypothesis, and random matrices}",
    eprint = "2209.02131",
    archivePrefix = "arXiv",
    primaryClass = "hep-th",
    doi = "10.1103/PhysRevD.108.066015",
    journal = "Phys. Rev. D",
    volume = "108",
    number = "6",
    pages = "066015",
    year = "2023"
}

@article{Rosso:2020wir,
    author = "Rosso, Felipe",
    title = "{$T\bar{T}$ deformation of random matrices}",
    eprint = "2012.11714",
    archivePrefix = "arXiv",
    primaryClass = "hep-th",
    doi = "10.1103/PhysRevD.103.126017",
    journal = "Phys. Rev. D",
    volume = "103",
    number = "12",
    pages = "126017",
    year = "2021"
}

@article{Chen:2022ozy,
    author = "Chen, Heng-Yu and Hikida, Yasuaki",
    title = "{Three-Dimensional de Sitter Holography and Bulk Correlators at Late Time}",
    eprint = "2204.04871",
    archivePrefix = "arXiv",
    primaryClass = "hep-th",
    reportNumber = "YITP-22-37",
    doi = "10.1103/PhysRevLett.129.061601",
    journal = "Phys. Rev. Lett.",
    volume = "129",
    number = "6",
    pages = "061601",
    year = "2022"
}

@article{Chen:2022xse,
    author = "Chen, Heng-Yu and Chen, Shi and Hikida, Yasuaki",
    title = "{Late-time correlation functions in dS$_{3}$/CFT$_{2}$ correspondence}",
    eprint = "2210.01415",
    archivePrefix = "arXiv",
    primaryClass = "hep-th",
    reportNumber = "YITP-22-99",
    doi = "10.1007/JHEP02(2023)038",
    journal = "JHEP",
    volume = "02",
    pages = "038",
    year = "2023"
}

@article{Hikida:2022ltr,
    author = "Hikida, Yasuaki and Nishioka, Tatsuma and Takayanagi, Tadashi and Taki, Yusuke",
    title = "{CFT duals of three-dimensional de Sitter gravity}",
    eprint = "2203.02852",
    archivePrefix = "arXiv",
    primaryClass = "hep-th",
    reportNumber = "YITP-22-20, IPMU22-0006",
    doi = "10.1007/JHEP05(2022)129",
    journal = "JHEP",
    volume = "05",
    pages = "129",
    year = "2022"
}

@article{Hikida:2021ese,
    author = "Hikida, Yasuaki and Nishioka, Tatsuma and Takayanagi, Tadashi and Taki, Yusuke",
    title = "{Holography in de Sitter Space via Chern-Simons Gauge Theory}",
    eprint = "2110.03197",
    archivePrefix = "arXiv",
    primaryClass = "hep-th",
    reportNumber = "YITP-21-105; IPMU21-0059",
    doi = "10.1103/PhysRevLett.129.041601",
    journal = "Phys. Rev. Lett.",
    volume = "129",
    number = "4",
    pages = "041601",
    year = "2022"
}

@article{Jafferis:2022uhu,
    author = "Jafferis, Daniel Louis and Kolchmeyer, David K. and Mukhametzhanov, Baur and Sonner, Julian",
    title = "{Matrix Models for Eigenstate Thermalization}",
    eprint = "2209.02130",
    archivePrefix = "arXiv",
    primaryClass = "hep-th",
    doi = "10.1103/PhysRevX.13.031033",
    journal = "Phys. Rev. X",
    volume = "13",
    number = "3",
    pages = "031033",
    year = "2023"
}

@article{Guica:2019nzm,
    author = "Guica, Monica and Monten, Ruben",
    title = "{$T\bar T$ and the mirage of a bulk cutoff}",
    eprint = "1906.11251",
    archivePrefix = "arXiv",
    primaryClass = "hep-th",
    doi = "10.21468/SciPostPhys.10.2.024",
    journal = "SciPost Phys.",
    volume = "10",
    number = "2",
    pages = "024",
    year = "2021"
}

@article{Apolo:2023vnm,
    author = "Apolo, Luis and Hao, Peng-Xiang and Lai, Wen-Xin and Song, Wei",
    title = "{Glue-on AdS holography for $ T\overline{T} $-deformed CFTs}",
    eprint = "2303.04836",
    archivePrefix = "arXiv",
    primaryClass = "hep-th",
    doi = "10.1007/JHEP06(2023)117",
    journal = "JHEP",
    volume = "06",
    pages = "117",
    year = "2023"
}

@article{Chattopadhyay:2024pdj,
    author = "Chattopadhyay, Arghya and Malvimat, Vinay and Mitra, Arpita",
    title = "{Krylov complexity of deformed conformal field theories}",
    eprint = "2405.03630",
    archivePrefix = "arXiv",
    primaryClass = "hep-th",
    doi = "10.1007/JHEP08(2024)053",
    journal = "JHEP",
    volume = "08",
    pages = "053",
    year = "2024"
}

@article{Sheikh-Jabbari:2025kjd,
    author = "Sheikh-Jabbari, M. M. and Taghiloo, V.",
    title = "{AdS$_3$ Freelance Holography, A Detailed Analysis}",
    eprint = "2510.10692",
    archivePrefix = "arXiv",
    primaryClass = "hep-th",
    month = "10",
    year = "2025"
}

@article{Blommaert:2024ymv,
    author = "Blommaert, Andreas and Mertens, Thomas G. and Papalini, Jacopo",
    title = "{The dilaton gravity hologram of double-scaled SYK}",
    eprint = "2404.03535",
    archivePrefix = "arXiv",
    primaryClass = "hep-th",
    doi = "10.1007/JHEP06(2025)050",
    journal = "JHEP",
    volume = "06",
    pages = "050",
    year = "2025"
}

@article{Heller:2025ddj,
    author = "Heller, Michal P. and Ori, Fabio and Papalini, Jacopo and Schuhmann, Tim and Wang, Meng-Ting",
    title = "{De Sitter holographic complexity from Krylov complexity in DSSYK}",
    eprint = "2510.13986",
    archivePrefix = "arXiv",
    primaryClass = "hep-th",
    month = "10",
    year = "2025"
}

@article{Erdmenger:2023wjg,
    author = "Erdmenger, Johanna and Jian, Shao-Kai and Xian, Zhuo-Yu",
    title = "{Universal chaotic dynamics from Krylov space}",
    eprint = "2303.12151",
    archivePrefix = "arXiv",
    primaryClass = "hep-th",
    doi = "10.1007/JHEP08(2023)176",
    journal = "JHEP",
    volume = "08",
    pages = "176",
    year = "2023"
}

@article{Caputa:2023vyr,
    author = "Caputa, Pawel and Magan, Javier M. and Patramanis, Dimitrios and Tonni, Erik",
    title = "{Krylov complexity of modular Hamiltonian evolution}",
    eprint = "2306.14732",
    archivePrefix = "arXiv",
    primaryClass = "hep-th",
    doi = "10.1103/PhysRevD.109.086004",
    journal = "Phys. Rev. D",
    volume = "109",
    number = "8",
    pages = "086004",
    year = "2024"
}

@article{Xu:2024hoc,
    author = "Xu, Jiuci",
    title = "{Von Neumann Algebras in Double-Scaled SYK}",
    eprint = "2403.09021",
    archivePrefix = "arXiv",
    primaryClass = "hep-th",
    month = "3",
    year = "2024"
}

@article{Okuyama:2023aup,
    author = "Okuyama, Kazumi and Suyama, Takao",
    title = "{Solvable limit of ETH matrix model for double-scaled SYK}",
    eprint = "2311.02846",
    archivePrefix = "arXiv",
    primaryClass = "hep-th",
    doi = "10.1007/JHEP04(2024)094",
    journal = "JHEP",
    volume = "04",
    pages = "094",
    year = "2024"
}

@article{Anninos:2024wpy,
    author = "Anninos, Dionysios and Galante, Dami\'an A. and Maneerat, Chawakorn",
    title = "{Cosmological observatories}",
    eprint = "2402.04305",
    archivePrefix = "arXiv",
    primaryClass = "hep-th",
    doi = "10.1088/1361-6382/ad5824",
    journal = "Class. Quant. Grav.",
    volume = "41",
    number = "16",
    pages = "165009",
    year = "2024"
}

@article{Okuyama:2024eyf,
    author = "Okuyama, Kazumi",
    title = "{Baby universe operators in double-scaled SYK}",
    eprint = "2408.03726",
    archivePrefix = "arXiv",
    primaryClass = "hep-th",
    month = "8",
    year = "2024"
}

@article{Berkooz:2018qkz,
    author = "Berkooz, Micha and Narayan, Prithvi and Simon, Joan",
    title = "{Chord diagrams, exact correlators in spin glasses and black hole bulk reconstruction}",
    eprint = "1806.04380",
    archivePrefix = "arXiv",
    primaryClass = "hep-th",
    doi = "10.1007/JHEP08(2018)192",
    journal = "JHEP",
    volume = "08",
    pages = "192",
    year = "2018"
}

@article{FarajiAstaneh:2024fpv,
    author = "Faraji Astaneh, Amin",
    title = "{Quantum Complexity of $T\bar{T}$-deformation and Its Implications}",
    eprint = "2408.06055",
    archivePrefix = "arXiv",
    primaryClass = "hep-th",
    month = "8",
    year = "2024"
}

@article{Susskind:2014rva,
    author = "Susskind, Leonard",
    title = "{Computational Complexity and Black Hole Horizons}",
    eprint = "1403.5695",
    archivePrefix = "arXiv",
    primaryClass = "hep-th",
    doi = "10.1002/prop.201500092",
    journal = "Fortsch. Phys.",
    volume = "64",
    pages = "24--43",
    year = "2016",
    note = "[Addendum: Fortsch.Phys. 64, 44--48 (2016)]"
}

@article{Balasubramanian:2001nb,
    author = "Balasubramanian, Vijay and de Boer, Jan and Minic, Djordje",
    title = "{Mass, entropy and holography in asymptotically de Sitter spaces}",
    eprint = "hep-th/0110108",
    archivePrefix = "arXiv",
    reportNumber = "VPI-IPPAP-01-01, UPR-964-T",
    doi = "10.1103/PhysRevD.65.123508",
    journal = "Phys. Rev. D",
    volume = "65",
    pages = "123508",
    year = "2002"
}

@article{Aguilar-Gutierrez:2024rka,
    author = "Aguilar-Gutierrez, Sergio E. and Baiguera, Stefano and Zenoni, Nicol\'o",
    title = "{Holographic complexity of the extended Schwarzschild-de Sitter space}",
    eprint = "2402.01357",
    archivePrefix = "arXiv",
    primaryClass = "hep-th",
    reportNumber = "OU-HET-1220",
    month = "2",
    year = "2024"
}

@article{Cotler:2016fpe,
    author = "Cotler, Jordan S. and Gur-Ari, Guy and Hanada, Masanori and Polchinski, Joseph and Saad, Phil and Shenker, Stephen H. and Stanford, Douglas and Streicher, Alexandre and Tezuka, Masaki",
    title = "{Black Holes and Random Matrices}",
    eprint = "1611.04650",
    archivePrefix = "arXiv",
    primaryClass = "hep-th",
    reportNumber = "SU-ITP-16-19, SU-ITP-16/19, YITP-16-124",
    doi = "10.1007/JHEP05(2017)118",
    journal = "JHEP",
    volume = "05",
    pages = "118",
    year = "2017",
    note = "[Erratum: JHEP 09, 002 (2018)]"
}

@article{Rabinovici:2023yex,
    author = "Rabinovici, E. and S\'anchez-Garrido, A. and Shir, R. and Sonner, J.",
    title = "{A bulk manifestation of Krylov complexity}",
    eprint = "2305.04355",
    archivePrefix = "arXiv",
    primaryClass = "hep-th",
    doi = "10.1007/JHEP08(2023)213",
    journal = "JHEP",
    volume = "08",
    pages = "213",
    year = "2023"
}

@article{Maldacena:2015waa,
    author = "Maldacena, Juan and Shenker, Stephen H. and Stanford, Douglas",
    title = "{A bound on chaos}",
    eprint = "1503.01409",
    archivePrefix = "arXiv",
    primaryClass = "hep-th",
    doi = "10.1007/JHEP08(2016)106",
    journal = "JHEP",
    volume = "08",
    pages = "106",
    year = "2016"
}

@article{Maldacena:2016hyu,
    author = "Maldacena, Juan and Stanford, Douglas",
    title = "{Remarks on the Sachdev-Ye-Kitaev model}",
    eprint = "1604.07818",
    archivePrefix = "arXiv",
    primaryClass = "hep-th",
    doi = "10.1103/PhysRevD.94.106002",
    journal = "Phys. Rev. D",
    volume = "94",
    number = "10",
    pages = "106002",
    year = "2016"
}

@article{Lin:2023trc,
    author = "Lin, Henry W. and Stanford, Douglas",
    title = "{A symmetry algebra in double-scaled SYK}",
    eprint = "2307.15725",
    archivePrefix = "arXiv",
    primaryClass = "hep-th",
    doi = "10.21468/SciPostPhys.15.6.234",
    journal = "SciPost Phys.",
    volume = "15",
    number = "6",
    pages = "234",
    year = "2023"
}

@article{Erdos:2014zgc,
    author = {Erd\H{o}s, L\'aszl\'o and Schr\"oder, Dominik},
    title = "{Phase Transition in the Density of States of Quantum Spin Glasses}",
    eprint = "1407.1552",
    archivePrefix = "arXiv",
    primaryClass = "math-ph",
    doi = "10.1007/s11040-014-9164-3",
    journal = "Math. Phys. Anal. Geom.",
    volume = "17",
    number = "3-4",
    pages = "441--464",
    year = "2014"
}

@article{Parisi_1994,
   title={D-dimensional arrays of Josephson junctions, spin glasses and q-deformed harmonic oscillators},
   volume={27},
   ISSN={1361-6447},
   url={http://dx.doi.org/10.1088/0305-4470/27/23/007},
   DOI={10.1088/0305-4470/27/23/007},
   number={23},
   journal={Journal of Physics A: Mathematical and General},
   publisher={IOP Publishing},
   author={Parisi, G},
   year={1994},
   month=dec, pages={7555–7568} }

@article{Berkooz:2024evs,
    author = "Berkooz, Micha and Brukner, Nadav and Jia, Yiyang and Mamroud, Ohad",
    title = "{From Chaos to Integrability in Double Scaled SYK}",
    eprint = "2403.01950",
    archivePrefix = "arXiv",
    primaryClass = "hep-th",
    month = "3",
    year = "2024"
}

@article{Berkooz:2024ofm,
    author = "Berkooz, Micha and Brukner, Nadav and Jia, Yiyang and Mamroud, Ohad",
    title = "{A Path Integral for Chord Diagrams and Chaotic-Integrable Transitions in Double Scaled SYK}",
    eprint = "2403.05980",
    archivePrefix = "arXiv",
    primaryClass = "hep-th",
    month = "3",
    year = "2024"
}

@article{Anegawa:2024yia,
    author = "Anegawa, Takanori and Watanabe, Ryota",
    title = "{Krylov complexity of fermion chain in double-scaled SYK and power spectrum perspective}",
    eprint = "2407.13293",
    archivePrefix = "arXiv",
    primaryClass = "hep-th",
    reportNumber = "KUNS-3008",
    month = "7",
    year = "2024"
}

@article{Rahman:2024iiu,
    author = "Rahman, Adel A. and Susskind, Leonard",
    title = "{$p$-Chords, Wee-Chords, and de Sitter Space}",
    eprint = "2407.12988",
    archivePrefix = "arXiv",
    primaryClass = "hep-th",
    month = "7",
    year = "2024"
}

@article{Susskind:2022bia,
    author = "Susskind, Leonard",
    title = "{De Sitter Space, Double-Scaled SYK, and the Separation of Scales in the Semiclassical Limit}",
    eprint = "2209.09999",
    archivePrefix = "arXiv",
    primaryClass = "hep-th",
    month = "9",
    year = "2022"
}

@article{Susskind:2023rxm,
    author = "Susskind, Leonard",
    title = "{A Paradox and its Resolution Illustrate Principles of de Sitter Holography}",
    eprint = "2304.00589",
    archivePrefix = "arXiv",
    primaryClass = "hep-th",
    month = "4",
    year = "2023"
}

@article{Rahman:2022jsf,
    author = "Rahman, Adel A.",
    title = "{dS JT Gravity and Double-Scaled SYK}",
    eprint = "2209.09997",
    archivePrefix = "arXiv",
    primaryClass = "hep-th",
    month = "9",
    year = "2022"
}

@article{Lin:2022nss,
    author = "Lin, Henry and Susskind, Leonard",
    title = "{Infinite Temperature's Not So Hot}",
    eprint = "2206.01083",
    archivePrefix = "arXiv",
    primaryClass = "hep-th",
    month = "6",
    year = "2022"
}

@article{Rahman:2023pgt,
    author = "Rahman, Adel A. and Susskind, Leonard",
    title = "{Comments on a Paper by Narovlansky and Verlinde}",
    eprint = "2312.04097",
    archivePrefix = "arXiv",
    primaryClass = "hep-th",
    month = "12",
    year = "2023"
}

@article{Araujo-Regado:2022gvw,
    author = "Araujo-Regado, Goncalo and Khan, Rifath and Wall, Aron C.",
    title = "{Cauchy slice holography: a new AdS/CFT dictionary}",
    eprint = "2204.00591",
    archivePrefix = "arXiv",
    primaryClass = "hep-th",
    doi = "10.1007/JHEP03(2023)026",
    journal = "JHEP",
    volume = "03",
    pages = "026",
    year = "2023"
}

@article{Araujo-Regado:2022jpj,
    author = "Araujo-Regado, Goncalo",
    title = "{Holographic Cosmology on Closed Slices in 2+1 Dimensions}",
    eprint = "2212.03219",
    archivePrefix = "arXiv",
    primaryClass = "hep-th",
    month = "12",
    year = "2022"
}

@article{An:2021fcq,
    author = "An, Zhongshan and Anderson, Michael T.",
    title = "{The initial boundary value problem and quasi-local Hamiltonians in General Relativity}",
    eprint = "2103.15673",
    archivePrefix = "arXiv",
    primaryClass = "gr-qc",
    doi = "10.1088/1361-6382/ac0a86",
    month = "3",
    year = "2021"
}

@article{Berkooz:2024lgq,
    author = "Berkooz, Micha and Mamroud, Ohad",
    title = "{A cordial introduction to double scaled SYK}",
    eprint = "2407.09396",
    archivePrefix = "arXiv",
    primaryClass = "hep-th",
    doi = "10.1088/1361-6633/ada889",
    journal = "Rept. Prog. Phys.",
    volume = "88",
    number = "3",
    pages = "036001",
    year = "2025"
}

@article{Anninos:2022ujl,
    author = {Anninos, Dionysios and Galante, Dami\'an A. and M\"uhlmann, Beatrix},
    title = "{Finite features of quantum de Sitter space}",
    eprint = "2206.14146",
    archivePrefix = "arXiv",
    primaryClass = "hep-th",
    doi = "10.1088/1361-6382/acaba5",
    journal = "Class. Quant. Grav.",
    volume = "40",
    number = "2",
    pages = "025009",
    year = "2023"
}

@article{Anninos:2020cwo,
    author = "Anninos, Dionysios and Galante, Dami\'an A.",
    title = "{Constructing AdS$_{2}$ flow geometries}",
    eprint = "2011.01944",
    archivePrefix = "arXiv",
    primaryClass = "hep-th",
    doi = "10.1007/JHEP02(2021)045",
    journal = "JHEP",
    volume = "02",
    pages = "045",
    year = "2021"
}

@article{Aguilar-Gutierrez:2023odp,
    author = "Aguilar-Gutierrez, Sergio E. and Bahiru, Eyoab and Esp\'\i{}ndola, Ricardo",
    title = "{The centaur-algebra of observables}",
    eprint = "2307.04233",
    archivePrefix = "arXiv",
    primaryClass = "hep-th",
    doi = "10.1007/JHEP03(2024)008",
    journal = "JHEP",
    volume = "03",
    pages = "008",
    year = "2024"
}

@article{kramers1926wellenmechanik,
  title={Wellenmechanik und halbzahlige Quantisierung},
  author={Kramers, Hendrik Anthony},
  journal={Zeitschrift f{\"u}r Physik},
  volume={39},
  number={10},
  pages={828--840},
  year={1926},
  publisher={Springer}
}

@article{Shenker:2013yza,
    author = "Shenker, Stephen H. and Stanford, Douglas",
    title = "{Multiple Shocks}",
    eprint = "1312.3296",
    archivePrefix = "arXiv",
    primaryClass = "hep-th",
    reportNumber = "SU-ITP-13-24",
    doi = "10.1007/JHEP12(2014)046",
    journal = "JHEP",
    volume = "12",
    pages = "046",
    year = "2014"
}

@article{Aguilar-Gutierrez:2026jjv,
    author = "Aguilar-Gutierrez, Sergio E. and Das, Rathindra Nath and Erdmenger, Johanna and Xian, Zhuo-Yu",
    title = "{Probing the Chaos to Integrability Transition in Double-Scaled SYK}",
    eprint = "2601.09801",
    archivePrefix = "arXiv",
    primaryClass = "hep-th",
    month = "1",
    year = "2026"
}

@article{Faruk:2023uzs,
    author = "Faruk, Mir Mehedi and Morvan, Edward and van der Schaar, Jan Pieter",
    title = "{Static sphere observers and geodesics in Schwarzschild-de Sitter spacetime}",
    eprint = "2312.06878",
    archivePrefix = "arXiv",
    primaryClass = "gr-qc",
    doi = "10.1088/1475-7516/2024/05/118",
    journal = "JCAP",
    volume = "05",
    pages = "118",
    year = "2024"
}

@article{brillouin1926mecanique,
  title={La m{\'e}canique ondulatoire de Schr{\"o}dinger; une m{\'e}thode g{\'e}n{\'e}rale de r{\'e}solution par approximations successives},
  author={Brillouin, L{\'e}on},
  journal={CR Acad. Sci},
  volume={183},
  number={11},
  pages={24--26},
  year={1926}
}

@article{wentzel1926verallgemeinerung,
  title={Eine verallgemeinerung der quantenbedingungen f{\"u}r die zwecke der wellenmechanik},
  author={Wentzel, Gregor},
  journal={Zeitschrift f{\"u}r Physik},
  volume={38},
  number={6},
  pages={518--529},
  year={1926},
  publisher={Springer}
}

@article{Chapman:2022mqd,
    author = "Chapman, Shira and Galante, Dami{\'a}n A. and Harris, Eleanor and Sheorey, Sameer U. and Vegh, David",
    title = "{Complex geodesics in de Sitter space}",
    eprint = "2212.01398",
    archivePrefix = "arXiv",
    primaryClass = "hep-th",
    doi = "10.1007/JHEP03(2023)006",
    journal = "JHEP",
    volume = "03",
    pages = "006",
    year = "2023"
}

@article{Chapman:2021eyy,
    author = "Chapman, Shira and Galante, Dami\'an A. and Kramer, Eric David",
    title = "{Holographic complexity and de Sitter space}",
    eprint = "2110.05522",
    archivePrefix = "arXiv",
    primaryClass = "hep-th",
    doi = "10.1007/JHEP02(2022)198",
    journal = "JHEP",
    volume = "02",
    pages = "198",
    year = "2022"
}

@article{Banihashemi:2024yye,
    author = "Banihashemi, Batoul and Shaghoulian, Edgar and Shashi, Sanjit",
    title = "{Flat space gravity at finite cutoff}",
    eprint = "2409.07643",
    archivePrefix = "arXiv",
    primaryClass = "hep-th",
    month = "9",
    year = "2024"
}

@article{Coleman:2020jte,
    author = "Coleman, Evan and Shyam, Vasudev",
    title = "{Conformal boundary conditions from cutoff AdS$_{3}$}",
    eprint = "2010.08504",
    archivePrefix = "arXiv",
    primaryClass = "hep-th",
    doi = "10.1007/JHEP09(2021)079",
    journal = "JHEP",
    volume = "09",
    pages = "079",
    year = "2021"
}

@article{Aguilar-Gutierrez:2024nst,
    author = "Aguilar-Gutierrez, Sergio E. and Svesko, Andrew and Visser, Manus R.",
    title = "{$ \textrm{T}\overline{\textrm{T}} $ deformations from AdS$_{2}$ to dS$_{2}$}",
    eprint = "2410.18257",
    archivePrefix = "arXiv",
    primaryClass = "hep-th",
    doi = "10.1007/JHEP01(2025)120",
    journal = "JHEP",
    volume = "01",
    pages = "120",
    year = "2025"
}

@article{Berkooz:2020uly,
    author = "Berkooz, Micha and Narovlansky, Vladimir and Raj, Himanshu",
    title = "{Complex Sachdev-Ye-Kitaev model in the double scaling limit}",
    eprint = "2006.13983",
    archivePrefix = "arXiv",
    primaryClass = "hep-th",
    doi = "10.1007/JHEP02(2021)113",
    journal = "JHEP",
    volume = "02",
    pages = "113",
    year = "2021"
}

@article{Davison:2016ngz,
    author = "Davison, Richard A. and Fu, Wenbo and Georges, Antoine and Gu, Yingfei and Jensen, Kristan and Sachdev, Subir",
    title = "{Thermoelectric transport in disordered metals without quasiparticles: The Sachdev-Ye-Kitaev models and holography}",
    eprint = "1612.00849",
    archivePrefix = "arXiv",
    primaryClass = "cond-mat.str-el",
    doi = "10.1103/PhysRevB.95.155131",
    journal = "Phys. Rev. B",
    volume = "95",
    number = "15",
    pages = "155131",
    year = "2017"
}

@article{Sachdev:2015efa,
    author = "Sachdev, Subir",
    title = "{Bekenstein-Hawking Entropy and Strange Metals}",
    eprint = "1506.05111",
    archivePrefix = "arXiv",
    primaryClass = "hep-th",
    doi = "10.1103/PhysRevX.5.041025",
    journal = "Phys. Rev. X",
    volume = "5",
    number = "4",
    pages = "041025",
    year = "2015"
}

@article{Basu:2025exh,
    author = "Basu, Debarshi and Chandra, Ashish and Wen, Qiang",
    title = "{Butterfly effect and $\textrm{T}\overline{\textrm{T}}$-deformation}",
    eprint = "2505.14331",
    archivePrefix = "arXiv",
    primaryClass = "hep-th",
    month = "5",
    year = "2025"
}

@article{Caputa:2020fbc,
    author = "Caputa, Pawel and Kruthoff, Jorrit and Parrikar, Onkar",
    title = "{Building Tensor Networks for Holographic States}",
    eprint = "2012.05247",
    archivePrefix = "arXiv",
    primaryClass = "hep-th",
    doi = "10.1007/JHEP05(2021)009",
    journal = "JHEP",
    volume = "05",
    pages = "009",
    year = "2021",
    note = "[Erratum: JHEP 09, 112 (2022)]"
}

@article{Zhang:2025dgm,
    author = "Zhang, Ming and Tan, Wen-Di and Lu, Mengqi and Bhattacharya, Dyuman and Yang, Jiayue and Mann, Robert B.",
    title = "{Finite-cutoff holographic thermodynamics}",
    eprint = "2507.01010",
    archivePrefix = "arXiv",
    primaryClass = "hep-th",
    doi = "10.1103/f94c-37sk",
    journal = "Phys. Rev. Res.",
    volume = "8",
    number = "1",
    pages = "013208",
    year = "2026"
}

@article{Batra:2024qju,
    author = "Batra, Gauri",
    title = "{Timelike boundaries in de Sitter JT gravity and the Gao-Wald theorem}",
    eprint = "2407.08913",
    archivePrefix = "arXiv",
    primaryClass = "hep-th",
    month = "7",
    year = "2024"
}

@article{Gu:2019jub,
    author = "Gu, Yingfei and Kitaev, Alexei and Sachdev, Subir and Tarnopolsky, Grigory",
    title = "{Notes on the complex Sachdev-Ye-Kitaev model}",
    eprint = "1910.14099",
    archivePrefix = "arXiv",
    primaryClass = "hep-th",
    doi = "10.1007/JHEP02(2020)157",
    journal = "JHEP",
    volume = "02",
    pages = "157",
    year = "2020"
}

@article{Rahman:2024vyg,
    author = "Rahman, Adel A. and Susskind, Leonard",
    title = "{Infinite Temperature is Not So Infinite: The Many Temperatures of de Sitter Space}",
    eprint = "2401.08555",
    archivePrefix = "arXiv",
    primaryClass = "hep-th",
    month = "1",
    year = "2024"
}

@article{Lin:2022rbf,
    author = "Lin, Henry W.",
    title = "{The bulk Hilbert space of double scaled SYK}",
    eprint = "2208.07032",
    archivePrefix = "arXiv",
    primaryClass = "hep-th",
    doi = "10.1007/JHEP11(2022)060",
    journal = "JHEP",
    volume = "11",
    pages = "060",
    year = "2022"
}

@article{Susskind:2023hnj,
    author = "Susskind, Leonard",
    title = "{De Sitter Space has no Chords. Almost Everything is Confined.}",
    eprint = "2303.00792",
    archivePrefix = "arXiv",
    primaryClass = "hep-th",
    doi = "10.22128/jhap.2023.661.1043",
    journal = "JHAP",
    volume = "3",
    number = "1",
    pages = "1--30",
    year = "2023"
}

@article{Miyashita:2025rpt,
    author = "Miyashita, Shoichiro and Sekino, Yasuhiro and Susskind, Leonard",
    title = "{DSSYK at Infinite Temperature: The Flat-Space Limit and the 't Hooft Model}",
    eprint = "2506.18054",
    archivePrefix = "arXiv",
    primaryClass = "hep-th",
    month = "6",
    year = "2025"
}

@article{Sekino:2025bsc,
    author = "Sekino, Yasuhiro and Susskind, Leonard",
    title = "{Double-Scaled SYK, QCD, and the Flat Space Limit of de Sitter Space}",
    eprint = "2501.09423",
    archivePrefix = "arXiv",
    primaryClass = "hep-th",
    month = "1",
    year = "2025"
}

@article{Deng:2023pjs,
    author = "Deng, Feiyu and Wang, Zhi and Zhou, Yang",
    title = "{End of the world brane meets $ T\overline{T} $}",
    eprint = "2310.15031",
    archivePrefix = "arXiv",
    primaryClass = "hep-th",
    doi = "10.1007/JHEP07(2024)036",
    journal = "JHEP",
    volume = "07",
    pages = "036",
    year = "2024"
}

@article{Okuyama:2023kdo,
    author = "Okuyama, Kazumi",
    title = "{Discrete analogue of the Weil-Petersson volume in double scaled SYK}",
    eprint = "2306.15981",
    archivePrefix = "arXiv",
    primaryClass = "hep-th",
    doi = "10.1007/JHEP09(2023)133",
    journal = "JHEP",
    volume = "09",
    pages = "133",
    year = "2023"
}

@article{Soni:2024aop,
    author = "Soni, Ronak M. and Wall, Aron C.",
    title = "{A New Covariant Entropy Bound from Cauchy Slice Holography}",
    eprint = "2407.16769",
    archivePrefix = "arXiv",
    primaryClass = "hep-th",
    month = "7",
    year = "2024"
}

@article{Okuyama:2023yat,
    author = "Okuyama, Kazumi",
    title = "{Matter correlators through a wormhole in double-scaled SYK}",
    eprint = "2312.00880",
    archivePrefix = "arXiv",
    primaryClass = "hep-th",
    doi = "10.1007/JHEP02(2024)147",
    journal = "JHEP",
    volume = "02",
    pages = "147",
    year = "2024"
}

@article{Bhattacharjee:2022ave,
    author = "Bhattacharjee, Budhaditya and Nandy, Pratik and Pathak, Tanay",
    title = "{Krylov complexity in large q and double-scaled SYK model}",
    eprint = "2210.02474",
    archivePrefix = "arXiv",
    primaryClass = "hep-th",
    reportNumber = "YITP-22-106",
    doi = "10.1007/JHEP08(2023)099",
    journal = "JHEP",
    volume = "08",
    pages = "099",
    year = "2023"
}

@article{Mertens:2020hbs,
    author = "Mertens, Thomas G. and Turiaci, Gustavo J.",
    title = "{Liouville quantum gravity -- holography, JT and matrices}",
    eprint = "2006.07072",
    archivePrefix = "arXiv",
    primaryClass = "hep-th",
    doi = "10.1007/JHEP01(2021)073",
    journal = "JHEP",
    volume = "01",
    pages = "073",
    year = "2021"
}

@article{Fan:2021bwt,
    author = "Fan, Yale and Mertens, Thomas G.",
    title = "{From quantum groups to Liouville and dilaton quantum gravity}",
    eprint = "2109.07770",
    archivePrefix = "arXiv",
    primaryClass = "hep-th",
    doi = "10.1007/JHEP05(2022)092",
    journal = "JHEP",
    volume = "05",
    pages = "092",
    year = "2022"
}

@article{Kyono:2017pxs,
    author = "Kyono, Hideki and Okumura, Suguru and Yoshida, Kentaroh",
    title = "{Comments on 2D dilaton gravity system with a hyperbolic dilaton potential}",
    eprint = "1704.07410",
    archivePrefix = "arXiv",
    primaryClass = "hep-th",
    reportNumber = "KUNS-2672",
    doi = "10.1016/j.nuclphysb.2017.07.013",
    journal = "Nucl. Phys. B",
    volume = "923",
    pages = "126--143",
    year = "2017"
}

@article{Suzuki:2021zbe,
    author = "Suzuki, Kenta and Takayanagi, Tadashi",
    title = "{JT gravity limit of Liouville CFT and matrix model}",
    eprint = "2108.12096",
    archivePrefix = "arXiv",
    primaryClass = "hep-th",
    reportNumber = "YITP-21-88, IPMU21-0054",
    doi = "10.1007/JHEP11(2021)137",
    journal = "JHEP",
    volume = "11",
    pages = "137",
    year = "2021"
}

@article{Goel:2020yxl,
    author = "Goel, Akash and Iliesiu, Luca V. and Kruthoff, Jorrit and Yang, Zhenbin",
    title = "{Classifying boundary conditions in JT gravity: from energy-branes to $\alpha$-branes}",
    eprint = "2010.12592",
    archivePrefix = "arXiv",
    primaryClass = "hep-th",
    doi = "10.1007/JHEP04(2021)069",
    journal = "JHEP",
    volume = "04",
    pages = "069",
    year = "2021"
}

@article{Collier:2023cyw,
    author = "Collier, Scott and Eberhardt, Lorenz and Muehlmann, Beatrix and Rodriguez, Victor A.",
    title = "{The Virasoro minimal string}",
    eprint = "2309.10846",
    archivePrefix = "arXiv",
    primaryClass = "hep-th",
    doi = "10.21468/SciPostPhys.16.2.057",
    journal = "SciPost Phys.",
    volume = "16",
    number = "2",
    pages = "057",
    year = "2024"
}

@article{Stanford:2014jda,
    author = "Stanford, Douglas and Susskind, Leonard",
    title = "{Complexity and Shock Wave Geometries}",
    eprint = "1406.2678",
    archivePrefix = "arXiv",
    primaryClass = "hep-th",
    doi = "10.1103/PhysRevD.90.126007",
    journal = "Phys. Rev. D",
    volume = "90",
    number = "12",
    pages = "126007",
    year = "2014"
}

@article{Iliesiu:2020zld,
    author = "Iliesiu, Luca V. and Kruthoff, Jorrit and Turiaci, Gustavo J. and Verlinde, Herman",
    title = "{JT gravity at finite cutoff}",
    eprint = "2004.07242",
    archivePrefix = "arXiv",
    primaryClass = "hep-th",
    doi = "10.21468/SciPostPhys.9.2.023",
    journal = "SciPost Phys.",
    volume = "9",
    pages = "023",
    year = "2020"
}

@article{Anninos:2022qgy,
    author = "Anninos, Dionysios and Galante, Dami\'an A. and Sheorey, Sameer U.",
    title = "{Renormalisation group flows of deformed SYK models}",
    eprint = "2212.04944",
    archivePrefix = "arXiv",
    primaryClass = "hep-th",
    doi = "10.1007/JHEP11(2023)197",
    journal = "JHEP",
    volume = "11",
    pages = "197",
    year = "2023"
}

@article{Susskind:2021dfc,
    author = "Susskind, Leonard",
    title = "{Black Holes Hint towards De Sitter Matrix Theory}",
    eprint = "2109.01322",
    archivePrefix = "arXiv",
    primaryClass = "hep-th",
    doi = "10.3390/universe9080368",
    journal = "Universe",
    volume = "9",
    number = "8",
    pages = "368",
    year = "2023"
}

@article{Shaghoulian:2021cef,
    author = "Shaghoulian, Edgar",
    title = "{The central dogma and cosmological horizons}",
    eprint = "2110.13210",
    archivePrefix = "arXiv",
    primaryClass = "hep-th",
    doi = "10.1007/JHEP01(2022)132",
    journal = "JHEP",
    volume = "01",
    pages = "132",
    year = "2022"
}

@article{Shaghoulian:2022fop,
    author = "Shaghoulian, Edgar and Susskind, Leonard",
    title = "{Entanglement in De Sitter space}",
    eprint = "2201.03603",
    archivePrefix = "arXiv",
    primaryClass = "hep-th",
    doi = "10.1007/JHEP08(2022)198",
    journal = "JHEP",
    volume = "08",
    pages = "198",
    year = "2022"
}

@article{Berkooz:2022mfk,
    author = "Berkooz, Micha and Isachenkov, Misha and Isachenkov, Mikhail and Narayan, Prithvi and Narovlansky, Vladimir",
    title = "{Quantum groups, non-commutative AdS$_{2}$, and chords in the double-scaled SYK model}",
    eprint = "2212.13668",
    archivePrefix = "arXiv",
    primaryClass = "hep-th",
    doi = "10.1007/JHEP08(2023)076",
    journal = "JHEP",
    volume = "08",
    pages = "076",
    year = "2023"
}

@article{Cooper:2013ffa,
    author = "Cooper, Patrick and Dubovsky, Sergei and Mohsen, Ali",
    title = "{Ultraviolet complete Lorentz-invariant theory with superluminal signal propagation}",
    eprint = "1312.2021",
    archivePrefix = "arXiv",
    primaryClass = "hep-th",
    doi = "10.1103/PhysRevD.89.084044",
    journal = "Phys. Rev. D",
    volume = "89",
    number = "8",
    pages = "084044",
    year = "2014"
}

@article{Yuan:2024utc,
    author = "Yuan, Haiming and Ge, Xian-Hui and Kim, Keun-Young",
    title = "{Pole skipping in two-dimensional de Sitter spacetime and double-scaled SYK model}",
    eprint = "2408.12330",
    archivePrefix = "arXiv",
    primaryClass = "hep-th",
    doi = "10.1103/f3cb-kmnc",
    journal = "Phys. Rev. D",
    volume = "112",
    number = "2",
    pages = "026022",
    year = "2025"
}

@article{Brown:1992br,
    author = "Brown, J. David and York, Jr., James W.",
    title = "{Quasilocal energy and conserved charges derived from the gravitational action}",
    eprint = "gr-qc/9209012",
    archivePrefix = "arXiv",
    reportNumber = "IFP-423-UNC, TAR-009-UNC",
    doi = "10.1103/PhysRevD.47.1407",
    journal = "Phys. Rev. D",
    volume = "47",
    pages = "1407--1419",
    year = "1993"
}

@article{Sachdev_1993,
   title={Gapless spin-fluid ground state in a random quantum Heisenberg magnet},
   volume={70},
   ISSN={0031-9007},
   url={http://dx.doi.org/10.1103/PhysRevLett.70.3339},
   DOI={10.1103/physrevlett.70.3339},
   number={21},
   journal={Physical Review Letters},
   publisher={American Physical Society (APS)},
   author={Sachdev, Subir and Ye, Jinwu},
   year={1993},
   month=may, pages={3339–3342} }

@article{JACKIW1985343,
title = {Lower dimensional gravity},
journal = {Nuclear Physics B},
volume = {252},
pages = {343-356},
year = {1985},
issn = {0550-3213},
doi = {https://doi.org/10.1016/0550-3213(85)90448-1},
url = {https://www.sciencedirect.com/science/article/pii/0550321385904481},
author = {Roman Jackiw}
}

@article{TEITELBOIM198341,
title = {Gravitation and hamiltonian structure in two spacetime dimensions},
journal = {Physics Letters B},
volume = {126},
number = {1},
pages = {41-45},
year = {1983},
issn = {0370-2693},
doi = {https://doi.org/10.1016/0370-2693(83)90012-6},
url = {https://www.sciencedirect.com/science/article/pii/0370269383900126},
author = {Claudio Teitelboim}
}

@article{Liu:2024ymn,
    author = "Liu, Xiaoyi and Santos, Jorge E. and Wiseman, Toby",
    title = "{New Well-Posed boundary conditions for semi-classical Euclidean gravity}",
    eprint = "2402.04308",
    archivePrefix = "arXiv",
    primaryClass = "hep-th",
    doi = "10.1007/JHEP06(2024)044",
    journal = "JHEP",
    volume = "06",
    pages = "044",
    year = "2024"
}

@article{Aguilar-Gutierrez:2024nau,
    author = "Aguilar-Gutierrez, Sergio E.",
    title = "{Towards complexity in de Sitter space from the double-scaled Sachdev-Ye-Kitaev model}",
    eprint = "2403.13186",
    archivePrefix = "arXiv",
    primaryClass = "hep-th",
    doi = "10.1007/JHEP10(2024)107",
    journal = "JHEP",
    volume = "10",
    pages = "107",
    year = "2024"
}

@article{Tsolakidis:2024wut,
    author = "Tsolakidis, Evangelos",
    title = "{Massive gravity generalization of $ T\overline{T} $ deformations}",
    eprint = "2405.07967",
    archivePrefix = "arXiv",
    primaryClass = "hep-th",
    doi = "10.1007/JHEP09(2024)167",
    journal = "JHEP",
    volume = "09",
    pages = "167",
    year = "2024"}

@article{Almheiri:2024xtw,
    author = "Almheiri, Ahmed and Goel, Akash and Hu, Xu-Yao",
    title = "{Quantum gravity of the Heisenberg algebra}",
    eprint = "2403.18333",
    archivePrefix = "arXiv",
    primaryClass = "hep-th",
    doi = "10.1007/JHEP08(2024)098",
    journal = "JHEP",
    volume = "08",
    pages = "098",
    year = "2024"
}

@article{Blommaert:2023opb,
    author = "Blommaert, Andreas and Mertens, Thomas G. and Yao, Shunyu",
    title = "{Dynamical actions and q-representation theory for double-scaled SYK}",
    eprint = "2306.00941",
    archivePrefix = "arXiv",
    primaryClass = "hep-th",
    doi = "10.1007/JHEP02(2024)067",
    journal = "JHEP",
    volume = "02",
    pages = "067",
    year = "2024"
}

@article{Bhattacharyya:2025gvd,
    author = "Bhattacharyya, Arpan and Ghosh, Saptaswa and Pal, Sounak and Vinod, Anandu",
    title = "{Wormholes in finite cutoff JT gravity: A study of baby universes and (Krylov) complexity}",
    eprint = "2502.13208",
    archivePrefix = "arXiv",
    primaryClass = "hep-th",
    month = "2",
    year = "2025"
}

@article{Nandy:2024evd,
    author = "Nandy, Pratik and Matsoukas-Roubeas, Apollonas S. and Mart{\'\i}nez-Azcona, Pablo and Dymarsky, Anatoly and del Campo, Adolfo",
    title = "{Quantum dynamics in Krylov space: Methods and applications}",
    eprint = "2405.09628",
    archivePrefix = "arXiv",
    primaryClass = "quant-ph",
    reportNumber = "RIKEN-iTHEMS-Report-24",
    doi = "10.1016/j.physrep.2025.05.001",
    journal = "Phys. Rept.",
    volume = "1125-1128",
    pages = "1--82",
    year = "2025"
}

@article{Baiguera:2025dkc,
    author = "Baiguera, Stefano and Balasubramanian, Vijay and Caputa, Pawel and Chapman, Shira and Haferkamp, Jonas and Heller, Michal P. and Halpern, Nicole Yunger",
    title = "{Quantum complexity in gravity, quantum field theory, and quantum information science}",
    eprint = "2503.10753",
    archivePrefix = "arXiv",
    primaryClass = "hep-th",
    reportNumber = "YITP-25-39",
    month = "3",
    year = "2025"
}

@article{Rabinovici:2025otw,
    author = "Rabinovici, Eliezer and S{\'a}nchez-Garrido, Adri{\'a}n and Shir, Ruth and Sonner, Julian",
    title = "{Krylov Complexity}",
    eprint = "2507.06286",
    archivePrefix = "arXiv",
    primaryClass = "hep-th",
    reportNumber = "CERN-TH-2025-128",
    month = "7",
    year = "2025"
}

@article{Aguilar-Gutierrez:2025mxf,
    author = "Aguilar-Gutierrez, Sergio E. and Xu, Jiuci",
    title = "{Geometry of Chord Intertwiner, Multiple Shocks and Switchback in Double-Scaled SYK}",
    eprint = "2506.19013",
    archivePrefix = "arXiv",
    primaryClass = "hep-th",
    month = "6",
    doi = "10.1007/JHEP02(2026)246",
    journal = "JHEP",
    volume = "02",
    pages = "246",
    year = "2026"
}

@article{Aguilar-Gutierrez:2025hty,
    author = "Aguilar-Gutierrez, Sergio E.",
    title = "{Symmetry sectors in chord space and relational holography in the DSSYK. Lessons from branes, wormholes, and de Sitter space}",
    eprint = "2506.21447",
    archivePrefix = "arXiv",
    primaryClass = "hep-th",
    doi = "10.1007/JHEP10(2025)044",
    journal = "JHEP",
    volume = "10",
    pages = "044",
    year = "2025"
}

@article{AliAhmad:2025kki,
    author = "Ali Ahmad, Shadi and Almheiri, Ahmed and Lin, Simon",
    title = "{$T\overline{T}$ and the black hole interior}",
    eprint = "2503.19854",
    archivePrefix = "arXiv",
    primaryClass = "hep-th",
    month = "3",
    year = "2025"
}

@article{Goel:2018ubv,
    author = "Goel, Akash and Lam, Ho Tat and Turiaci, Gustavo J. and Verlinde, Herman",
    title = "{Expanding the Black Hole Interior: Partially Entangled Thermal States in SYK}",
    eprint = "1807.03916",
    archivePrefix = "arXiv",
    primaryClass = "hep-th",
    doi = "10.1007/JHEP02(2019)156",
    journal = "JHEP",
    volume = "02",
    pages = "156",
    year = "2019"
}

@article{Segal1947IrreducibleRO,
  title={Irreducible representations of operator algebras},
  author={Irving Ezra Segal},
  journal={Bulletin of the American Mathematical Society},
  year={1947},
  volume={53},
  pages={73-88},
  url={https://api.semanticscholar.org/CorpusID:32218216}
}

@article{Cao:2025pir,
    author = "Cao, Xuchen and Gao, Ping",
    title = "{Single-Sided Black Holes in Double-Scaled SYK Model and No Man's Island}",
    eprint = "2511.01978",
    archivePrefix = "arXiv",
    primaryClass = "hep-th",
    month = "11",
    year = "2025"
}

@misc{kitaevTalks1, 
author         = "Kitaev, A.",
title= "{Hidden Correlations in the Hawking Radiation and Thermal Noise}",
howpublished = {\url{https://www.youtube.com/watch?v=OQ9qN8j7EZI}},
note = {Talk at 2015 Breakthrough Prize Fundamental Physics Symposium at Stanford University, U.S.A.} ,
month=" November",
year="2015"
  }

@misc{kitaevTalks2, 
author         = "Kitaev, A.",
title= "{A simple model of quantum holography (part 1)}",
howpublished = {\url{http://online.kitp.ucsb.edu/online/entangled15/kitaev/}},
note = {Talk at "Entanglement in Strongly-Correlated Quantum Matter" at the Kavli Institute for Theoretical Physics (KITP) at UC Santa Barbara, U.S.A.} ,
month=" April",
year="2015"
  }

@article{Aguilar-Gutierrez:2025pqp,
    author = "Aguilar-Gutierrez, Sergio E.",
    title = "{Building the holographic dictionary of the DSSYK from chords, complexity {\&} wormholes with matter}",
    eprint = "2505.22716",
    archivePrefix = "arXiv",
    primaryClass = "hep-th",
    doi = "10.1007/JHEP10(2025)221",
    journal = "JHEP",
    volume = "10",
    pages = "221",
    year = "2025"
}

@article{Ambrosini:2025hvo,
    author = "Ambrosini, Marco and Rabinovici, Eliezer and Sonner, Julian",
    title = "{Holography of K-complexity: Switchbacks and Shockwaves}",
    eprint = "2510.17975",
    archivePrefix = "arXiv",
    primaryClass = "hep-th",
    reportNumber = "CERN-TH-2025-206",
    month = "10",
    year = "2025"
}

@article{Almheiri:2024ayc,
    author = "Almheiri, Ahmed and Popov, Fedor K.",
    title = "{Holography on the quantum disk}",
    eprint = "2401.05575",
    archivePrefix = "arXiv",
    primaryClass = "hep-th",
    doi = "10.1007/JHEP06(2024)070",
    journal = "JHEP",
    volume = "06",
    pages = "070",
    year = "2024"
}

@article{Schouten:2025tvn,
    author = "Schouten, Koen and Isachenkov, Mikhail",
    title = "{The von Neumann algebraic quantum group $\mathrm{SU}_q(1,1)\rtimes \mathbb{Z}_2$ and the DSSYK model}",
    eprint = "2512.10101",
    archivePrefix = "arXiv",
    primaryClass = "math-ph",
    month = "12",
    year = "2025"
}

@article{vanderHeijden:2025zkr,
    author = "van der Heijden, Jeremy and Verlinde, Erik and Xu, Jiuci",
    title = "{Quantum Symmetry and Geometry in Double-Scaled SYK}",
    eprint = "2511.08743",
    archivePrefix = "arXiv",
    primaryClass = "hep-th",
    month = "11",
    year = "2025"
}

@misc{kitaevTalks3, 
author         = "Kitaev, A.",
title= "{A simple model of quantum holography (part 2)}",
howpublished = {\url{ http://online.kitp.ucsb.edu/online/entangled15/kitaev2/}},
note = {Talk at "Entanglement in Strongly-Correlated Quantum Matter" at the Kavli Institute for Theoretical Physics (KITP) at UC Santa Barbara, U.S.A.} ,
month=" April",
year="2015"
  }

@article{Forste:2025gng,
    author = "Forste, Stefan and Kruse, Yannic and Natu, Saurabh",
    title = "{Grand Canonical vs Canonical Krylov Complexity in Double-Scaled Complex SYK Model}",
    eprint = "2512.07715",
    archivePrefix = "arXiv",
    primaryClass = "hep-th",
    reportNumber = "BONN--TH--2025--35",
    month = "12",
    year = "2025"
}

@article{Gubankova:2025gbx,
    author = "Gubankova, Elena and Sachdev, Subir and Tarnopolsky, Grigory",
    title = "{Scaling limits of complex Sachdev-Ye-Kitaev models and holographic geometry}",
    eprint = "2512.05294",
    archivePrefix = "arXiv",
    primaryClass = "hep-th",
    month = "12",
    year = "2025"
}

@mastersthesis{Arundine:2025mcu,
    author = "Arundine, Mattia",
    title = "{The DSSYK Model: Charge and Holography}",
    eprint = "2512.21366",
    archivePrefix = "arXiv",
    primaryClass = "hep-th",
    type = "Other thesis",
    month = "12",
    year = "2025",
school="Università di Pisa"
}

@article{Ahn:2025exp,
    author = "Ahn, Yongjun and Grozdanov, Sa{\v{s}}o and Jeong, Hyun-Sik and Pedraza, Juan F.",
    title = "{Cosmological pole-skipping, shock waves and quantum chaotic dynamics of de Sitter horizons}",
    eprint = "2508.15589",
    archivePrefix = "arXiv",
    primaryClass = "hep-th",
    reportNumber = "IFT-UAM/CSIC-25-88, APCTP Pre2025 - 019",
    month = "8",
    year = "2025"
}

@article{Fu:2025kkh,
    author = "Fu, Yichao and Jeong, Hyun-Sik and Kim, Keun-Young and Pedraza, Juan F.",
    title = "{Toward Krylov-based holography in double-scaled SYK}",
    eprint = "2510.22658",
    archivePrefix = "arXiv",
    primaryClass = "hep-th",
    reportNumber = "IFT-UAM/CSIC-25-105, APCTP Pre2025 - 020",
    month = "10",
    year = "2025"
}

@article{Maldacena:1997re,
    author = "Maldacena, Juan Martin",
    title = "{The Large $N$ limit of superconformal field theories and supergravity}",
    eprint = "hep-th/9711200",
    archivePrefix = "arXiv",
    reportNumber = "HUTP-97-A097, HUTP-98-A097",
    doi = "10.4310/ATMP.1998.v2.n2.a1",
    journal = "Adv. Theor. Math. Phys.",
    volume = "2",
    pages = "231--252",
    year = "1998"
}

@article{Lin:2018xkj,
    author = "Lin, Jennifer",
    title = "{Entanglement entropy in Jackiw-Teitelboim Gravity}",
    eprint = "1807.06575",
    archivePrefix = "arXiv",
    primaryClass = "hep-th",
    month = "7",
    year = "2018"
}

@article{Lin:2017uzr,
    author = "Lin, Jennifer",
    title = "{Ryu-Takayanagi Area as an Entanglement Edge Term}",
    eprint = "1704.07763",
    archivePrefix = "arXiv",
    primaryClass = "hep-th",
    month = "4",
    year = "2017"
}

@article{Blommaert:2018iqz,
    author = "Blommaert, Andreas and Mertens, Thomas G. and Verschelde, Henri",
    title = "{Fine Structure of Jackiw-Teitelboim Quantum Gravity}",
    eprint = "1812.00918",
    archivePrefix = "arXiv",
    primaryClass = "hep-th",
    doi = "10.1007/JHEP09(2019)066",
    journal = "JHEP",
    volume = "09",
    pages = "066",
    year = "2019"
}

@article{Lin:2021tlr,
    author = "Lin, Jennifer",
    title = "{Entanglement entropy in Jackiw-Teitelboim gravity with matter}",
    eprint = "2107.11872",
    archivePrefix = "arXiv",
    primaryClass = "hep-th",
    month = "7",
    year = "2021"
}

@article{Mertens:2022ujr,
    author = "Mertens, Thomas G. and Sim{\'o}n, Joan and Wong, Gabriel",
    title = "{A proposal for 3d quantum gravity and its bulk factorization}",
    eprint = "2210.14196",
    archivePrefix = "arXiv",
    primaryClass = "hep-th",
    doi = "10.1007/JHEP06(2023)134",
    journal = "JHEP",
    volume = "06",
    pages = "134",
    year = "2023"
}

@book{Rangamani:2016dms,
    author = "Rangamani, Mukund and Takayanagi, Tadashi",
    title = "{Holographic Entanglement Entropy}",
    eprint = "1609.01287",
    archivePrefix = "arXiv",
    primaryClass = "hep-th",
    reportNumber = "YITP-16-106, YITP-16-106",
    doi = "10.1007/978-3-319-52573-0",
    publisher = "Springer",
    volume = "931",
    year = "2017"
}

@article{Chen:2019lcd,
    author = "Chen, Bin",
    title = "{Holographic Entanglement Entropy: A Topical Review}",
    doi = "10.1088/0253-6102/71/7/837",
    journal = "Commun. Theor. Phys.",
    volume = "71",
    number = "7",
    pages = "837",
    year = "2019"
}

@article{Harlow:2014yka,
    author = "Harlow, Daniel",
    title = "{Jerusalem Lectures on Black Holes and Quantum Information}",
    eprint = "1409.1231",
    archivePrefix = "arXiv",
    primaryClass = "hep-th",
    doi = "10.1103/RevModPhys.88.015002",
    journal = "Rev. Mod. Phys.",
    volume = "88",
    pages = "015002",
    year = "2016"
}

@article{Witten:2018zxz,
    author = "Witten, Edward",
    title = "{APS Medal for Exceptional Achievement in Research: Invited article on entanglement properties of quantum field theory}",
    eprint = "1803.04993",
    archivePrefix = "arXiv",
    primaryClass = "hep-th",
    doi = "10.1103/RevModPhys.90.045003",
    journal = "Rev. Mod. Phys.",
    volume = "90",
    number = "4",
    pages = "045003",
    year = "2018"
}

@article{Sorce:2023fdx,
    author = "Sorce, Jonathan",
    title = "{Notes on the type classification of von Neumann algebras}",
    eprint = "2302.01958",
    archivePrefix = "arXiv",
    primaryClass = "hep-th",
    reportNumber = "MIT-CTP/5527",
    doi = "10.1142/S0129055X24300024",
    journal = "Rev. Math. Phys.",
    volume = "36",
    number = "02",
    pages = "2430002",
    year = "2024"
}

@article{Casini:2022rlv,
    author = "Casini, Horacio and Huerta, Marina",
    title = "{Lectures on entanglement in quantum field theory}",
    eprint = "2201.13310",
    archivePrefix = "arXiv",
    primaryClass = "hep-th",
    doi = "10.22323/1.403.0002",
    journal = "PoS",
    volume = "TASI2021",
    pages = "002",
    year = "2023"
}

@inproceedings{Liu:2025krl,
    author = "Liu, Hong",
    title = "{Lectures on entanglement, von Neumann algebras, and emergence of spacetime}",
    booktitle = "{Theoretical Advanced Study Institute in Elementary Particle Physics 2023}: {Aspects of Symmetry}",
    eprint = "2510.07017",
    archivePrefix = "arXiv",
    primaryClass = "hep-th",
    reportNumber = "MIT-CTP/5938",
    month = "10",
    year = "2025"
}

@article{Doi:2022iyj,
    author = "Doi, Kazuki and Harper, Jonathan and Mollabashi, Ali and Takayanagi, Tadashi and Taki, Yusuke",
    title = "{Pseudoentropy in dS/CFT and Timelike Entanglement Entropy}",
    eprint = "2210.09457",
    archivePrefix = "arXiv",
    primaryClass = "hep-th",
    reportNumber = "YITP-22-121, IMPU22-0052",
    doi = "10.1103/PhysRevLett.130.031601",
    journal = "Phys. Rev. Lett.",
    volume = "130",
    number = "3",
    pages = "031601",
    year = "2023"
}

@article{Roberts:2018mnp,
    author = "Roberts, Daniel A. and Stanford, Douglas and Streicher, Alexandre",
    title = "{Operator growth in the SYK model}",
    eprint = "1802.02633",
    archivePrefix = "arXiv",
    primaryClass = "hep-th",
    doi = "10.1007/JHEP06(2018)122",
    journal = "JHEP",
    volume = "06",
    pages = "122",
    year = "2018"
}

@article{Conti:2018jho,
    author = "Conti, Riccardo and Iannella, Leonardo and Negro, Stefano and Tateo, Roberto",
    title = "{Generalised Born-Infeld models, Lax operators and the $ \mathrm{T}\overline{\mathrm{T}} $ perturbation}",
    eprint = "1806.11515",
    archivePrefix = "arXiv",
    primaryClass = "hep-th",
    doi = "10.1007/JHEP11(2018)007",
    journal = "JHEP",
    volume = "11",
    pages = "007",
    year = "2018"
}

@article{Conti:2018tca,
    author = "Conti, Riccardo and Negro, Stefano and Tateo, Roberto",
    title = "{The $ \mathrm{T}\overline{\mathrm{T}} $ perturbation and its geometric interpretation}",
    eprint = "1809.09593",
    archivePrefix = "arXiv",
    primaryClass = "hep-th",
    doi = "10.1007/JHEP02(2019)085",
    journal = "JHEP",
    volume = "02",
    pages = "085",
    year = "2019"
}

@article{Jafari:2019qns,
    author = "Jafari, Ghadir and Naseh, Ali and Zolfi, Hamed",
    title = "{Path Integral Optimization for $T\bar{T}$ Deformation}",
    eprint = "1909.02357",
    archivePrefix = "arXiv",
    primaryClass = "hep-th",
    doi = "10.1103/PhysRevD.101.026007",
    journal = "Phys. Rev. D",
    volume = "101",
    number = "2",
    pages = "026007",
    year = "2020"
}

@article{Araujo:2018rho,
    author = "Araujo, Thiago and Colg{\'a}in, E. {\'O} and Sakatani, Yuho and Sheikh-Jabbari, M. M. and Yavartanoo, Hossein",
    title = "{Holographic integration of $T \bar{T}$ {\textbackslash}{\&} $J \bar{T}$ via $O(d,d)$}",
    eprint = "1811.03050",
    archivePrefix = "arXiv",
    primaryClass = "hep-th",
    reportNumber = "IPM/P-2018/076",
    doi = "10.1007/JHEP03(2019)168",
    journal = "JHEP",
    volume = "03",
    pages = "168",
    year = "2019"
}

@article{Chandrasekaran:2022eqq,
    author = "Chandrasekaran, Venkatesa and Penington, Geoff and Witten, Edward",
    title = "{Large N algebras and generalized entropy}",
    eprint = "2209.10454",
    archivePrefix = "arXiv",
    primaryClass = "hep-th",
    journal = "JHEP",
    volume = "04",
    pages = "009",
    year = "2023",
    doi = "10.1007/JHEP04(2023)009",
}

@article{Wootters:1984wfv,
    author = "Wootters, William K.",
    title = "{{\textquotedblleft}Time{\textquotedblright} replaced by quantum correlations}",
    doi = "10.1007/BF02214098",
    journal = "Int. J. Theor. Phys.",
    volume = "23",
    number = "8",
    pages = "701--711",
    year = "1984"
}

@article{Page:1983uc,
    author = "Page, Don N. and Wootters, William K.",
    title = "{EVOLUTION WITHOUT EVOLUTION: DYNAMICS DESCRIBED BY STATIONARY OBSERVABLES}",
    reportNumber = "PRINT-83-0279 (PENN-STATE)",
    doi = "10.1103/PhysRevD.27.2885",
    journal = "Phys. Rev. D",
    volume = "27",
    pages = "2885",
    year = "1983"
}

@article{Engelhardt:2014gca,
    author = "Engelhardt, Netta and Wall, Aron C.",
    title = "{Quantum Extremal Surfaces: Holographic Entanglement Entropy beyond the Classical Regime}",
    eprint = "1408.3203",
    archivePrefix = "arXiv",
    primaryClass = "hep-th",
    doi = "10.1007/JHEP01(2015)073",
    journal = "JHEP",
    volume = "01",
    pages = "073",
    year = "2015"
}

@book{Henneaux:1994lbw,
    author = "Henneaux, Marc and Teitelboim, Claudio",
    title = "{Quantization of Gauge Systems}",
    isbn = "978-0-691-03769-1, 978-0-691-21386-6",
    publisher = "Princeton University Press",
    month = "8",
    year = "1994"
}

@book{dirac2013lectures,
  title={Lectures on quantum mechanics},
  author={Dirac, Paul AM},
  year={2013},
  publisher={Courier Corporation}
}

@article{Jafferis:2019wkd,
    author = "Jafferis, Daniel Louis and Kolchmeyer, David K.",
    title = "{Entanglement Entropy in Jackiw-Teitelboim Gravity}",
    eprint = "1911.10663",
    archivePrefix = "arXiv",
    primaryClass = "hep-th",
    month = "11",
    year = "2019"
}

@article{Lewkowycz:2013nqa,
    author = "Lewkowycz, Aitor and Maldacena, Juan",
    title = "{Generalized gravitational entropy}",
    eprint = "1304.4926",
    archivePrefix = "arXiv",
    primaryClass = "hep-th",
    doi = "10.1007/JHEP08(2013)090",
    journal = "JHEP",
    volume = "08",
    pages = "090",
    year = "2013"
}

@article{Aguilar-Gutierrez:2025otq,
    author = "Aguilar-Gutierrez, Sergio E.",
    title = "{Cosmological Entanglement Entropy from the von Neumann Algebra of Double-Scaled SYK {\&} Its Connection with Krylov Complexity}",
    eprint = "2511.03779",
    archivePrefix = "arXiv",
    primaryClass = "hep-th",
    month = "11",
    year = "2025"
}

@article{Galante:2025tnt,
    author = "Galante, Dami{\'a}n A. and Maneerat, Chawakorn and Svesko, Andrew",
    title = "{Conformal boundaries near extremal black holes}",
    eprint = "2504.14003",
    archivePrefix = "arXiv",
    primaryClass = "hep-th",
    month = "4",
    year = "2025"
}

@article{Arnowitt:1959eec,
    author = "Arnowitt, R. and Deser, S.",
    title = "{Quantum Theory of Gravitation: General Formulation and Linearized Theory}",
    doi = "10.1103/PhysRev.113.745",
    journal = "Phys. Rev.",
    volume = "113",
    pages = "745--750",
    year = "1959"
}

@article{Arnowitt:1960es,
    author = "Arnowitt, Richard L. and Deser, Stanley and Misner, Charles W.",
    title = "{Canonical variables for general relativity}",
    doi = "10.1103/PhysRev.117.1595",
    journal = "Phys. Rev.",
    volume = "117",
    pages = "1595--1602",
    year = "1960"
}

@article{Arnowitt:1960zzc,
    author = "Arnowitt, R. and Deser, S. and Misner, C. W.",
    title = "{Energy and the Criteria for Radiation in General Relativity}",
    doi = "10.1103/PhysRev.118.1100",
    journal = "Phys. Rev.",
    volume = "118",
    pages = "1100--1104",
    year = "1960"
}

@article{Arnowitt:1960zza,
    author = "Arnowitt, Richard L. and Deser, Stanley and Misner, Charles W.",
    title = "{Gravitational-electromagnetic coupling and the classical self-energy problem}",
    doi = "10.1103/PhysRev.120.313",
    journal = "Phys. Rev.",
    volume = "120",
    pages = "313--320",
    year = "1960"
}

@article{Arnowitt:1960zzb,
    author = "Arnowitt, R. and Deser, S. and Misner, C. W.",
    title = "{Interior Schwarzschild solutions and interpretation of source terms}",
    doi = "10.1103/PhysRev.120.321",
    journal = "Phys. Rev.",
    volume = "120",
    pages = "321",
    year = "1960"
}

@article{Griguolo:2025kpi,
    author = "Griguolo, Luca and Papalini, Jacopo and Russo, Lorenzo and Seminara, Domenico",
    title = "{A new perspective on dilaton gravity at finite cutoff}",
    eprint = "2512.21774",
    archivePrefix = "arXiv",
    primaryClass = "hep-th",
    month = "12",
    year = "2025"
}

@article{Guica:2025jkq,
    author = "Guica, Monica",
    title = "{From black holes to solvable irrelevant deformations and back}",
    eprint = "2512.23620",
    archivePrefix = "arXiv",
    primaryClass = "hep-th",
    month = "12",
    year = "2025"
}

@article{Arnowitt:1961zz,
    author = "Arnowitt, Richard L. and Deser, Stanley and Misner, Charles W.",
    title = "{Coordinate invariance and energy expressions in general relativity}",
    doi = "10.1103/PhysRev.122.997",
    journal = "Phys. Rev.",
    volume = "122",
    pages = "997",
    year = "1961"
}

@article{Arnowitt:1961zza,
    author = "Arnowitt, Richard L. and Deser, Stanley and Misner, Charles W.",
    title = "{Wave zone in general relativity}",
    doi = "10.1103/PhysRev.121.1556",
    journal = "Phys. Rev.",
    volume = "121",
    pages = "1556",
    year = "1961"
}

@article{Banihashemi:2025qqi,
    author = "Banihashemi, Batoul and Shaghoulian, Edgar and Shashi, Sanjit",
    title = "{Thermal effective actions from conformal boundary conditions in gravity}",
    eprint = "2503.17471",
    archivePrefix = "arXiv",
    primaryClass = "hep-th",
    month = "3",
    year = "2025"
}

@article{Galante:2025emz,
    author = "Galante, Dami{\'a}n A. and Myers, Robert C. and Zikopoulos, Themistocles",
    title = "{Conformal Boundary Conditions and Higher Curvature Gravity}",
    eprint = "2512.10930",
    archivePrefix = "arXiv",
    primaryClass = "hep-th",
    month = "12",
    year = "2025"
}

@article{Anninos:2025zgr,
    author = "Anninos, Dionysios and Galante, Dami{\'a}n A. and Georgescu, Silvia and Maneerat, Chawakorn and Svesko, Andrew",
    title = "{The Stretched Horizon Limit}",
    eprint = "2512.16738",
    archivePrefix = "arXiv",
    primaryClass = "hep-th",
    month = "12",
    year = "2025"
}

@article{Kudler-Flam:2023qfl,
    author = "Kudler-Flam, Jonah and Leutheusser, Samuel and Satishchandran, Gautam",
    title = "{Generalized black hole entropy is von Neumann entropy}",
    eprint = "2309.15897",
    archivePrefix = "arXiv",
    primaryClass = "hep-th",
    doi = "10.1103/PhysRevD.111.025013",
    journal = "Phys. Rev. D",
    volume = "111",
    number = "2",
    pages = "025013",
    year = "2025"
}

@article{Aguilar-Gutierrez:2025sqh,
    author = "Aguilar-Gutierrez, Sergio E.",
    title = "{Evolution With(out) Time: Relational Holography {\&} BPS Complexity Growth in $\mathcal{N}=2$ Double-Scaled SYK}",
    eprint = "2510.11777",
    archivePrefix = "arXiv",
    primaryClass = "hep-th",
    month = "10",
    doi = "10.1007/JHEP02(2026)229",
    journal = "JHEP",
    volume = "02",
    pages = "229",
    year = "2026"
}

@article{gelfand1943imbedding,
  title={On the imbedding of normed rings into the ring of operators in Hilbert space},
  author={Gelfand, Israel and Neumark, Mark},
  journal={Mathematical collection},
  volume={12},
  number={2},
  pages={197--217},
  year={1943},
  publisher={Russian Academy of Sciences, Steklov Mathematical Institute of the Russian Academy of Sciences}
}

@article{Witten:2021unn,
    author = "Witten, Edward",
    title = "{Gravity and the crossed product}",
    eprint = "2112.12828",
    archivePrefix = "arXiv",
    primaryClass = "hep-th",
    doi = "10.1007/JHEP10(2022)008",
    journal = "JHEP",
    volume = "10",
    pages = "008",
    year = "2022"
}
\end{document}